\documentclass[11pt, a4paper]{article}
\usepackage{jheparxiv}
\usepackage{wasysym}
\usepackage[utf8]{inputenc}
\usepackage{amsmath}
\usepackage{amsfonts}
\usepackage{amssymb}
\usepackage{latexsym}
\usepackage{mathrsfs}
\usepackage{mathtools}
\usepackage{braket}		
\usepackage{graphicx}
\usepackage{color}
\usepackage{empheq}
\usepackage{xcolor}
\usepackage{slashed}
\usepackage{twistor}
\usepackage[all]{xy}
\usepackage{tikz-cd}
\usepackage{bbm}
\usetikzlibrary{arrows.meta,decorations.pathmorphing,calc,shadings,shapes.misc}
\usepackage{physics}
\usepackage{blkarray}
\usepackage[most]{tcolorbox}
\usepackage{environ}
\usepackage{booktabs}
\usepackage{enumitem}

\usepackage{framed}

\usetikzlibrary{arrows.meta}

\newtcolorbox{highlighteddisplay}{
  enhanced,
  width=\linewidth,
  colback=brown!12,
  frame hidden,
  boxrule=0pt,
  sharp corners,
  boxsep=0pt,
  left=6pt,
  right=6pt,
  top=4pt,
  bottom=4pt,
  before skip balanced=\abovedisplayskip,
  after skip balanced=\belowdisplayskip,
  ams nodisplayskip
}

\NewEnviron{highlightedmultline}{%
  \begin{highlighteddisplay}
    \setlength{\abovedisplayskip}{0pt}%
    \setlength{\belowdisplayskip}{0pt}%
    \begin{multline}
      \BODY
    \end{multline}
  \end{highlighteddisplay}
}

\newtcolorbox{higheq}{
  enhanced,
  ams equation,
  width=\linewidth,
  colback=brown!12,
  frame hidden,
  boxrule=0pt,
  sharp corners,
  boxsep=0pt,
  left=6pt,
  right=6pt,
  top=6pt,
  bottom=6pt,
  before skip=\abovedisplayskip,
  after skip=\belowdisplayskip
}
\newtcolorbox{higheqtopless}{
  enhanced,
  ams equation,
  width=\linewidth,
  colback=brown!12,
  frame hidden,
  boxrule=0pt,
  sharp corners,
  boxsep=0pt,
  left=6pt,
  right=6pt,
  top=0pt,
  bottom=10pt,
  before skip=\abovedisplayskip,
  after skip=\belowdisplayskip
}

\newcommand{\vphi}{\varphi}
\renewcommand{\d}{\mathrm{d}}
\newcommand{\br}[1]{\overline{#1}}

\newcommand{\veps}{\varepsilon}

\newcommand{\til}[1]{\widetilde{#1}}
\newcommand{\Str}{{\mathrm{Str}}}

\newcommand{\gl}{{\mathfrak{gl}}}

\newcommand{\CT}{{\mathscr{T}}}

\newcommand{\sg}{{\mathsf{g}}}

\newcommand{\db}{{\dot\beta}}

\newcommand{\YM}{{\text{YM}}}

\renewcommand{\sl}{{\mathfrak{sl}}}

\newcommand{\psl}{{\mathfrak{psl}}}

\newcommand{\PE}{{\mathrm{PE}}}
\renewcommand{\Ber}{{\mathrm{Ber}}}
\newcommand{\Tot}{{\mathrm{Tot}}}
\newcommand{\Sym}{{\mathrm{Sym}}}
\newcommand{\fR}{{\mathfrak{R}}}
\newcommand{\fJ}{{\mathfrak{J}}}
\newcommand{\fL}{{\mathfrak{L}}}
\newcommand{\cS}{{\mathcal{S}}}
\newcommand{\cK}{{\mathcal{K}}}
\newcommand{\sh}{{\mathsf{h}}}

\renewcommand{\sc}{{\mathsf{c}}}
\renewcommand{\sb}{{\mathsf{b}}}
\newcommand{\sk}{{\mathsf{k}}}
\newcommand{\CE}{\mathscr{E}}
\newcommand{\bau}{{\boldsymbol{\tau}}}
\newcommand{\fx}{{\mathbf{x}}}
\newcommand{\fy}{{\mathbf{y}}}
\newcommand{\fz}{{\mathbf{z}}}

\title{Thermal Giant Graviton Expansion}

\author[a]{Kasia Budzik,}
\author[a,b]{Atul Sharma}

\affiliation[a]{Center for the Fundamental Laws of Nature, Harvard University,\\
Cambridge, MA, 02138, USA \vspace{0.1cm}}
\affiliation[b]{Black Hole Initiative, Harvard University,\\
Cambridge, MA, 02138, USA \vspace{0.1cm}}
\emailAdd{kbudzik@fas.harvard.edu}
\emailAdd{atulsharma@fas.harvard.edu}

\abstract{We construct giant graviton expansions of the thermal and $(-1)^F$-graded partition functions of $\cN=4$ and $\cN=1$ super Yang-Mills theories at zero 't Hooft coupling. These observables are generically non-BPS and do not admit a direct bulk interpretation in a supergravity regime. Instead, we exploit the corresponding self-dual theories, which arise as chiral zero coupling limits and admit twistor string holographic duals. We apply a quiver generalization of Murthy's expansion motivated by twistor strings. Notably, we obtain expansions at both finite and infinite $N$. We reproduce the terms in these expansions from bulk closed string and D-brane sectors, identifying the nonperturbative corrections with D-instanton/anti-D-instanton pairs. At infinite $N$, resumming the D-instanton sectors reproduces the Hagedorn temperatures of the free gauge theories, providing a microscopic bulk origin for their Hagedorn singularities.}

\begin{document}

\maketitle

\section{Introduction}
\label{sec:intro}

Holography at finite $N$ gives us a glimpse into the deeply quantum regime of gravity in the bulk. The presence of trace relations truncates the spectrum of the boundary gauge theory and gives rise to the stringy exclusion principle inducing a similar truncation in the gravitational dual \cite{Maldacena:1998bw,Susskind:1998dq,Evans:1998qu,McGreevy:2000cw,Grisaru:2000zn}. Recent years have witnessed substantial progress in our understanding of finite-$N$ AdS/CFT through the development of giant graviton expansions \cite{Arai:2019xmp,Arai:2019wgv,Arai:2019aou,Arai:2020qaj,Arai:2020uwd,Imamura:2021ytr,Imamura:2022aua,Gaiotto:2021xce,Murthy:2022ien,Lee:2022vig,Liu:2022olj,Eniceicu:2023uvd,Chen:2024cvf,Eleftheriou:2023jxr,Deddo:2024liu,Ezroura:2024wmp,Imamura:2024lkw,Budzik:2025vzr,Beccaria:2023sph,Beccaria:2023cuo,Hayashi:2024aaf,Chen:2024erz,Dorey:2025qht,Chen:2026fpe,Choi:2022ovw,Kim:2024ucf,Beccaria:2024szi,Eleftheriou:2025lac,Deddo:2025lfm,Purkayastha:2025whi,Imamura:2024zvw,Fujiwara:2023bdc}\footnote{See also \cite{Lee:2025veh} for a novel finite-$N$ expansion over bulk saddles in
AdS$_3$/CFT$_2$.} and the discovery of fortuitous states \cite{Choi:2022caq,Choi:2023vdm,Choi:2023znd,deMelloKoch:2024pcs,Kim:2025vup,Chang:2022mjp,Chang:2024zqi,Chang:2024lxt,Chang:2025rqy,Chang:2025mqp,Chang:2025wgo,Budzik:2023vtr,Chen:2025sum,Behan:2025hbx,Belin:2025hsg,Choi:2025bhi,deMelloKoch:2025cec,Hughes:2025tdy}. Trace relations play a central role in both phenomena and have themselves been the subject of extensive recent study \cite{Lee:2023iil,Lee:2024hef,Caputa:2025ikn,Eleftheriou:2026yct,deMelloKoch:2025ngs, deMelloKoch:2025eqt, deMelloKoch:2025qeq, deMelloKoch:2025rkw, deMelloKoch:2026pck, Ahn:2026ohj, deMelloKoch:2026plq}.

Much of the progress has been concerned with protected quantities, such as superconformal indices and supercharge cohomology, which can be studied at weak 't Hooft coupling and continued to strong coupling by virtue of supersymmetry. By contrast, finite-$N$ effects in \emph{non-BPS} observables can be accessed perturbatively at weak coupling, but their holographic interpretation is less direct. Unlike protected quantities, their weak-coupling values cannot generally be extrapolated to the supergravity regime. Consequently, the bulk origin of finite-$N$ truncations in the non-BPS spectrum remains less understood.

Tensionless holography provides one setting in which weak-coupling (BPS and non-BPS) observables can be studied holographically. The $\al'\to\infty$ tensionless regime is the opposite of the $\al'\to0$ supergravity regime of string theory in AdS. Hence, it is expected to be holographically dual to a weakly coupled boundary CFT. To date, this is best understood in the setting of AdS$_3$/CFT$_2$. The minimal tension limit of string theory in AdS$_3\times S^3$ is often found to be dual to the symmetric product orbifold of a free 2d CFT \cite{Eberhardt:2017pty,Eberhardt:2018ouy,Eberhardt:2019qcl,Eberhardt:2019niq,Gaberdiel:2023dxt,Aharony:2024fid}. As a result, BPS and non-BPS spectra, partition functions and correlators in such free symmetric product CFTs can be explicitly matched with the states, partition functions and worldsheet amplitudes of minimal tension strings \cite{Lunin:2000yv,Lunin:2001pw,Gaberdiel:2018rqv,Eberhardt:2019ywk,Eberhardt:2020bgq,Eberhardt:2021jvj,Dei:2019osr,Dei:2019iym,Dei:2020zui,Dei:2021xgh,Dei:2021yom,Dei:2022pkr,Knighton:2020kuh,Gaberdiel:2025rtu,Eberhardt:2025sbi}. Progress has also been made toward a tensionless description of AdS$_5$/CFT$_4$ \cite{Haggi-Mani:2000dxu,Sundborg:2000wp,Sezgin:2001zs,Mikhailov:2002bp,Sezgin:2002rt,Dhar:2003fi,Bianchi:2003wx,Beisert:2004di,Alday:2005kq,Gopakumar:2003ns,Gopakumar:2004qb,Gopakumar:2005fx,Furuuchi:2005qm,
Aharony:2006th,Aharony:2007fs,Gopakumar:2024jfq,Berkovits:2007zk,Berkovits:2007rj,Berkovits:2008qc,Berkovits:2019ulm,Bae:2016hfy,Bae:2017spv}. Most recently, an explicit worldsheet theory for strings on AdS$_5\times S^5$ dual to free $\cN=4$ sYM was proposed in \cite{Gaberdiel:2021qbb,Gaberdiel:2021jrv,Bhat:2021dez}, with similar topological string constructions also appearing in \cite{Berkovits:2025xok,Berkovits:2026ifu}.

In this work, we will take a different route to weak-coupling holography in four dimensions that is provided by the twistor string descriptions of \emph{self-dual} gauge theories. These are alternative zero coupling limits of 4d gauge theories in which the interactions are turned off in a chiral manner \cite{Siegel:1992wd,Siegel:1992xp,Chalmers:1996rq}. The self-dual sectors of certain 4d gauge theories admit anomaly-free holomorphic uplifts to twistor space \cite{Costello:2021bah,Costello:2022wso,Costello:2022upu,Costello:2023hmi, Costello:2023vyy, Fernandez:2023abp,Fernandez:2024qnu,Bittleston:2024efo,Dixon:2024mzh,Dixon:2024tsb,Morales:2025alm,Charanya:2026pnh}. Some of these six-dimensional uplifts have been geometrically engineered in the topological B-model. The self-dual\footnote{The self-dual theory can be perturbatively deformed into full $\cN=4$ sYM by insertions of the non-self-dual interaction Lagrangian. In the twistor string description, these insertions can be realized holographically by Witten's D-instantons. These are D9-branes in the setup of \cite{Sharma:2025ntb}.} sector of $\U(N)$ $\cN=4$ sYM is engineered by a stack of $N$ D5-branes wrapping twistor space $\PT$ inside a noncompact Calabi-Yau 7-fold $X=\Tot(\CO(-1)^{\oplus 4}\to\PT)$ \cite{Witten:2003nn,Boels:2006ir,Bittleston:2024efo,Sharma:2025ntb,Jarov:2025qhz}. Similarly, the self-dual sectors of $\U(N)$ $\cN=1$ and $\cN=2$ sYM are engineered by $N$ D5-branes wrapping $\PT$ inside related Calabi-Yau geometries \cite{Popov:2004nk,Giombi:2004xv, Bittleston:2024efo,Jarov:2025qhz,Balisa:2026icv}. Backreacting by the twistorial D5-branes gives rise to the closed string holographic duals of said self-dual theories \cite{Sharma:2025ntb,Jarov:2025qhz}. We refer to holographic dualities of this type as \emph{chiral holography}.\footnote{Twistorial rewritings of tensionless strings have also appeared in the AdS$_3$ case \cite{McStay:2024dtk}.} These constructions are closely related to the Berkovits-Witten twistor string \cite{Witten:2003nn,Berkovits:2004hg,Berkovits:2004jj} and are twistorial analogues of twisted holography \cite{Bonetti:2016nma,Ishtiaque:2018str,Costello:2018zrm,Costello:2020jbh, Fernandez:2024tue,Gaiotto:2024dwr}.

These exciting advances are opening the doors to new non-BPS precision tests of holography that directly operate at weak 't Hooft coupling. In this work, we study finite-$N$ effects in the simplest non-BPS observables of 4d supersymmetric gauge theories: their partition functions. We consider the thermal partition function and its Grassmann parity $(-1)^F$ graded counterpart, computed in the zero 't Hooft coupling limit. These partition functions receive contributions from both BPS and non-BPS states and therefore detect finite-$N$ trace relations in the full free theory spectrum. Such thermal observables have long been used to probe the phase structure of weakly coupled gauge theories on compact spaces, including Hagedorn behavior and confinement/deconfinement transitions \cite{Sundborg:1999ue,Sundborg:2000wp,Aharony:2003sx,Aharony:2005bq,Spradlin:2004pp, Kristensson:2020nly}. In their graded counterparts, cancellations between bosons and fermions can alter these transitions \cite{Unsal:2007fb,Basar:2013sza,Basar:2014jua,Basar:2015asd,Cherman:2015rpc,Cherman:2018mya,Cherman:2019ecx,Cherman:2020zea}.

We consider the thermal and graded partition functions of pure super Yang-Mills theories with $\cN=4$ and $\cN=1$ supersymmetry in their free limits. Working in radial quantization on $S^3$, we show that these free theory partition functions agree with the corresponding partition functions of $\cN=4$ and $\cN=1$ self-dual Yang-Mills.\footnote{This is essentially a consequence of the fact that the partition functions of both the free and self-dual theories are one-loop exact, and the linearization of the self-dual theories are the free theories.} This equality allows us to study the original free theory observables using twistor space methods. In the B-model description, the 4d graded partition functions lift to graded partition functions of the corresponding 6d D5-brane worldvolume theories, while the thermal partition functions lift to graded partition functions twisted by $\Z_2$ subgroups of the R-symmetry. The twistor uplift therefore provides a setting for studying the nonperturbative bulk effects dual to the finite-$N$ effects present in these non-BPS observables.

Our main result is the determination of finite-$N$ and infinite-$N$ expansions of the thermal/graded partition functions of free $\cN=4$ and $\cN=1$ sYM, aka \emph{the giant graviton expansions}, in terms of bulk ingredients that are explicitly realized in chiral holography. In particular, we compute both the closed string contributions and the brane contributions in the bulk, with the latter acting as nonperturbative corrections in the expansion. Thus, even in this intrinsically weakly coupled setting with no supergravity description, the terms in our expansions admit a direct bulk interpretation. As a corollary of our expansions, we will also obtain a bulk derivation of the Hagedorn temperatures of these free sYM theories.

In the rest of this Introduction, we first recall conventional giant graviton expansions, including the formulation by Murthy \cite{Murthy:2022ien} that we primarily use. Then we present an overview of our results as well as their interpretation in the formalism of chiral holography.


\subsection{Giant graviton expansions}

Recently, several finite-$N$ expansions of BPS indices, known as giant graviton expansions (GGE), have been proposed
\cite{Arai:2020qaj, Imamura:2021ytr,Gaiotto:2021xce,Lee:2022vig}. They express the finite-$N$ index in terms of its $N\to\infty$ limit and an infinite series of corrections. The main example is the GGE for the superconformal index of $\mathcal{N}=4$ sYM and its various specializations, see section \ref{ssec:standardGGE}.

In holographic examples, the infinite-$N$ contribution is reproduced by the index of multi-gravitons in the supergravity dual, while the finite-$N$ corrections are associated with analytically continued indices of giant graviton D-branes. In AdS$_5\times S^5$, these giants are D3-branes wrapping contractible $S^3\subset S^5$ \cite{McGreevy:2000cw,Grisaru:2000zn} and are dual to determinant operators in $\cN=4$ sYM \cite{Balasubramanian:2001nh}.

Beyond protected quantities, a generalization of giant graviton expansions to thermal and graded partition functions at zero 't Hooft coupling was proposed using Koszul-Tate resolutions \cite{Lee:2023iil}. The terms were conjecturally identified as contributions from D-brane excitations in the tensionless limit
\cite{Lee:2022vig,Lee:2023iil}. An explicit example is a single adjoint fermion
\cite{Gaiotto:2021xce,Lee:2022vig,Lee:2023iil}. It would be interesting to understand the relation between this construction and the expansions studied here.

A different exact expansion was proved by Murthy for a broad class of unitary matrix integrals, including many supersymmetric indices and free partition functions \cite{Murthy:2022ien}. We will refer to it as an `MGGE'. It takes the form
\begin{align}
Z_N &= Z_\infty \bigg( 1 + \sum_{m=1}^{\infty}G_N^{(m)}\bigg) \, .
\end{align}
Integral representations for the nonperturbative corrections $G_N^{(m)}$ were obtained in \cite{Liu:2022olj,Eniceicu:2023uvd,Chen:2024cvf} and take the form of $\U(m|m)$ supermatrix integrals \cite{Eniceicu:2023uvd}. A holographic interpretation of Murthy's GGE was proposed for BPS indices which can be engineered as worldvolume indices of D-branes in the topological B-model \cite{Budzik:2025vzr}. The corrections are associated with indices of brane/antibrane stacks completely transverse to the original stack of $N$ branes, called \emph{Koszul-dual} branes.

Murthy's identity itself does not rely on supersymmetry and can be applied directly to non-BPS free partition functions. However, a holographic realization in terms of branes, analogous to that proposed for BPS indices in
\cite{Budzik:2025vzr}, has not yet been established. To obtain one, we employ a quiver generalization of Murthy's expansion \cite{Purkayastha:2025whi}. In this generalization, $m$ is replaced by a vector of ranks $\boldsymbol{m}=(m_a)$, with one supergroup $\U(m_a|m_a)$ associated with each quiver node. For the thermal and graded free partition functions studied here, the relevant quiver will have two nodes.

Under their respective twistor uplifts, the 4d graded partition functions of $\cN=4$ and $\cN=1$ sYM become the graded partition functions of D5-brane worldvolume theories in a CY 7-fold and a CY 5-fold, respectively, while their thermal counterparts become ``twisted'' graded partition functions. These realizations in chiral holography allow us to construct the corresponding MGGEs directly from bulk ingredients:
\begin{itemize}
    \item the graded/twisted graded partition function of the BCOV theory\footnote{This is the string field theory of closed twistor strings \cite{Berkovits:2004jj,Sharma:2025ntb}, built along the lines of \cite{Bershadsky:1993cx,Costello:2012cy}.} on the CY7 in case of $\cN=4$ sYM, and CY5 in case of $\cN=1$ sYM;
    \item the graded/twisted graded partition functions of brane/antibrane stacks, transverse to the original stack of D5-branes (relative to the twistorial $\P^1\subset\PT$). These are D9-branes\footnote{These transverse branes are avatars of the D-instantons employed by Witten \cite{Witten:2003nn}, Berkovits \cite{Berkovits:2004hg}, and Roiban, Spradlin and Volovich \cite{Roiban:2004vt,Roiban:2004yf} to compute scattering amplitudes in $\cN=4$ sYM.} in CY7 in case of $\cN=4$ sYM, and D5-branes in CY5 in case of $\cN=1$ sYM.
\end{itemize}

By themselves, such brane/antibrane pairs will not be dual to determinant operators in the boundary theory. Instead, we will show that open string tachyon condensation on their worldvolume can reproduce the more familiar kinds of giant gravitons that are dual to determinant operators. In the $\cN=4$ self-dual Yang-Mills case studied in \cite{Sharma:2025ntb}, building on the work of \cite{Caron-Huot:2023wdh}, giant gravitons dual to $1/2$-BPS scalar determinants like $\det(y^{IJ}\Phi_{IJ})$ were found to be branes that wrap half of the directions transverse to twistor space. We will discuss the precise tachyon condensation pattern that gives rise to these giants.


\subsection{Overview of results}
\label{ssec:summary}

In the rest of this introduction, we briefly summarize our findings for the GGE. We mainly present here the results for the $\cN=4$ sYM case, and point to the relevant sections for their $\cN=1$ analogs at the end. We also specialize to a single fugacity $t$ that is conjugate to (twice) the conformal dimension and focus on the GGE for the graded partition function. The GGE for the thermal partition function follows from the simple substitution $t\mapsto -t$. Multi-fugacity generalizations will be discussed in the main text. There, we explain how the relation between the graded and thermal partition functions is implemented by a central R-symmetry twist and how the thermal GGE retains its holographic interpretation in terms of ``twisted graded" partition functions of closed and open strings in the bulk.

Consider $\U(N)$ $\cN=4$ sYM on $\R\times S^3$ in the limit of zero 't Hooft coupling. Its thermal and graded partition functions are defined as\footnote{A more common definition uses with $q=t^2=\e^{-\beta}$. Fermionic operators in gauge theories often have half-integer conformal dimensions, so we use $t$ as our expansion parameter to avoid working with half-integer powers of $q$. We are choosing the branch of the square root so that $0<t<1$, but other branches may be useful for accessing phase transitions.}
\begin{align}
    Z_N^{\mathrm{gr}}(t) =\Tr_{\cH_N} (-1)^F t^{2\Delta} \, , \qquad Z_N^{\mathrm{th}}(t) = \Tr_{\cH_N}\!t^{2\Delta} \, ,
\end{align}
where $t=\e^{-\beta/2}$, with $\beta=1/T$ being the inverse temperature, $\Delta$ denotes conformal dimension, and $\cH_N$ is the Hilbert space of states in radial quantization on $S^3$. As pure dilatations do not commute with any supercharge of the 4d $\cN=4$ superconformal algebra, both of these partition functions are not BPS. In the zero coupling limit, the Hilbert space $\cH_N$ is simply spanned by products of $\mathrm{u}(N)$-valued free field creation operators modulo the constraint of \emph{global} $\U(N)$ invariance \cite{Sundborg:1999ue,Aharony:2003sx}. On the \emph{free} $\cN=4$ spectrum, bosonic states have even $2\Delta$ and fermionic states have odd $2\Delta$. It follows that
\be
Z_N^{\mathrm{th}}(t) = Z_N^{\mathrm{gr}}(-t) \, .
\ee
Hence, from now on, we focus on the graded partition function and write
\be
Z_N(t) \vcentcolon= Z_N^{\mathrm{gr}}(t) 
\ee
unless stated otherwise.

Following \cite{Aharony:2003sx,Kinney:2005ej}, $Z_N$ can be expressed as a unitary matrix integral,
\be
Z_N = \int_{\U(N)}\d U\;\PE\qty[f \Tr(U)\Tr(U^{-1}) ] \,.
\ee
Here, the plethystic exponential in the integrand depends on the single-letter graded partition function
\be
f = f_\text{b} - f_\text{f} = \frac{6t^2+2t^3}{(1+t)^3}\,.
\ee
This is obtained using the results for the (unsigned) single-letter partition functions of bosonic and fermionic letters of the $\cN=4$ vector multiplet listed in \cite{Sundborg:1999ue,Aharony:2003sx},
\be
f_\text{b} = \frac{6t^2+12t^4-2t^6}{(1-t^2)^3}\,,\qquad f_\text{f} = \frac{16t^{3}}{(1-t^2)^3} \, .
\ee
The multi-fugacity generalizations of $Z_N$ were studied in \cite{Yamada:2006rx,Harmark:2006di,Harmark:2007px} and are reviewed in section \ref{sec:partition}.

At this stage, Murthy's original expansion \cite{Murthy:2022ien} can already be applied to the unitary matrix integral defining $Z_N$. It expresses $Z_N$ as $Z_\infty$ times a series of corrections $G^{(m)}_N$, $m\in\Z_{\geq0}$, which are $\U(m|m)$ supermatrix integrals with single-letter partition functions \cite{Eniceicu:2023uvd,Budzik:2025vzr}
\be
f^! = -\frac{f}{1-f} = -\frac{6t^2+2t^3}{(1-t)(1+4t+t^2)}\,.
\ee
Unfortunately, unlike the case of BPS indices \cite{Budzik:2025vzr}, this $f^!$ does not admit an obvious interpretation as the single-letter graded partition function of any brane worldvolume gauge theory known to us. To obtain an expansion that has a bulk interpretation, we look for a suitable generalization of MGGE.

The physically meaningful expansion that we eventually match with bulk computations takes the form of a double sum
\begin{higheq}\label{gget}
Z_N = Z_{\text{BCOV}}\sum_{m_0,m_1=0}^\infty G_N^{(m_0,m_1)}\,.
\end{higheq}
\noindent{}The gravitational prefactor $Z_{\text{BCOV}}$ is given by a plethystic exponential,
\be
Z_{\text{BCOV}} = \PE\bigg[\frac{8t+4t^2}{1-t^2}\bigg] = 1+8t+40t^2+160t^3+552t^4+1712t^5+\cdots\,.
\ee
Its precise origin as a closed string graded partition function will be described in section \ref{sec:bcov}, including its refinement \eqref{Zbcov4} by a maximal number of fugacities. But we remark that unlike other examples of GGEs, $Z_{\text{BCOV}}$ is \emph{not} the infinite-$N$ graded partition function $Z_\infty$ by itself.\footnote{In fact, the series for $Z_\infty$,
\be
Z_\infty = \prod_{n=1}^\infty\frac{1}{1-f(t^n)} = 1+6t^2-16t^3+72t^4-240t^5+\cdots\,,
\ee
does not appear to admit any compact expression as the plethystic exponential of a rational function (though of course, one can always write it as the PE of an infinite series in $t$).} Instead, we will find that $Z_{\text{BCOV}}$ equals $Z_{\infty,\infty}$, the graded partition function of a two-node quiver gauge theory in the limit in which both node ranks are taken to infinity.

More interesting are the summands $G_N^{(m_0,m_1)}$. These take the form of graded partition functions of $\U(m_0|m_0)\times\U(m_1|m_1)$ quiver gauge theories,
\be
G_N^{(m_0,m_1)} = \int\limits_{\U(m_0|m_0)}\hspace{-1em}\d \cW_0\int\limits_{\U(m_1|m_1)}\hspace{-1em}\d \cW_1\,(\Ber\,\cW_0)^N\,\PE\bigg[\sum_{a,b\in\{0,1\}}{\bf f}^!_{ab}\,\Str(\cW_a)\,\Str(\cW_b^{-1})\bigg]\,,
\ee
with the single-letter data given by the $2\times2$ matrix
\be
{\bf f}^! = \begin{pmatrix}
    {\displaystyle-\frac{8t+2t^2+2t^3}{(1-t)^3}}&&{\displaystyle\frac{4t+6t^2+2t^3}{(1-t)^3}}\\[1.5em]
    {\displaystyle\frac{2+6t+4t^2}{(1-t)^3}}&&{\displaystyle-\frac{8t+2t^2+2t^3}{(1-t)^3}}
\end{pmatrix}\,.
\ee
Here, `$\Str$' denotes the supertrace of the supermatrices $\cW_a,\cW_a^{-1}\in\U(m_a|m_a)$, $\Ber\,\cW_0$ is the Berezinian of $\cW_0$, and the contour prescription for evaluating the matrix integrals is given in section \ref{sec:ggemain}. The multi-fugacity generalization of ${\bf f}^!$ is obtained in \eqref{eq:fk4r}.

We experimentally find that the series expansions of $G_N^{(m_0,m_1)}$ in $t$ start at order $\max((N+m_0)m_0,m_1^2)$ or higher. Notice that the dependence on the original rank $N$ enters entirely through the $\U(m_0|m_0)$ node. From this perspective, the sectors with $m_1>0$ may be viewed as ``instanton corrections'' that dress the $m_1=0$ expansion. This intuition is also bolstered by the result we find for the large-$N$ limit of this expansion. In this limit, only the $m_0=0$ terms contribute and one recovers the desired large-$N$ partition function,
\begin{higheq}\label{Zinfexp}
Z_\infty = Z_\text{BCOV} \bigg(1+ \sum_{m_1=1}^\infty G_{\infty}^{(0,m_1)} \bigg)\,.
\end{higheq}
\noindent{}Hence, we find that the ``instanton corrections'' $G_{\infty}^{(0,m_1)}$ deform the closed string answer $Z_\text{BCOV}$ into the true large-$N$ answer $Z_\infty$.

Replacing $t$ by $-t$ in \eqref{Zinfexp} will generate a similar expansion for the large-$N$ thermal partition function of free $\cN=4$ sYM $Z^\text{th}_\infty$. It was found in \cite{Sundborg:1999ue,Aharony:2003sx} that $Z^\text{th}_\infty$ first becomes singular within the physical range $t\in(0,1)$ at the value $t=2-\sqrt{3}$. This corresponds to the Hagedorn temperature $T_H = 1/\log(7+4\sqrt{3})$ at which free sYM on $S^3$ deconfines. We find that $Z_\mathrm{BCOV}(-t)$ is non-singular for all $t\in(0,1)$, while the Hagedorn singularity is reproduced purely from the resummation of the instanton corrections $G^{(0,m_1)}_\infty(-t)$.


\subsection{Chiral holography}

The key idea that allows us to identify the gravitational origins of both $Z_\text{BCOV}$ and the corrections $G^{(m_0,m_1)}_N$ is \emph{chiral holography} \cite{Sharma:2025ntb}. This is an example of holography for twistor string theory. 

The twistor space of Euclidean $\R^4$ is denoted $\PT$ and given by the total space of the rank-2 holomorphic vector bundle $\CO(1)^{\oplus2}\to\P^1$ (see \cite{Adamo:2017qyl} for a review). As a smooth manifold, $\PT$ is diffeomorphic to $\R^4\times\P^1$, and compactification of holomorphic theories on $\PT$ along the $\P^1$ gives rise to self-dual gauge and gravitational theories on $\R^4$ \cite{Mason:2005zm,Boels:2006ir,Costello:2021bah}. Let $X$ denote the total space of the rank-4 vector bundle $\CO(-1)^{\oplus4}\to\PT$. Following \cite{Witten:2003nn,Bittleston:2024efo}, the self-dual sector of $\U(N)$ $\cN=4$ sYM on $\R^4$ can be engineered in the B-model on $X$ as the worldvolume theory of $N$ D5-branes wrapping the zero section of $X$. Backreacting by these branes furnishes a holographic duality between $\cN=4$ self-dual Yang-Mills and the B-model on $X-\PT$ in the presence of the backreaction \cite{Sharma:2025ntb, Jarov:2025qhz}. This has been termed chiral holography, as the boundary gauge theory is self-dual.

One of our key results will be to identify $Z_\text{BCOV}$ with the graded partition function of closed twistor strings in our bulk dual. Thereafter, we propose that the summands $G_N^{(m_0,m_1)}$ of our GGE encode nonperturbative corrections arising from D-branes that systematically correct the closed string answer $Z_\text{BCOV}$ to the gauge theory answer $Z_N$.

To characterize the branes that can appear in our GGE, we look at the category of branes in the B-model on $X$. This is given by $D^b\text{Coh}(X)$, the bounded derived category of coherent sheaves on $X$ \cite{Kontsevich:1994dn,Douglas:2000gi,Aspinwall:2004jr}. On the total spaces of vector bundles like $p:X\to\P^1$, this category admits a finite basis of `generators': the pullbacks by $p$ of $\CO_{\P^1}$ and $\CO_{\P^1}(1)$. We denote them by $\CO_X$ and $\CO_X(1)$ respectively. In physical terms, this means that every brane in the B-model on $X$ can be obtained by tachyon condensation from stacks of spacefilling branes $\CO_X,\CO_X(1)$, together with their parity-reversed partners $\Pi\CO_X$ and $\Pi\CO_X(1)$, which we refer to as \emph{antibranes} following standard B-model terminology.\footnote{More precisely, B-branes are represented by bounded $\mathbb Z$-graded complexes of coherent sheaves (see \cite{Aspinwall:2004jr} for a review). The notation $\Pi$ retains only the parity of this grading: copies placed in even and odd degrees are interpreted as branes and antibranes respectively.} To obtain branes with zero total K-theory charge, one takes equal number of copies of $\CO_X$ and $\Pi\CO_X$, and equal number of copies of $\CO_X(1)$ and $\Pi\CO_X(1)$.

Motivated by \cite{Budzik:2025vzr}, we further focus on branes/antibranes that are transverse to the backreacting stack of $N$ D5-branes wrapping $\PT$. Every point on $\R^4$ corresponds to a copy of $\P^1$ inside $\PT\simeq\R^4\times\P^1$, known as its \emph{twistor line}. Local operators in the gauge theory on $\R^4$ are engineered as non-local operators smeared along the twistor lines. So we look for transverse branes that intersect $\PT$ in a $\P^1$ (instead of at a point) and wrap the fibers of $\CO(-1)^{\oplus4}$ above it. 

To be precise, let $Y\subset X$ be the total space of $\CO(-1)^{\oplus4}\to\P^1$, with $\P^1$ being the twistor line of the origin of $\R^4$ (around which we radially quantize). We work with D9-branes that wrap the complex 5-fold $Y$. In \cite{Sharma:2025ntb}, these were identified with the D-instantons of twistor string theory \cite{Witten:2003nn,Roiban:2004yf}. A basis of the category of B-model branes supported within $Y$ is given by the generators $\CO_Y,\CO_Y(1)$. All possible ``giant graviton'' branes in chiral holography can thus be generated by tachyon condensation on stacks of these elementary D9-branes/antibranes. \emph{This} is the origin of the double sum in our expansion \eqref{gget}.

In the spirit of \cite{Budzik:2025vzr}, we expect the partial sum
$\sum_{i_0=0}^{m_0}\sum_{i_1=0}^{m_1}G_N^{(i_0,i_1)}$
to be associated with $m_0|m_0$ branes each supporting the Chan-Paton bundle
$\CO_Y$, and $m_1|m_1$ branes each supporting the Chan-Paton bundle $\CO_Y(1)$, in the presence of $N$ units of flux. The full double sum
$\sum_{m_0,m_1\geq0}G_N^{(m_0,m_1)}$ corresponds to taking both stack
sizes to infinity, formally associated with
\begin{align}
\CO_Y^{\oplus\,\infty|\infty} \oplus \CO_Y(1)^{\oplus\,\infty|\infty}\,.
\end{align}
We verify by an independent bulk calculation that ${\bf f}^!$ is the matrix of single-letter graded partition functions counting open string states on the branes $\Pi\CO_Y\oplus\CO_Y(1)$ (after an appropriate fugacity involution). The diagonal components ${\bf f}^!_{00},{\bf f}^!_{11}$ describe strings stretching from $\Pi\CO_Y$ to itself and $\CO_Y(1)$ to itself, while the off-diagonal components ${\bf f}^!_{01},{\bf f}^!_{10}$ capture strings stretching between $\Pi\CO_Y$ and $\CO_Y(1)$.

The expressions for the summands $G_N^{(m_0,m_1)}$ reduce to contour integrals over the eigenvalues of $\cW_0,\cW_1$. The resulting residues correspond to the tachyon condensates of a stack of $m_0|m_0$ $\CO_Y$ branes and $m_1|m_1$ $\CO_Y(1)$ branes. Thus, the sum over $m_0,m_1$ captures the contributions of all possible giant graviton-type branes that may appear in any GGE of $Z_N$. For example, we will see that the half-BPS giants found in \cite{Sharma:2025ntb}, dual to the determinant operators of \cite{Caron-Huot:2023wdh}, are engineered by the complex of sheaves displayed in equation \eqref{cmpx2}. The maps in this complex correspond to tachyon vevs, and these vevs will map to a specific residue of the contour integrals present in the correction $G_N^{(1,2)}$.

\subsubsection*{Outline} 
We end this introduction by presenting an outline of the sections that follow.
\begin{enumerate}
    \item[{\hyperref[sec:ggereview]{\underline{$\S$.2.}}}] We begin with a review of giant graviton expansions in section \ref{sec:ggereview}. We first recall the more conventional GGEs of superconformal indices, in which the finite-$N$ corrections are interpreted as indices of giant graviton worldvolume theories. We then review Murthy's giant graviton expansion and its generalization to quiver gauge theories with adjoint and bifundamental matter. These expansions are naturally organized in terms of contributions from stacks of Koszul-dual branes and antibranes. We also explain how one may recover the ordinary one-node MGGE by ``integrating out'' one of the nodes from our two-node MGGE.
    \vskip 0.25cm

    \item[{\hyperref[sec:partition]{\underline{$\S$.3.}}}] In section \ref{sec:partition}, we introduce the non-BPS observables studied in this work: the thermal and graded partition functions of free $\cN=4$ and $\cN=1$ sYM on $S^3$. We review the derivation of the associated single-letter partition functions and unitary matrix integrals, refined by a maximal number of fugacities corresponding to the Cartan subalgebras of the Lorentz and R-symmetries. We also explain how the thermal partition functions can be obtained as R-symmetry twists of the graded ones.

    \item[{\hyperref[sec:chiralholography]{\underline{$\S$.4.}}}] In section \ref{sec:chiralholography}, we review the twistor uplifts and holographic duals of the self-dual limits of super Yang-Mills theories. These are weak coupling limits, so the holographic duals do not take the form of supergravity descriptions in AdS spaces. Instead, we describe their origins in twistor strings. This was studied for the $\cN=4$ sYM case in \cite{Sharma:2025ntb,Jarov:2025qhz}, and is generalized to $\cN=1$ sYM in section \ref{ssec:ch1} in the present work.

    \item[{\hyperref[sec:sdpartition]{\underline{$\S$.5.}}}] In section \ref{sec:sdpartition}, we compute the graded partition functions of $\cN=4$ and $\cN=1$ self-dual Yang-Mills (sdYM) directly from their twistor uplifts. Using the BV-BRST description of the open string states, we show that their single-particle spectra coincide with those of the corresponding ordinary free theories, while the residual BRST differential imposes global $\U(N)$ invariance. This establishes that the self-dual and ordinary free partition functions agree, allowing us to study the observables of section \ref{sec:partition} using their twistor string duals.

    \item[{\hyperref[sec:bcov]{\underline{$\S$.6.}}}] Next, we study the closed string sector of these holographic duals. In section \ref{sec:bcov}, we compute the graded partition functions $Z_{\mathrm{BCOV}}$ of the BCOV theories describing closed twistor strings in the $\cN=4$ and $\cN=1$ constructions. Quite surprisingly, these partition functions fail to reproduce the large-$N$ limits of the corresponding boundary partition functions. This signals that closed strings alone are insufficient to reproduce the large-$N$ boundary spectrum at weak coupling. The resolution of this puzzle will emerge from the brane contributions in our GGEs.

    \item[{\underline{$\S$.7.}}] In section 7, we turn to open strings and the category of B-model branes relevant to our construction. We review how the branes supported on twistor space, as well as the branes supported on the geometries transverse to it, can be generated by a pair of `elementary' branes $\CO(n),\CO(n+1)$ for any $n\in\Z$. We develop a general method for counting the open strings between these generators and use it to reproduce the matrices of single-letter partition functions that will enter the quiver MGGE. We also describe the Koszul-dual transverse branes/antibranes and show how ordinary giant gravitons can arise from them by tachyon condensation.

    \item[{\hyperref[sec:ggemain]{\underline{$\S$.8.}}}] In section \ref{sec:ggemain}, we bring together the ingredients from the previous sections to construct the giant graviton expansion in chiral holography. We apply the two-node quiver MGGE to the worldvolume theory of the D5-brane stack $\CO_\PT^{\oplus N_0}\oplus\CO_\PT(-1)^{\oplus N_1}$. Its $N_0,N_1\to\infty$ partition function is identified with the BCOV closed string partition function, while the transformed single-letter data $\mathbf f^!$ are identified with the open string spectra of Koszul-dual brane/antibrane systems $\Pi\CO_Y^{\oplus\,m_0|m_0}\oplus\CO_Y(1)^{\oplus\,m_1|m_1}$ transverse to twistor space. Our MGGE comes with Berezinian insertions $(\Ber\,\cW_0)^{N_0}(\Ber\,\cW_1)^{N_1}$. We show that they arise from counting bifundamental strings stretched between the original and transverse branes. In this process, the transverse branes carrying $\CO_Y(1)$ are identified with Witten's D-instantons \cite{Witten:2003nn}.

    \item[{\hyperref[sec:thermalgge]{\underline{$\S$.9.}}}] In section \ref{sec:thermalgge}, we remove the ``scaffolding node'' $\CO_\PT(-1)^{\oplus N_1}$ by setting $N_1=0$ along with $N_0=N$. This yields our main finite-$N$ expansions for the graded and thermal partition functions of free self-dual $\cN=4$ and $\cN=1$ sYM. The resulting MGGE remains a double sum over two types of transverse brane/antibrane pairs, with the second type giving ``D-instanton corrections'' that resum to the conventional one-node Murthy expansion. We illustrate these formulae explicitly in the unrefined theories and also take initial steps toward relating the Murthy-type expansion to conventional GGEs by extracting the contributions of simple giant gravitons from residues of the supermatrix integrals.

    \item[{\hyperref[sec:hagedorn]{\underline{$\S$.10.}}}] In section \ref{sec:hagedorn}, we use the large-$N$ limit of our expansions to give a bulk derivation of the Hagedorn temperatures of free $\cN=4$ and $\cN=1$ sYM. Remarkably, the BCOV closed string partition functions remain regular at the Hagedorn temperatures. Instead, the Hagedorn poles are reproduced by the resummed contribution of Witten's D-instanton/anti-D-instanton pairs. This identifies the microscopic bulk degrees of freedom responsible for the familiar Hagedorn singularities of the free gauge theories.

    \item[{\hyperref[sec:discussion]{\underline{$\S$.11.}}}] In section \ref{sec:discussion}, we end with a discussion of the lessons learned as well as possible directions for future work. These include the search for a bulk description of the deconfinement/Hawking-Page transition at weak 't Hooft coupling, the interpretation of the scaffolding node, standard GGEs and non-BPS trace relations, 't Hooft coupling corrections, and potential extensions of our techniques to non-supersymmetric gauge theories that admit twistor uplifts.
\end{enumerate}


\section{Giant graviton expansions}
\label{sec:ggereview}

We begin by reviewing the notion of giant graviton expansions (GGEs). In section \ref{ssec:standardGGE}, we briefly discuss ``standard" GGEs of finite-$N$ superconformal indices. These are expansions in which the finite-$N$ corrections to the infinite-$N$ answer are related to indices of worldvolume theories on a stack of giant graviton branes in the holographic dual.

In section \ref{ssec:murthy}, we review Murthy's giant graviton expansion
(MGGE). Although Murthy's expansion is commonly referred to as a GGE, its finite-$N$ corrections are naturally related to ``Koszul-dual" brane/antibrane systems, rather than to ordinary giant gravitons \cite{Budzik:2025vzr}. We also review the generalization of MGGE to quiver gauge theories with gauge groups $\U(N_1)\times\cdots\times\U(N_k)$ \cite{Purkayastha:2025whi}. The two-node quiver MGGE will arise naturally in our chiral holography setting, where the relevant category of branes has two generators. We also show that, after setting the rank of the second node to zero, the two-node MGGE can be resummed into the standard one-node MGGE. We refer to this procedure as ``integrating out the scaffolding node''.


\subsection{Giant graviton expansions of superconformal indices}
\label{ssec:standardGGE}

We begin by reviewing giant graviton expansions of superconformal indices
\cite{Imamura:2021ytr,Gaiotto:2021xce,Lee:2022vig}. These expansions express a finite-$N$ index as its infinite-$N$ limit multiplied by an infinite series of corrections.

The simplest example is the $1/2$-BPS subsector of $\U(N)$ $\cN=4$ sYM. Its index\footnote{In this simple sector, the index equals the free partition function because all contributing states are bosonic and there are no cancellations from $(-1)^F$.} counts gauge invariant polynomials in a single complex adjoint scalar
$X$, with fugacity $q$ for its R-charge (also equal to its conformal dimension $R=\Delta$):
\begin{align}
Z_N(q) &= \Tr_{\cH_N^{1/2}} (-1)^F q^R =\prod_{n=1}^N\frac{1}{1-q^n} \, .
\end{align}
The finite-$N$ index admits the exact expansion \cite{Gaiotto:2021xce}
\begin{align}
Z_N(q)&= Z_\infty(q) \sum_{k=0}^\infty q^{Nk} \, \hat Z_k(q) \,, \label{eq:12bps}
\end{align}
where
\begin{align}
    Z_\infty(q) =\prod_{n=1}^\infty\frac{1}{1-q^n} 
\end{align}
is the infinite-$N$ limit, and
\begin{align}
\hat{Z}_k(q) = (-1)^k\frac{q^{k(k+1)/2}}{\prod_{n=1}^k(1-q^n)} 
\end{align}
are the corrections.

This expansion admits a natural holographic interpretation. The factor $Z_\infty(q)$ is the multi-particle index of perturbative $1/2$-BPS Kaluza-Klein gravitons in AdS$_5\times S^5$ \cite{Kinney:2005ej}. The corrections $\hat{Z}_k(q)$ are associated with giant gravitons, which are $1/2$-BPS D3-branes wrapping an $S^3\subset S^5$ in AdS$_5\times S^5$ \cite{McGreevy:2000cw,Grisaru:2000zn}, and dual to determinant operators $\det X$ \cite{Balasubramanian:2001nh}. The integer $k$ counts the number of giant gravitons, each carrying R-charge $N$, so that $q^{kN}$ is their classical contribution. The corrections $\hat Z_k(q)$ are the analytically continued $1/2$-BPS indices of the $\U(k)$ worldvolume theory on the stack of $k$ giants \cite{Gaiotto:2021xce,Lee:2022vig,Eleftheriou:2023jxr}:
\begin{align}
\hat{Z}_k(q)=Z_k(q^{-1}) =
\prod_{n=1}^k\frac{1}{1-q^{-n}} = (-1)^k\frac{q^{k(k+1)/2}}{\prod_{n=1}^k(1-q^n)} \, .
\end{align}

Finally, we note that the difference between $Z_N$ and $Z_\infty$ originates from finite-$N$ trace relations. The giant graviton expansion is expected to implement these cancellations. In the $1/2$-BPS subsector, this statement has been made precise by identifying the giant graviton contributions with the characters of the terms in a Koszul-Tate resolution of the finite-$N$ Hilbert space \cite{Lee:2023iil}.

Next we consider the fully refined superconformal index of $\U(N)$ $\cN=4$ sYM \cite{Kinney:2005ej}. In a convenient basis of charges, it is
\begin{align}\label{16index}
Z_N(p,q,x,y,z) &= \Tr_{\cH_N^{1/16}}(-1)^F p^{\Delta-j_1+j_2} q^{\Delta-j_1-j_2} x^{-R_2-R_3} y^{-R_3-R_1} z^{-R_1-R_2}\,,
\end{align}
subject to the constraint $xyz=pq$ on the fugacities. Here $\cH_N^{1/16}$ denotes the $1/16$-BPS Hilbert space of states annihilated by both $Q\vcentcolon=Q_{1-}$ and $Q^\dagger=S^{1-}$. These states saturate the BPS bound
\begin{align}
2\{Q,Q^\dagger\} =\Delta-2j_1-R_1-R_2-R_3 = 0 \, . \label{eq:116bound}
\end{align}
Here $\Delta$ denotes the conformal dimension, while $j_1,j_2$ and $R_1, R_2, R_3$ are the Cartan charges of the Lorentz and $\SO(6)_R$ symmetry groups, respectively. In these conventions the three complex scalars $X,Y,Z$ carry weights $x,y,z$.

The superconformal index is protected and may therefore be evaluated in the zero coupling limit, where it admits the Molien-Weyl representation
\begin{align}
Z_N(p,q,x,y,z) &= \int_{\U(N)} \dd U \, \exp\qty [ \sum_{n=1}^\infty \frac{1}{n} f(p^n,q^n,x^n,y^n,z^n) \Tr(U^n)\Tr(U^{-n}) ] \, .
\end{align}
Here $f$ is the single-letter index
\begin{align}
f(p,q,x,y,z) &= 1-\frac{(1-x)(1-y)(1-z)}{(1-p)(1-q)}\,, \label{eq:116f}
\end{align}
which counts the $1/16$-BPS letters of $\cN=4$ sYM satisfying \eqref{eq:116bound}. The gauge invariance is imposed by a $\U(N)$ integral. Writing $u_i$ for the eigenvalues of $U$, the normalized Haar measure is
\begin{align}
\d U = \frac{1}{N!} \prod_{i=1}^N\frac{\d u_i}{2\pi i u_i} \prod_{i\neq j}\qty(1 - \frac{u_i}{u_j}),
\qquad |u_i|=1 \, . \label{eq:dU}
\end{align}
In particular, the large-$N$ index is \cite{Kinney:2005ej}
\begin{align}
Z_\infty(p,q,x,y,z)
=
\prod_{n=1}^\infty
\frac{1}{1-f(p^n,q^n,x^n,y^n,z^n)} = \prod_{n=1}^\infty \frac{(1-p^n)(1-q^n)}{(1-x^n)(1-y^n)(1-z^n)} \,.
\end{align}
This is obtained by a saddle-point evaluation of the eigenvalue integrals.

The proposed giant graviton expansion of the full index is \cite{Imamura:2021ytr}
\begin{align}
Z_N(p,q,x,y,z)
= Z_\infty(p,q,x,y,z) \sum_{k_1,k_2,k_3=0}^\infty x^{Nk_1}y^{Nk_2}z^{Nk_3} \hat Z_{k_1,k_2,k_3}(p,q,x,y,z) \, .
\label{116GGE}
\end{align}
The integers $k_1,k_2,k_3$ count stacks of giant graviton D3-branes wrapping three differently oriented maximal $S^3\subset S^5$, dual to the determinant operators $\det X$, $\det Y$ and $\det Z$. The factors $x^{Nk_1}y^{Nk_2}z^{Nk_3}$ are their classical fugacity weights, while $\hat Z_{k_1,k_2,k_3}(p,q,x,y,z)$ is the analytically continued index of the $\U(k_1)\times\U(k_2)\times\U(k_3)$ worldvolume quiver theory, evaluated with an appropriate contour or pole selection rule  \cite{Imamura:2021ytr,Lee:2022vig}. It includes both open strings within each stack and open strings stretched between different stacks.\footnote{Although each giant graviton D3-brane is $1/2$-BPS, their worldvolume excitations might preserve less supersymmetry.}

Unlike the $1/2$-BPS identity \eqref{eq:12bps}, the expansion \eqref{116GGE} remains conjectural and has been checked through total wrapping number $k_1+k_2+k_3=3$ for low rank $N$ \cite{Imamura:2021ytr}.

Other examples of the giant graviton expansion include other BPS subsectors of $\cN=4$ sYM as well as other BPS indices and M-theory cases \cite{Arai:2020qaj, Lee:2022vig, Arai:2020uwd}.\footnote{The expansion of Gaiotto-Lee \cite{Gaiotto:2021xce} can be thought of as a GGE expansion in a preferred fugacity. See the discussion in \cite{Lee:2022vig}. Since the corrections are still indices of giant gravitons, we will group it together with the reviewed expansions, to differentiate from the Murthy-type expansion \cite{Murthy:2022ien}.} In general, branes of different types or different wrapping cycles may contribute, leading to a sum over multiple integers \cite{Lee:2022vig}.


\subsection{Murthy's expansion and quiver generalization}
\label{ssec:murthy}

We now turn to a second type of giant graviton expansion, which we call Murthy-type, or MGGE. Murthy's expansion was proved \cite{Murthy:2022ien} as an exact identity for a broad class of unitary matrix integrals of the form
\begin{align}
Z_N[f_\bullet] = \int_{\U(N)}\dd U\, \exp\left[
\sum_{n=1}^\infty\frac{1}{n}\,
f_n
\Tr(U^n)\Tr(U^{-n})
\right] \, ,
\end{align}
where $f_\bullet=\qty{f_n}_{n\geq 1}$ denotes a sequence of winding data. To keep
the notation light, we will usually denote the dependence on the full sequence simply by $Z_N[f]$. In the case of superconformal indices, the winding data are generated from a single-letter index $f(\qty{q_i})$ by $f_n=f(\qty{q_i^n})$, i.e., raising every fugacity to the $n^\text{th}$ power.

Using the Hubbard-Stratonovich transformation and a result of Borodin and Okounkov \cite{borodin2000fredholm}, Murthy proved the exact expansion \cite{Murthy:2022ien}
\begin{align}
Z_N[f] = Z_\infty[f] \sum_{m=0}^\infty G_N^{(m)}[f] \,. \label{eq:murthy}
\end{align}
Here, $Z_\infty[f]$ is again the infinite-$N$ limit of $Z_N[f]$, 
\begin{align}
Z_\infty[f] = \prod_{n=1}^\infty\frac{1}{1-f_n} \label{eq:Zinftyagain} \, .
\end{align}
Integral representations for the corrections $G_N^{(m)}[f]$ were obtained in \cite{Liu:2022olj,Eniceicu:2023uvd,Chen:2024cvf}. In particular, they can be written as $\U(m|m)$ supermatrix integrals \cite{Eniceicu:2023uvd},\footnote{These
supermatrix integrals bear a close resemblance to those arising in resurgent analyses of Hermitian matrix models
\cite{Marino:2022rpz,Schiappa:2023ned}.}
\begin{align}
G_N^{(m)}[f] = \int_{\U(m|m)}\d \cW\,(\text{Ber}\,\cW)^N\, \exp\left[
\sum_{n=1}^\infty\frac{1}{n}\,
f^!_n \, 
\Str(\cW^{n})\,\Str(\cW^{-n})
\right] \, , \label{eq:G}
\end{align}
where $\text{Ber}\,\cW$ is the Berezinian, $\Str\,\cW$ is the supertrace, and $\d\cW$ is the $\U(m|m)$ supermatrix measure. Writing $u_i$ and $v_j$ respectively for the eigenvalues of the two bosonic $\U(m)$ blocks, we have
\begin{align}
    \dd\cW = \frac{1}{(m!)^2} \prod_{i=1}^m \frac{\dd u_i}{2\pi i u_i} \prod_{j=1}^m \frac{\dd v_j}{2\pi i v_j} \frac{\prod_{i\neq j} \qty(1-\frac{u_i}{u_j}) \qty(1-\frac{v_i}{v_j}) }{\prod_{i,j} \qty(1-\frac{u_i}{v_j}) \qty(1-\frac{v_j}{u_i}) } \, .
\end{align}
The standard unit contours are modified to $|u_i|=r$ and $|v_j|=R$, with $0<r<R$ to avoid the poles of the super-Vandermonde factor. For given fugacities, the ratio $r/R$ is chosen sufficiently close to $1$, so that \eqref{eq:G} converges (e.g. $|t|<r/R$ in a single fugacity case).

Finally, the transformed winding data $f^!_n$ are related to the original winding data $f_n$ by
\begin{align}
    f^!_n = -\frac{f_n}{1-f_n} \, .
\end{align}
When $f_n=f(\{q_i^n\})$, these are generated by an ``effective" single-letter index
\begin{align}
f^! = -\frac{f}{1-f} \label{eq:fk}
\end{align}
of a distinct putative gauge theory.

Motivated by the relation \eqref{eq:fk}, a B-model holographic interpretation of the MGGE for certain superconformal indices was proposed in \cite{Budzik:2025vzr}. If $f$ is the single-letter graded partition function of the worldvolume theory on a B-model D-brane in $\C^d$, then $f^{!}$ is the single-letter graded partition function of a D-brane wrapping all directions transverse to the original brane, called its \emph{Koszul-dual} brane. Under this identification, the partial sum
\begin{align}
\sum_{i=0}^m G_N^{(i)}[f]
\end{align}
is conjectured to compute the graded partition function of a system of $m$ Koszul-dual branes and $m$ antibranes, in the presence of $N$ units of flux.\footnote{The supermatrix integrals studied in \cite{Budzik:2025vzr, berele2025invariants} correspond to \eqref{eq:G} with $N=0$, so the interpretation for $N>0$ is conjectural.} The worldvolume theory of this system has gauge supergroup $\U(m|m)$ and single-letter graded partition function $f^!$.

In the two examples considered above:
\begin{itemize}
    \item The $1/2$-BPS index, $f=x$, is the single-letter graded partition function of D$(-1)$-branes in $\C$, while $f^!=-x/(1-x)$ is the single-letter graded partition function of Koszul-dual D1-branes filling $\C$.
    \item The $1/16$-BPS superconformal index $f$, given by \eqref{eq:116f}, is the single-letter graded partition function of D3-branes in $\C^5$, while $f^!=-f/(1-f)$ is the single-letter graded partition function of Koszul-dual D5-branes in $\C^5$, transverse to the original D3-branes.
\end{itemize}
See appendix C of \cite{Budzik:2025vzr} for more examples. In this work, we study MGGEs for more intricate brane systems and bulk geometries, providing more evidence for this holographic interpretation.

As discussed above, Murthy's expansion is naturally organized in terms of Koszul-dual brane/antibrane systems \cite{Budzik:2025vzr}. In the B-model, lower-dimensional branes can be obtained by tachyon condensation from brane/antibrane pairs \cite{Douglas:2000gi,Aspinwall:2001pu,Aspinwall:2004jr}. Ordinary giant graviton branes can therefore arise as tachyon condensates within the Koszul-dual brane/antibrane systems appearing in Murthy's expansion. This gives a mechanism relating Murthy's expansion to standard GGEs. We will show examples of this mechanism in our chiral holography setting. We will construct a complex engineering a giant graviton brane from tachyon condensation on a collection of Koszul-dual branes and antibranes in section \ref{ssec:GGfromtachyons}. We will also show how the corresponding giant graviton graded partition function can be extracted by taking residues in the MGGE in section \ref{ssec:interpretations}. We leave the complete matching of the two types of expansions for future work.

We emphasize that the derivation of Murthy's GGE is insensitive to whether $f_n$ originates from a superconformal index. It therefore also applies to free partition functions, including the thermal and graded free partition functions studied here. However, the holographic interpretation of $f^!_n$ in terms of Koszul-dual branes, as in the BPS examples above, can be far from obvious. For our thermal/graded partition functions, we will obtain such a holographic realization by using a two-node quiver generalization of MGGE and embedding the system in chiral holography.

\paragraph{Quiver generalization.}  We now review the generalization of Murthy's GGE to quiver gauge theories with adjoint and bifundamental matter obtained in \cite{Purkayastha:2025whi} (allowing independent ranks at different nodes). The starting point is a matrix integral with gauge group
$\U(N_1)\times\cdots\times\U(N_k)$ of the form
\be\label{ZvecN}
Z_{N_1,\dots,N_k}[\mathbf{f}] = \int_{\U(N_1)}\d U_1\cdots\int_{\U(N_k)}\d U_k\,\exp\bigg(\sum_{n=1}^\infty\frac{1}{n}\sum_{a,b=1}^k ({\bf f}_{n})_{ab} \Tr(U_a^{n})\Tr(U_b^{-n})\bigg)\,.
\ee
The single-letter data is now assembled in the form of $k\times k$ matrices ${\bf f}_{n}$. In case of an index of a quiver gauge theory, ${\bf f}_n$ equals ${\bf f}(\{q_i^n\})$ for some matrix of single-letter indices ${\bf f}(\{q_i\})$. The entry $f_{ab}$ counts letters that transform as elements of $\Hom(E_b,E_a)=E_a\otimes E_b^\vee$, where $E_a$ denotes the $\U(N_a)$ bundle at the $a^\text{th}$ node of the quiver.

The $k$-node Murthy-type expansion takes the form \cite{Purkayastha:2025whi}
\be\label{quivermurthy}
Z_{N_1,\dots,N_k}[\mathbf{f}] = Z_{\infty,\dots,\infty}[\mathbf{f}]\sum_{m_1,\dots,m_k=0}^\infty G_{N_1,\dots,N_k}^{(m_1,\dots,m_k)}[\mathbf{f}]\,.
\ee

The prefactor $Z_{\infty,\dots,\infty}[\mathbf{f}]$ can be identified with the simultaneous $N_a\to\infty$ limit of $Z_{N_1,\dots,N_k}[\mathbf{f}]$:
\be\label{Zinfinf}
Z_{\infty,\dots,\infty}[\mathbf{f}] = \prod_{n=1}^\infty\frac{1}{\det(1-{\bf f}_n)}\,.
\ee

The summands $G_{N_1,\dots,N_k}^{(m_1,\dots,m_k)}[\mathbf{f}]$ take the form of $\U(m_1|m_1)\times\cdots\times\U(m_k|m_k)$ unitary supermatrix integrals
\begin{multline}
G_{N_1,\dots,N_k}^{(m_1,\dots,m_k)}[\mathbf{f}] = \int_{\U(m_1|m_1)}\d \cW_1\,(\Ber\,\cW_1)^{N_1}\cdots\int_{\U(m_k|m_k)}\d \cW_k\,(\Ber\,\cW_k)^{N_k}\\
\times\exp\bigg(\sum_{n=1}^\infty\frac{1}{n}\sum_{a,b=1}^k ({\bf f}^!_{n})_{ab}\,\Str(\cW_a^{n})\,\Str(\cW_b^{-n})\bigg)\, ,
\end{multline}
governed by a transformed set of matrices
\be\label{bfshk}
{\bf f}^!_n = -\qty[{\bf f}_n( \mathbf{1}-{\bf f}_n)^{-1}]^{\mathrm{T}}\,.
\ee
At each node $a$, the $|u_{a,i}|=r_a$ contours are again inside the $|v_{a,j}|=R_a$ contours. Off-diagonal zeromodes can require additional nesting between nodes, as will be the case in our expansions in sections \ref{sec:ggemain} and \ref{sec:thermalgge}. We review the quiver generalization of Murthy's proof in appendix \ref{app:proofs}.

In analogy with the one-node MGGE explained above, the original quiver in our applications will arise as the worldvolume theory on a collection of generator branes in the B-model. We will show that the transformed matrix
\begin{align}
   \mathbf{f}^!=-[\mathbf{f}(\mathbf{1}-\mathbf{f})^{-1}]^{\mathrm{T}} 
\end{align}
is the matrix of single-letter graded partition functions of the worldvolume theory on a second collection of generator branes supported in the transverse geometry (after an appropriate fugacity involution).

\paragraph{Integrating out the scaffolding node.} We will mostly use the two-node specialization of the quiver MGGE reviewed above. We relabel the nodes as $\U(N_0)$ and $\U(N_1)$, and set $N_0=N$ and $N_1=0$:
\begin{align}
Z_{N,0}[\mathbf{f}] &= Z_{\infty,\infty} [\mathbf{f}]
\sum_{m_0,m_1=0}^\infty
G_{N,0}^{(m_0,m_1)} [\mathbf{f}] \, . \label{eq:2nodeMGEE}
\end{align}

Since the second node has rank zero, the matrix integral $Z_{N,0}[\mathbf{f}]$ depends only on the $f_{00}$ component and $Z_{N,0}[\mathbf{f}]=Z_N[f_{00}]$. It therefore also admits the standard one-node Murthy expansion \eqref{eq:murthy}:
\begin{align}
Z_N[f_{00}] &= Z_\infty [f_{00}]
\sum_{m_0=0}^\infty \widetilde{G}_N^{(m_0)} [f_{00}] \,. \label{eq:1nodeMGGE}
\end{align}

The one-node expansion \eqref{eq:1nodeMGGE} can be recovered from the two-node expansion \eqref{eq:2nodeMGEE} by summing over all sectors associated with the second ``scaffolding" node at fixed $m_0$. Below, we list the important relations between both expansions and delegate the proofs to appendix \ref{app:proofs}.

First, we consider the $m_0=0$ terms. Since these are independent of $N$ we denote them as $G^{(0,m_1)}_{\infty,0}[\mathbf{f}]$. They resum to:
\begin{align}
    \sum_{m_1=0}^\infty G^{(0,m_1)}_{\infty,0}[\mathbf{f}] = Z_{\infty}[(\mathbf{f}^!)_{11}] \, . \label{eq:m00sum}
\end{align}
For $m_0>0$, the Berezinian factor suppresses each fixed $m_0$ term as $N\rightarrow\infty$, so that only $m_0=0$ terms contribute. Taking the infinite-$N$ limit of the two-node expansion \eqref{eq:2nodeMGEE} then gives
\begin{align}
Z_\infty[f_{00}] &= Z_{\infty,\infty}[\mathbf{f}]
\sum_{m_1=0}^\infty G_{\infty,0}^{(0,m_1)}[\mathbf{f}] = Z_{\infty,\infty}[\mathbf{f}] \, Z_{\infty}[(\mathbf{f}^!)_{11}] \, . \label{eq:Zinftyresum}
\end{align}
More generally, resumming the scaffolding terms at fixed $m_0>0$ reproduces the corresponding one-node correction after dividing by the $m_0=0$ sum:
\begin{align}
\widetilde{G}_N^{(m_0)}[f_{00}]
&= \frac{\sum_{m_1=0}^\infty
G_{N,0}^{(m_0,m_1)}[\mathbf{f}] }
{\sum_{m_1=0}^\infty
G_{\infty,0}^{(0,m_1)}[\mathbf{f}] } = \frac{Z_{\infty,\infty}[\mathbf{f}]}{Z_\infty[f_{00}]} \sum_{m_1=0}^\infty
G_{N,0}^{(m_0,m_1)}[\mathbf{f}] \, . \label{eq:GGrelation}
\end{align}
For the thermal/graded partition functions considered later in this paper, the standard MGGE \eqref{eq:1nodeMGGE} effectively
resums the brane/antibrane sectors associated with the scaffolding node. The quiver expansion \eqref{eq:2nodeMGEE} keeps these sectors explicit and will admit a direct holographic interpretation.


\section{Partition functions of free gauge theories}
\label{sec:partition}

Having reviewed the giant graviton expansions of BPS indices, in this section, we will introduce the non-BPS observables we are interested in. We will study the thermal and $(-1)^F$-graded partition functions of $\cN=4$ and $\cN=1$ super Yang-Mills (sYM) theories in their ordinary free limits. In each case, we will work with their refinements by a maximal number of fugacities. Much of this discussion builds on the classic studies \cite{Sundborg:1999ue,Aharony:2003sx} of free gauge theory partition functions on $S^1\times S^3$. In later sections, our free theory partition functions will be identified with the corresponding partition functions of $\cN=4$ and $\cN=1$ self-dual Yang-Mills (sdYM) theories. This will allow us to construct their giant graviton expansions from bulk ingredients in the framework of chiral holography.


\subsection{Free $\cN=4$ sYM}
\label{ssec:N=4}

Partition functions of even the familiar supersymmetric theories are generically non-BPS and can usually only be studied in perturbation theory at weak 't Hooft coupling. We will be interested in computing such partition functions in a trivial background and to zeroth order in the 't Hooft coupling, i.e., in the free limit. They will count states built from the creation operators of linearized gauge theories, modulo the constraint of global $\U(N)$ invariance.

Because free theories are usually conformally invariant (in the absence of mass terms), we can study the partition functions in radial quantization on $\R\times S^3$ and also introduce a state-operator correspondence.\footnote{When studying 't Hooft coupling corrections to the partition functions, one should replace the dilatation operator or any other fugacities that we use by fugacities conjugate to symmetries that remain unbroken. We will not study such loop corrections in this work.} Such partition functions can be computed as the Gaussian path integrals of free gauge theories on $S^1\times S^3$, up to an integral over the holonomies along $S^1$ that projects onto global $\U(N)$ invariants. A detailed discussion of these points can be found in \cite{Sundborg:1999ue,Aharony:2003sx}. Because the theories we study are free, all the nontriviality of the partition functions comes from their dependence on $N$. So they are the perfect observables to study finite-$N$ holography in a controlled yet non-BPS setting.

\paragraph{Definitions of partition functions.} In this section, we will review the counting of operators in free $\cN=4$ sYM. The $\cN=1$ case will be handled in the next section. In what follows, let $\al,\dal=0,1$ denote the left- and right-handed spinor indices of the Euclidean isometry group $\SO(4)$, and let $I=1,2,3,4$ denote the index of the fundamental representation of $\SU(4)_\text{R}$. 

Over the course of this work, we will study three kinds of partition functions. The first is the thermal partition function, which corresponds to imposing antiperiodic boundary conditions on the fermions. We define it as a trace over the full Hilbert space $\cH_N$,
\begin{multline}\label{Zth4}
Z_N^\text{th}(\fx,\fy,\fz) = \Tr_{\cH_N}\!x_1^{\frac{R_1+R_2+R_3}{2}}x_2^{\frac{R_1-R_2-R_3}{2}}x_3^{\frac{R_2-R_3-R_1}{2}}x_4^{\frac{R_3-R_1-R_2}{2}}\\\times y_1^{\frac{\Delta}{2}+j_2}y_2^{\frac{\Delta}{2}-j_2}z_1^{-\frac{\Delta}{2}-j_1}z_2^{-\frac{\Delta}{2}+j_1}\,,
\end{multline}
where $R_1,R_2,R_3$ generate a Cartan subalgebra of the $\mathfrak{su}(4)$ R-symmetry algebra, and $\Delta,j_1,j_2$ are the conformal dimension and left-, right-moving angular momenta respectively. This is a function of eight fugacities that we collect into vectors $\fx = (x_1,x_2,x_3,x_4)$, $\fy=(y_1,y_2)$ and $\fz=(z_1,z_2)$ for convenience. We will often index them as $x_I$, $y_\dal$ and $z_\al$. Thermal partition functions with chemical potentials for the
$\SU(4)_R$ Cartan were studied in \cite{Yamada:2006rx,Harmark:2006di}, while \cite{Harmark:2007px} included chemical potentials for both angular momenta and all three R-charges. The parametrization we use is a symmetric repackaging of their definitions \cite{Bianchi:2006ti,Gromov:2010vb}, see also appendix C.2 of \cite{Harmark:2021qma} for a helpful review.

The second type of partition function we consider is the graded partition function, which corresponds to imposing periodic boundary conditions on the fermions,
\begin{multline}\label{Zgr4}
Z_N^\text{gr}(\fx,\fy,\fz) = \Tr_{\cH_N}(-1)^Fx_1^{\frac{R_1+R_2+R_3}{2}}x_2^{\frac{R_1-R_2-R_3}{2}}x_3^{\frac{R_2-R_3-R_1}{2}}x_4^{\frac{R_3-R_1-R_2}{2}}\\\times y_1^{\frac{\Delta}{2}+j_2}y_2^{\frac{\Delta}{2}-j_2}z_1^{-\frac{\Delta}{2}-j_1}z_2^{-\frac{\Delta}{2}+j_1}\,.
\end{multline}
This can often display better high temperature properties due to the cancellations between bosons and fermions induced by the Grassmann parity signs $(-1)^F$ \cite{Cherman:2020zea}. We stress that regardless of the presence or absence of $(-1)^F$, neither of these is supersymmetry-protected.

The third kind of partition function is a twisted graded partition function in which, in addition to $(-1)^F$, we also grade by a $\Z_2$ in the center of the $\SU(4)$ R-symmetry group,
\begin{multline}\label{Ztw4}
Z_N^\text{tw}(\fx,\fy,\fz) = \Tr_{\cH_N}\!\bau_R\,(-1)^Fx_1^{\frac{R_1+R_2+R_3}{2}}x_2^{\frac{R_1-R_2-R_3}{2}}x_3^{\frac{R_2-R_3-R_1}{2}}x_4^{\frac{R_3-R_1-R_2}{2}}\\\times y_1^{\frac{\Delta}{2}+j_2}y_2^{\frac{\Delta}{2}-j_2}z_1^{-\frac{\Delta}{2}-j_1}z_2^{-\frac{\Delta}{2}+j_1}\,.
\end{multline}
For our twist, we choose the $\Z_2$ generator (with the convention that $R_1,R_2,R_3\in\frac12\Z$)
\be
\bau_R = (-1)^{2(R_1+R_2+R_3)}\,.
\ee
This corresponds to the central element $-\mathbbm{1}_4\in\SU(4)$. At present, this does not define an independent partition function. This is because the parity and R-charges are related,
\be
(-1)^F = (-1)^{2(R_1+R_2+R_3)}\,,
\ee
hence the thermal and twisted graded partition functions coincide,
\be\label{Ztwtoth}
Z_N^\text{th}(\fx,\fy,\fz) = Z_N^\text{tw}(\fx,\fy,\fz)\,.
\ee
So we will mainly present our results for the thermal and graded cases. However, the giant graviton expansion constructed in this paper will involve other twisted graded partition functions.

The graded and twisted graded partition functions are also related to each other. This is because the twist $\bau_R$ is an element of the Cartan subalgebra of $\SU(4)_R$. So grading by the twist is already included within grading by the Cartan generators. To see this, note that
\begin{multline}
    (-1)^{2(R_1+R_2+R_3)}x_1^{\frac{R_1+R_2+R_3}{2}}x_2^{\frac{R_1-R_2-R_3}{2}}x_3^{\frac{R_2-R_3-R_1}{2}}x_4^{\frac{R_3-R_1-R_2}{2}}\\= (x_1\e^{3\pi i})^{\frac{R_1+R_2+R_3}{2}}(x_2\e^{-\pi i})^{\frac{R_1-R_2-R_3}{2}}(x_3\e^{-\pi i})^{\frac{R_2-R_3-R_1}{2}}(x_4\e^{-\pi i})^{\frac{R_3-R_1-R_2}{2}}\,. 
\end{multline}
This motivates the definition of an involution on the $x_I$'s,
\be
\bau_R(\fx) = (x_1\e^{3\pi i},x_2\e^{-\pi i},x_3\e^{-\pi i},x_4\e^{-\pi i}) \, ,
\ee
which in particular preserves the product $x_1x_2x_3x_4$. In terms of this, we see that
\be\label{Ztwtogr}
Z^\text{tw}_N(\fx,\fy,\fz) = Z^\text{gr}_N(\bau_R(\fx),\fy,\fz)\,.
\ee
Additionally, combining this with \eqref{Ztwtoth} shows that the thermal and graded partition functions are related to each other. This is a specialization of the continuous R-symmetry interpolation between thermal and graded partition functions studied in \cite{Cherman:2020zea}. In later sections, we will first derive a GGE for the graded partition function, then translate it to a GGE for the twisted graded partition function using \eqref{Ztwtogr}, and thereafter to one for the thermal partition function using \eqref{Ztwtoth}.

\paragraph{Fugacity constraints.} We have expressed these partition functions as functions of eight fugacities, which a priori would have corresponded to the Cartan subalgebra of a $\mathfrak{gl}(4|4,\C)$ (complexified) superconformal symmetry. But since the true superconformal algebra of $\cN=4$ sYM is $\mathfrak{psl}(4|4,\C)$ which has a six-dimensional Cartan subalgebra, the eight fugacities need to be cut down to six. To do this, first impose the constraint\footnote{We will later obtain this constraint as the requirement that the volume form on supertwistor space have fugacity $1$.}
\be\label{xyzeq}
x_1x_2x_3x_4=y_1y_2z_1z_2\,.
\ee 
This brings us down from $\gl(4|4,\C)$ to $\sl(4|4,\C)$. To further reduce this to $\mathfrak{psl}(4|4,\C)$, notice that our partition functions are invariant under a simultaneous rescaling of all fugacities, so the fugacities can be taken up to projective rescalings:
\be
[x_1:x_2:x_3:x_4:y_1:y_2:z_1:z_2] \in \mathbb{P}^7\,. \label{eq:proj4}
\ee
Here, we are treating the fugacities to be complex. In practical calculations, we will gauge-fix this redundancy and restrict the fugacities to a suitable real slice on which the partition sums converge.

The simplest examples of ``gauge-fixed'' partition functions will be the unrefined thermal and graded partition functions. They are obtained by setting
\be\label{tspec}
x_1=x_2=x_3=x_4 = t\,,\qquad y_1=y_2=t^2\,,\qquad z_1=z_2=1\,,
\ee
where $t$ is a fugacity conjugate to dilatations. With this specialization, we obtain
\be
\begin{split}
    Z_N^\text{th}(t) &= \Tr_{\cH_N}\!t^{2\Delta}\,,\\
     Z_N^\text{gr}(t) &= \Tr_{\cH_N}(-1)^Ft^{2\Delta}\,. \\
    Z_N^\text{tw}(t) &= \Tr_{\cH_N}\!\bau_R\,(-1)^Ft^{2\Delta}
\end{split}
\ee
These will count local operators in free $\cN=4$ sYM graded purely by their conformal dimension. One can typically take $t$ to be complex as long as $|t|<1$ so as to ensure convergence. A standard choice is to take $t\in(0,1)$.

\paragraph{Single-letter partition functions.} The field content of $\cN=4$ sYM consists of a gauge field $A_{\al\dal}$, four left-handed and four right-handed Weyl fermions $\Psi^I_\al$, $\til\Psi_{I\dal}$, and six scalars $\Phi_{IJ}=-\Phi_{JI}$. In the free theory, they obey the zero rest mass free field equations
\be
\begin{split}
    &\p^{\al\dal}\til F_{\dal\db} = 0\,,\\
    &\p^{\al\dal}\til\Psi_{I\dal} = 0\,,\\
    &\p^{\al\dal}\p_{\al\dal}\Phi_{IJ} = 0\,,\\
    &\p^{\al\dal}\Psi^I_\al = 0\,,\\
    &\p^{\al\dal}F_{\al\beta} = 0\,,
\end{split}
\ee
where $\til F_{\dal\db}$ and $F_{\al\beta}$ are the self-dual and anti-self-dual parts of the linearized field strength of $A$. Local operators of free $\cN=4$ sYM are built from $\U(N)$-invariant products of the letters given by $F_{\al\beta}$, $\til F_{\dal\db}$, $\Psi^I_\al$, $\til\Psi_{I\dal}$, $\Phi_{IJ}$ and their $\p_{\al\dal}$-derivatives, modulo these equations of motion. So the first step in counting the local operators is to build partition functions that count these elementary letters.
\begin{table}[t]
    \centering
    \renewcommand{\arraystretch}{1.6}
    \begin{tabular}{@{}|c|c|c|c|c|c|c|c|@{}}
        \hline
        Letter
        & $\hspace{0.3cm}\Delta\hspace{0.3cm}$
        & $j_1$
        & $j_2$
        & $\hspace{0.2cm}R_1\hspace{0.2cm}$
        & $\hspace{0.2cm}R_2\hspace{0.2cm}$
        & $\hspace{0.2cm}R_3\hspace{0.2cm}$
        & Fugacity weight\\
        \hline
        \rule{0pt}{4.5ex}$\widetilde F_{\dot\alpha\dot\beta}$
        & $2$
        & $0$
        & $\widetilde m_{\dot\alpha}+\widetilde m_{\dot\beta}$
        & $0$
        & $0$
        & $0$
        & $\displaystyle
           \frac{y_{\dot\alpha}y_{\dot\beta}}{z_1z_2}$
        \\[3mm]
        $\widetilde\Psi_{I\dot\alpha}$
        & $\displaystyle\frac32$
        & $0$
        & $\widetilde m_{\dot\alpha}$
        & $r_I^1$
        & $r_I^2$
        & $r_I^3$
        & $\displaystyle
           \frac{x_Iy_{\dot\alpha}}{z_1z_2}$
        \\[3mm]
        $\Phi_{IJ}$
        & $1$
        & $0$
        & $0$
        & $r_I^1+r_J^1$
        & $r_I^2+r_J^2$
        & $r_I^3+r_J^3$
        & $\displaystyle
           \frac{x_Ix_J}{z_1z_2}$
        \\[3mm]
        $\Psi^I_{\alpha}$
        & $\displaystyle\frac32$
        & $m_\alpha$
        & $0$
        & $-r_I^1$
        & $-r_I^2$
        & $-r_I^3$
        & $\displaystyle
           \frac{y_1y_2}{x_Iz_\alpha}$
        \\[3mm]
        $F_{\alpha\beta}$
        & $2$
        & $m_\alpha+m_\beta$
        & $0$
        & $0$
        & $0$
        & $0$
        & $\displaystyle
           \frac{y_1y_2}{z_\alpha z_\beta}$
        \\[3mm]
        $\partial_{\alpha\dot\alpha}$
        & $1$
        & $m_\alpha$
        & $\widetilde m_{\dot\alpha}$
        & $0$
        & $0$
        & $0$
        & $\displaystyle
           \frac{y_{\dot\alpha}}{z_\alpha}$
        \\[3mm]
        \hline
    \end{tabular}
    \caption{Cartan charges and fugacity assignments for the letters of free
    $\mathcal N=4$ sYM, where
    $m_\alpha=\widetilde m_{\dot\alpha}=(\frac12,-\frac12)$, and
    $\mathbf r_I=(r_I^1,r_I^2,r_I^3)$ are the weights of the
    $\mathbf 4$ of $\SU(4)_R$: $\mathbf r_1=\frac12(1,1,1)$, $\mathbf r_2=\frac12(1,-1,-1)$, $\mathbf r_3=\frac12(-1,1,-1)$ and $\mathbf r_4=\frac12(-1,-1,1)$. The fugacity assignments have been simplified on the support of $x_1x_2x_3x_4=y_1y_2z_1z_2$.}
    \label{fugacities}
\end{table}

Because of the equations of motion, the only independent derivatives are those that are totally symmetrized in the dotted indices and separately in the undotted indices,
\begingroup
\allowdisplaybreaks
\begin{align}\label{letts}
    &\p_{(\al}^{(\dal}\cdots\p_{\beta)}^{\dot\beta}\til F^{\dot\gamma\dot\delta)}\,,\nonumber\\
    &\p_{(\al}^{(\dal}\cdots\p_{\beta)}^{\dot\beta}\til \Psi_I^{\dot\gamma)}\,,\nonumber\\
    &\p_{(\al}^{(\dal}\cdots\p_{\beta)}^{\dot\beta)}\Phi_{IJ}\,,\\
    &\p_{(\al}^{(\dal}\cdots\p_{\beta}^{\dot\beta)}\Psi^I_{\gamma)}\,,\nonumber\\
    &\p_{(\al}^{(\dal}\cdots\p_{\beta}^{\dot\beta)}F_{\gamma\delta)}\,.\nonumber
\end{align}
\endgroup
To build a Hilbert series counting these, we use the fugacity assignments of the elementary fields listed in table \ref{fugacities}. These correspond to a standard convenient choice of the Cartan subalgebra of the superconformal algebra, and have been simplified on the support of the constraint \eqref{xyzeq}.

With these assignments, we can build single-letter partition functions counting each of the series of derivatives in the five lines of \eqref{letts}. In the order in which they appear, the unsigned Hilbert series of the letters listed in \eqref{letts} are given by
\begingroup
\allowdisplaybreaks
\begin{align}
    f_{\til F} &= \frac{1}{z_1z_2}\sum_{n=0}^\infty\sum_{k=0}^{n+2}\sum_{\ell=0}^n\frac{y_1^ky_2^{n+2-k}}{z_1^\ell z_2^{n-\ell}}\,,\\
    f_{\til\Psi} &= \frac{x_1+x_2+x_3+x_4}{z_1z_2}\sum_{n=0}^\infty\sum_{k=0}^{n+1}\sum_{\ell=0}^n\frac{y_1^ky_2^{n+1-k}}{z_1^\ell z_2^{n-\ell}}\,,\\
    f_{\Phi} &= \frac{x_1x_2+x_1x_3+x_1x_4+x_2x_3+x_2x_4+x_3x_4}{z_1z_2}\sum_{n=0}^\infty\sum_{k=0}^{n}\sum_{\ell=0}^n\frac{y_1^ky_2^{n-k}}{z_1^\ell z_2^{n-\ell}}\,,\\
    f_{\Psi} &= y_1y_2\left(\frac{1}{x_1}+\frac{1}{x_2}+\frac{1}{x_3}+\frac{1}{x_4}\right)\sum_{n=0}^\infty\sum_{k=0}^{n}\sum_{\ell=0}^{n+1}\frac{y_1^ky_2^{n-k}}{z_1^\ell z_2^{n+1-\ell}}\,,\\
    f_{F} &= y_1y_2\sum_{n=0}^\infty\sum_{k=0}^{n}\sum_{\ell=0}^{n+2}\frac{y_1^ky_2^{n-k}}{z_1^\ell z_2^{n+2-\ell}}\,.
\end{align}
\endgroup
Each of these may be formally summed up into rational expressions, but eventually we will only need their signed total. Upon simplification on the support of \eqref{xyzeq}, this produces the single-letter graded partition function
\be\label{fN=4}
    f = f_{\til F} - f_{\til\Psi} + f_\Phi - f_\Psi + f_F = 1 - \frac{z_1P(z_1)-z_2P(z_2)}{z_1-z_2}\,,
\ee
where we have introduced the rational function
\be\label{Pprod}
P(z) = \frac{(1-z^{-1}x_1)(1-z^{-1}x_2)(1-z^{-1}x_3)(1-z^{-1}x_4)}{(1-z^{-1}y_1)(1-z^{-1}y_2)}\,.
\ee
The apparent singularity at $z_1=z_2$ is spurious. As a further consistency check, setting
\be
\begin{split}
    &x_1=\frac{1}{xyz}\,,\quad x_2=x\,,\quad x_3=y\,,\quad x_4=z\,,\\
    &y_1=p\,,\quad y_2=q\,,\quad z_1=1\,,\quad z_2=\frac{1}{pq}\,,\quad xyz=pq
\end{split}
\ee
restricts \eqref{Zgr4} to the $1/16$-BPS index \eqref{16index} and recovers the single-letter index \eqref{eq:116f}. 

One is also interested in the ``thermal'' single-letter partition function which is the unsigned sum
\be
f^\text{th} = f_{\til F} + f_{\til\Psi} + f_\Phi + f_\Psi + f_F\,.
\ee
But one easily finds that this is related to the signed sum $f$ by
\be\label{fthfm}
f^\text{th}(\fx,\fy,\fz) = f(-\fx,\fy,\fz)\,.
\ee
In what follows, all our results will be expressed in terms of the single-letter graded partition function $f$. A minimal six-fugacity parametrization of these refined single-letter partition functions may be found in \cite{Harmark:2007px}.

The unrefined expressions are also illuminating. Using the specialization \eqref{tspec}, we find the unrefined single-letter partition functions
\be\label{fspec}
f(t) = \frac{6t^2+2t^3}{(1+t)^3}\,,\qquad f^\text{th}(t) = \frac{6t^2-2t^3}{(1-t)^3}\,.
\ee
In particular, we see that
\be
f^\text{th}(t) = f(-t)\,.
\ee
These unrefined results were originally obtained in \cite{Sundborg:1999ue,Aharony:2003sx}.

\paragraph{Projection onto $\U(N)$ invariants.} We can obtain the thermal and graded partition functions by first constructing the multi-particle versions of these single-letter partition functions and then projecting onto $\U(N)$-invariants using a Molien-Weyl integral. This is the same technique one uses to compute indices, as reviewed in section \ref{ssec:standardGGE}.

Let $f_\text{b}(\fx,\fy,\fz)$ and $f_\text{f}(\fx,\fy,\fz)$ respectively denote the unsigned single-letter Hilbert series of bosonic and fermionic letters,
\be
f_\text{b} = \frac{f^\text{th}+f}{2}\,,\qquad f_\text{f} = \frac{f^\text{th}-f}{2}\,.
\ee
It is well-known that in terms of $f_\text{b}$, $f_\text{f}$, the thermal and graded partition functions are given by the unitary matrix integrals \cite{Aharony:2003sx,Kristensson:2020nly}
\begingroup
\allowdisplaybreaks
\begin{align}
    Z_N^\text{th} &= \int_{\U(N)}\d U\;\exp\bigg(\sum_{n=1}^\infty\frac{1}{n}\,\big(f_\text{b}(\fx^n,\fy^n,\fz^n)+(-1)^{n+1}f_\text{f}(\fx^n,\fy^n,\fz^n)\big)\,\chi_\text{adj}(U^n)\bigg)\,,\\
    Z_N^\text{gr} &= \int_{\U(N)}\d U\;\exp\bigg(\sum_{n=1}^\infty\frac{1}{n}\,\big(f_\text{b}(\fx^n,\fy^n,\fz^n)-f_\text{f}(\fx^n,\fy^n,\fz^n)\big)\,\chi_\text{adj}(U^n)\bigg)\,,
\end{align}
\endgroup
where $f_\text{b}(\fx^n,\fy^n,\fz^n)\equiv f_\text{b}(x_1^n,x_2^n,x_3^n,x_4^n,y_1^n,y_2^n,z_1^n,z_2^n)$, etc., and $\chi_\text{adj}(U)=\Tr(U)\Tr(U^{-1})$ is the character of the adjoint representation of $\U(N)$.

But because of the simple relation \eqref{fthfm}, we find that both of these can be written purely in terms of the graded sum $f(\fx,\fy,\fz)$,
\begingroup
\allowdisplaybreaks
\begin{align}
    Z_N^\text{th} &= \int_{\U(N)}\d U\;\exp\bigg(\sum_{n=1}^\infty\frac{f((-\fx)^n,\fy^n,\fz^n)}{n}\,\Tr(U^n)\Tr(U^{-n})\bigg)\,,\label{Zth4result}\\
    Z_N^\text{gr} &= \int_{\U(N)}\d U\;\exp\bigg(\sum_{n=1}^\infty\frac{f(\fx^n,\fy^n,\fz^n)}{n}\,\Tr(U^n)\Tr(U^{-n})\bigg)\,.\label{Zgr4result}
\end{align}
\endgroup
where $f((-\fx)^n,\fy^n,\fz^n)\equiv f((-x_1)^n,(-x_2)^n,(-x_3)^n,(-x_4)^n,y_1^n,y_2^n,z_1^n,z_2^n)$. In particular, we recover the relation implied by the R-symmetry twist discussed above,
\be
Z_N^\text{th}(\fx,\fy,\fz) = Z_N^\text{gr}(-\fx,\fy,\fz)\,,
\ee
that holds for the partition functions of the free theory. This relation is very convenient, as it is often easier to work with the graded partition function due to its similarity to BPS indices. It also tells us that the graded partition function counts \emph{all} operators and does not exhibit any cancellations order by order in the fugacities. 

For explicit evaluation, it is helpful to focus on the unrefined partition functions. In this specialization, one obtains the relation
\be
Z_N^\text{th}(t) = Z_N^\text{gr}(-t)\,.
\ee
In the limit $N\to\infty$, the matrix integrals can be evaluated in the saddle point approximation to yield \cite{Aharony:2003sx}
\be
Z_\infty^\text{gr}(t) = \prod_{n=1}^\infty\frac{1}{1-f(t^n)}\,,\qquad Z_\infty^\text{th}(t) = \prod_{n=1}^\infty\frac{1}{1-f((-t)^n)}\,.
\ee
For instance, the large-$N$ series expansion of the graded partition function is found to be
\be\label{Zinf4}
\begin{split}
    Z_\infty^\text{gr}(t) &= 1+6 t^2-16 t^3+72 t^4-240 t^5+928t^6-3456 t^7+12954 t^8\\
    &\hspace{2.5cm}-48112t^9+179712t^{10}-671088
   t^{11}+2504576
   t^{12} + \mathrm{O}(t^{13})\,.
\end{split}
\ee
This gives a signed count of all single-trace and multi-trace operators, including non-BPS ones in the free limit. The leading $t$ dependent term $6t^2$ counts the $6$ dimension $1$ operators $\Tr\Phi_{IJ}$. The next term $-16t^3$ counts the $16$ dimension $3/2$ operators $\Tr\Psi^I_\al$, $\Tr\til\Psi_{I\dal}$. The next term $72t^4$ counts dimension $2$ operators, including the $24$ operators $\Tr(\p_{\al\dal}\Phi_{IJ})$, the $6$ operators $\Tr F_{\al\beta}, \Tr F_{\dal\dot\beta}$, the $21$ operators $\Tr(\Phi_{IJ}\Phi_{KL})$, as well as the $21$ double-trace operators $\Tr(\Phi_{IJ})\Tr(\Phi_{KL})$. In particular, it counts the famously non-BPS Konishi operator $\Tr(\Phi^{IJ}\Phi_{IJ})$, which is not visible to any index.


\subsection{Free $\cN=1$ sYM}
\label{ssec:N=1}

To exhibit the general applicability of our weak-coupling techniques, we will also study the partition functions of $\cN=1$ super Yang-Mills. This is the theory of a single $\cN=1$ vector multiplet with a $\U(N)$ gauge group. Its elementary fields are a gauge field $A_{\al\dal}$ and a pair of Weyl fermions of either chirality $\Psi_\al,\til\Psi_\dal$. Much of the technology of the previous section can be imported verbatim to develop its thermal and graded partition functions in the free limit. The partition functions of this theory at large $N$ have previously been studied as the $N_f=1$ specialization of adjoint QCD \cite{Basar:2014jua}.

In the presence of interactions, the classical $\U(1)$ R-symmetry becomes anomalous in the quantum theory. But in the free limit, we can still use it to grade states. Denoting its generator by $R$, we can define the thermal and graded partition functions
\begin{align}
Z_N^\text{th}(\fx,\fy,\fz) &= \Tr_{\cH_N}\!x_1^{-\frac{3R}{4}}x_2^{\frac{R}{4}}y_1^{\frac{\Delta}{2}+j_2}y_2^{\frac{\Delta}{2}-j_2}z_1^{-\frac{\Delta}{2}-j_1}z_2^{-\frac{\Delta}{2}+j_1}\,,\label{Zth1}\\
Z_N^\text{gr}(\fx,\fy,\fz) &= \Tr_{\cH_N}(-1)^Fx_1^{-\frac{3R}{4}}x_2^{\frac{R}{4}}y_1^{\frac{\Delta}{2}+j_2}y_2^{\frac{\Delta}{2}-j_2}z_1^{-\frac{\Delta}{2}-j_1}z_2^{-\frac{\Delta}{2}+j_1}\,.\label{Zgr1}
\end{align}
These are functions of six fugacities that may be arranged as $\fx =(x_1,x_2)$, $\fy=(y_1,y_2)$ and $\fz=(z_1,z_2)$. We sometimes index these as $x_i$, $y_\dal$, $z_\al$. Furthermore, \eqref{Zth1} and \eqref{Zgr1} are invariant under the projective rescalings
\be
(x_1,x_2,y_1,y_2,z_1,z_2) \sim (\al x_1,\al^3 x_2,\al y_1,\al y_2,\al z_1,\al z_2)\quad \forall\;\al\in\C^\times\,, \label{eq:proj1}
\ee
which cuts down the independent fugacities to five. We also impose the relation
\be\label{xyzeq1}
x_1x_2=y_1y_2z_1z_2\,,
\ee
which cuts them down to four independent fugacities, precisely the amount needed to grade states by the four-dimensional Cartan subalgebra generated by $R,\Delta,j_1,j_2$.

We also introduce the twisted graded partition function
\be\label{Ztw1}
Z_N^\text{tw}(\fx,\fy,\fz) = \Tr_{\cH_N}\!\bau_R\,(-1)^Fx_1^{-\frac{3R}{4}}x_2^{\frac{R}{4}}y_1^{\frac{\Delta}{2}+j_2}y_2^{\frac{\Delta}{2}-j_2}z_1^{-\frac{\Delta}{2}-j_1}z_2^{-\frac{\Delta}{2}+j_1}\,.
\ee
In the $\cN=1$ case, we take the twist $\bau_R$ to be
\be
\bau_R = (-1)^R\,.
\ee
This generates a $\Z_2\subset\U(1)_R$, which in fact remains non-anomalous even in the interacting theory \cite{Shifman:1999mv}. The gauge field has no R-charge, and the gaugino components $\Psi_\al$, $\til\Psi_\dal$ have R-charges $+1$, $-1$ respectively. So $\bau_R$ again coincides with $(-1)^F$, and we find
\be
Z^\text{th}_N(\fx,\fy,\fz) = Z_N^\text{tw}(\fx,\fy,\fz)\,.
\ee
Additionally, defining the involution
\be
\bau_R(\fx) = (x_1\e^{-\pi i},x_2\e^{\pi i})\,,
\ee
and splitting $(-1)^R = \e^{3\pi i R/4}\e^{\pi i R/4}$, we observe that
\be
Z_N^\text{tw}(\fx,\fy,\fz) = Z_N^\text{gr}(\bau_R(\fx),\fy,\fz)\,.
\ee
Again, we will mainly state our results for $Z^\text{gr}_N$.

We can also study the unrefined limits of these partition functions. A natural specialization that respects \eqref{xyzeq1} is given by
\be\label{tspec1}
x_1=t \,, \qquad x_2= t^3 \,,\qquad y_1=y_2=t^2\,,\qquad z_1=z_2=1\,.
\ee
Just as in the $\cN=4$ case, the definitions \eqref{Zth1} and \eqref{Zgr1} reduce to partition functions graded purely by conformal dimension,
\be
\begin{split}
    Z_N^\text{th}(t) &= \Tr_{\cH_N}\!t^{2\Delta}\,,\\
    Z_N^\text{gr}(t) &= \Tr_{\cH_N}(-1)^Ft^{2\Delta}\,,\\
    Z_N^\text{tw}(t) &= \Tr_{\cH_N}\!\bau_R\,(-1)^Ft^{2\Delta}\,.
\end{split}
\ee
We may now compute these by counting letters of the free theory.

\begin{table}[t]
    \centering
    \renewcommand{\arraystretch}{1.6}
    \begin{tabular}{@{}|c|c|c|c|c|c|@{}}
        \hline
        Letter
        & $\hspace{0.3cm}\Delta\hspace{0.3cm}$
        & $j_1$
        & $j_2$
        & $\hspace{0.3cm}R\hspace{0.3cm}$
        & Fugacity weight
        \\
        \hline
        \rule{0pt}{4.5ex}$\widetilde F_{\dot\alpha\dot\beta}$
        & $2$
        & $0$
        & $\widetilde m_{\dot\alpha}+\widetilde m_{\dot\beta}$
        & $0$
        & $\displaystyle
           \frac{y_{\dot\alpha}y_{\dot\beta}}{z_1z_2}$
        \\[3mm]
        $\widetilde\Psi_{\dot\alpha}$
        & $\displaystyle\frac32$
        & $0$
        & $\widetilde m_{\dot\alpha}$
        & $-1$
        & $\displaystyle
           \frac{x_1y_{\dot\alpha}}{z_1z_2}$
        \\[3mm]
        $\Psi_{\alpha}$
        & $\displaystyle\frac32$
        & $m_\alpha$
        & $0$
        & $+1$
        & $\displaystyle
           \frac{y_1y_2}{x_1z_\alpha}$
        \\[3mm]
        $F_{\alpha\beta}$
        & $2$
        & $m_\alpha+m_\beta$
        & $0$
        & $0$
        & $\displaystyle
           \frac{y_1y_2}{z_\alpha z_\beta}$
        \\[3mm]
        $\partial_{\alpha\dot\alpha}$
        & $1$
        & $m_\alpha$
        & $\widetilde m_{\dot\alpha}$
        & $0$
        & $\displaystyle
           \frac{y_{\dot\alpha}}{z_\alpha}$
        \\[3mm]
        \hline
    \end{tabular}
    \caption{Cartan charges and fugacity assignments for the letters of free
    $\mathcal N=1$ sYM, where
    $m_\alpha=\widetilde m_{\dot\alpha}=(\frac12,-\frac12)$.
    The fugacity weights have been simplified on the support of
    $x_1x_2=y_1y_2z_1z_2$.}
    \label{fugacities-N1}
\end{table}

The fugacity assignments of the field content of the $\cN=1$ vector multiplet are listed in table \ref{fugacities-N1}. The free theory equations of motion are
\be
\begin{split}
    &\p^{\al\dal}\til F_{\dal\db} = 0\,,\\
    &\p^{\al\dal}\til\Psi_{\dal} = 0\,,\\
    &\p^{\al\dal}\Psi_\al = 0\,,\\
    &\p^{\al\dal}F_{\al\beta} = 0\,.
\end{split}
\ee
On their support, the independent letters generating the operator content of free $\cN=1$ sYM are given by the derivatives
\be\label{letts1}
\begin{split}
    &\p_{(\al}^{(\dal}\cdots\p_{\beta)}^{\dot\beta}\til F^{\dot\gamma\dot\delta)}\,,\\
    &\p_{(\al}^{(\dal}\cdots\p_{\beta)}^{\dot\beta}\til \Psi^{\dot\gamma)}\,,\\
    &\p_{(\al}^{(\dal}\cdots\p_{\beta}^{\dot\beta)}\Psi_{\gamma)}\,,\\
    &\p_{(\al}^{(\dal}\cdots\p_{\beta}^{\dot\beta)}F_{\gamma\delta)}\,.
\end{split}
\ee
The unsigned Hilbert series counting these letters read
\begingroup
\allowdisplaybreaks
\begin{align}
    f_{\til F}
    &= \frac{1}{z_1z_2}
    \sum_{n=0}^\infty
    \sum_{k=0}^{n+2}
    \sum_{\ell=0}^{n}
    \frac{y_1^k y_2^{n+2-k}}
         {z_1^\ell z_2^{n-\ell}}
    \,,
    \\
    f_{\til\Psi}
    &= \frac{x_1}{z_1z_2}
    \sum_{n=0}^\infty
    \sum_{k=0}^{n+1}
    \sum_{\ell=0}^{n}
    \frac{y_1^k y_2^{n+1-k}}
         {z_1^\ell z_2^{n-\ell}}
    \,,
    \\
    f_{\Psi}
    &= \frac{y_1y_2}{x_1}
    \sum_{n=0}^\infty
    \sum_{k=0}^{n}
    \sum_{\ell=0}^{n+1}
    \frac{y_1^k y_2^{n-k}}
         {z_1^\ell z_2^{n+1-\ell}}
    \,,
    \\
    f_F
    &= y_1y_2
    \sum_{n=0}^\infty
    \sum_{k=0}^{n}
    \sum_{\ell=0}^{n+2}
    \frac{y_1^k y_2^{n-k}}
         {z_1^\ell z_2^{n+2-\ell}}
    \,.
\end{align}
\endgroup
Computing their signed sum and simplifying using \eqref{xyzeq1} gives the single-letter graded partition function
\be\label{fN=1}
f = f_{\til F} - f_{\til\Psi} - f_\Psi + f_F = 1-\frac{z_1Q(z_1)-z_2Q(z_2)}{z_1-z_2}\,,
\ee
expressed in terms of another rational function
\be\label{Qprod}
Q(z) = \frac{(1-z^{-1}x_1)(1-z^{-3}x_2)}{(1-z^{-1}y_1)(1-z^{-1}y_2)}\,.
\ee
Similarly, the thermal analog is given by the unsigned sum
\be
f^\text{th} = f_{\til F} + f_{\til\Psi} + f_\Psi + f_F\,.
\ee
This is found to be expressible in terms of the signed sum as
\be
f^\text{th}(\fx,\fy,\fz) = f(-\fx,\fy,\fz)\,,
\ee
closely mirroring the $\cN=4$ case.

As a result, the graded partition function is given by the matrix integral
\be\label{Zgr1result}
Z_N^\text{gr} = \int_{\U(N)}\d U\;\exp\bigg(\sum_{n=1}^\infty\frac{f(\fx^n,\fy^n,\fz^n)}{n}\,\Tr(U^n)\Tr(U^{-n})\bigg)\,,
\ee
and the thermal partition function can again be obtained by changing the sign of $x_i$,
\be
Z_N^\text{th}(\fx,\fy,\fz) = Z_N^\text{gr}(-\fx,\fy,\fz)\,.
\ee
To illustrate their series expansions, let us again study their unrefined specializations. Substituting \eqref{tspec1} into our single-letter partition functions generates
\be\label{fspec1}
f(t) = -\frac{4t^3+2t^4}{(1-t)(1+t)^3}\,,\qquad f^\text{th}(t) = \frac{4t^3-2t^4}{(1+t)(1-t)^3}\,.
\ee
Once again, notice that $f^\text{th}(t)=f(-t)$. With this data, we can compute the large-$N$ graded partition function,
\be\label{Zinf1}
\begin{split}
    Z_\infty^\text{gr}(t) &= \prod_{n=1}^\infty\frac{1}{1-f(t^n)}\\
    &= 1-4 t^3+6 t^4-12 t^5+28 t^6-72t^7+168 t^8\\
    &\hspace{2.5cm}-364 t^9+828t^{10}-1956 t^{11}+4508t^{12} + \mathrm{O}(t^{13})\,.
\end{split}
\ee
The thermal answer is given by $Z_\infty^\text{th}(t) = Z_\infty^\text{gr}(-t)$, i.e., by making the replacement $t\mapsto-t$ term by term in this expansion. The numbers themselves remain unchanged.

Notice that since this theory contains no dimension 1 scalars, the $t$-dependence starts at cubic order, with the $-4t^3$ counting the dimension $3/2$ operators $\Tr\Psi_\al$, $\Tr\til\Psi_\dal$, the $6t^4$ counting the dimension $2$ operators $\Tr F_{\al\beta},\Tr F_{\dal\dot\beta}$, the $-12t^5$ counting the dimension $5/2$ operators $\Tr\p_{(\al|\dal}\Psi_{|\beta)}$, $\Tr\p_{\al(\dal}\til\Psi_{\dot\beta)}$, and so on. The first double-trace operators show up at order $t^6$: the $28t^6$ counts the $22=8+8+4+1+1$ single-trace operators $\Tr\p_{(\al}^\dal F_{\beta\gamma)}$, $\Tr\p^\al_{(\dal}F_{\dot\beta\dot\gamma)}$, $\Tr\textstyle(\Psi_\al\til\Psi_\dal)$, $\Tr(\Psi_\al\Psi^\al)$ and $\Tr\textstyle(\til\Psi_\dal\til\Psi^\dal)$, along with the $6=4+1+1$ double-trace operators $\Tr(\Psi_\al)\Tr\textstyle(\til\Psi_\dal)$, $\Tr(\Psi_\al)\Tr(\Psi^\al)$ and $\Tr\textstyle(\til\Psi_\dal)\Tr\textstyle(\til\Psi^\dal)$.

\section{Holography at zero 't Hooft coupling}
\label{sec:chiralholography}

We would like to construct giant graviton expansions of the free partition functions of the previous section. But the bulk duals of free gauge theories are famously hard to find. Recently, this problem has been attacked from the viewpoint of \emph{self-dual} gauge theories \cite{Sharma:2025ntb,Jarov:2025qhz,Balisa:2026icv}. These provide alternative zero-coupling limits of gauge theories that can often be embedded in twistor string theories \cite{Witten:2003nn,Bittleston:2024efo}. This fact can be used to engineer their gravity duals.

In this section, we review the twistor string holographic dual of $\cN=4$ self-dual Yang-Mills (sdYM) constructed in \cite{Sharma:2025ntb,Jarov:2025qhz}. We also describe a generalization of this construction to $\cN=1$ self-dual Yang-Mills. For each case, we review its brane construction in twistor string theory, along with the backreaction of said branes. In later sections, we will argue that the partition functions of the self-dual and free theories agree with each other. Thereafter, we will use the gravity duals of the self-dual theories to discover the GGEs of our free partition functions.

\subsection{Self-duality and twistor theory}
\label{ssec:sdreview}

Let us begin by briefly reviewing self-dual gauge theory and its twistor uplift. Consider $\U(N)$ Yang-Mills theory on Euclidean $\R^4$. Its self-dual subsector describes Yang-Mills connections $A$ whose field strength $F=\d A+A^2$ obeys the self-duality equations
\be
*F = F\,.
\ee
Let $\al,\beta=0,1$ and $\dal,\dot\beta=\dot0,\dot1$ be the left- and right-handed 2-spinor indices of $\SL_2(\C)$ that may be raised or lowered using the Levi Civita symbol. We follow the conventions of \cite{Adamo:2017qyl} for spinor contractions and angle and square brackets $\la\lambda\kappa\ra=\lambda^\al\kappa_\al$, $[\mu\rho]=\mu^\dal\rho_\dal$, etc. In spinor form, the action of self-dual Yang-Mills is given by
\be\label{sdym}
S_\text{sdYM} = \int_{\R^4}\d^4x\,\Tr\,(B^{\al\beta}F_{\al\beta})\,,
\ee
where $F_{\al\beta} = \p_{(\al|\dal}A^\dal_{|\beta)} + A_{(\al|\dal}A^\dal_{|\beta)}$ is the spinor equivalent of the anti-self-dual (asd) field strength $\frac12\,(F-*F)$, and $B_{\al\beta}$ is a $\mathfrak{u}(N)$-valued Lagrange multiplier. 

It is helpful to think of sdYM as a \emph{chiral} zero coupling limit of Yang-Mills theory. Deforming the BF-type action in \eqref{sdym} by a $B^2$ interaction results in an action that is perturbatively equivalent to Yang-Mills,
\be
S_\text{CS} = \int_{\R^4}\d^4x\,\Tr\left(B^{\al\beta}F_{\al\beta}+\frac{\sg_\text{YM}^2}{4}B^{\al\beta}B_{\al\beta}\right)\,.
\ee
This is the Chalmers-Siegel action for pure gauge theory \cite{Chalmers:1996rq}. Integrating out the Lagrange multiplier $B_{\al\beta}$ reproduces the standard Yang-Mills action including a theta term with a complex theta angle $\theta = -8\pi^2i/\sg_\YM^2$. 

If one takes an ordinary zero coupling limit $\sg_\YM\to0$ of the standard Yang-Mills action, one gets a non-chiral, abelian gauge theory with gauge group $\U(1)^{N^2}$. On the other hand, taking the zero coupling limit $\sg_\YM\to0$ of the Chalmers-Siegel action gives sdYM, which remains non-abelian. We refer to this as the \emph{self-dual limit}. On trivial gauge bundles, perturbation theory in $\sg_\YM$ around the standard free limit is equivalent to perturbation theory around the self-dual limit.\footnote{This fact played a key role in the development of twistor string formulae for scattering amplitudes \cite{Witten:2003nn,Berkovits:2004hg,Roiban:2004vt,Roiban:2004yf}, CSW recursion relations \cite{Cachazo:2004kj}, twistor action techniques \cite{Mason:2005zm,Boels:2006ir}, and weak coupling proofs of the amplitude/Wilson loop duality \cite{Alday:2007hr,Mason:2010yk,Adamo:2011dq}.} In this work, we will use the same idea to compute the partition functions of super Yang-Mills theories at zeroth order in perturbation theory using their self-dual limits.

The tool that brings forth the hidden integrability of the self-dual limit is \emph{twistor theory} \cite{Penrose:1967wn}. The \emph{twistor space} of four-dimensional flat space is a rank 2 holomorphic vector bundle over the Riemann sphere,
\be
\PT = \Tot(\CO(1)\oplus\CO(1)\to\P^1)\,.
\ee
Let $\lambda_\al=(\lambda_0,\lambda_1)$ be homogeneous coordinates on the $\P^1$ base of this fibration, and let $\mu^\dal=(\mu^{\dot0},\mu^{\dot1})$ be the coordinates along the two $\CO(1)$ fibers. Taken together, they furnish a twistor: $Z^A=(\mu^\dal,\lambda_\al)\in\PT$. Points $x^{\al\dal}$ in complexified flat space are in one-to-one correspondence with \emph{twistor lines} $\P^1_x\subset\PT$ via the incidence relations
\be
\P^1_x\;:\quad\mu^\dal = x^{\al\dal}\lambda_\al\,.
\ee
When the spacetime points are restricted to the Euclidean slice $\R^4\subset\C^4$, the incidence relations exhibit an actual diffeomorphism $\PT\simeq\R^4\times\P^1$ \cite{Woodhouse:1985id}. This allows one to obtain many 4d integrable theories on $\R^4$ by compactification of 6d holomorphic theories on $\PT$ along the $\P^1$ factor of $\R^4\times\P^1$ \cite{Mason:2005zm,Costello:2021bah}.

The twistor uplift of sdYM is given by a holomorphic BF theory,
\be\label{hbf}
S_\text{sdYM} = \int_{\PT}\D^3Z\wedge\Tr\,b\,(\dbar a+a^2)\,.
\ee
Here, $\D^3Z = \frac16\,\veps_{ABCD}\,Z^A\,\d Z^B\,\d Z^C\,\d Z^D$ is a weight $+4$ holomorphic volume form on $\PT$. The field $a$ is a partial connection on a $\GL_N(\C)$ vector bundle $E\to\PT$. We will be interested in computing partition functions of gauge theories on the trivial $\U(N)$ bundle on $\R^4$, which corresponds to working with the trivial bundle $E=\CO_{\PT}^{\oplus N}$. The other field $b$ is a weight $-4$ Lagrange multiplier whose equation of motion imposes the flatness condition $\dbar a+a^2=0$. A theorem of Ward \cite{Ward:1977ta,Ward:1990vs} tells us that if the restriction of $E$ to every twistor line is trivial -- as is the case for $E=\CO_\PT^{\oplus N}$ -- then the on-shell data $a$ descends to give a self-dual connection\footnote{A priori, $A$ is a $\GL_N(\C)$ connection on $\R^4$. It becomes a $\U(N)$ connection if $E$ is equipped with a positive real form compatible with the Euclidean reality conditions on twistor space \cite{Ward:1990vs}.} $A$ on $\R^4$. This is known as the \emph{Penrose-Ward transform}. 

It was later shown by Mason that this reduction also works off-shell. Compactification of the holomorphic BF action \eqref{hbf} along the twistor lines reproduces the BF action of sdYM theory \cite{Mason:2005zm}. This statement is classically true, but can fail in the quantum theory due to gauge anomalies of the holomorphic BF path integral. The anomalies can be cancelled through a combination of Green-Schwarz mechanisms and judicious choices of matter content \cite{Costello:2021bah,Costello:2023vyy}. They can also be cancelled by turning on any amount of supersymmetry, which is the route we will take.

\subsection{$\cN=4$ sdYM}
\label{ssec:ch4}

The self-dual limit of $\cN=4$ super Yang-Mills was first studied by Siegel \cite{Siegel:1992xp,Siegel:1992wd}. We will refer to it as $\cN=4$ sdYM. Upon integrating in the field $B_{\al\beta}$, the $\cN=4$ vector multiplet consists of the gauge theory fields $A_{\al\dal}$, $B_{\al\beta}$, four left-handed Weyl fermions $\Psi^I_\al$, four right-handed Weyl fermions $\til\Psi_{I\dal}$, and six scalars $\Phi_{IJ}=-\Phi_{JI}$, where $I,J$ are indices of the fundamental representation of the $\SU(4)$ R-symmetry. The R-symmetry indices can be raised as $\Phi^{IJ} = \frac12\,\veps^{IJKL}\Phi_{KL}$.

After a suitable rescaling of the fields by powers of $\sg_\YM$, Witten found that the action of $\cN=4$ sYM splits as a sum of the action of $\cN=4$ sdYM and a term containing the non-self-dual interactions \cite{Witten:2003nn},\footnote{This decomposition may be organized using Witten's auxiliary $\U(1)_S$ grading, under which
$(A,\widetilde\Psi,\Phi,\Psi,B)$ carry $S$-charges $(0,1,2,3,4)$, respectively.
The terms in $S^{\cN=4}_{\mathrm{sd}}$ then have total $S$-charge $4$, while those in $I^{\cN=4}_{\mathrm{int}}$ have total $S$-charge $8$. For bookkeeping one may assign $S$-charge $-4$ to $\sg_\YM^2$.}
\be
S_{\cN=4} = S^{\cN=4}_\text{sd} + \frac{\sg_\YM^2}{4}\,I^{\cN=4}_\text{int}\,.
\ee
The self-dual action is given by
\be\label{N=4sd}
S^{\cN=4}_\text{sd} = \int_{\R^4}\d^4x\;\Tr\left(B^{\al\beta}F_{\al\beta} + \til\Psi_{I\dal}D^{\al\dal}\Psi^I_\al - \frac18\,D_{\al\dal}\Phi_{IJ}D^{\al\dal}\Phi^{IJ} - \Phi^{IJ}\til\Psi_{I\dal}\til\Psi_J^\dal\right)\,,
\ee
where $D=\d+A$ is the gauge-covariant derivative. The non-self-dual interactions read
\be\label{intN4}
I^{\cN=4}_\text{int} = \int_{\R^4}\d^4x\;\Tr\left(B^{\al\beta}B_{\al\beta} - \Phi_{IJ}\Psi^I_\al\Psi^{J\al} - \frac{1}{16}\,[\Phi_{IJ},\Phi_{KL}][\Phi^{IJ},\Phi^{KL}]\right)\,.
\ee
The sd action does not contain the quartic interactions as well as half the Yukawa interactions. This fact underlies its helpful simplicity and integrability, and ensures that its partition function is one-loop exact.

The twistor uplift of $\U(N)$ $\cN=4$ sdYM is a $\GL_N(\C)$ holomorphic Chern-Simons theory on $\cN=4$ supertwistor space,
\be\label{N=4tac}
S^{\cN=4}_\text{sd} = \int_{\PT^{3|4}}\D^{3|4}\cZ\;\Tr\left(\frac12\,\cA\,\dbar\cA + \frac13\,\cA^3\right)\,.
\ee
Here, $\cZ^\mathrm{a} = (Z^A\,|\,\eta^I)$ is a supertwistor, defined up to the rescaling $\cZ^\mathrm{a} \sim \al\cZ^\mathrm{a}\;\forall\;\al\in\C^\times$. The superfield $\cA = \cA_0+\cA_1+\cA_2+\cA_3$ is a sum of component fields $\cA_p$ valued in antiholomorphic $(0,p)$-forms on $\PT$ for $p=0,1,2,3$. In the BV formalism, the $(0,1)$-part $\cA_1$ contains the physical fields, the $0$-form part $\cA_0$ contains ghosts of gauge symmetries, while $\cA_2,\cA_3$ contain their respective antifields. 

Each of these fields admits a superfield expansion in the $\eta^I$. The superfield expansion of $\cA_1$ contains the twistor representatives of the fields of the $\cN=4$ vector multiplet,
\be
\cA_1 = a + \til\psi_I\eta^I + \frac12\,\vphi_{IJ}\eta^I\eta^J + \psi^I\eta^3_I + b\,\eta^4\,,
\ee
where $\eta^3_L\equiv\frac16\,\veps_{IJKL}\eta^I\eta^J\eta^K$ and $\eta^4 = \frac{1}{24}\,\veps_{IJKL}\eta^I\eta^J\eta^K\eta^L$. The fields $a,\til\psi_I,\vphi_{IJ},\psi^I,b$ are $(0,1)$-forms on $\PT$ valued in the sheaves $\CO_\PT(n)$ for $n=0,-1,-2,-3,-4$, respectively. Their Penrose transforms give rise to the spacetime fields $A_{\al\dal},\til\Psi_{I\dal},\Phi_{IJ},\Psi^I_\al,B_{\al\beta}$ respectively \cite{Eastwood:1981jy}, and compactification of \eqref{N=4tac} along the $\P^1$ factor of $\PT\simeq\R^4\times\P^1$ reproduces the action \eqref{N=4sd} of $\cN=4$ sdYM \cite{Boels:2006ir}.

\paragraph{B-model construction.} This holomorphic Chern-Simons theory can be engineered as the worldvolume theory of $N$ D5-branes wrapping $\PT$ in a B-model on the Calabi-Yau 7-fold
\be\label{X}
X = \Tot(\CO(-1)^{\oplus4}\to\PT)\,.
\ee
Let $W_I$ be homogeneous coordinates on the fibers of $\CO(-1)^{\oplus4}$. Under $Z^A\mapsto\al Z^A$, they rescale as $W_I\mapsto\al^{-1}W_I$. So $X$ admits the manifestly weightless Calabi-Yau volume form\footnote{We often abuse notation by abbreviating this to just $\D^3Z\,\d^4W$, which is all one needs for going to affine patches where one of the $Z^A$ coordinates is set to $1$. But this more generally valid expression can be obtained by contracting the Euler vector field $\Upsilon=Z^A\p/\p Z^A-W_I\p/\p W_I$ into the non-projective form $\d^4Z\,\d^4W$.}
\be\label{omegaX}
\Omega_X = \D^3Z\,\d^4W - \D^3W\,\d^4Z\,,
\ee
where, similarly to $\D^3Z$, we have defined $\D^3W=\frac16\,\veps^{IJKL}W_I\,\d W_J\,\d W_K\,\d W_L$. The D5-branes wrap the locus $W_I=0$.

In addition to the open strings on the D5-branes, the B-model also contains closed strings. The closed string field theory of the B-model is known as BCOV theory \cite{Bershadsky:1993ta,Bershadsky:1993cx,Costello:2012cy}. Its field content consists of $(p,q)$-forms $\xi_{p,q}$ on $X$ for $p,q=0,\dots,7$ satisfying the off-shell constraint $\p\xi_{p,q} = 0$, i.e.\footnote{It is also common to drop the $p=0$ forms $\xi_{0,q}$ from the spectrum, as they happen to be essentially topological degrees of freedom. This is useful because one often trivializes the constraint $\p\xi_{p,q}=0$ by taking $\xi_{p,q}$ to be in the image of $\p$. We will do this in section \ref{ssec:bcov4}.}
\begin{align}
\ker\partial \subset \bigoplus_{p,q=0}^{7}\Omega^{p,q}(X) \, .
\end{align}

To construct a holographic dual of $\cN=4$ sdYM, one simply backreacts by the $N$ D5-branes. This procedure replaces the D5-branes by their 7-form flux which obeys\footnote{The full BCOV equations of motion are nonlinear, but the nonlinear term vanishes for the flux constructed below. We give more detail on the nonlinear eoms in the next section for $\cN=1$ sdYM.}
\be
\dbar\xi_{4,3} = N\,\bar\delta^4(W)\,\d^4W\,.
\ee
The source on the right hand side arises from the backreacting branes at $W_I=0$.

We can write an expression for the flux that solves this in open patches. In a patch $A\cdot Z\neq0$ for some reference twistor $A_A$, it is given by
\be\label{xi43}
\xi_{4,3} = \frac{N\,\D^3\br W}{\|W\|^8}\wedge\left(\d^4W - \D^3W\wedge\d\log(A\cdot Z)\right)\,.
\ee
Here, $\br W^I$ are the hermitian conjugates of $W_I$ with respect to the flat metric that preserves the $\SU(4)$ R-symmetry, and $\D^3\br W = \frac16\,\veps_{IJKL}\,\br W^I\d\br W^J\d\br W^K\d\br W^L$. The factor of $\D^3\br W/\|W\|^8$ is the Bochner-Martinelli kernel for the dbar operator on the $\C^4$ fibers of $\CO(-1)^{\oplus 4}$. This is accompanied by a factor of $\d^4W - \D^3W\wedge\d\log(A\cdot Z)$, which invokes the choice of patch so as to ensure that $\xi_{4,3}$ contracts to zero against the Euler vector field $\Upsilon = Z^A\p/\p Z^A-W_I\p/\p W_I$. 

The near-horizon limit remains trivial in the B-model, just as in twisted holography \cite{Costello:2018zrm}. So this procedure allows us to conjecture that $\U(N)$ $\cN=4$ sdYM is holographically dual to the B-model on $X$ minus its zero section equipped with this backreaction. The zero section that has been removed is precisely the copy of $\PT$ sitting at $W_I=0$ where the backreaction is singular. More details of chiral holography may be found in  \cite{Sharma:2025ntb,Jarov:2025qhz}. We will also describe the $\cN=1$ case in slightly greater detail in the next section, building on the D5-brane realization discussed in \cite{Popov:2004nk, Bittleston:2024efo} and giving an explicit expression for the backreaction. The reader interested in the $\cN=4$ case may skip it on first reading.

\subsection{$\cN=1$ sdYM}
\label{ssec:ch1}

Four-dimensional $\cN=1$ super Yang-Mills is the theory of a single $\cN=1$ vector multiplet. In the Chalmers-Siegel formalism of \cite{Chalmers:1996rq}, it can be written in terms of the gauge theory fields $A_{\al\dal},B_{\al\beta}$ along with a pair of Weyl fermions $\Psi_\al,\til\Psi_\dal$. Just as with the $\cN=4$ case, we can expand $\cN=1$ sYM perturbatively around its self-dual sector \cite{Chalmers:1996rq,Boels:2006ir},
\be
S_{\cN=1} = S^{\cN=1}_\text{sd} + \frac{\sg_\YM^2}{4}\,I^{\cN=1}_\text{int}\,.
\ee
In this case, the self-dual action consists purely of the covariantized kinetic terms,
\be\label{N=1sd}
S^{\cN=1}_\text{sd} = \int_{\R^4}\d^4x\;\Tr\big(B^{\al\beta}F_{\al\beta} + \til\Psi_{\dal}D^{\al\dal}\Psi_\al\big)\,,
\ee
while the non-self-dual interactions now amount to just the $B^2$ deformation
\be\label{intN1}
I^{\cN=1}_\text{int} = \int_{\R^4}\d^4x\;\Tr\,(B^{\al\beta}B_{\al\beta})\,.
\ee
The spacetime action \eqref{N=1sd} can again be uplifted to a twistor action.

We denote the supertwistor space on which its twistor uplift lives as $\PT^{3|2}$. This is obtained by starting with weighted projective superspace $\P(1,1,1,1|1,3)$ with homogeneous coordinates $\cZ^\mathrm{a} = (Z^A\,|\,\eta^i) = (\mu^\dal,\lambda_\al\,|\,\eta^i)$, $i=1,2$, and removing the locus $\{\lambda_\al=0\}$. The Grassmann coordinates $\eta^1,\eta^2$ now have weights $1,3$ respectively, so that $\PT^{3|2}$ admits the weightless volume form $\D^{3|2}\cZ = \D^3Z\,\d^2\eta$. The twistor action is now given by a holomorphic super Chern-Simons theory on $\PT^{3|2}$,
\be\label{N=1tac}
S^{\cN=1}_\text{sd} = \int_{\PT^{3|2}}\D^{3|2}\cZ\;\Tr\left(\frac12\,\cA\,\dbar\cA + \frac13\,\cA^3\right)\,.
\ee
As in the $\cN=4$ sdYM case, the superfield $\mathcal A=\mathcal A_0+\mathcal A_1+\mathcal A_2+\mathcal A_3$ has components $\mathcal A_p$ that are $(0,p)$-forms on $\PT$.
The component $\mathcal A_1$ contains the physical fields, $\mathcal A_0$ contains the ghost fields, and $\mathcal A_2,\mathcal A_3$ contain their respective antifields.

The expansion of $\mathcal A_1$ in the Grassmann coordinates $\eta^I$ is
\be\label{N=1A1}
\cA_1 = a + \til\psi\,\eta^1 + \psi\,\eta^2 + b\,\eta^1\eta^2\,.
\ee
The fields $a,\widetilde\psi,\psi,b$ are $(0,1)$-forms on $\PT$ valued in $\CO_\PT(n)$ for $n=0,-1,-3,-4$, respectively. Expanding the Chern-Simons functional and performing the Grassmann integrals, the action for the physical fields is found to be
\be
\int_{\PT}\D^3Z\wedge\Tr\big(b\,f_a + \til\psi\,\dbar_a\psi\big)\,,
\ee
where $\dbar_a = \dbar+a$ is Ward's partial connection, and $f_a=\dbar a+a^2$ is its curvature. The $\P^1$ compactification of this theory reproduces the $\cN=1$ sdYM action \eqref{N=1sd}. The twistor space fields $a,\til\psi,\psi,b$ Penrose transform to the spacetime fields $A_{\al\dal},\til\Psi_\dal,\Psi_\al,B_{\al\beta}$ respectively.

\paragraph{B-model construction.} By virtue of being a Chern-Simons type theory, we can again engineer \eqref{N=1tac} through branes in the B-model. The relevant twistor string construction has been described in \cite{Popov:2004nk} in the past, and was recently revisited in \cite{Bittleston:2024efo} in the context of celestial holography. One starts with the B-model on the Calabi-Yau 5-fold
\be\label{M}
M = \Tot(\CO(-1)\oplus\CO(-3)\to\PT)\,.
\ee
Let $(Z^A,w_i)$ be homogeneous coordinates on its base and fibers respectively and wrap a stack of $N$ D5-branes along the copy of $\PT$ given by the zero section $w_i = 0$. For the trivial gauge bundle $\CO_{\PT}^{\oplus N}$, the space of open string fields on the D5 worldvolume is given by \cite{Katz:2002gh}
\be
\bigoplus_{p,q}\Omega^{0,p}(\PT,\wedge^q\cN_{\PT/M}\otimes\mathrm{End}\,\CO_{\PT}^{\oplus N})\,,
\ee
where $\cN_{\PT/M}$ is the normal bundle of $\PT$ in $M$. We can identify $\eta^1\equiv\p_{w_1},\eta^2\equiv\p_{w_2}$ as a frame of the normal bundle, and the superfield expansion in $\eta^i$ as a component expansion of fields valued in the wedge algebra $\wedge^*\cN_{\PT/M}$. Thus, we reproduce the field content \eqref{N=1A1} of the $\cN=1$ vector multiplet. Their Chern-Simons action functional can also be reproduced through the open string field theory of B-model branes \cite{Witten:1992fb}.

To construct a holographic dual of $\cN=1$ sdYM, we simply backreact by these D5-branes. The backreaction generates a closed string flux, so let us briefly review the closed string B-model on $M$. The closed string field theory on $M$ is five-dimensional BCOV theory, which also makes an appearance in twisted holography for AdS$_3\times S^3$ \cite{Costello:2020jbh,Fernandez:2024tue}.

The field content of this theory consists of all $\p$-closed $(p,q)$-forms on $M$,
\be
\text{ker}\,\p\subset\bigoplus_{p,q=0}^5\Omega^{p,q}(M)\,.
\ee
Denote the field valued in $\Omega^{p,q}(M)$ by $\xi_{p,q}$. Using the Calabi-Yau volume form on $M$,
\be
\Omega_M = \D^3Z\;\d^2w-(w_1\d w_2-3\,w_2\d w_1)\,\d^4Z\,,
\ee
we can dualize $\xi_{p,q}$ into $(0,q)$-forms $\beta_{5-p,q}$ valued in polyvectors of degree $(5-p,0)$,
\be
\xi_{p,q} = \beta_{5-p,q}\,\lrcorner\,\Omega_M\,,\qquad\beta_{p,q}\in\Omega^{0,q}(M,\wedge^p T_M)\,.
\ee
It is convenient to collect the field content in a ``superfield''
\be
\boldsymbol{\beta} = \sum_{p,q}\beta_{p,q}\,.
\ee
The classical field equation obeyed by the closed string fields is the Maurer-Cartan equation
\be
\dbar\boldsymbol{\beta} + \frac12\,[\boldsymbol{\beta},\boldsymbol{\beta}] = 0\,,
\ee
where $[-,-]$ denotes the Schouten-Nijenhuis bracket.\footnote{The Schouten-Nijenhuis bracket combines the wedge product on the antiholomorphic form factors with the graded extension of the Lie bracket to polyvector fields. A local coordinate expression for this bracket may be found in equation (3.12) of \cite{Costello:2018zrm}.} It is possible to write an action functional for this field equation \cite{Bershadsky:1993cx,Costello:2012cy,Costello:2019jsy}, but it happens to be slightly nonlocal due to the off-shell constraint $\p\xi_{p,q}=0$ and will not be needed in what follows.

The presence of the $N$ D5-branes wrapping $\PT$ adds a source to this field equation,
\be
\dbar\boldsymbol{\beta} + \frac12\,[\boldsymbol{\beta},\boldsymbol{\beta}] = N\delta_{\PT}\,.
\ee
We can solve this using the Bochner-Martinelli kernel. The resulting flux is best expressed in terms of the $(p,q)$-forms $\xi_{p,q}$,
\be\label{xi21}
\xi_{2,1} = \frac{N\,\|\lambda\|^4\,\D\br w}{(|w_1|^2+\|\lambda\|^4|w_2|^2)^2}\wedge(\d^2w-\D w\wedge\d\log(A\cdot Z))\,,
\ee
in an open patch $A\cdot Z\neq0$ for some reference twistor $A_A$. Here, $\br w^i$ is the hermitian conjugate of $w_i$, and
\be
\begin{split}
    \D w &\vcentcolon= w_1\d w_2-3\,w_2\d w_1\,,\\
    \D\br w &\vcentcolon= \br w^1\d\br w^2-\br w^2\d\br w^1 + \br w^1\br w^2\,\dbar\log\|\lambda\|^4\,.
\end{split}
\ee
We are being imprecise about numerical factors like $1/2\pi i$, which will not affect our analysis. As always in holography, the crucial point is that the flux is proportional to $N$. 

Note that $\|\lambda\|^2\equiv|\lambda_0|^2+|\lambda_1|^2\neq0$ because $[\lambda_0:\lambda_1]$ are homogeneous coordinates on the twistor $\P^1$ over which both $\PT$ and the larger space $M$ fiber. So $\xi_{2,1}$ is singular precisely along the copy of $\PT$ situated at $w_1=w_2=0$. Consequently, we conjecture that $\U(N)$ $\cN=1$ sdYM on $\R^4$ is holographically dual to the B-model on $M-\PT$ in the presence of the backreaction \eqref{xi21}.


\section{Partition functions of self-dual gauge theories}
\label{sec:sdpartition}

The non-BPS graded partition functions reviewed in section \ref{sec:partition} do not admit a holographic dual in the supergravity limit of AdS/CFT. This is because they were computed in free gauge theories, while the supergravity regime is dual to strongly coupled gauge theories. Fortunately, on trivial gauge bundles, the partition functions of the free limits of many 4d gauge theories coincide with the partition functions of their self-dual limits that do admit holographic descriptions. Over the course of this section and the next, we will establish this fact for $\cN=4$ and $\cN=1$ sYM by direct calculation. We expect it to hold for all twistorial theories.

The key principle underlying this simplification is the fact that, just as for the free theories, the partition functions of self-dual (sd) theories are 1-loop exact.\footnote{See \cite{Krasnov:2016emc} for a helpful review of perturbative calculations in self-dual theories.} Such partition functions are path integrals on $S^1\times S^3$ with suitable periodicity conditions. Due to the 1-loop exactness, the computation of these path integrals reduces to the computation of 1-loop determinants, which simply count states in the Fock space of the free theories obtained from dropping the sd interactions. After an integral over the $S^1$-holonomy of the gauge field, the resulting path integrals coincide with partition functions of the usual free limits of non-self-dual gauge theories.

The 1-loop exactness can also be seen on twistor space. One finds that the self-dual theories uplift to holomorphic BF theories (plus holomorphic matter) on $\PT$. This is clear for the non-supersymmetric action \eqref{hbf}, but also holds for the supersymmetric cases when the Chern-Simons type actions \eqref{N=4tac} and \eqref{N=1tac} are expanded in terms of component fields. The partition functions of holomorphic BF theories are 1-loop exact and can be reduced to the computation of invariants like the holomorphic Ray-Singer torsion, as explicitly seen in \cite{Bittleston:2022nfr} for twistorial theories.

In the rest of this section, we will work with the twistor uplifts. We compute their partition functions around the trivial background, so there is no classical action that contributes to the path integral. The 1-loop determinants count states in the Hilbert spaces obtained from quantizing the phase spaces of the classical sd theories. This counting problem can be systematically attacked by working in the BV formalism. 


\subsection{States of self-dual theories}
\label{ssec:sdstates}

Once a given self-dual theory has been uplifted to $\PT$, its Hilbert space can be realized as the cohomology of its BV-BRST charge $Q$. In both the $\cN=4$ and $\cN=1$ cases, the classical BRST differential can be compactly expressed through its action on the Chern-Simons gauge superfield $\cA$,
\be
Q\cA = \dbar\cA + \cA^2\,. \label{eq:QA}
\ee
This can be decomposed by form degree by expanding $\cA=\sum_{p=0}^3\cA_p$ as a sum of $(0,p)$-forms $\cA_p$,
\be
\begin{split}
    Q\cA_0 &= \cA_0^2\,,\\
    Q\cA_1 &= \dbar\cA_0 + [\cA_0,\cA_1]\,,\\
    Q\cA_2 &= \dbar\cA_1 + [\cA_1,\cA_1] + [\cA_0,\cA_2]\,,\\
    Q\cA_3 &= \dbar\cA_2 + [\cA_1,\cA_2] + [\cA_0,\cA_3]\,,
\end{split}
\ee
where all $\cA_p$ have odd Grassmann parity,\footnote{In this convention, odd differential forms also carry odd Grassmann parity. So for example, if a partial connection $a=a_{\bar A}\d\bar{Z}^{\bar A}$ is assigned odd Grassmann parity, it means that the coefficients $a_{\bar A}$ in its expansion are graded even as usual.} and $[-,-]$ denotes the graded Lie bracket. The cohomology problem can now be solved perturbatively.

We will only solve the problem of \emph{counting} this cohomology, as developing explicit representatives of the cohomology classes is a much more nontrivial task.\footnote{Moreover, the differential \eqref{eq:QA} could potentially receive higher-loop quantum corrections when acting on composite operators \cite{Budzik:2023xbr, Choi:2025bhi, Budzik:2025zvu, Kim:2026rnx}.} We will primarily focus on the graded partition function, which plays the role of an index for the D5-brane theory.\footnote{The sd theories have a second coupling constant which is usually omitted as it can be absorbed in the scale of $b$ and other 6d fields. The graded partition function is an index in the sense of being independent of this coupling.}

The cohomology can be computed by a spectral sequence argument. The classical differential splits as $Q=Q_0+Q_1$, where $Q_0\cA = \dbar\cA$ and $Q_1\cA = \cA^2$ are themselves differentials. We first pass to the cohomology of $Q_0$. This is just the Dolbeault cohomology of $\PT$. The $(0,p)$-Dolbeault cohomology valued in any sheaf on $\PT$ is supported purely in degrees $p=0,1$. Degree-$1$ cohomology is the usual data $\cA_1$ that gets mapped to states of physical fields by the Penrose transform \cite{Eastwood:1981jy}. Degree-$0$ cohomology $\cA_0$ gives rise to states of the ghosts of gauge symmetries.

Since one generically finds $H^{0,2}=H^{0,3}=0$ on $\PT$, on passing to Dolbeault cohomology, one can set the antifields $\cA_2,\cA_3$ to zero. Setting $\dbar\cA_0=\dbar\cA_1=0$, the residual action of the BRST charge is found to be
\be
Q\cA_0 = \cA_0^2\,,\qquad Q\cA_1 = [\cA_0,\cA_1]\,.
\ee
This action resembles the Chevalley-Eilenberg differential defining a Lie algebra cohomology problem. We now make this resemblance precise in the super Yang-Mills cases.

\paragraph{States of $\cN=4$ sdYM.} We reviewed the field content of $\cN=4$ sdYM in section \ref{ssec:ch4}. The open string field $\cA$ is a superfield in four Grassmann coordinates $\eta^I$ on supertwistor space $\PT^{3|4}$. Let us begin by expanding the ghosts $\cA_0$ in $\eta^I$,
\be
\cA_0 = c + f_I\eta^I + \frac12\,\chi_{IJ}\eta^I\eta^J + g^I\eta^3_I + \til c\,\eta^4\,.
\ee
Under rescalings of $Z^A$, the $\eta^I$ carry weight $1$. So the ghosts $c,f_I,\chi_{IJ},g^I,\til c$ are valued in $\CO_{\PT}(-n)$ for $n=0,1,2,3,4$. 

For all $n\geq1$, the zeroth cohomology of $\PT$ valued in $\CO_{\PT}(-n)$ vanishes. For $n=0$, the zeroth cohomology is nonzero but one-dimensional:
\be
H^0(\PT,\CO_\PT) = \C\,.
\ee
This is the constant mode $c=c_0$ of the $c$-ghost which imposes \emph{global} $\GL_N(\C)$ invariance. Therefore, after passing to Dolbeault cohomology, we are left with
\be
\cA_0 = c_0\,.
\ee
The residual BRST differential on the second page of our spectral sequence then acts as
\be\label{Qres}
Qc_0 = c_0^2\,,\qquad Q\cA_1 = [c_0,\cA_1]\,.
\ee
Since $c_0$ is a constant $\gl_N(\C)$ matrix, $Qc_0=c_0^2$ describes precisely the $\gl_N(\C)$ Lie algebra cohomology differential.

Next, recall the expansion of the $(0,1)$-part $\cA_1$ in physical fields of the $\cN=4$ vector multiplet,
\be
\cA_1 = a + \til\psi_I\eta^I + \frac12\,\vphi_{IJ}\eta^I\eta^J + \psi^I\eta^3_I + b\,\eta^4\,.
\ee
The relevant Dolbeault cohomology groups are
\be\label{H01}
H^{0,1}(\PT,\CO(-n)) = \bigoplus_{k=\max(0,2-n)}^\infty\Sym^{k+n-2}(\C^2)\otimes\Sym^{k}(\C^2)\,,
\ee
for $n=0,1,2,3,4$. Let $\hat\lambda^\al = (\bar\lambda_{\dot0},\bar\lambda_{\dot1})$ denote the hermitian conjugate of $\lambda_\al$. Harmonic representatives of the $k^\text{th}$ summand of this direct sum decomposition are given by
\be\label{harrep}
\frac{\hat\lambda_{\al_1}\cdots\hat\lambda_{\al_{k+n-2}}\mu^{\dal_1}\cdots\mu^{\dal_k}}{\la\hat\lambda\lambda\ra^{k+n-2}}\,\bar e^0\,,\qquad \bar e^0\equiv\frac{\la\hat\lambda\,\d\hat\lambda\ra}{\la\hat\lambda\lambda\ra^2}\,,
\ee
which are singular only along the ``line at infinity'' $\{\lambda_\al=0\}$ that is not contained in $\PT$. Every such representative can be Penrose transformed into a single-particle state of free $\cN=4$ sYM, giving rise to a ``positive frequency'' state in radial quantization on $\R^4-0$.

In total, we find that the ``single-particle'' spectrum is obtained from taking the quotient of the algebra $\C[\![\mu^\dal,\eta^I,\hat\lambda_\al/\la\hat\lambda\lambda\ra]\!]\,\bar e^0$ by the supertwistor rescaling. We can write this as an expansion of $\cA_1$ in ``positive frequency modes'' on twistor space,
\begin{multline}\label{N=4spec}
    \cA_1 = \sum_{p_1=0}^1\sum_{p_2=0}^1\sum_{p_3=0}^1\sum_{p_4=0}^1\sum_{q_1=0}^\infty\sum_{q_2=0}^\infty\sum_{r_1=0}^\infty\sum_{r_2=0}^\infty a^\dagger_{\bf{p},\bf{q},\bf{r}}\,\frac{\hat\lambda_0^{r_1}}{\la\hat\lambda\lambda\ra^{r_1}}\frac{\hat\lambda_1^{r_2}}{\la\hat\lambda\lambda\ra^{r_2}}\,(\mu^{\dot0})^{q_1}(\mu^{\dot1})^{q_2}\\\times(\eta^1)^{p_1}(\eta^2)^{p_2}(\eta^3)^{p_3}(\eta^4)^{p_4}\bar e^0\,\delta_{p_1+p_2+p_3+p_4+q_1+q_2,r_1+r_2+2}\,,
\end{multline}
where $a^\dagger_{{\bf p},{\bf q},{\bf r}}$ for ${\bf p}=(p_1,p_2,p_3,p_4)$, ${\bf q}=(q_1,q_2)$, ${\bf r} =(r_1,r_2)$ are the $\gl_N(\C)$-valued creation operators for the open string spectrum on $\PT$.\footnote{We remark that $\PT$ is usually constructed by taking $\P^3$ with homogeneous coordinates $Z^A=(\mu^\dal,\lambda_\al)$ and removing the line $\{\lambda_\al=0\}$. The data of annihilation operators and negative frequency states can also be recovered from twistor theory as the cohomology of the space obtained from removing the twistor line of the origin $\{\mu^\dal=0\}$ from $\P^3$. We suppress the annihilation operators in our discussion.} The Kronecker delta in the second line imposes the weightlessness constraint.

The Fock space of open strings can now be generated by second quantizing the single-string spectrum \eqref{N=4spec} and taking cohomology with respect to the residual differential \eqref{Qres}. In terms of the creation operators, the residual differential acts as
\be
Qc_0 = c_0^2\,,\qquad Q a^\dagger_{\bf{p},\bf{q},\bf{r}} = [c_0,a^\dagger_{\bf{p},\bf{q},\bf{r}}]\,.
\ee
Beautifully, this is precisely the calculation that one is prescribed to do in free gauge theories: construct the free field Fock space and take global $\U(N)$ invariants. Due to the Penrose transform, the single-string states \eqref{N=4spec} are in one-to-one correspondence with the single-particle states of the free $\cN=4$ theory. Hence, the associated Fock spaces as well as their $\U(N)$ invariant subspaces coincide, thereby ensuring that the partition functions of the self-dual and free theories also coincide.

\paragraph{States of $\cN=1$ sdYM.} The analysis of the open string spectrum of the $\cN=4$ case can easily be recycled for the $\cN=1$ case. Recall from section \ref{ssec:ch1} that the open string field $\cA$ is now a superfield on $\PT^{3|2}$, and can be expanded in its two Grassmann coordinates $\eta^1,\eta^2$ of weight $1,3$ respectively. The expansion of the 0-form part $\cA_0$ takes the form
\be
\cA_0 = c + f\,\eta^1 + g\,\eta^2 + \til c\,\eta^1\eta^2\,,
\ee
where ghosts $c,f,g,\til c$ are valued in $\CO_\PT(-n)$ for $n=0,1,3,4$ respectively. The classical BRST differential again takes the form $Q\cA = \dbar\cA+\cA^2$. Passing to Dolbeault cohomology, we can set $f=g=\til c=0$ and $c=c_0=\text{constant}$, leaving us with just the mode $\cA_0 = c_0$ that is the ghost of global $\GL_N(\C)$ rotations.

The physical spectrum is encoded in the $(0,1)$-part $\cA_1$,
\be
\cA_1 = a + \til\psi\,\eta^1 + \psi\,\eta^2 + b\,\eta^1\eta^2\,.
\ee
As mentioned in section \ref{ssec:ch1}, the fields $a,b,\psi,\til\psi$ Penrose transform to the fields $A_{\al\dal}$, $B_{\al\beta}$, $\Psi_\al$, $\til\Psi_\dal$ of the $\cN=1$ vector multiplet. The associated Dolbeault cohomology groups are the same as \eqref{H01} but now for just the $n=0,1,3,4$ cases. We can also recycle their harmonic representatives \eqref{harrep} to obtain an expansion of the open string spectrum,
\begin{multline}\label{N=1spec}
    \cA_1 = \sum_{p_1=0}^1\sum_{p_2=0}^1\sum_{q_1=0}^\infty\sum_{q_2=0}^\infty\sum_{r_1=0}^\infty\sum_{r_2=0}^\infty a^\dagger_{\bf{p},\bf{q},\bf{r}}\,\frac{\hat\lambda_0^{r_1}}{\la\hat\lambda\lambda\ra^{r_1}}\frac{\hat\lambda_1^{r_2}}{\la\hat\lambda\lambda\ra^{r_2}}\,(\mu^{\dot0})^{q_1}(\mu^{\dot1})^{q_2}\\\times(\eta^1)^{p_1}(\eta^2)^{p_2}\bar e^0\,\delta_{p_1+3p_2+q_1+q_2,r_1+r_2+2}\,,
\end{multline}
where the creation operators $a^\dagger_{\bf{p},\bf{q},\bf{r}}$ are now labeled by the multi-indices ${\bf p} = (p_1,p_2)$, ${\bf q}=(q_1,q_2)$ and ${\bf r}=(r_1,r_2)$. Note that the Kronecker delta now contains a $3p_2$ instead of just $p_2$, because the weight of $\eta^2$ is now $3$ instead of $1$. Finally, the $\cN=1$ sdYM Fock space is built from $\U(N)$ invariant products of creation operators $a^\dagger_{\bf{p},\bf{q},\bf{r}}$ acting on the vacuum.


\subsection{State counting}
\label{ssec:multifugacity}

In the previous section, we outlined the construction of the spectra of $\cN=4$ and $\cN=1$ sdYM theories. Let us finally count these spectra to obtain partition functions for each of these theories. We will show that the results match the free theory partition functions discussed in section \ref{sec:partition}.

\paragraph{Partition functions of $\cN=4$ sdYM.} The complexified symmetry algebra of $\cN=4$ sYM is $\psl(4|4,\C)$. We can grade states with respect to its Cartan subalgebra generated by the conformal dimension $\Delta$, the left- and right-handed angular momenta $j_1$, $j_2$, and three R-symmetry generators $R_1,R_2,R_3$. Each of these symmetries uplifts to supertwistor space $\PT^{3|4}$ as holomorphic vector fields (see \cite{Elvang:2013cua} for a pedagogical discussion). The dilatation generator becomes
\be\label{dils}
\Delta = \frac{1}{2}\left(\lambda_\al\frac{\p}{\p\lambda_\al}-\mu^\dal\frac{\p}{\p\mu^\dal}\right)\,.
\ee
The Lorentz generators are given by
\be
J_{\al\beta} = \lambda_{\al}\frac{\p}{\p\lambda^{\beta}}+\lambda_{\beta}\frac{\p}{\p\lambda^{\al}}\,,\qquad\til J_{\dal\dot\beta} = \mu_{\dal}\frac{\p}{\p\mu^{\dot\beta}}+\mu_{\dot\beta}\frac{\p}{\p\mu^{\dal}}\,,
\ee
and the Cartan can be chosen to be $j_1=-J_{01}$, $j_2=\til J_{\dot0\dot1}$. Similarly, the R-symmetry generators are given by
\be
R^I{}_J = \eta^I\frac{\p}{\p\eta^J} - \frac14\,\delta^I_J\,\eta^K\frac{\p}{\p\eta^K}
\ee
from which we pick out $R_{1} = R^1{}_1+R^2{}_2, R_{2} = R^1{}_1+R^3{}_3, R_{3}=R^1{}_1+R^4{}_4$ or any other convenient choice of a Cartan. All such vector fields are defined up to shifts by the Euler homogeneity vector
\be\label{euler}
\Upsilon = Z^A\frac{\p}{\p Z^A} + \eta^I\frac{\p}{\p\eta^I} \, ,
\ee
which generates the rescalings $(Z^A,\eta^I)\sim(\alpha Z^A,\alpha\eta^I)$ that have been quotiented out in the construction of $\PT^{3|4}$. States of twistor strings are constrained to be uncharged under the action of $\Upsilon$.

Instead of directly introducing six fugacities conjugate to the six charges $\Delta$, $j_1$, $j_2$, $R_1$, $R_2$, $R_3$, we can streamline the analysis by working with partition functions refined by eight fugacities. For instance, we can introduce the thermal, graded and twisted graded partition functions
\begin{align}
Z_N^\text{th} &= \Tr_{\cH_N}\!(-1)^\text{gh}x_1^{-\fR_1}x_2^{-\fR_2}x_3^{-\fR_3}x_4^{-\fR_4}y_1^{-\fJ_1}y_2^{-\fJ_2}z_1^{-\fL_1}z_2^{-\fL_2}\,.\label{Zth4twistor}\\
Z_N^\text{gr} &= \Tr_{\cH_N}(-1)^{\text{gh}+F}x_1^{-\fR_1}x_2^{-\fR_2}x_3^{-\fR_3}x_4^{-\fR_4}y_1^{-\fJ_1}y_2^{-\fJ_2}z_1^{-\fL_1}z_2^{-\fL_2}\,,\label{Zgr4twistor}\\
Z_N^\text{tw} &= \Tr_{\cH_N}(-1)^{\text{gh}+F}\bau_R\,x_1^{-\fR_1}x_2^{-\fR_2}x_3^{-\fR_3}x_4^{-\fR_4}y_1^{-\fJ_1}y_2^{-\fJ_2}z_1^{-\fL_1}z_2^{-\fL_2}\,.\label{Ztw4twistor}
\end{align}
Here, we have introduced four fugacities $x_1,x_2,x_3,x_4$ conjugate to the four vector fields
\be\label{Rs}
\mathfrak{R}_1 = \eta^1\frac{\p}{\p\eta^1}\,,\qquad\mathfrak{R}_2 = \eta^2\frac{\p}{\p\eta^2}\,,\qquad\mathfrak{R}_3 = \eta^3\frac{\p}{\p\eta^3}\,,\qquad\mathfrak{R}_4 = \eta^4\frac{\p}{\p\eta^4}\,,
\ee
two fugacities $y_1,y_2$ conjugate respectively to
\be\label{Js}
\mathfrak{J}_1 = \mu^{\dot0}\frac{\p}{\p\mu^{\dot0}}\,,\qquad\mathfrak{J}_2 = \mu^{\dot1}\frac{\p}{\p\mu^{\dot1}}\,,
\ee
and two fugacities $z_1,z_2$ conjugate respectively to
\be\label{Ls}
\mathfrak{L}_1 = \lambda_0\frac{\p}{\p\lambda_0}\,,\qquad\mathfrak{L}_2 = \lambda_1\frac{\p}{\p\lambda_1}\,.
\ee
The twisting by $-\mathbbm{1}_4\in\SU(4)_R$ is taken to be
\be
\bau_R = (-1)^{3\,\fR_1-\fR_2-\fR_3-\fR_4}\,.
\ee
This can be identified with the spacetime fermion number $F$,
\be
(-1)^F = \bau_R\,,
\ee
so that the thermal and twisted graded partition functions are again equal in this context. For BRST invariance, each of these partition functions has also been graded by an insertion of $(-1)^\text{gh}$, where `gh' denotes the BV-BRST ghost number.\footnote{BRST ghosts are always periodic on the thermal circle by virtue of representing gauge transformations. So gauge theory partition functions are necessarily graded by BRST ghost number. Most of the time, one directly works with the ghost-free Fock space and does not make this explicit.} On $\PT$, one assigns states of the $(0,p)$-form field $\cA_p$ a ghost number $\text{gh}=1-p$.

This representation is precisely analogous to the eight-fugacity partition functions \eqref{Zth4}, \eqref{Zgr4} and \eqref{Ztw4} of the free theory. The vectors in \eqref{Rs}, \eqref{Js} and \eqref{Ls} generate the Cartan of $\gl(4|4,\C)$, but the true R-symmetry is $\mathfrak{psl}(4|4,\C)$. One of the fugacities is again redundant because under a rescaling $(x_I,y_\dal,z_\al)\sim(\alpha x_I,\alpha y_\dal,\alpha z_\al)$, the partition functions scale as $\alpha^{-\Upsilon}$, which evaluates to $1$ on states that are uncharged under $\Upsilon$. Moreover, any symmetry of the B-model must leave invariant the volume form of its Calabi-Yau target. Requiring that the volume form of supertwistor space $\D^3Z\,\d^4\eta$ have fugacity 1 imposes the constraint $x_1x_2x_3x_4=y_1y_2z_1z_2$. This is precisely the constraint that we had encountered in the free theory case \eqref{xyzeq}.

For our purposes, it will suffice to compute the graded partition function. The single-letter graded partition function $f(\fx,\fy,\fz)$ can be obtained by counting the creation operators $a^\dagger_{{\bf p},{\bf q},{\bf r}}$. We have been slightly cavalier in working with \emph{holomorphic} vector fields generating \emph{complexified} symmetries, without imposing any reality conditions. The harmonic representatives \eqref{harrep} contain $\hat\lambda_\al$ and are adapted to Euclidean reality conditions. To act with the vector fields $\fL_1,\fL_2$ on these would require working with the real parts of $\fL_1,\fL_2$. Instead, we find it simpler to take a signature-agnostic approach by utilizing the \v{C}ech-Dolbeault isomorphism.

By this isomorphism, we can replace the $(0,1)$-form Dolbeault representatives of the cohomology \eqref{H01} by \v{C}ech representatives that are meromorphic functions on $\PT^{3|4}$ with singularities at both $\lambda_0=0$ and $\lambda_1=0$ but regular elsewhere. The expansion of $\cA_1$ in \v{C}ech representatives takes the form
\begin{multline}\label{N=4spec1}
    \cA_1^\text{\v{C}ech} = \sum_{p_1=0}^1\sum_{p_2=0}^1\sum_{p_3=0}^1\sum_{p_4=0}^1\sum_{q_1=0}^\infty\sum_{q_2=0}^\infty\sum_{r_1=0}^\infty\sum_{r_2=0}^\infty a^\dagger_{\bf{p},\bf{q},\bf{r}}\,\lambda_0^{-r_1-1}\lambda_1^{-r_2-1}(\mu^{\dot0})^{q_1}(\mu^{\dot1})^{q_2}\\\times(\eta^1)^{p_1}(\eta^2)^{p_2}(\eta^3)^{p_3}(\eta^4)^{p_4}\,\delta_{p_1+p_2+p_3+p_4+q_1+q_2,r_1+r_2+2}\,.
\end{multline}
This is equivalent to the expansion \eqref{N=4spec}. The vectors \eqref{Rs}, \eqref{Js}, \eqref{Ls} assign fugacities $z_1^{-1},z_2^{-1},y_1^{-1},y_2^{-1},x_1^{-1},x_2^{-1},x_3^{-1},x_4^{-1}$ to $\lambda_0,\lambda_1,\mu^{\dot0},\mu^{\dot1},\eta^1,\eta^2,\eta^3,\eta^4$ respectively. The creation operators come with the opposite powers of fugacities, so that $\cA_1^\text{\v{C}ech}$ has total fugacity zero. Hence, we obtain the fugacity assignments
\be
[a^\dagger_{\bf{p},\bf{q},\bf{r}}] = (-1)^{p_1+p_2+p_3+p_4}x_1^{p_1}x_2^{p_2}x_3^{p_3}x_4^{p_4}y_1^{q_1}y_2^{q_2}z_1^{-r_1-1}z_2^{-r_2-1}\,,
\ee
where the indices obey $p_1+p_2+p_3+p_4+q_1+q_2=r_1+r_2+2$, and the overall sign is the Grassmann parity $(-1)^F$. There is no ghost number sign as all of these states sit at zero ghost number.

The single-letter graded partition function is given by the sum over these fugacities,
\be
f_{\cN=4} = \sum_{p_1=0}^1\sum_{p_2=0}^1\sum_{p_3=0}^1\sum_{p_4=0}^1\sum_{q_1=0}^\infty\sum_{q_2=0}^\infty\sum_{r_1=0}^\infty\sum_{r_2=0}^\infty[a^\dagger_{\bf{p},\bf{q},\bf{r}}]\,\delta_{p_1+p_2+p_3+p_4+q_1+q_2,r_1+r_2+2}\,.
\ee
As usual, we do not include the ghost zero mode $c_0$ in our counting, as it would contribute to $f_{\cN=4}$ with fugacity $1$ whose $\PE$ would diverge. Performing the sums yields
\be\label{fN4}
f_{\cN=4} = 1 - \frac{z_1P(z_1)-z_2P(z_2)}{z_1-z_2}\,,
\ee
where $P(z)$ is the same rational function \eqref{Pprod} that we obtained in the free theory calculation. 

The result for the graded partition function is now identical to the free theory partition function \eqref{Zgr4result}. As is standard, one can show that the $\U(N)$ integral arises from the integral over the holonomy of the gauge field $A_{\al\dal}$ along the $S^1$. Because the sdYM path integral is 1-loop exact, the effective action $\sum_{n\geq1}\frac{1}{n}\,f_{\cN=4}(x^n,y^n,z^n)\Tr(U^n)\Tr(U^{-n})$ for the holonomy $U$ also remains unchanged from the free theory.

\paragraph{Partition functions of $\cN=1$ sdYM.} The $\cN=1$ supertwistor space $\PT^{3|2}$ carries only two Grassmann coordinates $\eta^1,\eta^2$. We introduce the corresponding vector fields
\be
\fR_1 = \eta^1\frac{\p}{\p\eta^1}\,,\qquad\fR_2 = \eta^2\frac{\p}{\p\eta^2}\,.
\ee
We can use these in addition to $\fJ_1,\fJ_2,\fL_1,\fL_2$ to construct a graded partition function,
\be
Z_N^\text{gr} = \Tr_{\cH_N}(-1)^{\text{gh}+F}x_1^{-\fR_1}x_2^{-\fR_2}y_1^{-\fJ_1}y_2^{-\fJ_2}z_1^{-\fL_1}z_2^{-\fL_2}\,.
\ee
Its thermal and twisted graded versions are defined analogously. All these vector fields are now taken modulo the Euler vector field
\be\label{euler1}
Z^A\frac{\p}{\p Z^A} + \eta^1\frac{\p}{\p\eta^1} + 3\,\eta^2\frac{\p}{\p\eta^2}
\ee
which generates the rescaling $(Z^A|\eta^1,\eta^2)\sim(\alpha Z^A|\alpha\eta^1,\alpha^3\eta^2)$ that is quotiented out in the construction of $\PT^{3|2}$.

On spacetime, both the free and self-dual limits of $\cN=1$ sYM only have one R-symmetry. As in the $\cN=4$ case, the fugacities are defined up to the projective rescalings $(x_1,x_2,y_\dal,z_\al)\sim(\alpha x_1,\alpha^3x_2,\alpha y_\dal,\alpha z_\al)$, and invariance of the volume form $\D^3Z\,\d^2\eta$ imposes the relation $x_1x_2=y_1y_2z_1z_2$. The single-letter graded partition function is obtained by counting the creation operators occurring in the \v{C}ech analog of the mode expansion \eqref{N=1spec},
\begin{multline}\label{N=1spec1}
    \cA_1^\text{\v{C}ech} = \sum_{p_1=0}^1\sum_{p_2=0}^1\sum_{q_1=0}^\infty\sum_{q_2=0}^\infty\sum_{r_1=0}^\infty\sum_{r_2=0}^\infty a^\dagger_{\bf{p},\bf{q},\bf{r}}\,\lambda_0^{-r_1-1}\lambda_1^{-r_2-1}(\mu^{\dot0})^{q_1}(\mu^{\dot1})^{q_2}\\\times(\eta^1)^{p_1}(\eta^2)^{p_2}\,\delta_{p_1+3p_2+q_1+q_2,r_1+r_2+2}\,.
\end{multline}
The creation operators $a^\dagger_{\bf{p},\bf{q},\bf{r}}$ now carry the fugacity assignments
\be
[a^\dagger_{\bf{p},\bf{q},\bf{r}}] = (-1)^{p_1+p_2}x_1^{p_1}x_2^{p_2}y_1^{q_1}y_2^{q_2}z_1^{-r_1-1}z_2^{-r_2-1}\,.
\ee
So the single-letter count is computed by the sum
\be
f_{\cN=1} = \sum_{p_1=0}^1\sum_{p_2=0}^1\sum_{q_1=0}^\infty\sum_{q_2=0}^\infty\sum_{r_1=0}^\infty\sum_{r_2=0}^\infty[a^\dagger_{\bf{p},\bf{q},\bf{r}}]\,\delta_{p_1+3p_2+q_1+q_2,r_1+r_2+2}\,.
\ee
This sum can again be performed in closed form to yield
\be\label{fN1}
f_{\cN=1} = 1 - \frac{z_1Q(z_1)-z_2Q(z_2)}{z_1-z_2}\,,
\ee
expressed in terms of the elementary product found in \eqref{Qprod}. 

Again, the resulting $\U(N)$ matrix integral agrees with the free theory result \eqref{Zgr1result}. This completes the definition of the refined graded partition function of $\cN=1$ sdYM.


\section{Partition functions of closed strings}
\label{sec:bcov}

As we briefly reviewed in sections \ref{ssec:ch4}, \ref{ssec:ch1}, closed twistor strings are governed by BCOV theory on a 7-fold in the $\cN=4$ case and on a 5-fold in the $\cN=1$ case. In this section, we compute the graded partition functions $Z_\text{BCOV}$ of such closed string states, as they are crucial ingredients that will enter our giant graviton expansions. Quite surprisingly, we will find that they \emph{fail} to match the infinite-$N$ graded partition functions of their sdYM duals. This suggests that even at large $N$, there must exist extra degrees of freedom in addition to closed twistor strings that are needed to reproduce the boundary multi-graviton spectrum.

We will find a beautiful resolution of this puzzle when we obtain our giant graviton expansions in section \ref{sec:ggemain}. The mismatch between $Z_{\mathrm{BCOV}}$ and the infinite-$N$ graded partition function $Z_\infty$ of the boundary sdYM theory will be cured by an infinite series of D-instanton/anti-D-instanton corrections. The D-instantons that accomplish this feat will precisely be Witten's D-instantons\footnote{We will only encounter D-instantons that are degree-1 in the terminology of twistor strings \cite{Witten:2003nn}. A degree-$d$ D-instanton is a brane that intersects $\PT$ in a degree-$d$ rational curve.} that also proved to be the origin of the simplicity of tree-level gluon amplitudes in \cite{Witten:2003nn}.

With this hindsight, in the rest of this section, we focus on computing the graded partition functions of BCOV theory. The fruit of our labor may be found in equation \eqref{Zbcov4} for the $\cN=4$ case and \eqref{Zbcov1} for the $\cN=1$ case. The twisted graded partition functions can again be obtained by R-symmetry shifts explained in section \ref{sec:sdpartition}.


\subsection{$\cN=4$ sdYM}
\label{ssec:bcov4}

To compute the partition function of closed twistor strings, we can work with their closed string field theory. In the $\U(N)$ $\cN=4$ sdYM case, this is BCOV theory on a 7-fold, in the presence of $N$ units of backreaction. Before backreaction, one works with the 7-fold $X=\Tot(\CO(-1)^{\oplus4}\to\PT)$, with base and fiber coordinates $Z^A$ and $W_I$ respectively. The flux sourced by $N$ D5-branes wrapping $\PT$ was reviewed in \eqref{xi43}. It is singular along the source locus $\PT\simeq\{W_I=0\}$. So, strictly speaking, the backreaction-coupled BCOV theory lives on the space $X^\times = X-\PT$.

For the purposes of computing the graded partition function we can consider $X^\times$ with the backreaction flux turned off. The backreaction deforms the BCOV differential by $\bar\partial\mapsto\bar\partial+\qty[\xi_{4,3},-]$. This can potentially lift cohomology classes \cite{Budzik:2023xbr} (or just deform them as in \cite{Costello:2018zrm}). However, any states lifted in this way occur in pairs of opposite Grassmann parity and identical fugacity weights. Their contributions therefore cancel in the graded partition function. This argument also applies to the twisted graded partition functions, since the twist commutes with the deformed differential. We therefore perform the counting with the undeformed $\bar\partial$ differential on $X^\times$. For completeness, we provide a sample computation of the deformation of a BCOV state in the presence of the backreaction in appendix \ref{app:backreaction}.

The field content of BCOV theory contains a field $\xi_{k,l}\in\Omega^{k,l}(X^\times)$ for every $k,l$, satisfying the constraint $\p\xi_{k,l} = 0$. One usually drops the ``topological degrees of freedom'' $\xi_{0,l}$ and takes the remaining $\xi_{k,l}$ to be $\p$-exact: $\xi_{k,l}=\p\gamma_{k-1,l}$, for some potentials $\gamma_{k-1,l}$ with $k=1,\dots,7$ and $l=0,1,\dots,7$. In the absence of backreaction, these fields are subject to the linearized field equations and gauge freedoms
\be\label{dbarxikl}
\dbar\xi_{k,l} = 0\,,\qquad\delta\xi_{k,l} = \dbar f_{k,l}\,.
\ee
So before turning on the flux, the linearized BCOV states are cohomology classes $\xi_{k,l}\in H^{0,l}(X,\Omega^{k,0}_\text{ex})$, where $\Omega^{k,0}_\text{ex}$ denotes the bundle of $\p$-exact $k$-forms on $X$. 

In holography, we look for states that are sourced by boundary operator insertions. Equivalently, we look for states that couple to operators in the worldvolume theory of the backreacting branes. Let us take the latter viewpoint. For instance, the bulk-brane coupling of a BCOV state $\xi$ to an open string operator $\mathcal{O}_{I_1\cdots I_n}$ on the D5-branes wrapping $\PT$ schematically takes the form
\be
\int_{\PT}\p^{I_1}\cdots\p^{I_n}\xi\;\mathcal{O}_{I_1\cdots I_n}\,,
\ee
where $I_r$ are R-symmetry indices and $\p^I = \p/\p W_I$ are derivatives in the $\CO(-1)^{\oplus4}$ directions normal to $\PT$. For this to give rise to a local operator insertion at a spacetime point $x\in\R^4$, we need $\xi$ to contain a delta function $\bar\delta^2(\mu-x\lambda)$ supported along the twistor line $\P^1_x$. Without loss of generality, we will focus on states dual to operators placed at $x=0$.

States with such distributional support are classified by the notion of \emph{local cohomology}. Let $E\to X$ be a holomorphic vector bundle. The $E$-valued local cohomology $H^{0,*}_C(X,E)$ of $X$ supported at a subspace $C\subset X$ measures the failure of sections of $E$ on $X-C$ to extend across $C$. Similar to twisted holography \cite{Costello:2018zrm}, instead of taking our states $\xi_{k,l}$ to be valued in ordinary Dolbeault cohomology $H^{0,l}(X,\Omega^{k,0}_\text{ex})$, we are instructed to take
\be
\xi_{k,l}\in H^{0,l}_Y(X,\Omega^{k,0}_\text{ex})\,,
\ee
where $Y = \{\mu^\dal=0\}$ is a copy of $\CO(-1)^{\oplus4}\to\P^1$. Such local cohomology will necessarily be supported in a formal neighborhood of $Y$, i.e., its representatives can be chosen to be proportional to derivatives of the delta function $(0,2)$-forms $\bar\delta^2(\mu)$,
\be
\xi_{k,l} = f_{k,l-2}(Z,W)\,\p_{\dot0}^r\p_{\dot1}^s\bar\delta^2(\mu)\,,\qquad r,s\geq0\,,
\ee
with coefficients $f_{k,l-2}$ given by holomorphic $(k,l-2)$-forms of weight $r+s+2$ under $(Z^A,W_I)\sim(\alpha\,Z^A,\alpha^{-1}W_I)$. In particular, one finds that our local cohomology is only nonzero in degrees $l\geq2$. These are the states that we will count.

As a simple example, consider the case $k=7$. Let $\Omega_X=\D^3Z\,\d^4W-\D^3W\,\d^4Z$ be the holomorphic top-form on $X$. The cohomology ring $H^{0,l}_Y(X,\Omega^{7,0}_\text{ex})$ is supported purely in degree $l=2$ and is spanned by states of the form
\be\label{72states}
\xi_{7,2}[{\bf p},{\bf q},{\bf r}] = \lambda_0^{r_1}\lambda_1^{r_2}W_1^{p_1}W_2^{p_2}W_3^{p_3}W_4^{p_4}\p_{\dot0}^{q_1}\p_{\dot1}^{q_2}\bar\delta^2(\mu)\,\Omega_X\,,
\ee
where $p_1,p_2,p_3,p_4,q_1,q_2,r_1,r_2\geq0$, and $r_1+r_2=p_1+p_2+p_3+p_4+q_1+q_2+2$ to ensure weightlessness. Because they are top holomorphic forms on $X$, they are automatically $\p$-closed. It is not hard to show that they are also $\p$-exact.

\paragraph{Sidestepping the $\p$-closure constraint.} More generally, although this local cohomology problem can be solved using standard techniques, imposing the $\p$-exactness constraint quickly grows quite laborious. Thankfully, if one is only interested in \emph{counting} the states, one can also work directly at the level of the bundles $\Omega^{k,0}$ without imposing $\p$-exactness. This can be accomplished through a set of tricks. 

Let us begin by introducing some notation. For each $k=1,\dots,7$, denote the local cohomology ring without the $\p$-exactness constraint by
\be
 \cC_k = \bigoplus_{l=0}^7H^{0,l}_Y(X,\Omega^{k,0})\,.
\ee
Since $\p$ commutes with taking $\dbar$-cohomology, it induces maps $\p:\cC_k\to\cC_{k+1}$. We denote the associated kernels and images by
\be
\begin{split}
    \cK_k &= \ker(\p:\cC_k\to\cC_{k+1})\,,\\
    \cB_k &= \text{im}(\p:\cC_{k-1}\to\cC_k)\,.
\end{split}
\ee
Our $\p$-exact BCOV states $\xi_{k,l}$ are computed precisely by the groups $\cB_k$.

To each $\cB_k$, we will associate a Hilbert series $\sb_k({\bf x},{\bf y},{\bf z})$ graded by the same fugacities that entered the boundary partition functions of section \ref{ssec:multifugacity}. Having done this, one could compute the single-particle graded partition function of BCOV theory,
\be\label{fbcovdef}
f_\text{BCOV}({\bf x},{\bf y},{\bf z}) = \sum_{k=1}^7(-1)^{k+1}\sb_k({\bf x},{\bf y},{\bf z})\,,
\ee
and obtain the multi-particle graded partition function through plethystic exponentiation,
\be
Z_\text{BCOV}(\mathbf{x},\mathbf{y},\mathbf{z}) = \PE[f_{\text{BCOV}}(\mathbf{x},\mathbf{y},\mathbf{z})]= \exp\qty(\sum_{n=1}^\infty \frac{1}{n}\, f_\text{BCOV}(\mathbf{x}^n,\mathbf{y}^n,\mathbf{z}^n))\,.
\ee
This is mostly doable, but can be greatly simplified by eliminating $\sb_k$ in terms of the Hilbert series of $\cC_k$.

To this end, introduce the $\p$-cohomology groups
\be\label{cHk}
\cH_k = \frac{\cK_k}{\cB_k}
\ee
which measure whether a $\p$-closed state is $\p$-exact or not. Let $\sk_k({\bf x},{\bf y},{\bf z})$, $\sc_k({\bf x},{\bf y},{\bf z})$ and $\sh_k({\bf x},{\bf y},{\bf z})$ denote the Hilbert series of $\cK_k$, $\cC_k$ and $\cH_k$ respectively. From the first isomorphism theorem, we have
\be
\cB_{k+1} = \frac{\cC_k}{\cK_k}\,.
\ee
At the level of Hilbert series, this yields the relation $\sb_{k+1} = \sc_k - \sk_k$. Similarly, from the definition of the cohomology groups \eqref{cHk}, we also obtain $\sh_k = \sk_k-\sb_k$. Eliminating $\sk_k$ from these two identities, we obtain the recursion
\be
\sb_k = \sc_k-\sh_k-\sb_{k+1}\,.
\ee
Using this, we can reduce the graded sum \eqref{fbcovdef} to
\be\label{fbcov1}
f_\text{BCOV} = \sum_{k=1}^7(-1)^{k+1}k\,\sc_k-\sum_{k=1}^7(-1)^{k+1}k\,\sh_k\,.
\ee
It turns out that $\cH_k=0$ for all $k\neq2,3$, and is very sparse even for $k=2,3$. This will allow us to compute $f_\text{BCOV}$ directly from $\sc_k$, effectively sidestepping the $\p$-closure constraint.

\paragraph{Fugacity assignments.} Having chosen a basis of states that are monomials in $Z^A,W_I$ and $\d Z^A,\d W_I$, it becomes straightforward to build their Hilbert series. On the boundary, we counted operators. In the bulk, one instead counts the BCOV states that they couple to. As with the boundary theory, we make the fugacity assignments
\be\label{fugass}
\begin{split}
    &[W_1] = x_1\,,\qquad[W_2] = x_2\,,\qquad[W_3] = x_3\,,\qquad[W_4] = x_4\,,\\
    &[\mu^{\dot0}] = y_1^{-1}\,,\qquad[\mu^{\dot1}] = y_2^{-1}\,,\qquad[\lambda_0] = z_1^{-1}\,,\qquad[\lambda_1] = z_2^{-1}\,.
\end{split}
\ee
These are taken modulo the relations \eqref{xyzeq} and \eqref{eq:proj4} as in section \ref{ssec:multifugacity}. The volume form $\Omega_X$ has fugacity $[\Omega_X]=x_1x_2x_3x_4/y_1y_2z_1z_2$, which is equal to $1$ but is better kept as it is for subsequent simplifications. 

Again, as a simple example, the states in \eqref{72states} get assigned the fugacities
\be
\big[\xi_{7,2}[{\bf p},{\bf q},{\bf r}]\big] = x_1^{p_1+1}x_2^{p_2+1}x_3^{p_3+1}x_4^{p_4+1}y_1^{q_1}y_2^{q_2}z_1^{-r_1-1}z_2^{-r_2-1}\,.
\ee
Summing over these, we obtain the Hilbert series of $\cC_7$,
\be
\begin{split}
    \sc_7 &= \sum_{p_1=0}^\infty\sum_{p_2=0}^\infty\sum_{p_3=0}^\infty\sum_{p_4=0}^\infty\sum_{q_1=0}^\infty\sum_{q_2=0}^\infty\sum_{r_1=0}^\infty\sum_{r_2=0}^\infty\big[\xi_{7,2}[{\bf p},{\bf q},{\bf r}]\big]\,\delta_{r_1+r_2,p_1+p_2+p_3+p_4+q_1+q_2+2}\\
    &= -\frac{x_1x_2x_3x_4}{z_1-z_2}\left(\frac{z_1^{-3}}{D(z_1)}-\frac{z_2^{-3}}{D(z_2)}\right)\,.
\end{split}
\ee
Here, we have defined the denominator
\be\label{d(z)}
D(z) = (1-z^{-1}x_1)(1-z^{-1}x_2)(1-z^{-1}x_3)(1-z^{-1}x_4)(1-z^{-1}y_1)(1-z^{-1}y_2)\,.
\ee
We will encounter this again in what follows.

\paragraph{Passing to non-projective forms.} For $k=7$, the computation of $(k,*)$-states was simplified by the fact that they were all proportional to the $(7,0)$-volume form $\Omega_X$. For other values of $k$, we need to contend with the issue that our states are generated by wedge products of $\d Z^A,\d W_I$, but such wedge products are not automatically projective. To be projective, a state $\xi_{k,l}$ must be invariant under the $\C^\times$-rescalings $(Z^A,W_I)\sim(\alpha\,Z^A,\alpha^{-1}W_I)$ and must also be annihilated by the Euler vector field
\be
\Upsilon = Z^A\frac{\p}{\p Z^A} - W_I\frac{\p}{\p W_I}\,.
\ee
The $\C^\times$-invariance is easy to impose, but vanishing under contractions with $\Upsilon$ is slightly more involved.

To make progress, it is helpful to work with arbitrary differential forms in $\d Z^A,\d W_I$. This can only be done by working on the ambient space $\til X = \C^8-\C^6$ obtained by taking $\C^8$ with \emph{affine} coordinates $Z^A,W_I$ and removing $\C^6=\{\lambda_\al=0\}$. Our 7-fold $X$ is the quotient of $\til X$ by the $\C^\times$-action $(Z,W)\sim(\alpha\,Z,\alpha^{-1}W)$. Let $\til{\Omega}^{k,0}$ denote the bundle of $(k,0)$-forms on $\til X$. Such forms no longer need vanish when contracted into $\Upsilon$. Also let $\til Y=\{\mu^\dal=0\}\subset\til X$. For $k=1,\dots,8$, denote by
\be
\til\cC_k = \bigoplus_{l=0}^7H^{0,l}_{\til Y}(\til X,\til\Omega^{k,0})^{\C^\times}
\ee
the $\C^\times$-invariant local cohomology of $\til X$ supported at $\til Y$ and valued in $\til\Omega^{k,0}$. We will count the $\C^\times$-invariant local cohomology annihilated by the interior product $\iota_\Upsilon\equiv\Upsilon\lrcorner$ through an inclusion-exclusion argument.

Because $\Upsilon$ is holomorphic, it commutes with taking $\dbar$-cohomology and induces maps on $\til\cC_k$. By definition, our projective local cohomologies are the kernels of the induced maps,
\be\label{Ckker}
\cC_k = \ker(\iota_\Upsilon:\til\cC_k\to\til\cC_{k-1})\,.
\ee
Because $\Upsilon$ is nowhere-vanishing on $\til X$, a $p$-form $\omega$ obeys $\iota_\Upsilon\omega = 0$ if and only if it can be written as $\omega = \iota_\Upsilon\eta$ for some $(p+1)$-form $\eta$. Hence, we can rewrite \eqref{Ckker} as
\be\label{Ckim}
\cC_k = \text{im}(\iota_\Upsilon:\til\cC_{k+1}\to\til\cC_k)\,.
\ee
By the first isomorphism theorem, we can express the image of $\iota_\Upsilon$ as
\be
\text{im}(\iota_\Upsilon:\til\cC_{k+1}\to\til\cC_k) = \frac{\til\cC_{k+1}}{\ker(\iota_\Upsilon:\til\cC_{k+1}\to\til\cC_k)}\,.
\ee
Combining this with \eqref{Ckker} and \eqref{Ckim} leads to
\be\label{CtoCtilde}
\cC_k = \frac{\til\cC_{k+1}}{\cC_{k+1}}\,.
\ee
At the level of Hilbert series, this yields a recursion relation for $\sc_k$,
\be
\sc_k = \til\sc_{k+1} - \sc_{k+1}\,,
\ee
in terms of the Hilbert series $\til\sc_k({\bf x},{\bf y},{\bf z})$ of the non-projective cohomology $\til\cC_k$. This can be solved to obtain
\be\label{ctoctilde}
\sc_k = \sum_{p=k+1}^8(-1)^{p-k-1}\til\sc_{p}\,.
\ee
Here, the range of $k$ is $k=1,2,\dots,7$, and we have used the fact that $\sc_8 = 0$ to truncate the recursion. Though note that $\til\sc_8\neq0$ because $\til X$ can support $(8,0)$-forms.

Using \eqref{ctoctilde}, we can eliminate $\sc_k$ in the graded sum \eqref{fbcov1} in favor of the series $\til\sc_k$. This gives us our main computational formula
\be\label{fbcov2}
f_\text{BCOV} = \frac12\sum_{k=1}^7(-1)^{k+1}k(k+1)\,\til\sc_{k+1}-\sum_{k=1}^7(-1)^{k+1}k\,\sh_k\,.
\ee
We may now count the $\C^\times$-invariant local cohomology of the ambient $\til X$ with nary a care for the Euler vector field or $\p$-exactness.

\paragraph{State counting.} A generic element $\til\xi_{k,l}$ of the non-projective cohomology ring $\til\cC_k$ is proportional to derivatives of $\bar\delta^2(\mu)$ and lies in the span of $k$-fold wedge products of $\d Z^A,\d W_I$. Because the $\CO(-1)$ fibers are affine, just like \eqref{72states}, we can work with a basis of states that are monomials in the $W_I$,
\begin{multline}\label{xitildekl}
    \til\xi_{k,l} = f_n(\lambda)W_1^{p_1}W_2^{p_2}W_3^{p_3}W_4^{p_4}\p_{\dot0}^{q_1}\p_{\dot1}^{q_2}\bar\delta^2(\mu)\\
    \times(\d W_1)^{r_1}(\d W_2)^{r_2}(\d W_3)^{r_3}(\d W_4)^{r_4}(\d Z^1)^{s_1}(\d Z^2)^{s_2}(\d Z^3)^{s_3}(\d Z^4)^{s_4}\,.
\end{multline}
Here, we have $n\in\Z$, $p_1,p_2,p_3,p_4,q_1,q_2\geq0$, and $r_1,r_2,r_3,r_4,s_1,s_2,s_3,s_4\in\{0,1\}$. Moreover, the powers of the differentials $\d W_I,\d Z^A$ must add up to
\be
r_1+r_2+r_3+r_4+s_1+s_2+s_3+s_4=k\,,
\ee
where $k$ can go up to $k=8$. It will be convenient to define
\be
r = r_1+r_2+r_3+r_4\,.
\ee
We also demand $\C^\times$-invariance. To do this, we take the coefficient $f_n(\lambda)$ to be a section of $H^{0,*}(\P^1,\CO(n))$ and fix $n$ by matching homogeneity,
\be\label{npqr}
n = p_1+p_2+p_3+p_4+q_1+q_2+2+2r-k\,.
\ee
When $n\geq0$, only the zeroth cohomology of $\P^1$ is nonvanishing. So in most cases, the coefficients $f_n(\lambda)$ are spanned by degree $n$ monomials in $\lambda_0,\lambda_1$. But because of the $-k$ in \eqref{npqr}, one occasionally encounters the first cohomology of $\P^1$ whose Dolbeault representatives are $(0,1)$-forms proportional to $\bar e^0=\la\hat\lambda\,\d\hat\lambda\ra/\|\lambda\|^4$. All in all, the antiholomorphic form-degree of $\til\xi_{k,l}$ can only take two values: $l=2,3$. Mostly, $\til\cC_k$ is supported in $l=2$.

To count such states, we first need to build the Hilbert series of the cohomology of $\P^1$. As we saw in the counting of the states of $\cN=4$ sdYM in section \ref{ssec:multifugacity}, it is easier to assign fugacities to \v{C}ech representatives. For $n\geq0$, $H^{0,1}(\P^1,\CO(n))=0$, whereas the \v{C}ech representatives of $H^0(\P^1,\CO(n))$ are given by
\be
\lambda_0^{n-k}\lambda_1^k\,,\qquad 0\leq k\leq n\,.
\ee
With the fugacity assignments \eqref{fugass}, the associated Hilbert series is computed as the refined Euler character
\be\label{chiP1pos}
\chi_{\P^1}(\CO(n)) = \sum_{k=0}^n\frac{1}{z_1^{n-k}z_2^k} = \frac{z_1^{-1-n}-z_2^{-1-n}}{z_1^{-1}-z_2^{-1}}\,,\qquad n\geq0\,.
\ee
For $n=-1$, one finds $H^0(\P^1,\CO(-1))=H^{0,1}(\P^1,\CO(-1))=0$, so the corresponding Euler character vanishes. For $n\leq-2$, we have the `dual' result $H^0(\P^1,\CO(n))=0$, whereas $H^{0,1}(\P^1,\CO(n))$ is generated by the \v{C}ech representatives
\be
\lambda_0^{n+k}\lambda_1^{-k}\,,\qquad 1\leq k\leq-n-1\,.
\ee
This is counted by the Euler character
\be\label{chiP1neg}
\chi_{\P^1}(\CO(n)) = -\sum_{k=1}^{-n-1}\frac{1}{z_1^{n+k}z_2^{-k}} = \frac{z_1^{-1-n}-z_2^{-1-n}}{z_1^{-1}-z_2^{-1}}\,,\qquad n\leq-2\,,
\ee
where the minus sign comes from these classes residing in degree $1$ cohomology. So we obtain a uniform formula for all $\CO(n)$ (notice that $z_1^{-1-n}-z_2^{-1-n}=1-1=0$ for $n=-1$, as desired). For every $n\in\Z$, this counts the choices of $f_n(\lambda)$ that can enter the states \eqref{xitildekl}. 

Next, we build a Hilbert series for the holomorphic form generators $(\d W)^{r_1}\cdots(\d Z^4)^{s_4}$ appearing in the second line of \eqref{xitildekl}. We can have $r$ $\d W_I$ factors and $k-r$ $\d Z^A$ factors, for the ranges $0\leq r\leq 4$ and $0\leq k-r\leq 4$. So in fact $r$ must fall within the range
\be
\max(0,k-4)\leq r\leq\min(4,k)\,.
\ee
For a given $k,r$, the choices of $k$-fold wedge products of $\d W_I,\d Z^A$ with $r$ $\d W_I$ factors is counted by the series
\be
\mathsf{e}_r(x_1,x_2,x_3,x_4)\,\mathsf{e}_{k-r}(y_1^{-1},y_2^{-1},z_1^{-1},z_2^{-1})\,,
\ee
where $\mathsf{e}_r(t_1,\dots,t_n)$ denotes the $r^\text{th}$ elementary symmetric polynomial in $n$ variables,
\be\label{esym}
\mathsf{e}_r(t_1,\dots,t_n) = \sum_{1\leq a_1<a_2<\cdots<a_r\leq n}t_{a_1}t_{a_2}\cdots t_{a_r}\,.
\ee
For example, when $k=r=1$, we just count the four generators $\d W_I$ whose fugacity sum is $x_1+x_2+x_3+x_4 = \mathsf{e}_1(x_1,x_2,x_3,x_4)$, etc.

Using these building blocks, we can reduce the Hilbert series of the states \eqref{xitildekl} to
\be\label{sckexp}
\til\sc_k = \sum_{r=\max(0,k-4)}^{\min(4,k)}\mathsf{e}_r(x_1,x_2,x_3,x_4)\,\mathsf{e}_{k-r}(y_1^{-1},y_2^{-1},z_1^{-1},z_2^{-1})\,A_{2r-k}({\bf x},{\bf y},{\bf z})\,.
\ee
Here, $A_{2r-k}({\bf x},{\bf y},{\bf z})$ counts states $\til\xi_{k,l}$ in the span of wedge products of $r$ $\d W$'s and $k-r$ $\d Z$'s. It is a special case of the more general count $A_m({\bf x},{\bf y},{\bf z})$ of monomials of the form
\be
f_n(\lambda)W_1^{p_1}W_2^{p_2}W_3^{p_3}W_4^{p_4}\p_{\dot0}^{q_1}\p_{\dot1}^{q_2}\bar\delta^2(\mu)
\ee
that carry weight $m$ under $(Z,W)\sim(\alpha\,Z,\alpha^{-1}W)$. Such monomials satisfy the constraint
\be
n = p_1+p_2+p_3+p_4+q_1+q_2+2+m
\ee
which of course reduces to \eqref{npqr} when one sets $m=2r-k$ as done in the formula \eqref{sckexp}.

One can easily build $A_m({\bf x},{\bf y},{\bf z})$ for generic $m\in\Z$ by using our result for the refined Euler character of $\P^1$,
\begin{multline}
    A_m = \sum_{p_1=0}^\infty\sum_{p_2=0}^\infty\sum_{p_3=0}^\infty\sum_{p_4=0}^\infty\sum_{q_1=0}^\infty\sum_{q_2=0}^\infty x_1^{p_1}x_2^{p_2}x_3^{p_3}x_4^{p_4}y_1^{q_1+1}y_2^{q_2+1}\\
    \times\chi_{\P^1}(\CO(p_1+p_2+p_3+p_4+q_1+q_2+2+m))\,.
\end{multline}
Plugging in the formula $\chi_{\P^1}(\CO(n)) = (z_1^{-1-n}-z_2^{-1-n})/(z_1^{-1}-z_2^{-1})$ valid for all $n\in\Z$ and performing the sums results in
\be\label{Am}
A_m = -\frac{y_1y_2z_1z_2}{z_1-z_2}\left(\frac{z_1^{-3-m}}{D(z_1)}-\frac{z_2^{-3-m}}{D(z_2)}\right)\,,
\ee
with $D(z)$ being the denominator that we previously encountered in \eqref{d(z)} while counting the states of $\xi_{7,2}$. Substituting this result for $A_m$ in \eqref{sckexp} completes the determination of the non-projective series $\til\sc_k$ that enter our master formula \eqref{fbcov2} for $f_\text{BCOV}$.

\paragraph{$\p$-cohomology.} The only other ingredients that go into \eqref{fbcov2} are the Hilbert series of the $\p$-cohomology groups \eqref{cHk}. Since we are computing $\p$-cohomology on $\dbar$-cohomology, this can be thought of as a spectral sequence computation of local de Rham cohomology. The result is very simple because our geometry is essentially affine, except for cohomology coming from the twistor $\P^1$.

Only two de Rham cohomology rings are nonzero, $\cH_2$ and $\cH_3$, and both of them have only one generator. The isolated generator of $\cH_2$ is just the delta function with support at $\mu^\dal=0$,
\be\label{xi22cl}
\xi_{2,2} = \bar\delta^2(\mu)\,\d^2\mu\,.
\ee
This is obviously $\p$-closed, but it fails to be $\p$-exact, which follows from the fact that it is the Poincar\'e dual of the singular locus $Y$. Moreover, it has fugacity $1$, so it definitely needs to be removed from $f_\text{BCOV}$ lest it cause $\PE[f_\text{BCOV}]$ to diverge. Similarly, the isolated generator of $\cH_3$ is
\be\label{xi33cl}
\xi_{3,3} = \bar\delta^2(\mu)\,\bar e^0\,\D^3Z\,.
\ee
Again, it is easily verified that this is $\p$-closed but not $\p$-exact. The \v{C}ech representative of $\bar e^0$ is $1/\lambda_0\lambda_1$. So we assign $\bar e^0$ the corresponding fugacity $[\bar e^0]=z_1z_2$. At the same time, $\D^3Z$ has fugacity $[\D^3Z]=1/y_1y_2z_1z_2$. So the net state again has fugacity $1$ and needs to be removed from $f_\text{BCOV}$. 

From these isolated states and their fugacities, we obtain\footnote{Both these states are fermionic, but we have already incorporated the signs from the holomorphic form grading as explicit signs in \eqref{fbcov2}. So the sign in $\sh_3$ as opposed to $\sh_2$ is just the sign coming from the antiholomorphic form grading.}
\be\label{hhs}
\sh_2=1\,,\quad\sh_3=-1\,,
\ee
and $\sh_k=0$ otherwise. The second sum on the right hand side of \eqref{fbcov2} subtracts precisely these two contributions and renders the plethystic exponential well-defined. In fact, the wavefunction \eqref{xi33cl} arises as the state that couples to the operator $\Tr(c_0)$ on the worldvolume of the backreacting  $\D5$'s. Removing $\Tr(c_0)$ from the partition function of the boundary sdYM theory corresponds to removing this closed-but-not-exact state from the bulk BCOV theory. It would be interesting to find a similar interpretation for the other state \eqref{xi22cl}.

\paragraph{Assembling $\boldsymbol{f}_\text{BCOV}$.} Substituting the $\p$-cohomology Hilbert series \eqref{hhs} in our master formula \eqref{fbcov2}, we obtain
\be
f_\text{BCOV}^{\cN=4} = \til\sc_2 - 3\,\til\sc_3 +6\,\til\sc_4 - 10\,\til\sc_5 + 15\,\til\sc_6 - 21\,\til\sc_7 + 28\,\til\sc_8 + 5\,.
\ee
The last step is to combine this with the results of our computation for $\til\sc_k$ in \eqref{sckexp}, \eqref{Am} and simplify the sum. When the dust settles, we obtain a remarkably simple answer for the signed count of single-particle BCOV states,
\be
f_\text{BCOV}^{\cN=4} = S(z_1) + S(z_2)\,,
\ee
presented in terms of the sum
\be\label{Sz}
S(z) = \frac{x_1}{z-x_1}+\frac{x_2}{z-x_2}+\frac{x_3}{z-x_3}+\frac{x_4}{z-x_4}-\frac{y_1}{z-y_1}-\frac{y_2}{z-y_2}\,.
\ee
From this, we obtain our main result for the graded partition function of closed twistor strings dual to $\cN=4$ sdYM,
\be\label{Zbcov4}
\begin{split}
Z_\text{BCOV}^{\cN=4}(\fx,\fy,\fz) &= \PE[S(z_1)]\,\PE[S(z_2)] \\
&= \prod_{n=1}^\infty \frac{(1-y_1^n/z_1^n)(1-y_2^n/z_1^n)}{(1-x_1^n/z_1^n)(1-x_2^n/z_1^n)(1-x_3^n/z_1^n)(1-x_4^n/z_1^n)} \\
& \hspace{1.5cm}\times\frac{(1-y_1^n/z_2^n)(1-y_2^n/z_2^n)}{(1-x_1^n/z_2^n)(1-x_2^n/z_2^n)(1-x_3^n/z_2^n)(1-x_4^n/z_2^n)} \, .
\end{split}
\ee
Naively, we would expect this to be dual to the $N\to\infty$ limit of the graded partition function $Z_N[f_{\cN=4}]$ of $\cN=4$ sdYM. But this expectation fails miserably!

When the fugacities are specialized to a single fugacity $t$ by means of \eqref{tspec}, we obtain
\be
f_\text{BCOV}^{\cN=4}(t) = \frac{8t+4t^2}{1-t^2}\,.
\ee
Its plethystic exponential gives the unrefined graded partition function
\be
\begin{split}
Z_\text{BCOV}^{\cN=4}(t) &= \PE\bigg[\frac{8t+4t^2}{1-t^2}\bigg]\\
&= 1+8 t+40 t^2+160 t^3+552 t^4+1712
   t^5+4896 t^6+13120 t^7+33320
   t^8\\
   &\hspace{2.4cm}+80872 t^9+188784
   t^{10}+425952 t^{11}+932640
   t^{12}+\mathrm{O}(t^{13})\,.
\end{split}
\ee
This manifestly fails to match the expansion of $Z_\infty[f_{\cN=4}]$ found in \eqref{Zinf4}. So what's going on? Does this signify a failure of the holographic duality conjectured in \cite{Sharma:2025ntb}?

It turns out that the duality \emph{does} hold, but unlike other instances of holography, even at infinite $N$, the bulk closed string spectrum is \emph{not} sufficient to reproduce the boundary operator spectrum. In section \ref{ssec:gge4}, as a corollary of our giant graviton expansion for $Z_N[f_{\cN=4}]$, we will find that the closed twistor string partition function receives instanton corrections. Our instantons happen to be precisely the degree-1 D-instantons that Witten encountered in the computation of scattering amplitudes from twistor strings \cite{Witten:2003nn}. The open strings on the D-instantons will systematically correct the closed string counting to precisely reproduce an accounting of the boundary spectrum.


\subsection{$\cN=1$ sdYM}
\label{ssec:bcov1}

All the tricks developed in the $\cN=4$ case can be recycled to count BCOV states in the twistor string dual of $\cN=1$ sdYM.

Recall from section \ref{ssec:ch1} that the bulk dual of $\cN=1$ sdYM is given by the B-model on the Calabi-Yau 5-fold $M=\text{Tot}(\CO(-1)\oplus\CO(-3)\to\PT)$ in presence of the backreaction \eqref{xi21}. The closed string field theory of this B-model is again BCOV theory on $M$. Its field content consists of $\p$-closed $(k,l)$-forms $\xi_{k,l}$ on $M$ with $k=1,\dots,5$ and $l=0,\dots,5$. 

Let $S$ denote the subspace $\{\mu^\dal=0\}\subset M$. It is a copy of $\CO(-1)\oplus\CO(-3)\to\P^1$, with the base being the twistor line of the origin of $\R^4$. In this section, let $\Omega^{k,0}$ be the bundle of $(k,0)$-forms on $M$, and $\Omega^{k,0}_\text{ex}$ the subbundle of $\p$-exact $(k,0)$-forms. The BCOV states dual to boundary operators can again be computed in perturbation theory around representatives of the local cohomology rings
\be
\xi_{k,l} \in \cB_k \vcentcolon= \bigoplus_{l=0}^5 H^{0,l}_S(M,\Omega^{k,0}_\text{ex})
\ee
that capture states in the absence of backreaction. We now repeat the steps of the previous section to obtain their Hilbert series.

The first step is to pass from the cohomology rings $\cB_k$ containing  $\p$-exact representatives to the unconstrained cohomology rings
\be
\cC_k = \bigoplus_{l=0}^5 H^{0,l}_S(M,\Omega^{k,0})\,.
\ee
We can obtain $\cB_k$ as the images of the maps $\p:\cC_{k-1}\to\cC_k$ induced by the holomorphic exterior derivative on $M$,
\be
\cB_k = \text{im}(\p:\cC_{k-1}\to\cC_k)\,.
\ee
Similarly, we define the kernels and $\p$-cohomology rings
\be
\cK_k = \ker(\p:\cC_k\to\cC_{k+1})\,,\qquad \cH_k = \frac{\cK_k}{\cB_k}\,.
\ee
Associate the Hilbert series $\sc_k$, $\sb_k$, $\sk_k$ and $\sh_k$ to $\cC_k$, $\cB_k$, $\cK_k$ and $\cH_k$ respectively.

As before, the definition of $\cH_k$ gives us the relation $\sh_k = \sk_k-\sb_k$; while an application of the first isomorphism theorem in the form of $\cB_{k+1} = \cC_k/\cK_k$ yields $\sb_{k+1} = \sc_k-\sk_k$. Eliminating $\sk_k$ from these yields $\sb_k = \sc_k-\sh_k-\sb_{k+1}$. Using this to recursively solve for $\sb_k$, we can compute the single-particle graded partition function of BCOV on $M$,
\be
    f_\text{BCOV} = \sum_{k=1}^5(-1)^{k+1}\sb_k = \sum_{k=1}^5(-1)^{k+1}k\,\sc_k-\sum_{k=1}^5(-1)^{k+1}k\,\sh_k\,.
\ee
Once again, we have dropped the ``topological degrees of freedom'' $\xi_{0,l}$ by not including a $k=0$ term in the sum.

The next step is to pass to non-projective forms. Let $w_1,w_2$ denote the coordinates on the $\CO(-1)$ and $\CO(-3)$ fibers respectively. Take $\til M = \C^6-\C^4$ to be the space obtained from $\C^6$ with affine coordinates $Z^A,w_i$ by removing the locus $\C^4 = \{\lambda_\al=0\}$, and let $\til S = \{\mu^\dal=0\}\subset\til M$. One obtains $M$ by quotienting $\til M$ by the $\C^\times$-action $(Z^A,w_1,w_2)\sim(\alpha\,Z^A,\alpha^{-1}w_1,\alpha^{-3}w_2)$. Differential forms on $M$ are equivalent to $\C^\times$-invariant forms on $\til M$ that are additionally in the kernel of contraction with the Euler vector field
\be
\Upsilon = Z^A\frac{\p}{\p Z^A} - w_1\frac{\p}{\p w_1} - 3\,w_2\frac{\p}{\p w_2}\,.
\ee
Instead of working with the kernel of the interior product $\iota_\Upsilon$, we will find it easier to work with the $\C^\times$-invariant, non-projective local cohomology of $\til M$,
\be
\til\cC_k = \bigoplus_{l=0}^5 H^{0,l}_{\til S}(\til M,\til\Omega^{k,0})^{\C^\times}\,,
\ee
where $\til\Omega^{k,0}$ is the bundle of $(k,0)$-forms on $\til M$, and $k$ can now run up to $6$.

Since $\Upsilon$ is nowhere-vanishing on $\til M$, a form $\omega$ in the kernel of $\iota_\Upsilon$ must necessarily be in the image of $\iota_\Upsilon$. So we obtain the same relation
\be
\begin{split}
    \cC_k &= \ker(\iota_\Upsilon:\til\cC_k\to\til\cC_{k-1}) = \text{im}(\iota_\Upsilon:\til\cC_{k+1}\to\til\cC_{k}) \\
    &= \frac{\til\cC_{k+1}}{\ker(\iota_\Upsilon:\til\cC_{k+1}\to\til\cC_{k})} = \frac{\til\cC_{k+1}}{\cC_{k+1}}
\end{split}
\ee
that we obtained in the $\cN=4$ case in equation \eqref{CtoCtilde}. This yields the recursion
\be
\sc_k = \til\sc_{k+1} - \sc_{k+1}\,,
\ee
which can be used to eliminate $\sc_k$ in favor of the Hilbert series $\til\sc_k$ of $\til\cC_k$. This results in
\be\label{fbcov3}
f_\text{BCOV} = \frac12\sum_{k=1}^5(-1)^{k+1}k(k+1)\,\til\sc_{k+1}-\sum_{k=1}^5(-1)^{k+1}k\,\sh_k\,,
\ee
which we can again evaluate by freely enumerating states that may or may not be projective or $\p$-exact.

A general element of the cohomology $\til\cC_k$ belongs to the span of the monomials
\be
\til\xi_{k,l} = f_n(\lambda)\,w_1^{p_1}w_2^{p_2}\p_{\dot0}^{q_1}\p_{\dot1}^{q_2}\bar\delta^2(\mu)\,(\d w_1)^{r_1}(\d w_2)^{r_2}(\d Z^1)^{s_1}(\d Z^2)^{s_2}(\d Z^3)^{s_3}(\d Z^4)^{s_4}\,,
\ee
where $n\in\Z$, $p_1,p_2,q_1,q_2\geq0$, and $r_1,r_2,s_1,s_2,s_3,s_4\in\{0,1\}$. The coefficients $f_n(\lambda)$ are cohomology classes in $H^{0,*}(\P^1,\CO(n))$. For this to be a $(k,l)$-form, $f_n(\lambda)$ must be a $(0,l-2)$-form, which requires $l=2,3$. The holomorphic form degree also fixes
\be
r_1+r_2+s_1+s_2+s_3+s_4=k\,,
\ee
where $0\leq k\leq 6$. Demanding $\C^\times$-invariance additionally fixes
\be
n = p_1+3p_2+q_1+q_2+2+2r_1+4r_2-k\,.
\ee
Modulo these constraints, we can enumerate states at fixed $k$ to obtain $\til\sc_k$.

We make the following fugacity assignments:
\be
[w_1] = x_1\,,\qquad [w_2] = x_2\,,
\ee
while the fugacities of $\mu^\dal,\lambda_\al$ are taken to be the same as in \eqref{fugass}. In the $\cN=1$ case, they obey the relations \eqref{eq:proj1} and \eqref{xyzeq1}. The fugacity polynomial that counts the $k$-fold wedge products of $\d w_i,\d Z^A$ is given by
\be\label{x12e}
x_1^{r_1}x_2^{r_2}\,\mathsf{e}_{k-r_1-r_2}(y_1^{-1},y_2^{-1},z_1^{-1},z_2^{-1})\,.
\ee
As in equation \eqref{esym}, for $0\leq r\leq n$, the expression $\mathsf{e}_{r}(t_1,\dots,t_n)$ denotes the $r^\text{th}$ elementary symmetric polynomial in $n$ variables. As a convenient piece of notation, we extend its definition to $r\in\Z$ by setting $\mathsf{e}_{r}(t_1,\dots,t_n) = 0$ for $r>n$ or $r<0$. Then the expression \eqref{x12e} is understood to be zero for $k>r_1+r_2+4$ or $k<r_1+r_2$.

The complete Hilbert series $\til\sc_k$ is given by
\be\label{ctilk}
\til\sc_k = \sum_{r_1=0}^1\sum_{r_2=0}^1x_1^{r_1}x_2^{r_2}\,\mathsf{e}_{k-r_1-r_2}(y_1^{-1},y_2^{-1},z_1^{-1},z_2^{-1})\,A_{2r_1+4r_2-k}\,,
\ee
where $A_m$ counts monomials of the form
\be
f_n(\lambda)\,w_1^{p_1}w_2^{p_2}\p_{\dot0}^{q_1}\p_{\dot1}^{q_2}\bar\delta^2(\mu)
\ee
that carry net weight $m$ under our $\C^\times$-action. This choice of net weight fixes the value of the weight of $f_n(\lambda)$ to be
\be
n = p_1+3p_2+q_1+q_2+2+m\,.
\ee
Therefore, $A_m$ is computed by the sum
\be
A_m = \sum_{p_1=0}^\infty\sum_{p_2=0}^\infty\sum_{q_1=0}^\infty\sum_{q_2=0}^\infty x_1^{p_1}x_2^{p_2}y_1^{q_1+1}y_2^{q_2+1}\chi_{\P^1}(\CO(p_1+3p_2+q_1+q_2+2+m))\,.
\ee
Finally, an application of $\chi_{\P^1}(\CO(n)) = (z_1^{-1-n}-z_2^{-1-n})/(z_1^{-1}-z_2^{-1})$ yields
\be
A_m = -\frac{y_1y_2z_1z_2}{z_1-z_2}\left(\frac{z_1^{-3-m}}{D(z_1)}-\frac{z_2^{-3-m}}{D(z_2)}\right)\,,
\ee
where $D(z)$ denotes the denominator
\be
D(z) = (1-z^{-1}x_1)(1-z^{-3}x_2)(1-z^{-1}y_1)(1-z^{-1}y_2)\,.
\ee
Substituting this into \eqref{ctilk} determines the non-projective Hilbert series $\til\sc_k$.

The $\p$-cohomology is spanned by the same two states \eqref{xi22cl}, \eqref{xi33cl} that we encountered in the $\cN=4$ case, yielding $\sh_k=0$ except for $\sh_2=1$ and $\sh_3=-1$. Hence, our formula \eqref{fbcov3} for the BCOV single-particle count expands out to give
\be
f_{\text{BCOV}}^{\cN=1} = \til\sc_2 - 3\,\til\sc_3 + 6\,\til\sc_4 - 10\,\til\sc_5 + 15\,\til\sc_6 + 5\,.
\ee
Plugging in our result \eqref{ctilk} for the various $\til\sc_k$ and simplifying, we obtain
\be
f_{\text{BCOV}}^{\cN=1} = \widetilde{S}(z_1)+\widetilde{S}(z_2)\,,
\ee
with $\widetilde{S}(z)$ now being the sum
\be\label{SzN=1}
\widetilde{S}(z) = \frac{x_1}{z-x_1}+\frac{x_2}{z^3-x_2}-\frac{y_1}{z-y_1}-\frac{y_2}{z-y_2}\,.
\ee
Thus, the graded partition function of closed twistor strings dual to $\cN=1$ sdYM is given by
\be\label{Zbcov1}
\begin{split}
Z_\text{BCOV}^{\cN=1}(\fx,\fy,\fz) &= \PE[\widetilde{S}(z_1)]\,\PE[\widetilde{S}(z_2)] \\
&= \prod_{n=1}^\infty \frac{(1-y_1^n/z_1^n)(1-y_2^n/z_1^n)}{(1-x_1^n/z_1^n)(1-x_2^n/z_1^{3n})} \frac{(1-y_1^n/z_2^n)(1-y_2^n/z_2^n)}{(1-x_1^n/z_2^n)(1-x_2^n/z_2^{3n})} \, ,
\end{split}
\ee
which takes the same general form as the $\cN=4$ case \eqref{Zbcov4}, but with a slightly different definition of $\widetilde{S}(z)$.

Applying the fugacity specialization \eqref{tspec1}, we obtain the unrefined result
\be
f_\text{BCOV}^{\cN=1}(t) = \frac{2t+2t^3}{(1+t)(1-t^3)}\,.
\ee
Computing its plethystic exponential gives
\be
\begin{split}
Z_\text{BCOV}^{\cN=1}(t) &= \PE\bigg[\frac{2t+2t^3}{(1+t)(1-t^3)}\bigg]\\
&= 1+2 t+t^2+4 t^3+6 t^4+2 t^5+12 t^6+16 t^7+5
   t^8\\
   &\hspace{3cm}+28 t^9+36 t^{10}+12 t^{11}+60
   t^{12}+\mathrm{O}(t^{13})\,.
\end{split}
\ee
Again, this fails to match the large $N$ graded partition function of $\cN=1$ sdYM found in \eqref{Zinf1}. In section \ref{ssec:gge1}, we will cure this mismatch by means of Witten-type D-instanton corrections. 


\section{Partition functions of open strings}
\label{sec:openstrings}

In this section, we study the D-branes relevant to the chiral holography realizations of $\cN=4$ and $\cN=1$ sdYM. As reviewed in section \ref{sec:chiralholography}, these theories are engineered by D5-branes wrapping $\PT$ in the B-model on suitable Calabi-Yau manifolds fibering over $\PT$.

We begin by discussing the categories of D-branes supported on $\PT$ and, more generally, on total spaces of holomorphic vector bundles over $\P^1$. Each such category admits two generators, with natural choices given by the pullbacks of $\CO_{\P^1}(s)$ and $\CO_{\P^1}(s+1)$, for any $s\in\Z$, to the total spaces of the bundles. We then develop a general method for computing the single-letter graded partition functions of open strings on these D-branes. Applying this method to D5-branes on $\PT$ with trivial Chan-Paton bundle, we recover the single-letter graded partition functions of free/self-dual $\cN=4$ and $\cN=1$ sYM, computed previously in sections \ref{sec:partition} and \ref{sec:sdpartition}. Including both types of generators, we obtain two-node matrices of single-letter graded partition functions $\mathbf{f}$.

Finally, motivated by Murthy's GGE reviewed in section \ref{ssec:murthy}, we turn to D-branes supported on submanifolds transverse to $\PT$ (relative to the twistorial $\P^1$) in our Calabi-Yau targets. These are D9-branes in the $\cN=4$ case and D5-branes in the $\cN=1$ case. They are closely related to Witten's D-instantons \cite{Witten:2003nn,Bittleston:2024efo,Sharma:2025ntb}. The categories of such transverse branes likewise admit two generators. Applying the same counting method, we compute their matrices of single-letter graded partition functions of open strings. These are the data $\mathbf{f}^!$ that (after an appropriate involution of the fugacities) will enter the corrections in the two-node generalization of Murthy's expansion discussed in section \ref{sec:ggemain}.


\subsection{The category of branes}
\label{ssec:category}

For any Calabi-Yau manifold $\cM$, the category of branes in the B-model on $\cM$ is the bounded derived category of coherent sheaves on $\cM$ \cite{Kontsevich:1994dn,Douglas:2000gi,Aspinwall:2004jr}, usually denoted $D^b\text{Coh}(\cM)$. Objects in the category of branes are represented by complexes of coherent sheaves. 

In such complexes, we refer to the terms in even degrees as branes and those in odd degrees as antibranes. As an abbreviation that suppresses degree, if $\CE$ is a sheaf representing a brane, we will represent its antibrane by the parity-reversed sheaf $\Pi\CE$. The maps between successive terms in a complex describe open string tachyon profiles, with the requirement that successive maps compose to zero being equivalent to the tachyon equations of motion. For example, a brane and its antibrane connected by the identity map form an exact complex and can annihilate completely. More generally, a complex describes a B-model brane obtained by tachyon condensation on its constituent branes and antibranes. Quasi-isomorphic complexes describe the same B-model brane. 

These facts are well-illustrated by the case $\cM=\C^d$. It is a standard result that $D^b\text{Coh}(\C^d)$ is generated by $\CO_{\C^d}$, the structure sheaf of $\C^d$. What this means in practice is that any brane in the B-model on $\C^d$ can be described as a complex of sheaves whose entries are direct sums of $\CO_{\C^d}$. Physically, these entries describe stacks of the spacefilling brane and its antibrane, with smaller branes arising from tachyon condensation and brane-antibrane annihilation.

When constructing our GGEs, we will also need the refined fact that for any closed, complex submanifold $\cS\subset\cM$, the branes contained in $\cS$ can be built from pushforwards to $\cM$ of complexes of coherent sheaves on $\cS$. The category of such branes is referred to as the derived category $D^b_\cS\text{Coh}(\cM)$ of coherent sheaves with support within $\cS$ \cite{Rouquier2008}. This is a full subcategory of $D^b\text{Coh}(\cM)$.

For the study of non-BPS partition functions, we are primarily interested in geometries given by total spaces of holomorphic vector bundles over $\P^1$. Recall from section \ref{ssec:ch4} that the chiral holographic setup for $\cN=4$ sdYM employed the Calabi-Yau 7-fold
\be
X = \Tot(\CO(-1)^{\oplus4}\oplus\CO(1)^{\oplus2}\to\P^1)\,,
\ee
which is just another way of expressing \eqref{X} by using the fact that $\PT$ is the total space of $\CO(1)^{\oplus2}\to\P^1$. Similarly, the chiral holographic setup for $\cN=1$ sdYM, from section \ref{ssec:ch1}, employed the Calabi-Yau 5-fold
\be
M = \Tot(\CO(-1)\oplus\CO(-3)\oplus\CO(1)^{\oplus2}\to\P^1)\,,
\ee
which is the same as \eqref{M}. In each of these cases, the sdYM theory is engineered by wrapping $N$ D5-branes along the isolated copy of $\PT$.

We can easily generalize the above discussion to the geometries $\cM=X,M$. In both of these cases, and more generally for the total space of any holomorphic vector bundle over $\P^1$, the category $D^b\text{Coh}(\cM)$ is generated by just two generators \cite{beilinson1978coherent,bondal2002generators}. Any consecutive pair of sheaves $\CO_{\cM}(s)$, $\CO_{\cM}(s+1)$ for some $s\in\Z$ provides a choice of generators, where $\CO_\cM(s)$ denotes the pullback of the degree-$s$ line bundle $\CO(s)\to\P^1$ by the fibration $\cM\to\P^1$. We will mostly work with a convenient choice like $s=0$ or $s=-1$.

This applies both to $\PT$ and to the transverse geometries considered below. In each case, the two generators provide the elementary branes from which more general branes are constructed. We will illustrate this with an explicit example of constructing giant graviton branes by tachyon condensation in section \ref{ssec:GGfromtachyons}.

This idea of generating all D-branes from a finite set of ``elementary'' branes has also appeared in previous physical studies. For example, \cite{Aspinwall:2002hg} used pullbacks of $\CO_{\P^1}$ and a skyscraper sheaf $\CO_p$ over a reference point $p\in\P^1$ as alternative generators of branes on geometries fibering over $\P^1$. Moreover, we will interpret the generators $\CO_{Y_x}(1)$ for the subspaces $Y_x=\{\mu^\dal=x^{\al\dal}\lambda_\al\}$ as examples of Witten's D-instantons that were employed in the original twistor string calculations of $\cN=4$ scattering amplitudes \cite{Witten:2003nn}.


\subsection{Counting open strings}
\label{ssec:counting}

In this section, we compute the single-letter graded partition functions of open strings between the generator branes. These will count the open string spectra in their linearized worldvolume theories. But as usual, since we are computing graded (or twisted graded) partition functions, it is enough to count (with signs) the linearized cohomology, just as in section \ref{sec:sdpartition}. 


It is very useful to work this out in some generality before restricting to Calabi-Yau's fibering over $\PT$, as we will find multiple occasions to recycle the result. Let us begin with a manifold $\cM$ given by the total space of the bundle
\be\label{gentarget}
\bigoplus_{\ell=1}^m\CO(a_\ell)\oplus\bigoplus_{r=1}^n\CO(b_r)\to\P^1\,,
\ee
and assume that the degrees obey $\sum_\ell a_\ell+\sum_rb_r=-2$ so as to ensure that $\cM$ is Calabi-Yau. Let $\mu^\ell,w_r$ be coordinates on the $\CO(a_\ell)$, $\CO(b_r)$ fibers respectively, and let $\lambda_\al$ be coordinates on the base $\P^1$ as usual. Consider a pair of branes along the complex submanifold
\be
\cS=\{w_r=0\}\subset\cM\,,
\ee
one with gauge bundle $\CO_\cS(d_1)$ and another with gauge bundle $\CO_\cS(d_2)$ for some $d_1,d_2\in\Z$. Let us count the open strings stretching from $\CO_\cS(d_1)$ to $\CO_\cS(\d_2)$.

Let $\cA$ denote the associated open string field. In linear theory, open string states of the B-model obey $\dbar\cA=0$ modulo the gauge symmetry $\delta\cA = \dbar\xi$. Hence, they are elements of Dolbeault cohomology. For open strings stretching from $\CO_\cS(d_1)$ to $\CO_\cS(d_2)$, the states $\cA$ are valued in the $\dbar$-cohomology \cite{Katz:2002gh}
\be
\cA\in\bigoplus_{k,l}H^{0,k}\big(\cS,\wedge^l\cN_{\cS/\cM}\otimes\CO_\cS(d_2-d_1)\big)\,.
\ee
As this only depends on the difference $d_2-d_1$, we only need to count the strings stretching from $\CO_\cS$ to $\CO_\cS(d)$ for arbitrary $d\in\Z$ and set $d=d_2-d_1$ at the end.

Let $\eta^r=\p/\p w_r$ be a frame of the normal bundle $\cN_{\cS/\cM}$. The cohomology ring
\be\label{H0*d}
\bigoplus_{k,l}H^{0,k}\big(\cS,\wedge^l\cN_{\cS/\cM}\otimes\CO_\cS(d)\big)
\ee
consists of polynomials in the Grassmann even coordinates $\mu^\ell$ and Grassmann odd coordinates $\eta^r$, with coefficients in the cohomology of various sheaves on $\P^1$. A typical state looks like
\be
\cA = a^\dagger_{f,\mathbf{p},\mathbf{q}}\,f(\lambda)\,\prod_{\ell=1}^m(\mu^\ell)^{p_\ell}\prod_{r=1}^n(\eta^r)^{q_r}\,,
\ee
where each $p_\ell\in\Z_{\geq0}$ and $q_r\in\{0,1\}$, and $f(\lambda)$ is an element of $\P^1$-cohomology,
\be\label{P1dpq}
f\in\bigoplus_{k=0,1}H^{0,k}\big(\P^1,\CO\big(d-\textstyle\sum_\ell p_\ell a_\ell+\textstyle\sum_r q_rb_r\big)\big)\,.
\ee
The weight of $f$ has been chosen such that the state $\cA$ has total weight $d$. Here, we have used the fact that $\mu^\ell$ has weight $a_\ell$ whereas $\eta^r$ has weight $-b_r$. The coefficients $a_{f,\mathbf{p},\mathbf{q}}^\dagger$ are creation operators for the Fock space of free open strings. They have been labeled by the choice of $f$ as well as vectors $\mathbf{p}=(p_\ell)_{1\leq\ell\leq m}$ and $\mathbf{q} = (q_r)_{1\leq r\leq n}$. A typical state of a single open string is obtained by acting with such a creation operator on the open string vacuum,
\be
a^\dagger_{f,\mathbf{p},\mathbf{q}}|0\ra\,.
\ee
We will count such single-string states.

Let's associate fugacity $y_\ell^{-1}$ to $\mu^\ell$, $x_r$ to $w_r$, and $z_1^{-1},z_2^{-1}$ to $\lambda_0,\lambda_1$ respectively. Because $\eta^r=\p/\p w_r$, it is assigned fugacity $x_r^{-1}$. As the coordinates are projective, the fugacities are only defined projectively:
\be
(x_r,y_\ell,z_\al)\sim(\alpha^{-b_r}x_r,\alpha^{a_\ell}y_\ell,\alpha z_\al)\quad\forall\;\alpha\in\C^\times\,.
\ee
Moreover, for a symmetry of the target space to be a symmetry of the B-model, it must leave invariant the volume form of the target space. The volume form of our target \eqref{gentarget} is $\D\lambda\,\d^m\mu\,\d^nw$, where $\D\lambda\equiv\lambda_0\d\lambda_1-\lambda_1\d\lambda_0$. This has fugacity
\be
[\D\lambda\,\d^m\mu\,\d^nw] = \frac{x_1x_2\cdots x_n}{y_1y_2\cdots y_mz_1z_2}\,.
\ee
Demanding that we only grade our states by symmetries under which the volume form is uncharged gives rise to the constraint
\be\label{fugrelsgen}
x_1x_2\cdots x_n=y_1y_2\cdots y_mz_1z_2
\ee
on the fugacities. This is a generalization of the constraints that we encountered in free as well as sd gauge theories.

To count single-string states, we simply need to count their creation operators. The total fugacity assigned to $\cA$ must be $1$, so the creation operator $a^\dagger_{f,\mathbf{p},\mathbf{q}}$ is assigned the inverse of the fugacity of its accompanying state $f(\lambda)(\mu^1)^{p_1}\cdots(\mu^m)^{p_m}(\eta^1)^{q_1}\cdots(\eta^n)^{q_n}$. Equivalently, we are computing the graded Hilbert series of the dual of the cohomology ring \eqref{H0*d}.  The refined graded Hilbert series counting these creation operators is thus given by the sum
\begin{multline}\label{fd}
\vphi_d = \sum_{p_1=0}^\infty\cdots\sum_{p_m=0}^\infty\sum_{q_1=0}^1\cdots\sum_{q_n=0}^1(-1)^{q_1+\cdots+q_n}y_1^{p_1}\cdots y_m^{p_m}x_1^{q_1}\cdots x_n^{q_n}\\
\times\chi^\text{op}_{\P^1}\big(\CO\big(d-\textstyle\sum_\ell p_\ell a_\ell+\textstyle\sum_r q_rb_r\big)\big)\,.
\end{multline}
In this expression, the second line contains the Euler characteristic of the dual of the $\P^1$-cohomology \eqref{P1dpq} refined by $z_1,z_2$. This counts the possible values that $f(\lambda)$ can take for a given $\mathbf{p},\mathbf{q}$. The remaining sum ranges over all possible values of $\mathbf{p},\mathbf{q}$. The minus signs arise from the odd grading of the $\eta^r$'s.

We already computed the refined Euler character that counted the \v{C}ech representatives of the cohomology of $\P^1$ in equations \eqref{chiP1pos}, \eqref{chiP1neg}. Now suppose the \v{C}ech representative of $f(\lambda)$ is given by $f= \lambda_0^r\lambda_1^s$. The monomial $\lambda_0^r\lambda_1^s$ has fugacity $z_1^{-r}z_2^{-s}$, so it contributes a factor of $z_1^rz_2^s$ to the fugacity of $a^\dagger_{f,\mathbf{p},\mathbf{q}}$. Hence, the character that counts the creation operators can be obtained by simply replacing $z_1\mapsto z_1^{-1}$, $z_2\mapsto z_2^{-1}$ in \eqref{chiP1pos}, \eqref{chiP1neg}. For all $a\in\Z$, this gives us the uniform answer
\be
\chi^\text{op}_{\P^1}(\CO(a)) = \frac{z_1^{1+a}-z_2^{1+a}}{z_1-z_2}\,.
\ee
Consequently, plugging the result
\be
\chi^\text{op}_{\P^1}\big(\CO\big(d-{\textstyle\sum_\ell} p_\ell a_\ell+{\textstyle\sum_r} q_rb_r\big)\big) = \frac{z_1^{1+d-\sum_\ell p_\ell a_\ell+\sum_r q_rb_r}-z_2^{1+d-\sum_\ell p_\ell a_\ell+\sum_r q_rb_r}}{z_1-z_2}
\ee
into the sum \eqref{fd}, we can decouple and perform the $p_\ell,q_r$ sums. When the dust settles, we find
\be\label{phiS}
\vphi_d({\bf x},{\bf y},{\bf z}) = \frac{z_1^{1+d}P_{{\bf a},{\bf b}}({\bf x},{\bf y},z_1)-z_2^{1+d}P_{{\bf a},{\bf b}}({\bf x},{\bf y},z_2)}{z_1-z_2}\,,
\ee
where ${\bf x} = (x_1,\dots,x_n)$, ${\bf y}=(y_1,\dots,y_m)$, ${\bf z}=(z_1,z_2)$, ${\bf a}=(a_1,\dots,a_m)$, ${\bf b} = (b_1,\dots,b_n)$, and we encounter the products
\be\label{PS}
P_{{\bf a},{\bf b}}({\bf x},{\bf y},z) = \frac{\prod_{r=1}^n(1-z^{b_r}x_r)}{\prod_{\ell=1}^m(1-z^{-a_\ell}y_\ell)}\,.
\ee
The products found in \eqref{Pprod} and \eqref{Qprod} while counting the free $\cN=4$ and $\cN=1$ sdYM spectra are special cases of this general formula. Note that $\vphi_0$ contains a $1$ in its series expansion. This is the usual ghost zero mode that we will subtract.

Returning to the situation of interest, for any integer $s\in\Z$, we can take the generators of the category of branes in $\cM$ with support within $\cS$ to be the sheaves $\CO_\cS(s)$, $\CO_\cS(s+1)$. From the general result \eqref{phiS}, we can obtain the counts of open strings stretching between any pair of generators. Arranging this as a $2\times2$ matrix, we obtain
\begingroup
\setlength{\abovedisplayskip}{4pt}
\setlength{\belowdisplayskip}{4pt}
\setlength{\abovedisplayshortskip}{2pt}
\setlength{\belowdisplayshortskip}{2pt}
\be\label{fS}
\raisebox{-4pt}[0pt][0pt]{$\mathbf f =$}\;\,
\begin{blockarray}{cc@{\qquad}l}
{\scriptstyle \CO_{\cS}(s+1)}
&
{\scriptstyle \CO_{\cS}(s)}
&
\\[0.2em]
\begin{block}{(cc)@{\quad\;}l}
1-\vphi_0
&
-\vphi_{1}
&
{\scriptstyle \CO_{\cS}(s+1)}
\\
-\vphi_{-1}
&
1-\vphi_0
&
{\scriptstyle \CO_{\cS}(s)}
\\
\end{block}
\end{blockarray}
\,.
\ee
\endgroup
The entries of $\mathbf{f}$ count open strings viewed as maps between these sheaves with domain labeled by the column and target labeled by the row, and the minus signs indicate the overall odd Grassmann parity of the open string field. For later convenience, we have labeled the rows and columns by the two sheaves in the order $\{\CO_\cS(s+1),\CO_\cS(s)\}$. We will denote the components of this matrix as $f_{ab}$, with $a,b=0,1$. It will also be convenient to think of $\mathbf{f}$ as a single-letter graded partition function for the brane stack $\CO_\cS(s+1)\oplus\CO_\cS(s)$.


\subsection{Branes wrapping twistor space}

We now apply this general formula for single-letter graded partition functions to the D5-branes wrapping $\PT$ in the Calabi-Yau 7-fold $X$ and 5-fold $M$. With trivial Chan-Paton bundles, these branes engineer $\cN=4$ and $\cN=1$ sdYM, respectively.\footnote{More generally, the D5-brane theories engineer sd gauge theories on spacetime whenever the bundle on twistor space is trivial when restricted to all twistor lines.} Their single-letter graded partition functions reproduce the results of section \ref{sec:sdpartition}. 

To construct the GGE, we will study D5-branes with nontrivial Chan-Paton bundles. In this case, it turns out to be convenient to choose $\CO_\PT$ and $\CO_\PT(-1)$ as the two generators of branes wrapping $\PT$. Working with general bundles will correspond to working with quiver gauge theories on stacks of both types of generators. To such quiver gauge theories, we will apply the two-node quiver generalization of Murthy's expansion.

\paragraph{$\cN=4$ sdYM.}

Applying the counting formulae developed above to $\CO_{\PT}$ and $\CO_{\PT}(-1)$ in the CY 7-fold $X$ gives the following matrix of refined single-letter graded partition functions:
\begin{align}
    \mathbf{f}_{\cN=4}(\mathbf{x},\mathbf{y},\mathbf{z}) = \begin{pmatrix} \displaystyle 1-\frac{z_1P(z_1)-z_2P(z_2)}{z_1-z_2} && \displaystyle -\frac{z_1^2P(z_1)-z_2^2P(z_2)}{z_1-z_2} \\ \vspace{-0.4em} \\ \displaystyle -\frac{P(z_1)-P(z_2)}{z_1-z_2} && \displaystyle 1-\frac{z_1P(z_1)-z_2P(z_2)}{z_1-z_2} \end{pmatrix} \, , \label{eq:f4r}
\end{align}
where
\begin{align}
    P(z)=\frac{(1-z^{-1}x_1)(1-z^{-1}x_2)(1-z^{-1}x_3)(1-z^{-1}x_4)}{(1-z^{-1}y_1)(1-z^{-1}y_2)} \, .
\end{align}
Notice that whereas the diagonal entries are weightless under a simultaneous rescaling $(\fx,\fy,\fz)\sim(\al \fx,\al \fy,\al \fz)$, the off-diagonal entries carry the same homogeneities $\pm1$ as that of the strings they are counting. We can treat such entries as sections of line bundles over the $\P^7$ defined by the fugacities, in particular on the surface $x_1x_2x_3x_4=y_1y_2z_1z_2$. The unitary matrix integral that computes the associated $\U(N_0)\times\U(N_1)$-invariant graded partition function will be invariant under these rescalings.

Setting $x_I=t$, $y_1=y_2=t^2$ and $z_1=z_2=1$, one obtains the unrefined single-letter graded partition function,
\begin{higheq}
    \mathbf{f}_{\cN=4}(t) = \begin{pmatrix} \displaystyle \frac{6t^2+2t^3}{(1+t)^3}  && \displaystyle \frac{-2 - 2 t + 4 t^2}{(1+t)^3} \\ \vspace{-0.4em} \\ \displaystyle \frac{-4 t + 2 t^2 + 2 t^3}{(1+t)^3} && \displaystyle \frac{6t^2+2t^3}{(1+t)^3} \end{pmatrix} \, . \label{eq:f4u}
\end{higheq}
\noindent{}The upper-left entry counting strings from $\CO_\PT$ to itself reproduces the single-letter graded partition function $f_{\cN=4}$ of $\cN=4$ sdYM computed in \eqref{fN4}, and matches the free theory result \eqref{fN=4} computed directly on spacetime. The other entries are extra data that do not enter the sd gauge theories but will be crucial for the construction of our GGEs.

\paragraph{$\cN=1$ sdYM.} 

Applying the counting formulae to $\CO_{\PT}$ and $\CO_{\PT}(-1)$ in the CY 5-fold $M$ gives the following matrix of single-letter graded partition functions of open strings: 
\begin{align}
    \mathbf{f}_{\cN=1}(\mathbf{x},\mathbf{y},\mathbf{z}) = \begin{pmatrix} \displaystyle 1-\frac{z_1Q(z_1)-z_2Q(z_2)}{z_1-z_2} && \displaystyle -\frac{z_1^2Q(z_1)-z_2^2Q(z_2)}{z_1-z_2} \\ \vspace{-0.4em} \\ \displaystyle -\frac{Q(z_1)-Q(z_2)}{z_1-z_2} && \displaystyle 1-\frac{z_1Q(z_1)-z_2Q(z_2)}{z_1-z_2} \end{pmatrix} \, , \label{eq:f1r}
\end{align}
where
\begin{align}
    Q(z)=\frac{(1-z^{-1}x_1)(1-z^{-3}x_2)}{(1-z^{-1}y_1)(1-z^{-1}y_2)} \, .
\end{align}
The unrefined single-letter graded partition function simplifies to
\begin{higheq}
    \mathbf{f}_{\cN=1}(t) = \begin{pmatrix} \displaystyle -\frac{4t^3+2t^4}{(1-t) (1+t)^3}  && \displaystyle -\frac{2+3t+t^3}{(1-t) (1+t)^3} \\ \vspace{-0.4em} \\ \displaystyle -\frac{t+3t^3+2 t^4}{(1-t) (1+t)^3} && \displaystyle -\frac{4t^3+2t^4}{(1-t) (1+t)^3} \end{pmatrix} \, . \label{eq:f1u}
\end{higheq}
\noindent{}In this case, the upper-left entry counting strings from $\CO_\PT$ to itself reproduces the $\cN=1$ sdYM result \eqref{fN1} and matches the free theory result \eqref{fN=1}.


\subsection{Koszul-dual branes}
\label{ssec:koszul}

In this section, we identify candidate branes for a giant graviton expansion. Our choice is motivated by the categorical interpretation of Murthy's expansion \cite{Budzik:2025vzr}.

As we reviewed in section \ref{ssec:murthy}, Murthy proved an expansion that works in great generality. In particular, it provides an expansion for a variety of protected indices such as the $1/16$-BPS index or the Schur index of $\cN=4$ sYM. Many such indices can be viewed as graded partition functions of twists of their parent supersymmetric theories. Such twists can often be engineered as D-brane worldvolume theories in the topological B-model on flat space $\C^n$ \cite{Yoo:2025qlw}. For instance, the $1/16$-BPS sector of $\U(N)$ $\cN=4$ sYM can be encoded as the theory of $N$ D3-branes wrapping $\C^2$ in the B-model on $\C^5$, the Schur sector as the chiral algebra on $N$ D1-branes wrapping $\C$ in the B-model on $\C^3$ \cite{Costello:2018zrm}, etc. Backreacting by these brane stacks gives rise to self-contained holographic duals of the twisted theories \cite{Costello:2018zrm}.

Murthy's giant graviton expansion (MGGE) expresses a finite-$N$ supersymmetric index as its infinite-$N$ limit multiplied by a series of corrections.  If $f$ denotes the single-letter index of the original theory, these corrections involve the data
\begin{align}\label{fsh1node}
    f^!=-\frac{f}{1-f} \, .
\end{align}
In \cite{Budzik:2025vzr}, $f^!$ was recognized to be the single-letter index of D-branes wrapping the transverse plane to the original stack of D-branes described by $f$. For example, for the $1/16$-BPS index, these are D5-branes wrapping a $\C^3\subset\C^5$ that is transverse to the original D3-branes. A similar identification also holds for other twists like the Schur or half-BPS subsectors. It was therefore conjectured \cite{Budzik:2025vzr} that Murthy's GGE is naturally expressed in terms of brane/antibrane pairs that fill the transverse plane to the original stack of $N$ D-branes. Since the relation $(1-f^!)(1-f)=1$ is very generally obeyed by Hilbert series of associative algebras that are Koszul dual to each other \cite{Loday:2012ev}, we refer to the transverse branes as ``Koszul-dual branes''.

The standard giant graviton branes on the other hand wrap \emph{subspaces} of the plane transverse to the backreacting branes. The key fact that appears to underlie the success of Murthy's expansion is that all B-model branes contained in a plane $\C^n$ can be engineered from tachyon condensation on brane/antibrane pairs filling $\C^n$.

Here, we generalize this idea to the B-model on other target spaces. In our case, we are interested in B-model on the total spaces of vector bundles over $\P^1$. Our theories can be engineered as D5-branes wrapping $\PT$ in CY7 $X$ or CY5 $M$. As discussed in section \ref{ssec:category}, in these cases, the category of branes admits two generators. Stacks containing both types of generators on $\PT$ give a two-node quiver theory, with the single-letter data encoded in a $2\times 2$ matrix $\mathbf{f}$. We therefore use the two-node generalization of Murthy's expansion reviewed in section \ref{ssec:murthy}. This expansion is expressed in terms of another $2\times 2$ matrix
\begin{align}
    \mathbf{f}^! = - \qty[ \mathbf{f}(\mathbf{1}-\mathbf{f})^{-1} ]^{\mathrm{T}}
\end{align}
built as a generalization of the single-node relation \eqref{fsh1node}. The relation $(\mathbf{1}-\mathbf{f})(\mathbf{1}-\mathbf{f}^!)^{\mathrm{T}}=\mathbf{1}$ has the form of a matrix version of an identity obeyed by Hilbert series of Koszul-dual algebras \cite{Beilinson1996KoszulDP}.

Motivated by the identification in \cite{Budzik:2025vzr}, we expect the two-node MGGE to also admit a similar interpretation in terms of transverse Koszul-dual branes and antibranes in our geometries. In our setting, local operators on $\R^4$ correspond to non-local operators on $\PT$ smeared along the twistorial $\P^1$. We therefore expect the relevant Koszul-dual branes to wrap a $\P^1\subset\PT$, in addition to the fiber directions transverse to $\PT$. Without loss of generality, we can take the $\P^1$ to be the twistor line of the origin. Then, in the $\cN=4$ sdYM case, the geometry wrapped by the transverse branes is expected to be
\begin{align}
    Y=\qty{\mu^{\dot\alpha}=0} \simeq \Tot( \CO(-1)^4 \rightarrow \P^1 ) \, . \label{eq:Y}
\end{align}
We therefore identify the candidate Koszul-dual branes as the transverse D9-branes wrapping $Y\subset X$. Similarly, in the $\cN=1$ sdYM case, the transverse geometry (with respect to $\P^1\subset\PT$) is
\begin{align}
    S = \qty{\mu^{\dot\alpha}=0} \simeq \Tot(\CO(-1)\oplus\CO(-3)\rightarrow\P^1) \, , \label{eq:S}
\end{align}
and the candidate Koszul-dual branes are the transverse D5-branes wrapping $S\subset M$. In both these cases the category of branes admits two generators, just as for D5-branes wrapping $\PT$.

These generators and their antibranes can produce branes supported on smaller subspaces of $Y$ or $S$ through tachyon condensation. In section \ref{ssec:GGfromtachyons}, we illustrate this by using tachyon condensation to construct the giant graviton D5$'$-branes of \cite{Caron-Huot:2023wdh,Sharma:2025ntb} that are dual to determinant operators in $\cN=4$ sdYM.

We now compute the refined and unrefined single-letter graded partition functions, $\widetilde{\mathbf{f}}_{\cN=4}$ and $\widetilde{\mathbf{f}}_{\cN=1}$, of the open strings between the two generators in each geometry. We will relate them to the Koszul-dual data $\mathbf{f}^!_{\cN=4}$ and $\mathbf{f}^!_{\cN=1}$ in section \ref{sec:ggemain}.

\paragraph{$\cN=4$ sdYM.}

We apply the counting formulae developed in section \ref{ssec:counting} to the transverse D9-branes wrapping $Y$ in the CY 7-fold $X$. We choose $\Pi\CO_Y$ and $\CO_Y(1)$ as the two generators in that order, since this choice will be convenient for the GGE in section \ref{sec:ggemain}. The refined single-letter graded partition function of open strings is given by a matrix:
\begin{align}
    \widetilde{\mathbf{f}}_{\cN=4}(\mathbf{x},\mathbf{y},\mathbf{z}) = \begin{pmatrix} \displaystyle 1-\frac{z_1\widetilde{P}(z_1)-z_2\widetilde{P}(z_2)}{z_1-z_2} && \displaystyle \frac{\widetilde{P}(z_1)-\til P(z_2)}{z_1-z_2} \\ \vspace{-0.4em} \\ \displaystyle \frac{z_1^2 \widetilde{P}(z_1)- z_2^2 \widetilde{P}(z_2)}{z_1-z_2} && \displaystyle 1-\frac{z_1\widetilde{P}(z_1)-z_2\widetilde{P}(z_2)}{z_1-z_2} \end{pmatrix} \, , \label{eq:ft4r}
\end{align}
where
\begin{align}
    \widetilde{P}(z)=\frac{(1-z y_1^{-1})(1-z y_2^{-1})}{(1-z x_1^{-1})(1-z x_2^{-1})(1-z x_3^{-1})(1-z x_4^{-1})} \, .
\end{align}
The unrefined single-letter graded partition function simplifies to
\begin{higheq}
    \widetilde{\mathbf{f}}_{\cN=4}(t) = \begin{pmatrix} \displaystyle \frac{2+2t+8 t^2}{(1-t)^3}  && \displaystyle -\frac{2+6t+4 t^2}{(1-t)^3} \\ \vspace{-0.4em} \\ \displaystyle -\frac{4t+6t^2+2t^3}{(1-t)^3} && \displaystyle \frac{2+2t+8 t^2}{(1-t)^3} \end{pmatrix} \, . \label{eq:ft4u}
\end{higheq}

\paragraph{$\cN=1$ sdYM.}

We likewise apply the counting formulae developed above to the transverse D5-branes wrapping $S$ in the CY 5-fold $M$. We choose $\Pi\CO_S$ and $\CO_S(1)$ as the two generators in that order. The refined single-letter graded partition function of open strings is given by a matrix:
\begin{align}
    \widetilde{\mathbf{f}}_{\cN=1}(\mathbf{x},\mathbf{y},\mathbf{z}) = \begin{pmatrix} \displaystyle 1-\frac{z_1\widetilde{Q}(z_1)-z_2\widetilde{Q}(z_2)}{z_1-z_2} && \displaystyle \frac{\widetilde{Q}(z_1)-\widetilde{Q}(z_2)}{z_1-z_2} \\ \vspace{-0.4em} \\ \displaystyle \frac{z_1^2 \widetilde{Q}(z_1)- z_2^2 \widetilde{Q}(z_2)}{z_1-z_2} && \displaystyle 1-\frac{z_1\widetilde{Q}(z_1)-z_2\widetilde{Q}(z_2)}{z_1-z_2} \end{pmatrix} \, , \label{eq:ft1r}
\end{align}
where
\begin{align}
    \widetilde{Q}(z)=\frac{(1-z y_1^{-1})(1-z y_2^{-1})}{(1-z x_1^{-1})(1-z^3 x_2^{-1})} \, .
\end{align}
The unrefined single-letter graded partition function simplifies to
\begin{higheq}
    \widetilde{\mathbf{f}}_{\cN=1}(t) = \begin{pmatrix} \displaystyle \frac{2+4 t+3 t^2+t^3+2 t^4}{(1-t)
   (1+t+t^2)^2}  && \displaystyle -\frac{2+5 t+3
   t^2+t^3+t^4}{(1-t)(1+t+t^2)^2} \\ \vspace{-0.4em} \\ \displaystyle -\frac{t+t^2+3 t^3+5 t^4+2 t^5}{(1-t)
   (1+t+t^2)^2} && \displaystyle \frac{2+4 t+3 t^2+t^3+2 t^4}{(1-t)(1+t+t^2)^2} \end{pmatrix} \, . \label{eq:ft1u}
\end{higheq}


\subsection{Giant gravitons from tachyon condensation} 
\label{ssec:GGfromtachyons}

It is helpful to see the tachyon condensation idea in action by working out the complexes of sheaves that represent some known examples of giant gravitons. For simplicity, let us consider the giant gravitons dual to half-BPS determinants in $\cN=4$ sdYM. A scalar determinant like $\det(\Phi_{12})$ placed at the origin is dual to a brane in our 7-fold $X$ that wraps $\mu^\dal=W_1=W_2=0$. This was demonstrated in \cite{Sharma:2025ntb}, based on ideas from the earlier works \cite{Caron-Huot:2023wdh,Budzik:2021fyh,Jiang:2019xdz}. We will be interested in carving out this brane from tachyon condensation on the worldvolume of branes and antibranes wrapping $Y=\{\mu^\dal=0\}\subset X$, which is the locus transverse to $\PT$ relative to the base $\P^1$.

Just as $X$, the subspace $Y$ is a bundle over $\P^1$, namely $Y=\Tot(\CO(-1)^{\oplus4}\to\P^1)$, with the base $\P^1$ being identified with the twistor line of the origin. Hence, the bounded derived category of coherent sheaves on $X$ with support within $Y$ is generated by $\CO_Y, \CO_Y(1)$. Let $G\subset Y$ denote the locus $W_1=W_2=0$. Our giant is a brane that wraps $G$ and carries the trivial bundle, so it corresponds to the structure sheaf $\CO_G$. Because $G$ is codimension-two inside $Y$, we can use the Koszul resolution of $\CO_G$ to represent it as the cohomology of the complex
\be\label{cmpx1}
\begin{tikzcd}[
  column sep=2.2em,
  row sep=1.1em,
  every arrow/.append style={line width=0.6pt}
]
&&
\CO_Y(1)
  \arrow[dr, "W_1"{pos=0.35}]
  \arrow[dd, phantom, "\bigoplus" description]
&&
\\
0 \arrow[r]
&
\CO_Y(2)
  \arrow[ur, "-W_2"{pos=0.7}]
  \arrow[dr, "W_1"'{pos=0.7}]
&&
\CO_Y
  \arrow[r]
&
0
\\
&&
\CO_Y(1)
  \arrow[ur, "W_2"'{pos=0.35}]
&&
\end{tikzcd}\,.
\ee
The idea here is that there is no cohomology at $\CO_Y(2)$ or $\CO_Y(1)^{\oplus2}$. E.g., if $s$ is a local section of $\CO_Y(2)$, then by continuity, it is in the kernel of the map $s\mapsto(-W_2s,W_1s)$ if and only if $s=0$, etc. The only nonzero cohomology appears at the rightmost $\CO_Y$ and is given by the cokernel $\CO_Y/(W_1,W_2)\simeq\CO_G$. 

Each sheaf $\CO_Y(n)$ appearing in this complex can be thought of as a brane wrapping $Y$. More precisely, the sheaves $\CO_Y,\CO_Y(2)$ appearing in even degrees are branes, and the sheaves appearing in odd degrees such as $\CO_Y(1)^{\oplus2}$ are antibranes. We will sometimes represent the latter as $\Pi\CO_Y(1)^{\oplus2}$ when their degree in the complex is left implicit.

The Koszul resolution is one out of many possible resolutions of $\CO_G$. To find a resolution in terms of just the generator branes $\CO_Y,\CO_Y(1)$, we can use the fact that $\CO_Y(2)$ fits into the short exact sequence
\be\label{eulerseq}
0\longrightarrow\CO_Y\overset{\lambda_\al}{\longrightarrow}\CO_Y(1)^{\oplus2}\overset{\lambda^\al}{\longrightarrow}\CO_Y(2)\longrightarrow0\,.
\ee
Splicing this into \eqref{cmpx1} by composing the map $\CO_Y(1)^{\oplus2}\longrightarrow\CO_Y(2)$ in \eqref{eulerseq} with the map $\CO_Y(2)\longrightarrow\CO_Y(1)^{\oplus2}$ in \eqref{cmpx1}, we obtain
\be\label{cmpx2}
\begin{tikzcd}[
  column sep=1.8em,
  row sep=1.3em,
  every arrow/.append style={line width=0.6pt}
]
&&
\CO_Y(1)
  \arrow[rrr, "-\lambda_1W_2"{pos=0.45}]
  \arrow[
    ddrrr,
    "\lambda_1W_1"{pos=0.2, inner sep=0.6pt}
  ]
  \arrow[dd, phantom, "\bigoplus" description]
&&&
\CO_Y(1)
  \arrow[dr, "W_1"{pos=0.35}]
  \arrow[dd, phantom, "\bigoplus" description]
&&
\\
0 \arrow[r]
&
\CO_Y
  \arrow[ur, "\lambda_0"{pos=0.7}]
  \arrow[dr, "\lambda_1"'{pos=0.7}]
&&&&&
\CO_Y
  \arrow[r]
&
0
\\
&&
\CO_Y(1)
  \arrow[
    uurrr,
    crossing over,
    crossing over clearance=2pt,
    "\lambda_0W_2"'{pos=0.2, inner sep=0.6pt}
  ]
  \arrow[rrr, "-\lambda_0W_1"'{pos=0.45}]
&&&
\CO_Y(1)
  \arrow[ur, "W_2"'{pos=0.35}]
&&
\end{tikzcd}\,.
\ee
Since \eqref{eulerseq} is exact, the cohomology of this complex is again supported purely at the rightmost $\CO_Y$ and recovers $\CO_G$. Hence, we see that the giant graviton $\CO_G$ is just one example of a brane that can be constructed out of the generator branes $\CO_Y$ and $\CO_Y(1)$. Stacks of giants can be similarly obtained by starting with stacks of the generators. 

More generally, \emph{any} such bounded complex of sheaves gives rise to a brane that could act as a potential giant graviton. For our purposes, it will be sufficient to restrict attention to branes that do not carry any K-theory charge. The charge of a brane $\mathscr{E}$ formed from condensing $m_0,m_1,n_0,n_1$ branes of type $\CO_Y,\CO_Y(1),\Pi\CO_Y,\Pi\CO_Y(1)$ respectively is given by
\be
[\mathscr{E}] = (m_0-n_0)[\CO_Y] + (m_1-n_1)[\CO_Y(1)]\,.
\ee
For this to vanish, we separately require $m_0=n_0$ and $m_1=n_1$. Therefore, we will only include giants formed from condensing stacks of the form $\CO_Y^{\oplus\,m_0|m_0}\oplus\CO_Y(1)^{\oplus\,m_1|m_1}$. This zero-charge condition will arise naturally in the structure of our GGEs in the coming sections.


\section{Giant graviton expansion in chiral holography}
\label{sec:ggemain}

In section \ref{sec:partition}, we set up the unitary matrix integrals \eqref{Zgr4result}, \eqref{Zgr1result} that compute the graded partition functions of free $\cN=4$ and $\cN=1$ sYM respectively, incorporating refinements by a maximal number of fugacities. In section \ref{sec:sdpartition}, we showed that they coincided with the graded partition functions of the associated self-dual theories. The corresponding thermal partition functions were obtained by reversing the signs of the R-symmetry fugacities.

In section \ref{sec:chiralholography}, we described the twistorial uplifts of $\U(N)$ $\cN=4$ and $\cN=1$ sdYM and their realizations as worldvolume theories on $N$ D5-branes supported on $\PT$ inside CY7 $X$ and CY5 $M$, respectively. In both cases, the category of branes supported on $\PT$ is generated by two generators, which we take to be $\CO_{\PT}$ and $\CO_{\PT}(-1)$.

In this section, we consider two-node quiver gauge theories on systems of $N_0$ D5-branes of type $\CO_{\PT}$ and $N_1$ D5-branes of type $\CO_{\PT}(-1)$. We then apply the two-node quiver generalization of Murthy's expansion \eqref{quivermurthy} to the graded partition functions $Z_{N_0,N_1}$ of the two-node quiver gauge theories.

We first identify the prefactors $Z_{\infty,\infty}$, obtained in the simultaneous $N_0,N_1\to\infty$ limit, with the graded closed string partition functions $Z_{\mathrm{BCOV}}$ of the holographic duals. As discussed in previous sections, we expect Murthy's expansion to admit a holographic interpretation in terms of graded partition functions of brane/antibrane systems supported in the geometry transverse to the original D5-branes (relative to the twistor $\P^1\subset\PT$). As discussed in section \ref{ssec:koszul}, the generator branes for these systems are:
\begin{itemize}
    \item D9-branes wrapping $Y$ in CY7 $X$, with bundles $\Pi\CO_Y$ and $\CO_Y(1)$, for $\cN=4$ sdYM,
    \item D5-branes wrapping $S$ in CY5 $M$, with bundles $\Pi\CO_S$ and $\CO_S(1)$, for $\cN=1$ sdYM.
\end{itemize}
We show that their graded single-letter partition functions computed in section \ref{ssec:koszul}, after an appropriate fugacity involution, reproduce the two-node quiver data $\mathbf{f}^!$ entering Murthy's expansions. We also explain the Berezinian insertions $(\Ber\,\mathcal W_0)^{N_0}(\Ber\,\mathcal W_1)^{N_1}$ as contributions of open strings stretching between the original D5-branes and the transverse brane/antibrane stacks. These insertions encode the effect of the original branes backreaction on the transverse stacks. Combining these results with the closed string contributions of BCOV theory, we therefore construct the giant graviton expansions directly from bulk ingredients.

In section \ref{sec:thermalgge}, we will specialize to $N_0=N$ and $N_1=0$ and recover the graded partition functions of free $\cN=4$ and $\cN=1$ sYM. The two-node quiver MGGE also applies to twisted graded partition functions. In the $N_0=N$, $N_1=0$ specialization, reversing the signs of the R-symmetry fugacities will yield an MGGE for the thermal partition functions.


\subsection{Two-node quiver Murthy expansion}

We consider the worldvolume theory on a stack of $N_0$ D5-branes of type $\CO_{\PT}$ and $N_1$ D5-branes of type $\CO_{\PT}(-1)$, carrying the Chan-Paton bundle
\begin{align}
\CO_{\PT}^{\oplus N_0} \oplus \CO_{\PT}(-1)^{\oplus N_1} \, . \label{eq:CPbundle}
\end{align}
The resulting theory is a two-node quiver gauge theory with gauge group $\U(N_0)\times\U(N_1)$ and adjoint and bifundamental matter.\footnote{Strictly speaking, $\U(N_0)\times\U(N_1)$ is a choice of compact real form for the complexified automorphism group of the D5-brane system.} 

The construction is identical for the two theories once the matrix of single-letter graded partition functions is specified. For either theory, let ${\bf f}=(f_{ab})_{a,b=0,1}$ be the matrix of single-letter graded partition functions of open strings from the brane of type $b$ to the brane of type $a$. The refined and unrefined matrices were computed in \eqref{eq:f4r} and \eqref{eq:f4u} for the $\cN=4$ sdYM case, and in \eqref{eq:f1r} and \eqref{eq:f1u} for the $\cN=1$ sdYM case. The graded partition function of the corresponding two-node quiver gauge theory is
\begin{align}
Z_{N_0,N_1}[\mathbf f] &= \int_{\U(N_0)}\dd U_0 \int_{\U(N_1)}\dd U_1\, \exp\qty[ \sum_{n=1}^{\infty}\frac{1}{n} \sum_{a,b=0}^{1} (\mathbf f_n)_{ab} \Tr(U_a^{n})\Tr(U_b^{-n})] \, ,
\label{eq:two-node-Z}
\end{align}
where $\mathbf f_n$ is obtained by raising every fugacity appearing in $\mathbf f$ to the $n$th power. The $\pm1$ projective weights of the off-diagonal entries $f_{01},f_{10}$ discussed below \eqref{eq:f4r} drop out of \eqref{eq:two-node-Z}: integration over the central $\U(1)$ factors at the two nodes projects onto gauge invariant combinations, for which the opposite homogeneities carried by bifundamentals of opposite orientation cancel. 

One can similarly obtain the twisted graded partition function by replacing $\mathbf{x}\mapsto-\mathbf{x}$ in the refined case and $t\mapsto -t$ in the unrefined case. As discussed in section \ref{ssec:multifugacity}, since the twisting undoes the fermion parity, this is equal to the thermal partition function of the theory on the D5-brane stack \eqref{eq:CPbundle}.

Applying the quiver generalization of Murthy's expansion \eqref{quivermurthy} to our two-node quiver gauge theory, gives
\begin{align}
Z_{N_0,N_1}[\mathbf f] &= Z_{\infty,\infty}[\mathbf f] \sum_{m_0,m_1=0}^{\infty} G_{N_0,N_1}^{(m_0,m_1)}[\mathbf f] \, , \label{eq:two-node-MGGE}
\end{align}
where
\begin{align}
Z_{\infty,\infty}[\mathbf f] &= \prod_{n=1}^{\infty} \frac{1}{\det(\mathbf{1}-\mathbf f_n)} \, , \label{eq:Zinfty2}
\end{align}
and
\begin{multline}\label{eq:twonodeG}
G_{N_0,N_1}^{(m_0,m_1)}[\mathbf f] =
\int_{\U(m_0|m_0)}\dd\cW_0\, (\Ber\,\cW_0)^{N_0} \int_{\U(m_1|m_1)}\dd\cW_1\, (\Ber\,\cW_1)^{N_1}\\
\times \exp\qty[ \sum_{n=1}^{\infty}\frac{1}{n} \sum_{a,b=0}^{1} (\mathbf f_n^!)_{ab}\, \Str(\cW_a^{n})\,\Str(\cW_b^{-n}) ] \, ,
\end{multline}
where the measure and contours are explained in section \ref{ssec:murthy}. Due to off-diagonal zero modes we additionally nest the contours $r_1<R_1<r_0<R_0$.

The supermatrix integrals are governed by the ``Koszul-dual" single-letter graded partition functions
\begin{align}
\mathbf f^! = -\qty[\mathbf f(\mathbf{1}-\mathbf f)^{-1}]^{\mathrm{T}} \, . \label{eq:fk8}
\end{align}
The matrix $\mathbf{f}_n^!$ is obtained from $\mathbf{f}^!$ by raising every fugacity to the $n^\text{th}$ power. We now evaluate these matrices for both theories.

\paragraph{$\cN=4$ sdYM.} The Koszul-dual matrix associated to $\mathbf{f}_{\cN=4}$, computed in \eqref{eq:f4r}, is given by
\begin{align}
    \mathbf{f}^!_{\cN=4}(\mathbf{x},\mathbf{y},\mathbf{z}) = \begin{pmatrix} \displaystyle 1-\frac{z_1P(z_2)^{-1}-z_2P(z_1)^{-1}}{z_1-z_2} && \displaystyle \frac{P(z_2)^{-1}-P(z_1)^{-1}}{z_1-z_2} \\ \vspace{-0.4em} \\ \displaystyle \frac{z_1^2 P(z_2)^{-1}- z_2^2 P(z_1)^{-1}}{z_1-z_2} && \displaystyle 1-\frac{z_1P(z_2)^{-1}-z_2P(z_1)^{-1}}{z_1-z_2} \end{pmatrix} \, , \label{eq:fk4r}
\end{align}
expressed as before in terms of the product
\begin{align}
    P(z)=\frac{(1-z^{-1}x_1)(1-z^{-1}x_2)(1-z^{-1}x_3)(1-z^{-1}x_4)}{(1-z^{-1}y_1)(1-z^{-1}y_2)} \, .
\end{align}
Taking its unrefined specialization gives
\begin{higheq}
    \mathbf{f}^!_{\cN=4}(t) = \begin{pmatrix} \displaystyle -\frac{8t+2t^2+2t^3}{(1-t)^3}  && \displaystyle \frac{4t+6t^2+2t^3}{(1-t)^3} \\ \vspace{-0.4em} \\ \displaystyle \frac{2+6t+4t^2}{(1-t)^3} && \displaystyle -\frac{8t+2t^2+2t^3}{(1-t)^3} \end{pmatrix} \, . \label{eq:fk4u}
\end{higheq}

\paragraph{$\cN=1$ sdYM.} The Koszul-dual matrix associated to $\mathbf{f}_{\cN=1}$, computed in \eqref{eq:f1r}, is given by
\begin{align}
    \mathbf{f}^!_{\cN=1}(\mathbf{x},\mathbf{y},\mathbf{z}) = \begin{pmatrix} \displaystyle 1-\frac{z_1Q(z_2)^{-1}-z_2Q(z_1)^{-1}}{z_1-z_2} && \displaystyle \frac{Q(z_2)^{-1}-Q(z_1)^{-1}}{z_1-z_2} \\ \vspace{-0.4em} \\ \displaystyle \frac{z_1^2 Q(z_2)^{-1}- z_2^2 Q(z_1)^{-1}}{z_1-z_2} && \displaystyle 1-\frac{z_1Q(z_2)^{-1}-z_2Q(z_1)^{-1}}{z_1-z_2} \end{pmatrix} \, , \label{eq:fk1r}
\end{align}
where
\begin{align}
    Q(z)=\frac{(1-z^{-1}x_1)(1-z^{-3}x_2)}{(1-z^{-1}y_1)(1-z^{-1}y_2)} \, .
\end{align}
Taking the unrefined specialization gives
\begin{higheq}
    \mathbf{f}^!_{\cN=1}(t) = \begin{pmatrix} \displaystyle -\frac{2t+t^2+3t^3+4t^4+2t^5}{(1-t)
   (1+t+t^2)^2}  && \displaystyle \frac{t+t^2+
   3t^3+5t^4+2t^5}{(1-t)(1+t+t^2)^2} \\ \vspace{-0.4em} \\ \displaystyle \frac{2+5t+3 t^2+t^3+t^4}{(1-t)
   (1+t+t^2)^2} && \displaystyle -\frac{2t+t^2+3 t^3+4t^4+2t^5}{(1-t)(1+t+t^2)^2} \end{pmatrix} \, . \label{eq:fk1u}
\end{higheq}


\subsection{Matching the closed string contribution}

We now match the simultaneous infinite-rank prefactors $Z_{\infty,\infty}[\mathbf{f}]$ of the two-node MGGE with the graded closed string partition functions of the corresponding B-model bulk duals. The general expression for $Z_{\infty,\infty}[\mathbf{f}]$ was provided in \eqref{Zinfinf} and \eqref{eq:Zinfty2}. The closed string graded partition functions $Z_{\mathrm{BCOV}}^{\cN=4}$ and $Z_{\textrm{BCOV}}^{\cN=1}$ were computed in \eqref{Zbcov4} and \eqref{Zbcov1}, respectively.

\paragraph{$\cN=4$ sdYM.} For the single-letter matrix $\mathbf{f}_{\cN=4}$ given by \eqref{eq:f4r}, we obtain a beautiful simplification of the determinant:
\begin{align}
    \det(\mathbf{1}-\mathbf{f}_{\cN=4}) = P(z_1)P(z_2) \, .
\end{align}
Substituting this into the general formula \eqref{eq:Zinfty2} yields
\begin{align}
Z_{\infty,\infty}[\mathbf f_{\cN=4}] &= \prod_{n=1}^\infty \frac{(1-y_1^n/z_1^n)(1-y_2^n/z_1^n)}{(1-x_1^n/z_1^n)(1-x_2^n/z_1^n)(1-x_3^n/z_1^n)(1-x_4^n/z_1^n)} \\
& \hspace{1.5cm}\times\frac{(1-y_1^n/z_2^n)(1-y_2^n/z_2^n)}{(1-x_1^n/z_2^n)(1-x_2^n/z_2^n)(1-x_3^n/z_2^n)(1-x_4^n/z_2^n)} \, .
\label{eq:N4-BCOV-match}
\end{align}
This is precisely equal to the graded partition function of closed twistor strings in the CY7 $X$ derived in \eqref{Zbcov4}. Hence, we conclude our first main result,
\begin{higheqtopless}\label{Zbcov4match}
\vspace{-0.4em}
Z_{\infty,\infty}[\mathbf{f}_{\cN=4}] = Z_\text{BCOV}^{\cN=4} \, .
\end{higheqtopless}
\noindent{}The infinite-$N$ graded partition function of the original $\cN=4$ sdYM theory is obtained by setting $N_1=0$ and taking $N_0\rightarrow\infty$, giving $Z_{\infty,0}$. The BCOV graded partition function instead equals $Z_{\infty,\infty}$, which explains the previous mismatch with the naive answer.

\paragraph{$\cN=1$ sdYM.} Similarly, for the single-letter matrix $\mathbf{f}_{\cN=1}$ given by \eqref{eq:f1r} we get
\begin{align}
    \det(\mathbf{1}-\mathbf{f}_{\cN=1}) = Q(z_1)Q(z_2) \, .
\end{align}
Substituting this into the general formula \eqref{eq:Zinfty2} gives
\begin{align}
Z_{\infty,\infty}[\mathbf f_{\cN=1}] &= \prod_{n=1}^\infty \frac{(1-y_1^n/z_1^n)(1-y_2^n/z_1^n)}{(1-x_1^n/z_1^n)(1-x_2^n/z_1^{3n})} \frac{(1-y_1^n/z_2^n)(1-y_2^n/z_2^n)}{(1-x_1^n/z_2^n)(1-x_2^n/z_2^{3n})} \, .
\label{eq:N1-BCOV-match}
\end{align}
We again find that $Z_{\infty,\infty}$ matches the closed twistor string partition function \eqref{Zbcov1},
\begin{higheqtopless}\label{Zbcov1match}
\vspace{-0.4em}
Z_{\infty,\infty}[\mathbf{f}_{\cN=1}] = Z_\text{BCOV}^{\cN=1} \, .
\end{higheqtopless}
\noindent{}Having finally ascribed a holographic meaning to the closed string partition functions, we next study the nonperturbative brane/antibrane corrections.


\subsection{Matching the Koszul-dual brane contributions}

We now compare the matrices $\mathbf{f}^!$ entering the two-node MGGE with the single-letter graded partition functions of Koszul-dual branes $\widetilde{\mathbf{f}}$ computed in section \ref{ssec:koszul}. We first establish the match for the general geometry considered in section \ref{ssec:counting}, before specializing to the two theories.

Consider the original branes wrapping
\begin{align}
\mathcal{S} = \{w_r=0 \} \subset\cM \, ,
\end{align}
and the transverse branes wrapping
\be
\til{\mathcal{S}} = \{\mu^\ell=0 \}\subset\cM \, .
\ee
Both types of branes wrap the base $\P^1$, along which they intersect.

Recall from section \ref{ssec:counting} that the matrix of single-letter graded partition functions for the generators $\CO_\cS$ and $\CO_\cS(-1)$ is
\begingroup
\setlength{\abovedisplayskip}{4pt}
\setlength{\belowdisplayskip}{4pt}
\setlength{\abovedisplayshortskip}{2pt}
\setlength{\belowdisplayshortskip}{2pt}
\be\label{fS2}
\raisebox{-4pt}[0pt][0pt]{$\mathbf f =$}\;\,
\begin{blockarray}{cc@{\qquad}l}
{\scriptstyle \CO_{\cS}}
&
{\scriptstyle \CO_{\cS}(-1)}
&
\\[0.2em]
\begin{block}{(cc)@{\quad\;}l}
1-\vphi_0
&
-\vphi_{1}
&
{\scriptstyle \CO_{\cS}}
\\
-\vphi_{-1}
&
1-\vphi_0
&
{\scriptstyle \CO_{\cS}(-1)}
\\
\end{block}
\end{blockarray}
\,,
\ee
\endgroup
where the individual Hilbert series $\varphi_d$ were given in \eqref{phiS}, written in terms of the products  $P_{\mathbf{a},\mathbf{b}}$ given by \eqref{PS}. Using $\det(\mathbf{1}-\mathbf{f})=P_{\mathbf{a},\mathbf{b}}(z_1)P_{\mathbf{a},\mathbf{b}}(z_2)$, we obtain
\begin{align} 
    \mathbf{f}^! &= -\qty[\mathbf{f}(\mathbf{1}-\mathbf{f})^{-1}]^{\mathrm{T}} \\
    &= \begin{pmatrix} \displaystyle 1-\frac{z_1P_{\mathbf{a},\mathbf{b}}(z_2)^{-1}-z_2P_{\mathbf{a},\mathbf{b}}(z_1)^{-1}}{z_1-z_2} && \displaystyle \frac{P_{\mathbf{a},\mathbf{b}}(z_2)^{-1}-P_{\mathbf{a},\mathbf{b}}(z_1)^{-1}}{z_1-z_2} \\ \vspace{-0.4em} \\ \displaystyle \frac{z_1^2 P_{\mathbf{a},\mathbf{b}}(z_2)^{-1}- z_2^2 P_{\mathbf{a},\mathbf{b}}(z_1)^{-1}}{z_1-z_2} && \displaystyle 1-\frac{z_1P_{\mathbf{a},\mathbf{b}}(z_2)^{-1}-z_2P_{\mathbf{a},\mathbf{b}}(z_1)^{-1}}{z_1-z_2} \end{pmatrix} \, . \label{eq:bfk}
\end{align}
Because the off-diagonal entries in \eqref{eq:bfk} are not accompanied by the usual minus signs, we consider the transverse stack $\Pi\CO_{\til\cS}\oplus\CO_{\til\cS}(1)$. The parity reversal functor removes these signs by shifting the parity of the off-diagonal open string fields, while leaving the diagonal entries unchanged.

We use the same counting formulae from section \ref{ssec:counting} to obtain the single-letter graded partition functions $\widetilde{\mathbf{f}}$ for this stack. The on-shell states are now polynomials in $w_r$ and the normal bundle frame $\veps_\ell=\p/\p\mu^\ell$, with coefficients in the Dolbeault cohomology of $\P^1$. The fugacity of $w_r$ is $x_r$ as before, while that of $\veps_\ell$ is $y_\ell$. So we can obtain $\til{\bf f}$ by exchanging $y_\ell$ with $x_r^{-1}$ and $x_r$ with $y_\ell^{-1}$. The result is
\begingroup
\setlength{\abovedisplayskip}{4pt}
\setlength{\belowdisplayskip}{4pt}
\setlength{\abovedisplayshortskip}{2pt}
\setlength{\belowdisplayshortskip}{2pt}
\be\label{ftilgen}
\raisebox{-4pt}[0pt][0pt]{$\til{\mathbf f} =$}\;\,
\begin{blockarray}{cc@{\qquad}l}
{\scriptstyle \Pi\CO_{\cS}}
&
{\scriptstyle \CO_{\cS}(1)}
&
\\[0.2em]
\begin{block}{(cc)@{\quad\;}l}
1-\til\vphi_0
&
\til\vphi_{-1}
&
{\scriptstyle \Pi\CO_{\cS}}
\\
\til\vphi_{1}
&
1-\til\vphi_0
&
{\scriptstyle \CO_{\cS}(1)}
\\
\end{block}
\end{blockarray}
\,,
\ee
\endgroup
where the entries $\til\vphi_d$ are given by
\begin{align}
    \til\vphi_d({\bf x},{\bf y},{\bf z}) &= \frac{z_1^{1+d}\til P_{{\bf a},{\bf b}}({\bf x},{\bf y},z_1)-z_2^{1+d}\til P_{{\bf a},{\bf b}}({\bf x},{\bf y},z_2)}{z_1-z_2}\,,\label{phiStil}
\end{align}
and
\begin{align}
    \til P_{{\bf a},{\bf b}}({\bf x},{\bf y},z) &= \frac{\prod_{\ell=1}^m(1-z^{a_\ell}y_\ell^{-1})}{\prod_{r=1}^n(1-z^{-b_r}x_r^{-1})}\,.\label{PStil}
    \end{align}
Essentially, this has been obtained by simply exchanging the tangent and normal directions to the branes in the previous expressions \eqref{phiS} and \eqref{PS} (holding the $\P^1$ fixed).

At this point, we observe that $\til{\bf f}$ is related to ${\bf f}^!$ by the action of a suitable involution on the fugacities,
\be\label{shrieky}
{\bf f}^!(x_r,y_\ell,z_1,z_2) = \til{\bf f}\bigg(\frac{z_1^{-b_r}z_2^{-b_r}}{x_r},\frac{z_1^{a_\ell}z_2^{a_\ell}}{y_\ell},z_1,z_2\bigg)\,. 
\ee
Note that the involution employed here,
\be
x_r\mapsto\frac{(z_1z_2)^{-b_r}}{x_r}\,,\qquad y_\ell\mapsto\frac{(z_1z_2)^{a_\ell}}{y_\ell}\,,
\ee
preserves the fugacity constraint \eqref{fugrelsgen}. The fugacities are inverted because normal directions of the original branes become tangent directions of the transverse branes.

Thus, after the fugacity involution, the transverse-brane matrix $\widetilde{\mathbf{f}}$ reproduces the single-letter data $\mathbf{f}^!$ entering the two-node MGGE. Using the relation \eqref{shrieky}, we will be able to interpret the terms in our expansions as graded partition functions of branes transverse to twistor space within our bulk geometries.

\paragraph{$\cN=4$ sdYM.} We specialize to $\cM=X$, $\cS=\PT$ and transverse D9-branes wrapping $\widetilde{\cS}=Y$. 

The Koszul-dual single-letter graded partition function $\mathbf{f}^!$ appearing in the two-node MGGE was computed in \eqref{eq:fk4r} and \eqref{eq:fk4u}. The single-letter graded partition function of D9-branes with CP $\Pi\CO_Y$ and $\CO_Y(1)$ was computed in \eqref{eq:ft4r} and \eqref{eq:ft4u}.

We can easily check the matching \eqref{shrieky} reduces to
\be\label{eq:fshmatch4r}
{\bf f}^!_{\cN=4}(x_1,x_2,x_3,x_4,y_1,y_2,z_1,z_2) = \til{\bf f}_{\cN=4}\bigg(\frac{z_1z_2}{x_1},\frac{z_1z_2}{x_2},\frac{z_1z_2}{x_3},\frac{z_1z_2}{x_4},\frac{z_1z_2}{y_1},\frac{z_1z_2}{y_2},z_1,z_2\bigg) \,.
\ee
Under the unrefined specialization the fugacity involution reduces to
\begin{higheqtopless}
\vspace{-0.4em} \label{eq:fmatch4u}
{\bf f}^!_{\cN=4}(t) = \til{\bf f}_{\cN=4}(t^{-1}) \,.
\end{higheqtopless}
Along the lines of \cite{Budzik:2025vzr}, this allows us to identify the partial sums $\sum_{i_0,i_1=0}^{m_0,m_1} G_{N_0,N_1}^{(i_0,i_1)}$ as graded partition functions of the D9/anti-D9-brane pairs with gauge bundle $\Pi\CO_Y^{\oplus\,m_0|m_0}\oplus\CO_Y(1)^{\oplus\,m_1|m_1}$.\footnote{Because $\Pi\CO_Y^{\oplus\,m_0|m_0}$ is the same bundle as $\CO_Y^{\oplus\,m_0|m_0}$, we can equivalently think of the corrections in the MGGE as contributions from the D9/anti-D9 stacks $\CO_Y^{\oplus\,m_0|m_0}\oplus\CO_Y(1)^{\oplus\,m_1|m_1}$.}

\paragraph{$\cN=1$ sdYM.} Next, we specialize to $\cM=M$, $\cS=\PT$ and transverse D5-branes wrapping $\widetilde{\cS}=S$. 

The Koszul-dual single-letter graded partition function $\mathbf{f}^!$ appearing in the two-node MGGE was computed in \eqref{eq:fk1r} and \eqref{eq:fk1u}. The single-letter graded partition function of transverse D5-branes with CP bundles $\Pi\CO_S$ and $\CO_S(1)$ was computed in \eqref{eq:ft1r} and \eqref{eq:ft1u}.

We can again easily check the matching \eqref{shrieky} reduces to
\be\label{eq:fshmatch1r}
{\bf f}^!_{\cN=1}(x_1,x_2,y_1,y_2,z_1,z_2) = \til{\bf f}_{\cN=1}\bigg(\frac{z_1z_2}{x_1},\frac{(z_1z_2)^3}{x_2},\frac{z_1z_2}{y_1},\frac{z_1z_2}{y_2},z_1,z_2\bigg) \,.
\ee
Under the unrefined specialization the fugacity involution reduces to
\begin{higheqtopless}
\vspace{-0.4em} \label{eq:fmatch1u}
{\bf f}^!_{\cN=1}(t) = \til{\bf f}_{\cN=1}(t^{-1}) \,.
\end{higheqtopless}
\noindent{}This again allows us to identify the partial sums $\sum_{i_0,i_1=0}^{m_0,m_1} G_N^{(i_0,i_1)}$ as contributions of D5/anti-D5-brane pairs with gauge bundle $\Pi\CO_S^{\oplus\,m_0|m_0}\oplus\CO_S(1)^{\oplus\,m_1|m_1}$.


\subsection{Backreaction and Witten's D-instantons} 

The factors of $(\Ber\,\cW_0)^{N_0}$ and $(\Ber\,\cW_1)^{N_1}$ in
\eqref{eq:twonodeG} are the only places where $N_0,N_1$-dependence enters
the MGGE \eqref{eq:two-node-MGGE}. It is possible to understand these
factors as arising from the backreaction of the D5-branes wrapping $\PT$
on the transverse branes. Following the strategy of
\cite{Sharma:2025ntb}, we will compute this backreaction as the effect of
integrating out the strings stretching between the transverse branes and
the branes wrapping $\PT$. In this process, we will also find that the
transverse branes carrying the bundle $\CO(1)$ are Witten's
D-instantons.\footnote{This presentation of D-instantons was previously
used in \cite{Bittleston:2024efo} to construct top-down celestial
holograms.}

We work this out for the $\cN=4$ case, as the $\cN=1$ case is identical.
We are interested in computing the spectrum of bifundamental strings
between a pair of branes represented by bundles $\CE_\PT$, $\CE_Y$
supported on $\PT$ and $Y$ respectively. Such strings are confined to the
brane intersection $Y\cap\PT\simeq\P^1$, the twistor line of the origin. The states of strings stretching from $\CE_Y$ to $\CE_\PT$ are given by
\be
\text{9-5 strings}\;:\quad
\bigoplus_{k=0,1}
H^{0,k}(\P^1,\CE_\PT\otimes\CE_Y^\vee)\,,
\ee
whereas the states of strings stretching from $\CE_\PT$ to $\CE_Y$ are
given by
\be
\text{5-9 strings}\;:\quad
\bigoplus_{k=0,1}
H^{0,k}(\P^1,\CE_Y\otimes\CE_\PT^\vee\otimes K_{\P^1})\,.
\ee
Here, $K_{\P^1}\simeq\CO(-2)$ denotes the canonical bundle of $\P^1$, and we are abusing notation by continuing to denote the restriction $\CE_\PT|_{\P^1}$ by
$\CE_\PT$, etc.

When $\CE_\PT=\CO_\PT^{\oplus N}$ and $\CE_Y=\CO_Y(1)$, we have $\CE_\PT\otimes\CE_Y^\vee=\CE_Y\otimes\CE_\PT^\vee\otimes K_{\P^1} = \CO(-1)^{\oplus N}$. So both types of strings are valued in $\CO(-1)$. Their linearized action is given by
\be
\int_{\P^1}\D\lambda\,\beta\,\dbar\al\,,
\ee
where $\D\lambda=\lambda_\al\d\lambda^\al$ is the weight $2$ volume form
on $\P^1$, and
$\al,\beta\in\Omega^{0,*}(\P^1,\CO(-1))\otimes\Pi\C^N$ are fermions valued
in the fundamental and antifundamental representations of
$\GL(N,\C)$ respectively. This is the action of a degree-1 D-instanton
that Witten used in his formulation of twistor strings as a B-model on
$\P^{3|4}$ \cite{Witten:2003nn}. Correlation functions of such
D-instantons and their higher degree cousins give powerful tools for
computing gauge theory amplitudes \cite{Roiban:2004vt,Roiban:2004yf}. More generally, this allows us to think of the $\CO_Y(1)^{\oplus\,m_1|m_1}$ stack as being comprised of $m_1$ D-instanton/anti-D-instanton pairs.\footnote{For convenience, in this work, we also refer to branes carrying copies of the bundle $\CO_Y$ as D-instantons.}

Next, let us study the origins of the Berezinian factors. The
backreaction of the D5's wrapping $\PT$ can be understood from
open-closed duality as the effect of integrating out the open strings on
the D5's. In particular, the backreaction on our transverse D9 stacks
arises from integrating out the 5-9 and 9-5 bifundamental strings.
Equivalently, we can leave them unintegrated and think of the
backreaction-coupled D9-brane theory as just the total theory of the
9-9, 9-5 and 5-9 strings. The partition function of this total theory
should reproduce the partition function of the theory of 9-9 strings
coupled to backreaction.

We therefore consider the full D5-brane stack
\be
\CE_\PT
=
\CO_\PT^{\oplus N_0}
\oplus
\CO_\PT(-1)^{\oplus N_1}\,,
\ee
together with the transverse brane/antibrane stack
\be
\CE_Y
=
\Pi\CO_Y^{\oplus\,m_0|m_0}
\oplus
\CO_Y(1)^{\oplus\,m_1|m_1}\,.
\ee
The mixed open string spectra between the generator branes are
summarized in table \ref{tab:mixed-open-string-spectra}.
\begin{table}[t]
    \centering
    \renewcommand{\arraystretch}{1.4}
    \begin{tabular}{@{}|c|c|c|c|@{}}
        \hline
        $\CE_\PT$
        & $\CE_Y$
        & 9--5 strings
        & 5--9 strings
        \\
        \hline
        $\CO_\PT$
        & $\CO_Y$
        & $H^{0,0}(\P^1,\CO)\cong\mathbb C$
        & $H^{0,1}(\P^1,K_{\P^1})\cong\mathbb C$
        \\
        $\CO_\PT$
        & $\CO_Y(1)$
        & $0$
        & $0$
        \\
        $\CO_\PT(-1)$
        & $\CO_Y$
        & $0$
        & $0$
        \\
        $\CO_\PT(-1)$
        & $\CO_Y(1)$
        & $H^{0,1}(\P^1,K_{\P^1})\cong\mathbb C$
        & $H^{0,0}(\P^1,\CO)\cong\mathbb C$
        \\
        \hline
    \end{tabular}
    \caption{Mixed open string spectra between the generator branes
    on $\PT$ and $Y$, in the convention described in the text.
    All unlisted cohomology groups vanish.}
    \label{tab:mixed-open-string-spectra}
\end{table}
On-shell, the $\CO_\PT$ stack only couples to the transverse $\Pi\CO_Y$ stack, while the $\CO_\PT(-1)$ stack only couples to the transverse $\CO_Y(1)$ stack. This is precisely the diagonal pairing that is needed to generate the two Berezinian factors in \eqref{eq:twonodeG}.

To make this explicit, let $f_a^{(59)}$ and $f_a^{(95)}$ denote the
signed single-letter counts of 5-9 and 9-5 strings respectively, where $a=0$ labels the $\Pi\CO_Y^{\oplus\,m_0|m_0}$ node and $a=1$ the $\CO_Y(1)^{\oplus\,m_1|m_1}$ node. With
our conventions, a 5-9 string transforms in the fundamental of
$\U(m_a|m_a)$ and a 9-5 string in the antifundamental. Their contribution
to the transverse supermatrix integral is therefore
\be\label{bifundPE}
\PE\bigg[
\sum_{a=0}^1f_a^{(59)}\Str(\cW_a)
+
\sum_{a=0}^1f_a^{(95)}\Str(\cW_a^{-1})
\bigg]\,.
\ee
Let us first consider the $a=0$ node. For the unshifted brane $\CO_Y$,
table \ref{tab:mixed-open-string-spectra} gives a 9-5 state in
$H^{0,0}(\P^1,\CO)$ and a 5-9 state in
$H^{0,1}(\P^1,K_{\P^1})$. Since the mixed open string fields are
fermionic, the former state is fermionic whereas the latter is bosonic.
However, the actual generator appearing in our transverse stack is
$\Pi\CO_Y$, so their parities are reversed. Including the multiplicity
$N_0$ of the $\CO_\PT$ branes therefore gives
\be
f_0^{(59)}=-N_0\,,
\qquad
f_0^{(95)}=+N_0\,.
\ee
For the $a=1$ node, no parity reversal is present. Table
\ref{tab:mixed-open-string-spectra} gives a 9-5 state in
$H^{0,1}(\P^1,K_{\P^1})$ and a 5-9 state in
$H^{0,0}(\P^1,\CO)$. These are respectively bosonic and fermionic, so
including the multiplicity $N_1$ of the $\CO_\PT(-1)$ branes gives
\be
f_1^{(59)}=-N_1\,,
\qquad
f_1^{(95)}=+N_1\,.
\ee
Combining the two nodes, the entire mixed-string contribution is
therefore
\be
\begin{split}
Z_{59+95}&=\PE\bigg[\sum_{a=0}^1N_a\Big(\Str(\cW_a^{-1})-\Str(\cW_a)\Big)\bigg]\\
&=\prod_{a=0}^1\exp\bigg[\sum_{n=1}^\infty\frac{N_a}{n}\Big(\Str(\cW_a^{-n})-\Str(\cW_a^n)\Big)\bigg]=\prod_{a=0}^1\left(\frac{\Ber(1-\cW_a)}{\Ber(1-\cW_a^{-1})}\right)^{N_a}\\
&=(\Ber\,\cW_0)^{N_0}(\Ber\,\cW_1)^{N_1}\,.
\end{split}
\ee
In the last equality, we used
$\Ber(1-\cW^{-1})=\Ber(\cW)^{-1}\Ber(1-\cW)$, where the minus sign in
$1-\cW^{-1}=-\cW^{-1}(1-\cW)$ drops out because
$\Ber(-\mathbf{1}_{m|m})=1$.

This precisely reproduces the two Berezinian insertions appearing in
\eqref{eq:twonodeG}. Thus, all of the $N_0,N_1$-dependence of the
two-node MGGE can be interpreted as the effect of the backreaction of the
original D5-brane system on the transverse brane/antibrane stacks.


\section{Giant graviton expansion of free partition functions}
\label{sec:thermalgge}

In this section, we present and illustrate the giant graviton expansion of the non-BPS thermal and graded partition functions of free $\cN=4$ and $\cN=1$ sYM, using their identifications with the corresponding self-dual partition functions established in section \ref{sec:sdpartition}. To this end, we remove the $\CO_\PT(-1)^{\oplus\,N_1}$ branes introduced in the previous section. I.e., we specialize our two-node quiver gauge theories to the physical $\CO_\PT^{\oplus\,N_0}$ node which engineers the twistorial uplift of $\cN=4$ or $\cN=1$ sdYM. We will occasionally refer to the $\CO_\PT(-1)^{\oplus\,N_1}$ node as a \emph{scaffolding node}.

The crucial point is that even after setting $N_1=0$, the MGGE remains a \emph{double sum} over two types of Koszul-dual brane/antibrane pairs. The sectors associated with the transverse brane stack $\CO(1)^{\oplus\,m_1|m_1}$ are corrections involving Witten's D-instantons from the point of view of the original stack of $N_0=N$ D5-branes. Resumming these sectors recovers the standard one-node MGGE, as explained in section \ref{ssec:murthy}.

In this section, we focus on the thermal partition functions. As explained in sections \ref{sec:partition} and \ref{sec:sdpartition}, the thermal partition function is obtained from the graded partition function by an R-symmetry twist, which can be implemented by mapping the fugacities
\begin{align}
    \fx\mapsto -\fx , , \qquad  \fy\mapsto\fy , , \qquad  \fz\mapsto \fz \label{eq:twistr}
\end{align}
in the refined case, and
\begin{align}
    t \mapsto -t \label{eq:twistu}
\end{align}
in the unrefined case. The utility of identifying the thermal partition function with the twisted graded partition function is that it allows us to identify the ingredients appearing on the right-hand side of the MGGE with twisted graded partition functions of BCOV theory and the transverse branes. This is natural, since it is precisely these R-twisted graded bulk partition functions that are compatible with the thermal boundary conditions.

We illustrate the expansions for the unrefined thermal partition functions of free $\cN=4$ and $\cN=1$ sYM in sections \ref{ssec:gge4} and \ref{ssec:gge1}, respectively, by computing several corrections through order $t^{8}$ (or $t^9$). We also illustrate how the scaffolding node terms resum to the one-node GGE. We note that the underlying expansions are exact up to all orders by results of \cite{borodin2000fredholm,Murthy:2022ien,Purkayastha:2025whi}.

In the final section \ref{ssec:interpretations}, we also take initial steps in relating our Murthy-type expansion to the more conventional giant graviton expansions \cite{Imamura:2021ytr,Gaiotto:2021xce,Lee:2022vig} by identifying contributions of some simple giant graviton branes dual to ordinary determinant operators.


\subsection{Thermal GGE for free $\cN=4$ sYM}
\label{ssec:gge4}

Our main result is obtained by removing the scaffolding node. Setting $N_0=N$ and $N_1=0$ in the two-node expansion \eqref{eq:two-node-MGGE}, we find
\begin{higheq}\label{ggemain}
Z_N[f_{\cN=4}] = Z_\text{BCOV}^{\cN=4}\sum_{m_0=0}^\infty \sum_{m_1=0}^\infty G_{N}^{(m_0,m_1)}[\mathbf{f}_{\cN=4}]\,,
\end{higheq}
\noindent{}where $G_{N}^{(m_0,m_1)}\vcentcolon=G_{N,0}^{(m_0,m_1)}$ are obtained by setting $N_0=N$ and $N_1=0$ in \eqref{eq:twonodeG}:
\be\label{GNm}
G_{N}^{(m_0,m_1)}[\mathbf{f}] = \int\limits_{\U(m_0|m_0)}\d \cW_0\,(\Ber\,\cW_0)^N\!\int\limits_{\U(m_1|m_1)}\d \cW_1\,\PE\bigg[\sum_{a,b=0}^1{(\bf f}^!)_{ab}\,\Str(\cW_a)\,\Str(\cW_b^{-1})\bigg] .
\ee
The measures and contours are specified in sections \ref{ssec:murthy} and \ref{sec:ggemain}.

This gives a giant graviton expansion of the graded partition function of self-dual $\cN=4$ sYM, and hence of free $\cN=4$ sYM. The corresponding thermal expansion follows by applying the twist \eqref{eq:twistr} or \eqref{eq:twistu} to both sides of \eqref{ggemain}. In the graded case, the RHS consists of the graded partition function of BCOV theory, systematically corrected by the graded partition functions of transverse brane/antibrane pairs. In the thermal case, the RHS here is \emph{not} built from the ordinary thermal partition functions of the individual bulk theories.\footnote{In our conventions, these would instead be obtained by replacing $y\mapsto-y$, rather than $x\mapsto-x$, in their graded counterparts.} Instead, we encounter R-twisted graded partition functions in the bulk. This is not a contradiction: holography only requires the boundary thermal partition function to agree with a bulk path integral obeying the corresponding thermal boundary conditions. The R-twist precisely implements these boundary conditions for our bulk fields.

Notice that since we have set $N_1=0$, only the $\U(m_0|m_0)$ integral interacts with the data of the boundary rank $N_0=N$.  The $\U(m_1|m_1)$ integrals act as an additional layer of D-instanton corrections. By taking an $N\to\infty$ limit of our GGE, we obtain an expansion of $Z_\infty$ itself
\begin{higheq}\label{Dinstexp}
Z_\infty[f_{\cN=4}] = Z_\text{BCOV}^{\cN=4}\sum_{m_1=0}^\infty G_\infty^{(0,m_1)}[\mathbf{f}_{\cN=4}]\,.
\end{higheq}
\noindent{}The corrections with $m_0=0$ are independent of $N$ and we can denote them $G^{(0,m_1)}_\infty$. This makes manifest the first difference of our non-BPS GGE from the GGEs of BPS indices. The corrections in BPS GGEs drop out completely as $N\to\infty$, but this is not the case here. For us, the infinite-$N$ closed string answer itself receives corrections.

Resumming the scaffolding sectors at fixed $m_0$ recovers the standard one-node Murthy expansion reviewed in section \ref{ssec:murthy},
\begin{align}
    Z_N[f_{\cN=4}] &= Z_\infty[f_{\cN=4}] \sum_{m_0=0}^\infty \widetilde{G}^{(m_0)}_N[f_{\cN=4}] \,.
\end{align}
Here, the summands are given by
\be\label{GtNm}
\widetilde{G}_{N}^{(m_0)}[f] = \int\limits_{\U(m_0|m_0)}\d \cW_0\,(\Ber\,\cW_0)^N \,\PE\bigg[ f^!\, \,\Str\,\cW_0\,\Str\,\cW_0^{-1}\bigg]\,,
\ee
and in the $\cN=4$ case, $f^!$ is found to be
\be
f^!_{\cN=4} = -\frac{f_{\cN=4}}{1-f_{\cN=4}} = 1-\frac{z_1-z_2}{z_1P(z_1)-z_2P(z_2)}\,.
\ee
The one-node corrections are related to the two-node corrections by \eqref{eq:GGrelation}:
\begin{align}
    \widetilde{G}^{(m_0)}_N[f_{\cN=4}] &= \frac{Z_{\textrm{BCOV}}^{\cN=4}}{Z_\infty[f_{\cN=4}]} \sum_{m_1=0}^\infty G^{(m_0,m_1)}_N[\mathbf{f}_{\cN=4}] \, .
\end{align}
We remark that it is highly nontrivial to obtain this secondary expansion over $m_1$ and is the key place where the input of chiral holography was essential.

\paragraph{Numerical tests.} Here we present explicit expansions for the unrefined thermal partition function of free $\cN=4$ sYM. Throughout the rest of this section, $Z_N$, $Z_\infty$ denote their thermal versions and $Z_{\mathrm{BCOV}}$, $G$, $\widetilde{G}$ their twisted graded versions in the $\cN=4$ sYM case.

The infinite-$N$ thermal partition function is given by
\be
Z_\infty = 1+6 t^2+16 t^3+72 t^4+240 t^5+928t^6+3456 t^7+12954 t^8+48112t^9+\mathrm{O}(t^{10})\,.
\ee
Our corrections will systematically reduce this to the finite-$N$ answer $Z_N$. We also present the BCOV twisted graded partition function for completeness,
\begin{multline}
    Z_\text{BCOV} = 1-8 t+40 t^2-160 t^3+552 t^4-1712
   t^5+4896 t^6\\-13120 t^7+33320
   t^8-80872 t^9+\mathrm{O}(t^{10})\,.
\end{multline}
This is obtained by sending $t\mapsto-t$ in \eqref{Zbcov4}.

Generalizing a result of \cite{Liu:2022olj}, the corrections $G_{N_0,N_1}^{(m_0,m_1)}$ in the two-node GGE are expected to start at order $\mathrm{O}(t^L)$, where\footnote{Notice e.g.\ that four brane/antibrane pairs $(m_0,m_1)=(2,2)$ contribute earlier than certain three brane/antibrane pairs $(m_0,m_1)=(3,0)$.}
\begin{align}
    L=\max\qty[ (N_0+m_0)m_0 \, , (N_1+m_1)m_1 ] \, . \label{eq:L4}
\end{align}
We confirm this experimentally for the $N_1=0$ case by studying the expansions of $G^{(m_0,m_1)}_N$.

\begin{enumerate}[leftmargin=*]

\item $N=0$

\begin{table}[h!]
\centering
\small
\renewcommand{\arraystretch}{1.5}
\begin{tabular}{c|cccccccc}
$G^{(0,1)}_0$
& $+8t$ & $+30t^2$ & $+96t^3$ & $-368t^4$ & $-2832t^5$ & $+792t^6$ & $+44992t^7$ & $+14028t^8$ \\
$G^{(0,2)}_0$
& & & & $+736t^4$ & $+4272t^5$ & $+4536t^6$ & $-25312t^7$ & $+59622t^8$ \\
\hline
$G^{(1,0)}_0$
& $+8t$ & $+30t^2$ & $+96t^3$ & $-368t^4$ & $-2832t^5$ & $+792t^6$ & $+44992t^7$ & $+14028t^8$ \\
$G^{(1,1)}_0$
& $-8t$ & $-36t^2$ & $-160t^3$ & $+1096t^4$ & $+4752t^5$ & $-25928t^6$ & $-158944t^7$ & $+508128t^8$ \\
$G^{(1,2)}_0$
& & & & $-1072t^4$ & $-3312t^5$ & $+19904t^6$ & $+94336t^7$ & $-596028t^8$ \\
\hline
$G^{(2,0)}_0$
& & & & $+736t^4$ & $+4272t^5$ & $+4536t^6$ & $-25312t^7$ & $+59622t^8$ \\
$G^{(2,1)}_0$
& & & & $-1072t^4$ & $-3312t^5$ & $+19904t^6$ & $+94336t^7$ & $-596028t^8$ \\
$G^{(2,2)}_0$
& & & & $+336 t^4$ & $-960 t^5$ & $-24440 t^6$ & $-69024 t^7$ & $+536652 t^8$ \\
\end{tabular}
\caption{Two-node corrections for $N=0$ up to order $t^8$, with $G^{(0,0)}_{0}=1$. The next corrections $(m_0,m_1)=(0,3),\dots$ are expected at $\mathrm{O}(t^9)$.}
\label{tab:G00}
\end{table}

First, we sum the corrections from table \ref{tab:G00} at fixed $m_0$. After multiplying by $Z_{\mathrm{BCOV}}/Z_\infty$, we compare these sums with the one-node corrections:
\begingroup
\allowdisplaybreaks
\begin{align}
    \frac{Z_{\mathrm{BCOV}}}{Z_\infty} \sum_{m_1=0}^2 G^{(0,m_1)}_0 &= 1 + \mathrm{O}(t^9) \\
    \widetilde{G}^{(0)}_0 &= 1 \\
    \frac{Z_{\mathrm{BCOV}}}{Z_\infty} \sum_{m_1=0}^2 G^{(1,m_1)}_0 &= -6 t^2 - 16 t^3 - 36 t^4 - 48 t^5 - 24 t^6 - 288 t^8 + \mathrm{O}(t^9) \\
    \widetilde{G}^{(1)}_0 &= -6 t^2-16 t^3-36 t^4-48 t^5-24 t^6-288 t^8 + \mathrm{O}(t^9) \\
    \frac{Z_{\mathrm{BCOV}}}{Z_\infty} \sum_{m_1=0}^2 G^{(2,m_1)}_0 &= 246 t^8 + \mathrm{O}(t^9) \\
    \widetilde{G}^{(2)}_0 &= 246 t^8 + \mathrm{O}(t^9) \, .
\end{align}
\endgroup

Then, we sum all corrections from table \ref{tab:G00} and compare the result with the zero-rank partition function:
\begin{align}
    Z_{\textrm{BCOV}} \sum_{m_0}^2\sum_{m_1}^2 G^{(m_0,m_1)}_0 &= 1+\mathrm{O}(t^9) \, \\
   Z_0 &= 1 \, .
\end{align}

\item $N=1$

\begin{table}[h!]
\centering
\small
\renewcommand{\arraystretch}{1.5}
\begin{tabular}{c|cccccccc}
$G^{(0,1)}_1$
& $+8 t$ & $+30 t^2$ & $+96 t^3$ & $-368 t^4$ & $-2832 t^5$ & $+792 t^6$ & $+44992 t^7$ & $+14028 t^8$ \\
$G^{(0,2)}_1$
& & & & $+736 t^4$ & $+4272 t^5$ & $+4536 t^6$ & $-25312 t^7$ & $+ 59622t^8$ \\
\hline
$G^{(1,0)}_1$
& & $-36 t^2$ & $-128 t^3$ & $+99 t^4$ & $+2592 t^5$ & $+2036 t^6$ & $-38016 t^7$ & $-40887 t^8$ \\
$G^{(1,1)}_1$
& & $+36 t^2$ & $+128 t^3$ & $+596 t^4$ & $+1176 t^5$ & $+8116 t^6$ & $+35280 t^7$ & $-232314 t^8$ \\
$G^{(1,2)}_1$
& & & & $-716 t^4$ & $-4032 t^5$ & $-11926 t^6$ & $-6512 t^7$ & $+ 229830 t^8$ \\
\hline
$G^{(2,0)}_1$
& & & & & & $+10536 t^6$ & $+69120 t^7$ & $+34398 t^8$ \\
$G^{(2,1)}_1$
& & & & & & $-19412 t^6$ & $-106736 t^7$ & $+240972 t^8$ \\
$G^{(2,2)}_1$
& & & & & & $+8876 t^6$ & $+37616 t^7$ & $-275370 t^8$ \\
\end{tabular}
\caption{Two-node corrections for $N=1$ up to order $t^8$, with $G^{(0,0)}_{1}=1$. The next corrections $(m_0,m_1)=(0,3),\dots$ are expected at $\mathrm{O}(t^9)$.}
\label{tab:G10}
\end{table}

We sum the corrections from table \ref{tab:G10} at fixed $m_0$. After multiplying by $Z_{\mathrm{BCOV}}/Z_\infty$, we compare these sums with the one-node corrections:
\begingroup
\allowdisplaybreaks
\begin{align}
    \frac{Z_{\mathrm{BCOV}}}{Z_\infty} \sum_{m_1=0}^2 G^{(0,m_1)}_1 &= 1 + \mathrm{O}(t^9) \\
    \widetilde{G}^{(0)}_1 &= 1 \\
    \frac{Z_{\mathrm{BCOV}}}{Z_\infty} \sum_{m_1=0}^2 G^{(1,m_1)}_1 &= -21 t^4 - 96 t^5 - 376 t^6 - 1344 t^7 - 4395 t^8 + \mathrm{O}(t^9) \\
    \widetilde{G}^{(1)}_1 &= -21 t^4-96 t^5-376 t^6-1344 t^7-4395 t^8 + \mathrm{O}(t^9) \\
    \frac{Z_{\mathrm{BCOV}}}{Z_\infty} \sum_{m_1=0}^2 G^{(2,m_1)}_1 &= 0 + \mathrm{O}(t^9)\\
    \widetilde{G}^{(2)}_1 &= 0 + \mathrm{O}(t^9) \, .
\end{align}
\endgroup

Then, we sum all corrections from table \ref{tab:G10} and compare the result with the $N=1$ partition function:
\begin{align}
\begin{split}
    Z_{\textrm{BCOV}} \smash[b]{\sum_{m_0,m_1=0}^2} G^{(m_0,m_1)}_1 &= 1 + 6 t^2 + 16 t^3 + 51 t^4 + 144 t^5 \\
    &\hspace{1.5cm} + 426 t^6 + 1200 t^7 + 3255 t^8 + \mathrm{O}(t^9) \,
\end{split} \\
\begin{split}
   Z_1 &= 1+6 t^2+16 t^3+51 t^4+144 t^5 \\
   &\hspace{1.5cm} +426 t^6+1200 t^7+3255 t^8 + \mathrm{O}(t^9) \, .
\end{split}
\end{align}
The first difference from $Z_\infty$ appears at $\mathrm{O}(t^4)$. At $N=1$, the $21$ single-trace operators $\Tr(\Phi_{IJ}\Phi_{KL})$ are not independent from the $21$ double-trace operators $\Tr(\Phi_{IJ})\Tr(\Phi_{KL})$. So one gets a difference of $Z_1-Z_\infty=-21t^4+\mathrm{O}(t^5)$. The differences at higher orders in $t$ have a similar origin.

\item $N=2$

\begin{table}[h!]
\centering
\small
\renewcommand{\arraystretch}{1.5}
\begin{tabular}{c|cccccccc}
$G^{(0,1)}_2$
& $+8 t$ & $+30 t^2$ & $+96 t^3$ & $-368 t^4$ & $-2832 t^5$ & $+792 t^6$ & $+44992 t^7$ & $+14028 t^8$ \\
$G^{(0,2)}_2$
& & & & $+736 t^4$ & $+4272 t^5$ & $+4536 t^6$ & $-25312 t^7$ & $+59622 t^8$ \\
\hline
$G^{(1,0)}_2$
& & & $+120 t^3$ & $+324 t^4$ & $-1296 t^5$ & $-6428 t^6$ & $+16344 t^7$ & $+82458 t^8$ \\
$G^{(1,1)}_2$
& & & $-120 t^3$ & $-124 t^4$ & $-1080 t^5$ & $-25640 t^6$ & $-132896 t^7$ & $+255576 t^8$ \\
$G^{(1,2)}_2$
& & & & $-200 t^4$ & $+2376 t^5$ & $+32012 t^6$ & $+115768 t^7$ & $-344088 t^8$ \\
\hline
$G^{(2,0)}_2$
& & & & & & & & $+114696 t^8$ \\
$G^{(2,1)}_2$
& & & & & & & & $-224784 t^8$ \\
$G^{(2,2)}_2$
& & & & & & & & $+110088 t^8$ \\
\end{tabular}
\caption{Two-node corrections for $N=2$ up to order $t^8$, with $G^{(0,0)}_{2}=1$. The next corrections $(m_0,m_1)=(0,3),\dots$ are expected at $\mathrm{O}(t^9)$.}
\label{tab:G20}
\end{table}

We sum the corrections from table \ref{tab:G20} at fixed $m_0$. After multiplying by $Z_{\mathrm{BCOV}}/Z_\infty$, we compare these sums with the one-node corrections:
\begingroup
\allowdisplaybreaks
\begin{align}
    \frac{Z_{\mathrm{BCOV}}}{Z_\infty} \sum_{m_1=0}^2 G^{(0,m_1)}_2 &= 1 + \mathrm{O}(t^9) \\
    \widetilde{G}^{(0)}_2 &= 1 \\
    \frac{Z_{\mathrm{BCOV}}}{Z_\infty} \sum_{m_1=0}^2 G^{(1,m_1)}_2 &= -56 t^6 - 336 t^7 - 1686 t^8 + \mathrm{O}(t^9) \\
    \widetilde{G}^{(1)}_2 &= -56 t^6-336 t^7-1686 t^8 + \mathrm{O}(t^9) \\
    \frac{Z_{\mathrm{BCOV}}}{Z_\infty} \sum_{m_1=0}^2 G^{(2,m_1)}_2 &= 0 + \mathrm{O}(t^9) \\
    \widetilde{G}^{(2)}_2 &= 0 + \mathrm{O}(t^9) \, .
\end{align}
\endgroup

Then, we sum all corrections from table \ref{tab:G20} and compare the result with the $N=2$ partition function:
\begingroup
\allowdisplaybreaks
\begin{align}
\begin{split}
    Z_{\textrm{BCOV}} \smash[b]{\sum_{m_0,m_1=0}^2} G^{(m_0,m_1)}_2 &= 1 + 6 t^2 + 16 t^3 + 72 t^4 + 240 t^5 \\
    &\hspace{1.5cm} + 872 t^6 + 3120 t^7 + 10932 t^8 + \mathrm{O}(t^9) \, 
\end{split}\\
\begin{split}
   Z_2 &= 1+6 t^2+16 t^3+72 t^4+240 t^5 \\
   &\hspace{1.5cm} +872 t^6+3120 t^7+10932 t^8 + \mathrm{O}(t^9) \, .
\end{split}
\end{align}
\endgroup
The difference $Z_2-Z_\infty = -56t^6 + \mathrm{O}(t^7)$ here starts at $\mathrm{O}(t^6)$. As expected, the finite-$N$ answer systematically approaches the large-$N$ answer as $N$ is increased. More generally, we have experimentally checked up to $N=6$ that $Z_N-Z_\infty = -{N+6\choose 5}t^{2N+2}+\mathrm{O}(t^{2N+3})$.

\item $N=\infty$. 

\begin{table}[h!]
\centering
\small
\renewcommand{\arraystretch}{1.5}
\begin{tabular}{c|ccccccccc}
$G^{(0,1)}_\infty$
& $+8 t$ & $+30 t^2$ & $+96 t^3$ & $-368 t^4$ & $-2832 t^5$ & $+792 t^6$ & $+44992 t^7$ & $+14028 t^8$ & $-589752 t^9$ \\
$G^{(0,2)}_\infty$
& & & & $+736 t^4$ & $+4272 t^5$ & $+4536 t^6$ & $-25312 t^7$ & $+59622 t^8$ & $+159264 t^9$ \\
$G^{(0,3)}_\infty$
& & & & & & & & & $+706112 t^9$ \\
\end{tabular}
\caption{Instanton corrections for $N=\infty$ up to order $t^9$. The next corrections $(m_0,m_1)=(0,4),\dots$ are expected at $\mathrm{O}(t^{16})$.}
\label{tab:Zinf4}
\end{table}

Here we illustrate the instanton corrections \eqref{Dinstexp} to $Z_\infty$. We sum all corrections from table \ref{tab:Zinf4} and compare the result with the $N=\infty$ partition function:
\begingroup
\allowdisplaybreaks
\begin{align}
\begin{split}
    Z_{\textrm{BCOV}} \smash[b]{\sum_{m_1=0}^3} G^{(0,m_1)}_\infty &= 1 + 6 t^2 + 16 t^3 + 72 t^4 + 240 t^5 + 928 t^6 \\
    &\hspace{2cm} + 3456 t^7 + 
 12954 t^8 + 48112 t^9 + \mathrm{O}(t^{10})
\end{split} \\
\begin{split}
    Z_\infty &= 1+6 t^2+16 t^3+72 t^4+240 t^5+928 t^6 \\
    &\hspace{2cm} +3456 t^7+12954 t^8 + 48112 t^9 + \mathrm{O}(t^{10}) \, .
\end{split}
\end{align}
\endgroup
This explicitly resolves the mismatch between the infinite-$N$ gauge theory partition function and the bulk closed string partition function. The closed twistor string answer receives corrections from Witten's D-instanton/anti-D-instanton pairs.

\end{enumerate}


\subsection{Thermal GGE for free $\cN=1$ sYM}
\label{ssec:gge1}

In the $\cN=1$ case, we likewise set the scaffolding node to zero. Since the construction parallels the $\cN=4$ case, we present slightly less details in this section.

Setting $N_0=N$ and $N_1=0$ in the two-node expansion \eqref{eq:two-node-MGGE}, we find
\begin{higheq}\label{ggemain1}
Z_N[f_{\cN=1}] = Z_\text{BCOV}^{\cN=1}\sum_{m_0=0}^\infty \sum_{m_1=0}^\infty G_{N}^{(m_0,m_1)}[\mathbf{f}_{\cN=1}]\,,
\end{higheq}
\noindent{}where $G_{N}^{(m_0,m_1)}:=G_{N,0}^{(m_0,m_1)}$ are obtained by setting $N_0=N$ and $N_1=0$ in \eqref{eq:twonodeG}. 

This gives a giant graviton expansion of the graded partition function of self-dual $\cN=1$ sYM, and hence of free $\cN=1$ sYM. The corresponding thermal expansion follows by applying the twist \eqref{eq:twistr} or \eqref{eq:twistu}. The bulk derivation and holographic interpretation of the terms on the RHS were explained in section \ref{sec:ggemain}.

By taking an $N\to\infty$ limit of our GGE, we obtain an expansion of $Z_\infty$ itself
\begin{higheq}\label{Dinstexp1}
Z_\infty[f_{\cN=1}] = Z_\text{BCOV}^{\cN=1}\sum_{m_1=0}^\infty G_\infty^{(0,m_1)}[\mathbf{f}_{\cN=1}]\,.
\end{higheq}
\noindent{}The corrections with $m_0=0$ are independent of $N$ and we can denote them $G^{(0,m_1)}_\infty$. Just as with the $\cN=4$ case, the $\cN=1$ closed twistor string partition function receives D-instanton/anti-D-instanton corrections that reduce it to the correct $Z_\infty$. These are the D-instantons of $\cN=1$ twistor strings \cite{Popov:2004nk}.

Resumming the scaffolding sectors at fixed $m_0$ recovers the standard one-node Murthy expansion, reviewed in section \ref{ssec:murthy}:
\begin{align}
    Z_N[f_{\cN=1}] &= Z_\infty[f_{\cN=1}] \sum_{m_0=0}^\infty \widetilde{G}^{(m_0)}_N[f_{\cN=1}] \, .
\end{align}
The one-node corrections are related to the two-node corrections by \eqref{eq:GGrelation}:
\begin{align}
    \widetilde{G}^{(m_0)}_N[f_{\cN=1}] &= \frac{Z_{\textrm{BCOV}}^{\cN=1}}{Z_\infty[f_{\cN=1}]} \sum_{m_1=0}^\infty G^{(m_0,m_1)}_N[\mathbf{f}_{\cN=1}] \, .
\end{align}

\paragraph{Numerical tests.} Here we present the explicit expansions for the unrefined thermal partition function of free $\cN=1$ sYM. Throughout the rest of this section, $Z_N$, $Z_\infty$ denote their thermal versions and $Z_{\mathrm{BCOV}}$, $G$, $\widetilde{G}$ their twisted graded versions in the $\cN=1$ sYM case. Let us list the thermal $Z_\infty$ and the twisted graded $Z_\text{BCOV}$ for completeness,
\begin{align}
    Z_\infty &= 1+4 t^3+6 t^4+12 t^5+28 t^6+72t^7+168 t^8+364 t^9+\mathrm{O}(t^{10})\,,\\
    Z_\text{BCOV} &= 1-2 t+t^2-4 t^3+6 t^4-2 t^5+12 t^6-16 t^7+5
   t^8-28 t^9+\mathrm{O}(t^{10})\,.
\end{align}
These are used in our tests below.

\begin{enumerate}[leftmargin=*]

\item $N=0$

\begin{table}[h!]
\centering
\small
\renewcommand{\arraystretch}{1.5}
\begin{tabular}{c|cccccccc}
$G^{(0,1)}_0$
& $+2 t$ & $+ 3 t^2$ &  $+ 12 t^3$ & $+ 23 t^4$ & $+42 t^5$ & $+63 t^6$ & $+70 t^7$ & $- 9 t^8$ \\
$G^{(0,2)}_0$
& & & & $+6 t^4$ & $+ 18 t^5$ & $+ 78 t^6$ & $+ 266 t^7$ & $+ 789 t^8$ \\
\hline
$G^{(1,0)}_0$
& $+2 t$ & $+ 3 t^2$ & $+ 12 t^3$ & $+ 23 t^4$ & $+ 42 t^5$ & $+ 63 t^6$ & $+ 70 t^7$ & $- 9 t^8$ \\
$G^{(1,1)}_0$
& $-2 t$ & $- 3 t^2$ & $- 16 t^3$ & $- 25 t^4$ & $- 54 t^5$ & $- 79 t^6$ & $- 114 t^7$ & $- 81 t^8$ \\
$G^{(1,2)}_0$
& & & & $-12 t^4$ & $- 24 t^5$ & $- 86 t^6$ & $- 228 t^7$ & $- 588 t^8$ \\
\hline
$G^{(2,0)}_0$
& & & & $+6 t^4$ & $+ 18 t^5$ & $+ 78 t^6$ & $+ 266 t^7$ & $+ 789 t^8$ \\
$G^{(2,1)}_0$
& & & & $-12 t^4$ & $- 24 t^5$ & $- 86 t^6$ & $- 228 t^7$ & $- 588 t^8$ \\
$G^{(2,2)}_0$
& & & & $+6 t^4$ & $+ 6 t^5$ & $+ 8 t^6$ & $- 38 t^7$ & $- 201 t^8$ \\
\end{tabular}
\caption{Two-node corrections for $N=0$ up to order $t^8$, with $G^{(0,0)}_{0}=1$. The next corrections $(m_0,m_1)=(0,3),\dots$ are expected at $\mathrm{O}(t^9)$.}
\label{tab:G100}
\end{table}

First, we sum the corrections from table \ref{tab:G100} at fixed $m_0$. After multiplying by $Z_{\mathrm{BCOV}}/Z_\infty$, we compare these sums with the one-node corrections:
\begingroup
\allowdisplaybreaks
\begin{align}
    \frac{Z_{\mathrm{BCOV}}}{Z_\infty} \sum_{m_1=0}^2 G^{(0,m_1)}_0 &=1 +\mathrm{O}(t^9) \\
    \widetilde{G}^{(0)}_0 &= 1 \\
    \frac{Z_{\mathrm{BCOV}}}{Z_\infty} \sum_{m_1=0}^2 G^{(1,m_1)}_0 &= -4 t^3 - 6 t^4 - 12 t^5 - 12 t^6 - 24 t^7 - 36 t^8 + \mathrm{O}(t^9) \\
    \widetilde{G}^{(1)}_0 &= -4 t^3-6 t^4-12 t^5-12 t^6-24 t^7-36 t^8 + \mathrm{O}(t^9) \\
    \frac{Z_{\mathrm{BCOV}}}{Z_\infty} \sum_{m_1=0}^2 G^{(2,m_1)}_0 &= 0 + \mathrm{O}(t^9)\\
    \widetilde{G}^{(2)}_0 &= 0 + \mathrm{O}(t^9) \, .
\end{align}
\endgroup

Then, we sum all corrections from table \ref{tab:G100} and compare the result with the zero-rank partition function:
\begin{align}
    Z_{\textrm{BCOV}} \sum_{m_0}^2\sum_{m_1}^2 G^{(m_0,m_1)}_0 &= 1 + \mathrm{O}(t^9) \, \\
   Z_0 &= 1 \, .
\end{align}

\item $N=1$

\begin{table}[h!]
\centering
\small
\renewcommand{\arraystretch}{1.5}
\begin{tabular}{c|cccccccc}
$G^{(0,1)}_1$
& $+2 t$ & $+ 3 t^2$ & $+ 12 t^3$ & $+ 23 t^4$ & $+ 42 t^5$ & $+ 63 t^6$ & $+ 70 t^7$ & $- 9 t^8$ \\
$G^{(0,2)}_1$
& & & & $+6 t^4$ & $+ 18 t^5$ & $+ 78 t^6$ & $+ 266 t^7$ & $+ 789 t^8$ \\
\hline
$G^{(1,0)}_1$
& & $-3 t^2$ & $- 6 t^3$ & $- 21 t^4$ & $- 42 t^5$ & $- 69 t^6$ & $- 78 t^7$ & $- 30 t^8$ \\
$G^{(1,1)}_1$
& & $+3 t^2$ & $+ 6 t^3$ & $+ 27 t^4$ & $+ 60 t^5$ & $+ 150 t^6$ & $+ 342 t^7$ & $+ 837 t^8$ \\
$G^{(1,2)}_1$
& & & & $-6 t^4$ & $- 18 t^5$ & $- 87 t^6$ & $- 300 t^7$ & $- 942 t^8$ \\
\hline
$G^{(2,0)}_1$
& & & & & & $+10 t^6$ & $+ 36 t^7$ & $+ 171 t^8$ \\
$G^{(2,1)}_1$
& & & & & & $-20 t^6$ & $- 72 t^7$ & $- 324 t^8$ \\
$G^{(2,2)}_1$
& & & & & & $+10 t^6$ & $+ 36 t^7$ & $+ 153 t^8$ \\
\end{tabular}
\caption{Two-node corrections for $N=1$ up to order $t^8$, with $G^{(0,0)}_{1}=1$. The next corrections $(m_0,m_1)=(0,3),\dots$ are expected at $\mathrm{O}(t^9)$.}
\label{tab:G110}
\end{table}

We sum the corrections from table \ref{tab:G110} at fixed $m_0$. After multiplying by $Z_{\mathrm{BCOV}}/Z_\infty$, we compare these sums with the one-node corrections:
\begingroup
\allowdisplaybreaks
\begin{align}
    \frac{Z_{\mathrm{BCOV}}}{Z_\infty} \sum_{m_1=0}^2 G^{(0,m_1)}_1 &= 1 + \mathrm{O}(t^9) \\
    \widetilde{G}^{(0)}_1 &= 1 \\
    \frac{Z_{\mathrm{BCOV}}}{Z_\infty} \sum_{m_1=0}^2 G^{(1,m_1)}_1 &= -6 t^6 - 24 t^7 - 69 t^8 + \mathrm{O}(t^9) \\
    \widetilde{G}^{(1)}_1 &= -6 t^6-24 t^7-69 t^8 + \mathrm{O}(t^9) \\
    \frac{Z_{\mathrm{BCOV}}}{Z_\infty} \sum_{m_1=0}^2 G^{(2,m_1)}_1 &= 0 + \mathrm{O}(t^9) \\
    \widetilde{G}^{(2)}_1 &= 0 + \mathrm{O}(t^9) \, .
\end{align}
\endgroup

Then, we sum all corrections from table \ref{tab:G110} and compare the result with the $N=1$ partition function:
\begin{align}
\begin{split}
    Z_{\textrm{BCOV}} \smash[b]{\sum_{m_0,m_1=0}^2} G^{(m_0,m_1)}_1 &= 1 + 4 t^3 + 6 t^4 + 12 t^5 \\
    &\hspace{1.5cm} + 22 t^6 + 48 t^7 + 99 t^8 + \mathrm{O}(t^9) \, 
\end{split} \\
\begin{split}
   Z_1 &= 1+4 t^3+6 t^4+12 t^5 \\
   &\hspace{1.5cm} +22 t^6+48 t^7+99 t^8 + \mathrm{O}(t^9) \, .
\end{split}
\end{align}
Here, the difference $Z_1-Z_\infty = -6t^6+\mathrm{O}(t^7)$ signals that at $N=1$, the 6 single-trace operators $\Tr\textstyle(\Psi_\al\til\Psi_\dal)$, $\Tr(\Psi_\al\Psi^\al)$, $\Tr\textstyle(\til\Psi_\dal\til\Psi^\dal)$ are no longer independent of the 6 double-trace operators $\Tr(\Psi_\al)\Tr\textstyle(\til\Psi_\dal)$, $\Tr(\Psi_\al)\Tr(\Psi^\al)$, $\Tr\textstyle(\til\Psi_\dal)\Tr\textstyle(\til\Psi^\dal)$.

\item $N=2$

\begin{table}[h!]
\centering
\small
\renewcommand{\arraystretch}{1.5}
\begin{tabular}{c|cccccccc}
$G^{(0,1)}_2$
& $+2 t$ & $+ 3 t^2$ & $+ 12 t^3$ & $+ 23 t^4$ & $+ 42 t^5$ & $+ 63 t^6$ & $+ 70 t^7$ & $- 9 t^8$ \\
$G^{(0,2)}_2$
& & & & $+6 t^4$ & $+ 18 t^5$ & $+ 78 t^6$ & $+ 266 t^7$ & $+ 789 t^8$ \\
\hline
$G^{(1,0)}_2$
& & & $+4 t^3$ & $+ 9 t^4$ & $+ 30 t^5$ & $+ 61 t^6$ & $+ 96 t^7$ & $+ 96 t^8$ \\
$G^{(1,1)}_2$
& & & $-4 t^3$ & $- 9 t^4$ & $- 36 t^5$ & $- 75 t^6$ & $- 210 t^7$ & $- 606 t^8$ \\
$G^{(1,2)}_2$
& & & & & $+6 t^5 $ &$ +14 t^6$ & $+ 114 t^7$ & $+ 510 t^8$ \\
\hline
$G^{(2,0)}_2$
& & & & & & & & $+15 t^8$ \\
$G^{(2,1)}_2$
& & & & & & & & $-30 t^8$ \\
$G^{(2,2)}_2$
& & & & & & & & $+15 t^8$ \\
\end{tabular}
\caption{Two-node corrections for $N=2$ up to order $t^8$, with $G^{(0,0)}_{2}=1$. The next corrections $(m_0,m_1)=(0,3),\dots$ are expected at $\mathrm{O}(t^9)$.}
\label{tab:G120}
\end{table}

We sum the corrections from table \ref{tab:G120} at fixed $m_0$. After multiplying by $Z_{\mathrm{BCOV}}/Z_\infty$, we compare these sums with the one-node corrections:
\begingroup
\allowdisplaybreaks
\begin{align}
    \frac{Z_{\mathrm{BCOV}}}{Z_\infty} \sum_{m_1=0}^2 G^{(0,m_1)}_2 &= 1 + \mathrm{O}(t^9) \\
    \widetilde{G}^{(0)}_2 &= 1 \\
    \frac{Z_{\mathrm{BCOV}}}{Z_\infty} \sum_{m_1=0}^2 G^{(1,m_1)}_2 &= \mathrm{O}(t^9) \\
    \widetilde{G}^{(1)}_2 &= \mathrm{O}(t^9)\\
    \frac{Z_{\mathrm{BCOV}}}{Z_\infty} \sum_{m_1=0}^2 G^{(2,m_1)}_2 &= \mathrm{O}(t^9) \\
    \widetilde{G}^{(2)}_2 &= \mathrm{O}(t^9)\, .
\end{align}
\endgroup

Then, we sum all corrections from table \ref{tab:G120} and compare the result with the $N=2$ partition function:
\begin{align}
\begin{split}
    Z_{\textrm{BCOV}} \smash[b]{\sum_{m_0,m_1=0}^2} G^{(m_0,m_1)}_2 &= 1 + 4 t^3 + 6 t^4 + 12 t^5 \\
    &\hspace{1.5cm} + 28 t^6 + 72 t^7 + 168 t^8 + \mathrm{O}(t^9) \, 
\end{split} \\
\begin{split}
   Z_2 &= 1+4 t^3+6 t^4+12 t^5 \\
   &\hspace{1.5cm} +28 t^6+72 t^7+168 t^8 + \mathrm{O}(t^9) \, .
\end{split}
\end{align}

\item $N=\infty$. 

\begin{table}[h!]
\centering
\small
\renewcommand{\arraystretch}{1.5}
\begin{tabular}{c|ccccccccc}
$G^{(0,1)}_\infty$
& $+2 t$ & $+ 3 t^2$ & $+ 12 t^3$ & $+ 23 t^4$ & $+ 42 t^5$ & $+ 63 t^6$ & $+ 70 t^7$ & $- 9 t^8$ & $-222 t^9$ \\
$G^{(0,2)}_\infty$
& & & & $+6 t^4$ & $+ 18 t^5$ & $+ 78 t^6$ & $+ 266 t^7$ & $+ 789 t^8$ & $+ 1974 t^9$ \\
$G^{(0,3)}_\infty$
& & & & & & & & & $+20 t^9$ \\
\end{tabular}
\caption{Instanton corrections for $N=\infty$ up to order $t^9$. The next corrections $(m_0,m_1)=(0,4),\dots$ are expected at $\mathrm{O}(t^{16})$.}
\label{tab:Zinf1}
\end{table}

Finally, we illustrate the instanton corrections \eqref{Dinstexp1} to $Z_\infty$. We sum all corrections from table \ref{tab:Zinf1} and compare the result with the $N=\infty$ thermal partition function:
\begin{align}
    \hspace{-0.5cm}Z_{\textrm{BCOV}} \smash[b]{\sum_{m_1=0}^3} G^{(0,m_1)}_\infty &= 1 + 4 t^3 + 6 t^4 + 12 t^5 + 28 t^6 + 72 t^7 + 168 t^8 + 364 t^9+ \mathrm{O}(t^{10}) \\
    Z_\infty &= 1+4 t^3+6 t^4+12 t^5 +28 t^6+72 t^7+168 t^8 +364 t^9 + \mathrm{O}(t^{10}) \, .
\end{align}

\end{enumerate}


\subsection{Giant gravitons from residues}
\label{ssec:interpretations}

Having proved and numerically tested our MGGE for the thermal and graded partition functions of $\cN=4$ sdYM, we now illustrate its relation to conventional giant graviton expansions by identifying a residue associated with a half-BPS giant graviton \cite{Sharma:2025ntb,Caron-Huot:2023wdh}.

An important application of standard GGEs is the identification of bulk duals of trace relations. At finite $N$, the reason the single-trace spectrum of the boundary gauge theory truncates is because one can use trace relations like the Cayley–Hamilton theorem to eliminate traces of the form $\Tr X^{N+1}$ in terms of traces of smaller powers of $X\in\mathfrak{u}(N)$. Just as single-trace operators are dual to physical states of the bulk gravitational theory, the trace relations themselves can be interpreted as being dual to ghost-like states in the bulk \cite{Lee:2023iil}. These ghost states contribute negatively to the bulk degrees of freedom, thereby reducing the infinite-$N$ spectrum to the finite-$N$ spectrum. In \cite{Lee:2023iil,Lee:2024hef,Caputa:2025ikn}, by analyzing the GGEs of the various BPS indices, it was recognized that over-maximal giant gravitons furnish examples of such ghost-like states. This acts as a quantitative test of the stringy exclusion principle in holography. Using chiral holography, we should in principle be able to extend this identification beyond BPS sectors. We will not do so here in full generality, but content ourselves with studying half-BPS giants discussed in \cite{Sharma:2025ntb} that were found to be dual to scalar determinants $\det(y^{IJ}\Phi_{IJ})$ (with $y^{IJ}$ being a null 6-vector).

Let us explicitly check that our GGE incorporates contributions of such half-BPS giants. We will do this by picking out a specific residue in the contour integral representation of the summand $G_N^{(1,2)}$. The fact that the $m_0=1$, $m_1=2$ term contains the contribution of such a giant follows from their construction via the cohomology of complexes like \eqref{cmpx2}. The giant corresponding to \eqref{cmpx2} wraps $\mu^\dal=W_1=W_2=0$ and is dual to the determinant $\det(\Phi_{12})$. This complex manifestly involves exactly one copy of $\CO_Y\oplus\Pi\CO_Y$ and two copies of $\CO_Y(1)\oplus\Pi\CO_Y(1)$.

To pick out the associated residue, we must first organize our complex as a tachyon vev on $\mathscr{E}_{1,2}\equiv\CO_Y^{\oplus\,1|1}\oplus\CO_Y(1)^{\oplus\,2|2}$. Let us express the open string field on this stack as a $\gl(3|3,\C)$ matrix with rows and columns arranged in the order $\{\CO_Y,\CO_Y(1)^{\oplus2}\,|\,\Pi\CO_Y,\Pi\CO_Y(1)^{\oplus2}\}$. Taking the columns to label the domains and rows to label the targets of the various maps, we can express the complex \eqref{cmpx2} as the matrix
\be
\mathscr{T} = \left(
\begin{array}{ccc|ccc}
0&0&0&0&W_1&W_2\\
0&0&0&\lambda_0&0&0\\
0&0&0&\lambda_1&0&0\\
\hline
0&0&0&0&0&0\\
0&-\lambda_1W_2&\lambda_0W_2&0&0&0\\
0&\lambda_1W_1&-\lambda_0W_1&0&0&0
\end{array}
\right)\,.
\ee
Viewing this as an element of $\Ext^0(\mathscr{E}_{1,2},\mathscr{E}_{1,2})$, we can think of $\CT$ as an open string tachyon vev on our $\D9/\overline{\D9}$ stack. It obeys the open string field equation $\dbar\CT+\CT^2=0$ by virtue of obeying the linearized field equation $\dbar\CT=0$ and being nilpotent: $\CT^2=0$. The nilpotence condition is of course equivalent to \eqref{cmpx2} defining a differential complex.

To translate this into a residue, the first task is to assign this a matrix of fugacities. The fugacity for $W_I$ is $x_I$, but since the partition function counts creation operators, one must instead use the fugacity $1/x_I$. Due to \eqref{eq:fshmatch4r}, the $\D9/\overline{\D9}$ partition functions we actually encounter have fugacities $x_I$ mapped to $x_I\mapsto z_1z_2/x_I$. So to construct a fugacity assignment for $\CT$, we simply replace every occurrence of $W_I$ by $x_I/z_1z_2$. Similarly, the fugacities of $\lambda_0,\lambda_1$ are $z_1^{-1},z_2^{-1}$ and remain unchanged in \eqref{eq:fshmatch4r}, so the rule is to replace every occurrence of $\lambda_0,\lambda_1$ by $z_1,z_2$ respectively. Finally, we assign Chan-Paton fugacities
$v_0,u_{1,1},u_{1,2}$ to the first three rows and $u_0,v_{1,1},v_{1,2}$ to the last three rows, with their inverses assigned in the same order to the columns.

With these rules, we obtain the matrix of fugacity assignments for the entries of $\CT$,
\begin{align}
\left(
\begin{array}{cccccc}
 0 & 0 & 0 & 0 & \frac{v_{0} x_1}{v_{1,1} z_{1} z_2}
   & \frac{v_{0} x_2}{v_{1,2} z_1 z_2} \\
 0 & 0 & 0 & \frac{u_{1,1} z_1}{u_{0}} & 0 & 0 \\
 0 & 0 & 0 & \frac{u_{1,2}z_2}{u_{0}} & 0 & 0 \\
 0 & 0 & 0 & 0 & 0 & 0 \\
 0 & \frac{v_{1,1} x_2}{u_{1,1} z_1} &
   \frac{v_{1,1} x_2}{u_{1,2} z_2} & 0 & 0 & 0 \\
 0 & \frac{v_{1,2} x_1}{u_{1,1} z_1} &
   \frac{v_{1,2} x_1}{u_{1,2} z_2} & 0 & 0 & 0 \\
\end{array}\right) \, . \label{eq:matrix}
\end{align}

In the contour integral \eqref{GNm} for the case $m_0=1,m_1=2$, one encounters a pole for each entry of this matrix. There are 6 fugacities, and we can eliminate 5 by the following self-consistent residue prescription, leaving a trivial integral over the common overall fugacity:
\begin{align}
    v_ {1, 1} &= \frac {x_ 1} {z_ 1 z_ 2} v_ 0 \, \qquad v_ {1, 2} = \frac {x_ 2} {z_ 1 z_ 2} v_ 0 \, , \qquad u_{1,1} = \frac{x_2}{z_1} v_{1,1} = \frac{x_1 x_2}{z_1^2 z_2} v_0 \, , \\
    & \,\,\,\,\, u_{1,2} = \frac{x_1}{z_2} v_{1,2} = \frac{x_1 x_2}{z_1 z_2^2} v_0 \, , \qquad u_0 = z_1 u _{1,1}= \frac{x_1x_2}{z_1z_2} v_0 \, , \label{eq:poles}
\end{align}
where the residues are to be taken in the order they appear in this equation. These residues are within the contour specified in section \ref{sec:ggemain}. As a check, taking these residues sets all nonzero entries of \eqref{eq:matrix} to $1$.

Evaluating these residues, we recover the contribution
\begin{align}
    \PE[f'] = \exp\qty( \sum_{n=1}^\infty \frac{1}{n} f'_n  ) \, ,
\end{align}
where
\begin{align}
    f'=1-\frac{z_1P'(z_2)-z_2P'(z_1)}{z_1-z_2} \, , \quad P'(z)=\frac{(1-y_1/z)(1-y_2/z)(1-z/x_1)(1-z/x_2)}{(1-x_3/z)(1-x_4/z)} \, .
\end{align}
The Berezinian factor also gives
\begin{align}
    (\Ber\,\cW_0)^N \rightarrow \qty(\frac{x_1x_2}{z_1z_2})^N
\end{align}
This is precisely equal to the contribution of a half-BPS giant dual to a determinant $\det\Phi_{12}$ that we would expect in a ``standard" GGE, as reviewed in section \ref{ssec:standardGGE}. Its single-letter graded partition function $f'$ can be obtained using the prescription of section \ref{ssec:counting}, followed by the fugacity involution \eqref{eq:fshmatch4r}.


\section{Hagedorn temperature from twistor strings} 
\label{sec:hagedorn}

One of our main results is a holographic expansion even for the infinite-rank graded partition functions. In the case of free $\cN=4$ sYM, it takes the form
\begin{align}
Z_\infty[f_{\cN=4}]=Z_{\textrm{BCOV}}^{\cN=4}\sum_{m_1=0}^\infty G^{(0,m_1)}_\infty[\mathbf{f}_{\cN=4}] \, .
 \end{align}
We interpreted these remaining corrections as contributions from Witten's D-instanton/anti-D-instanton pairs. Using the identity \eqref{eq:m00sum}, they can be resummed into
\begin{align}
Z_\infty[f_{\cN=4}]=Z_{\textrm{BCOV}}^{\cN=4} \, Z_\infty[({\bf f}^!_{\cN=4})_{11}] \, .
\end{align}
These formulae also hold for the thermal/twisted graded partition functions after the fugacity shift $t\mapsto -t$ (in the unrefined case). 

The thermal partition function on the LHS and its first singularity in $t$ are given by
\begin{align}
    Z_\infty^{\cN=4,\mathrm{th}}(t) &= Z_\infty[f_{\cN=4}](-t) =  \prod_{n=1}^\infty \frac{(1 + (-t)^n)^3}{(1-(-t)^n) (1 + 4(-t)^n+t^{2n})} \, , \\
    t_H^{\cN=4} &= 2-\sqrt{3} \,.
\end{align}
The corresponding temperature obtained from using $t=\e^{-1/(2T)}$ is known as the Hagedorn temperature (measured in units of the $S^3$ radius),
\begin{align}
    T_H^{\cN=4}=-\frac{1}{2\log t_H^{\cN=4}} = \frac{1}{2\log(2+\sqrt{3})} \approx 0.380 \, .
\end{align}
This is the standard Hagedorn temperature of free large-$N$ $\cN=4$ sYM
on $S^1\times S^3$ \cite{Sundborg:1999ue,Aharony:2003sx}. At zero
't Hooft coupling, the Hagedorn singularity coincides with the
deconfinement transition of the large-$N$ unitary matrix model.

We can now reproduce this Hagedorn temperature from the bulk. On the RHS, the twisted graded closed string partition function
\begin{align}
  Z_{\mathrm{BCOV}}^{\cN=4}(-t) = \prod_{n=1}^\infty \frac{(1+(-t)^n)^4}{(1-(-t)^n)^4}
\end{align}
is regular for $0\leq t < 1$, including at $t=t_H^{\cN=4}$. So we find the surprising result that the Hagedorn singularity does not arise from the closed string spectrum. On the contrary, it is the resummed contribution of Witten's D-instanton brane/antibrane pairs that reproduces the Hagedorn pole at $1-4t+t^2=0$ from the bulk:
\begin{align}
    Z_\infty[({\bf f}^!_{\cN=4})_{11}](-t) &= \prod_{n=1}^\infty \frac{(1-(-t)^n)^3}{(1+(-t)^n)(1+4(-t)^n+t^{2n})} \, , \\
    t_H^{\cN=4} &= 2-\sqrt{3} \, . 
\end{align}
This signals the potential formation of a condensate of D-instantons at high temperatures.

A similar analysis holds for the free $\cN=1$ sYM case:
\begin{align}
Z_\infty[f_{\cN=1}] & =Z_{\textrm{BCOV}}^{\cN=1}\sum_{m_1=0}^\infty G^{(0,m_1)}_\infty[\mathbf{f}_{\cN=1}] \\
&= Z_{\textrm{BCOV}}^{\cN=1} \, Z_\infty[(\mathbf{f}_{\cN=1}^!)_{11}] \, . 
 \end{align}
The thermal partition function again has a singularity:
\begin{align}
    Z_\infty^{\cN=1,\mathrm{th}}(t)&=Z_\infty[f_{\cN=1}](-t) = \prod_{n=1}^\infty \frac{(1-(-t)^n)(1+(-t)^n)^3}{1+2(-t)^n+2(-t)^{3n}+t^{4n}}\, , \\
    t_H^{\cN=1} &=\frac{1+\sqrt{3}-\sqrt{2\sqrt{3}}}{2} \, ,
\end{align}
which gives the corresponding Hagedorn temperature
\begin{align}
    T_H^{\cN=1}=-\frac{1}{2\log t_H^{\cN=1}} = \frac{1}{\log(1+\sqrt{3}+\sqrt{3+2\sqrt{3}})} \approx 0.601 \, .
\end{align}
The twisted graded BCOV partition function on the RHS is again regular for $0\leq t<1$:
\begin{align}
    Z_{\textrm{BCOV}}^{\cN=1}(-t) = \prod_{n=1} \frac{(1+(-t)^n)^4}{(1+(-t)^n+t^{2n})^2} \, ,
\end{align}
while the resummed D-instanton brane/antibrane contribution shares the same Hagedorn pole:
\begin{align}
    Z_\infty[(\mathbf{f}_{\cN=1}^!)_{11}](-t) &= \prod_{n=1} \frac{(1-(-t)^n)(1+(-t)^n+t^{2n})^2}{(1+(-t)^n)(1+2(-t)^n+2(-t)^{3n}+t^{4n})}\,,\\
    t_H^{\cN=1}&=\frac{1+\sqrt{3}-\sqrt{2\sqrt{3}}}{2} \, .
\end{align}
We therefore identify the resummed D-instanton brane/antibrane sector as the source of the Hagedorn singularity in our bulk expansion.

Previous approaches to deriving Hagedorn physics from the bulk include reconstructing closed string worldsheets from thermal free-field correlators \cite{Furuuchi:2005qm}, as well as one-loop studies of proposed tensionless bulk duals of free Yang-Mills and $\cN=4$ sYM \cite{Bae:2016hfy,Bae:2017spv}. Harmark and Orselli matched Hagedorn temperatures directly between gauge theory and string theory in a near-critical chemical potential limit where both descriptions become weakly coupled \cite{Harmark:2006ta}. The Hagedorn temperature of planar $\cN=4$ sYM, with and without chemical potentials, has also been studied for arbitrary 't Hooft coupling using integrability and the quantum spectral curve \cite{Harmark:2017yrv,Harmark:2018red,Harmark:2021qma}, but its tensionless string origin at vanishing 't Hooft coupling has remained unclear.

What is distinct in the present construction is that our twistor string dual allows us to identify the microscopic bulk degrees of freedom responsible for the standard free sYM Hagedorn singularity: the BCOV closed string partition function remains regular at the Hagedorn temperature, while the Hagedorn pole is carried by Witten's D-instantons.


\section{Discussion}
\label{sec:discussion}

In this work, we studied the $N$-dependence of the non-BPS spectrum of weakly coupled gauge theories. We analyzed the non-BPS thermal and graded partition functions of $\cN=4$ and $\cN=1$ super Yang-Mills theories, refined by a maximal number of fugacities. The chiral zero coupling limits of these theories gave rise to $\cN=4$ and $\cN=1$ \emph{self-dual} Yang-Mills (sdYM) theories, using which we computed their partition functions to zeroth order in the 't Hooft coupling. 

$\cN=4$ sdYM has recently been shown to admit an explicit
twistor string dual \cite{Sharma:2025ntb, Jarov:2025qhz}, and in this work we outlined
the analogous construction for $\cN=1$ sdYM. By combining this perspective with Murthy's expansion \cite{Murthy:2022ien} and its holographic interpretation \cite{Budzik:2025vzr}, we derived the giant graviton expansions of our free partition functions and identified them as expansions in terms of closed strings and branes in the dual bulk theories. This furnishes a quantitative mechanism to potentially extend the stringy exclusion principle beyond BPS subsectors of supersymmetric theories.

Our results open up some very exciting avenues for future work, such as computing 't Hooft coupling corrections to giant graviton expansions, exploring Hawking-Page transitions, building the explicit holographic dictionary for chiral holography, and finding the gravity duals of genuinely non-supersymmetric 4d gauge theories. We end by briefly discussing these and related ideas.

\paragraph{Hawking-Page transitions.} In section \ref{sec:hagedorn}, we reproduced the Hagedorn temperatures of free $\cN=4$ and $\cN=1$ sYM theories on $S^1\times S^3$ from their twistor string bulk duals. Remarkably, the closed twistor string partition functions remain regular at the Hagedorn temperature, while the Hagedorn pole is carried by the resummed (twisted graded) partition functions of Witten's D-instanton/anti-D-instanton pairs. At large-$N$ and zero 't Hooft coupling, deconfinement transition occurs at the Hagedorn temperature \cite{Sundborg:1999ue,Aharony:2003sx}. At strong coupling, deconfinement is described holographically by the Hawking-Page transition between thermal AdS and an AdS black hole saddle \cite{Hawking:1982dh,Witten:1998zw}. It is natural to expect a similar change of dominant bulk saddle in our twistor string duals at weak coupling. Such a transition does take place in tensionless strings on AdS$_3\times S^3$ \cite{Eberhardt:2020bgq} and their symmetric orbifold duals \cite{Keller:2011xi}. Understanding its analog in chiral holography requires the construction of the thermal versions of our Calabi-Yau 7-fold and 5-fold geometries. We hope to return to this in future work.

\paragraph{Scaffold node.} Perhaps the most surprising aspect of our analysis is that our twistorial GGE for free partition functions is completely hidden on spacetime. The D5-brane theory on $\PT$ can only be interpreted as a self-dual Yang-Mills theory on $\R^4$ if the gauge bundle on the D5's is trivial along twistor lines. As this is not the case for $\CO_\PT(-1)$, the theory on the D5 stack $\CO_\PT^{\oplus N_0}\oplus\CO_\PT(-1)^{\oplus N_1}$ does not appear to admit a spacetime interpretation for $N_1>0$. So the scaffold $\U(N_1)$ node that we used to derive our GGE could only have been introduced on twistor space. It would be very interesting to find an inherently spacetime interpretation of the scaffolded partition function $Z_{N_0,N_1}$ that generalizes $Z_N \equiv Z_{N,0}$. In this regard, a better choice of generator branes might be $\CO_\PT$ and the structure sheaf $\CO_D$ of any divisor $D\subset\PT$ at fixed $\lambda_\al\in\P^1$ \cite{Aspinwall:2002hg}. Since $D\simeq\C^2\simeq\R^4$, the theory on $\CO_D^{\oplus N_1}$ could be interpreted as a holomorphic BF theory on spacetime, coupled to self-dual sYM.\footnote{We thank Kevin Costello for pointing out this possibility.}

\paragraph{Koszul-dual branes.} A key feature of our construction is that the thermal GGE is engineered from the graded GGE by an R-symmetry twist. In this description,
\begin{align}
    \mathbf f^! = -[\mathbf f(\mathbf 1-\mathbf f)^{-1}]^{\mathrm{T}} \label{eq:koszul}
\end{align}
continues to admit its interpretation as the twisted graded single-letter partition function of Koszul-dual branes (transverse D9-branes for $\cN=4$ and transverse D5-branes for $\cN=1$) \cite{Budzik:2025vzr}. These observations suggest that the MGGEs of more general thermal partition functions would retain their holographic interpretations in terms of Koszul-dual branes when $(-1)^F$ can be realized as a central element of a global symmetry. For $\cN=4$ sYM, the continuous subgroup of the R-symmetry of $\mathfrak{psl}(4|4)$ interpolates between the graded and thermal partition functions \cite{Cherman:2020zea}. For pure $\cN=1$ sYM, the analogous $\U(1)_R$ symmetry is anomalous, but its $\mathbb Z_2$ subgroup realizing $(-1)^F$ remains exact \cite{Shifman:1999mv}. Another example is the $\mathfrak{psl}(2|2)$ symmetry underlying the chiral algebra sector of $\cN=4$ sYM \cite{Beem:2013sza}, for which $(-1)^F$ can be realized by its $\SU(2)_R$ symmetry and the graded vacuum character is the Schur index.

\paragraph{Giant graviton expansions.} Standard GGEs \cite{Imamura:2021ytr, Gaiotto:2021xce, Lee:2022vig} originally derive their name from expansions of indices in which each correction is an index of giant graviton branes. Murthy's GGE provides an alternative construction wherein the individual terms are associated to brane/antibrane stacks in the B-model \cite{Eniceicu:2023uvd,Budzik:2025vzr}, while conventional giant graviton contributions can emerge from suitable residues. In this paper, we mainly used Murthy's GGE. Nevertheless, this raises the question of whether a standard GGE could also exist for non-BPS partition functions. Our preliminary results in sections \ref{ssec:GGfromtachyons} and \ref{ssec:interpretations} show that half-BPS giants dual to scalar determinants in $\cN=4$ sdYM do enter our GGEs through various residues. For instance, the contribution of a single half-BPS giant can be recovered by evaluating the contour integrals in $G_N^{(1,2)}$ as residues around subsets of poles like \eqref{eq:poles}. This is encouraging, as it might be possible to find a finite set of additional giant graviton-like branes that recast our expansion as a standard GGE. 

\paragraph{Non-BPS trace relations.} Finding a standard expansion remains a useful question because it is the (overmaximal) giant graviton branes that are usually viewed as being dual to trace relations in the gauge theory \cite{Lee:2023iil,Lee:2024hef,Caputa:2025ikn}. For instance, at $N=1$, the giant graviton dual to $\det(\Phi_{12})$ has overmaximal states that are dual to trace relations like $\Tr(\Phi_{12}^2) = (\Tr(\Phi_{12}))^2$, etc. Usually in the case of BPS subsectors, one only encounters trace relations built from BPS letters. However, our thermal GGEs also incorporate non-BPS trace relations such as the Konishi-type trace relation
\be
\Tr(\Phi_{12}\Phi_{34}) = \Tr(\Phi_{12})\Tr(\Phi_{34})\quad\text{at $N=1$}\,.
\ee
It will be very interesting to find the giant graviton in our twistorial bulk that is dual to this trace relation. More precisely, we anticipate that there will exist a specific tachyon condensation pattern on the Koszul-dual brane/antibrane stacks that gives rise to a giant graviton that carries such a non-BPS overmaximal open string state.

\paragraph{'t Hooft coupling corrections to GGE.} The GGEs of superconformal indices do not receive any 't Hooft coupling corrections by virtue of supersymmetry. But our thermal GGEs should receive such corrections. This provides a novel setting in which finite-$N$ corrections can be explored side-by-side with small $\lambda$ corrections. Computing the partition functions to order $\lambda^L$ involves $L$ insertions of the interaction Lagrangians \eqref{intN4}, \eqref{intN1}. They can be engineered by turning on $L$ copies of Witten's D-instantons in the bulk, integrated over their moduli. This could provide a bulk prediction for the boundary loop corrections, and in particular for the loop-corrected Hagedorn temperature of weakly coupled sYM \cite{Spradlin:2004pp}. It would also be interesting to compare such results derived through chiral holography to the integrability and strong coupling results such as \cite{Harmark:2017yrv}.

\paragraph{A dictionary for chiral holography.} One of the original aims of this project was to test chiral holography in a setting that does not require knowing its detailed holographic state/operator dictionary. At present, the construction of such an explicit dictionary has only been achieved in the half-BPS sector of $\cN=4$ sYM \cite{Sharma:2025ntb}, where the half-BPS determinants of \cite{Caron-Huot:2023wdh} have been mapped to giant gravitons wrapping two out of the four $\CO(-1)$ fibers in our twistor string target space $\CO(-1)^{\oplus4}\to\PT$. In equation \eqref{Dinstexp}, we found that both closed twistor strings and open strings on D-instantons are needed to reproduce the spectrum of $\cN=4$ sdYM. It will be very interesting to use this decomposition as a clue to mapping bulk closed strings and branes to the complete non-BPS large-$N$, and potentially even finite-$N$, spectrum.

\paragraph{Gravity duals of non-supersymmetric theories.} A new possibility that arises at weak coupling is that many \emph{non-supersymmetric} self-dual gauge theories can also be engineered through suitable orbifolds and orientifolds of twistor string theory \cite{Costello:2023vyy,Bittleston:2024efo}. We hope to use such constructions to develop the bulk duals of weakly coupled non-supersymmetric gauge theories. Our giant graviton expansions did not rely on supersymmetry, so we expect them to generalize to such settings. Similarly, it will be interesting to study the phase structures of such gauge theories through their gravity duals.


\acknowledgments

We would like to thank Roland Bittleston, Kevin Costello, Matthew Dodelson, Lorenz Eberhardt, Davide Gaiotto, Elliott Gesteau, Matthew Heydeman, Ji Hoon Lee, Juan Maldacena, Shiraz Minwalla, Marcus Spradlin, Andy Strominger and Anastasia Volovich for helpful conversations. The work of KB is supported by the Simons Collaboration on Celestial Holography. AS is supported by the Gordon and Betty Moore Foundation and the John Templeton Foundation via the Black Hole Initiative. We used ChatGPT for minor editing, reference searches, some technical discussions, and improving Mathematica code. The original research ideas and numerical results were produced by the authors.


\appendix

\section{MGGE generalizations}
\label{app:proofs}

In this appendix, we provide details on the generalization of Murthy's expansion to quiver gauge theories \eqref{quivermurthy}. We also complete the proof that integrating out the second node produces the standard one node expansion.

\subsection{Quiver generalization}
\label{app:quivergge}

Here we provide more detail on the derivation of the quiver expansion \eqref{quivermurthy} \cite{Purkayastha:2025whi}.

First, let us introduce a set of vectors assembling the fundamental and antifundamental characters,
\be
({\bf p}_n^+)_a = \Tr U_a^n\,,\qquad ({\bf p}_n^-)_a = \Tr U_a^{-n}\,.
\ee
Thinking of ${\bf p}_n^+$ as a $1\times k$ row vector and ${\bf p}_n^-$ as a $k\times 1$ column vector, we can reexpress \eqref{ZvecN} as
\be
Z_{N_1,\dots,N_k}[\mathbf{f}] = \int_{\U(N_1)\times\cdots\times\U(N_k)}\prod_{a=1}^k\d U_a\,\exp\bigg(\sum_{n=1}^\infty\frac{1}{n}\,{\bf p}_n^+{\bf f}_n{\bf p}_n^-\bigg)\,.
\ee
We now perform a Hubbard-Stratonovich transformation to decouple the nodes of the quiver. This is a formal Gaussian integral by means of which we write
\be
\exp\bigg(\frac1n\,{\bf p}_n^+{\bf f}_n{\bf p}_n^-\bigg) = \int\frac{\d^k{\bf t}^+_n\,\d^k{\bf t}^-_n}{n^k\det({\bf f}_n)}\,\exp\left(-\frac1n\,{\bf t}_n^-{\bf f}_n^{-1}{\bf t}_n^+ + \frac1n\,{\bf p}^+_n{\bf t}^+_n + \frac{1}{n}\,{\bf t}^-_n{\bf p}^-_n\right)\,,
\ee
where ${\bf f}_n^{-1}$ is the matrix inverse of ${\bf f}_n$, and $\det({\bf f}_n)$ its determinant. The integration variables consist of a $1\times k$ row vector ${\bf t}^-_n$ and a $k\times1$ column vector ${\bf t}^+_n$. We label their components as
\be
({\bf t}^\pm_n)_a = t_{n,a}^\pm\,.
\ee
We have also chosen to suppress the factors of $2\pi$ in the Gaussian integral by choosing an appropriate normalization of the measure $\d^k{\bf t}^+_n\,\d^k{\bf t}^-_n$.

Using this representation of the exponential factors, we can factorize the unitary matrix integrals in $Z_{N_1,\dots,N_k}$ to obtain
\be
Z_{N_1,\dots,N_k}[\mathbf{f}] = \int\prod_{n=1}^\infty\frac{\d^k{\bf t}^+_n\,\d^k{\bf t}^-_n}{n^{k}\det({\bf f}_n)}\,\exp\left(-\frac1n\,{\bf t}_n^-{\bf f}_n^{-1}{\bf t}_n^+\right)\prod_{a=1}^k\til Z_{N_a}(t^+_a,t^-_a) \, , \label{eq:ehhhh}
\ee
where $t^\pm_a$ collectively denote the integration variables $t_{n,a}^\pm$. The right hand side contains $k$ decoupled copies of the same auxiliary matrix integral
\be\label{hubbard}
\til Z_N(t^+,t^-) = \int_{\U(N)}\d U\,\exp\bigg(\sum_{n=1}^\infty\frac{1}{n}\,t_n^+\Tr U^n + \sum_{n=1}^\infty\frac{1}{n}\,t_n^-\Tr U^{-n}\bigg)\,.
\ee
This is a universal integral, as it does not depend on the details of ${\bf f}_n$.

At this point, Murthy derives an expansion for $\til Z_N$ using the results of \cite{borodin2000fredholm}. Over the course of many further studies \cite{Liu:2022olj,Eniceicu:2023uvd,Chen:2024cvf}, this expansion has been reduced to
\begin{multline}
\til Z_N(t^+,t^-) = \exp\bigg(\sum_{n=1}^\infty\frac{t^+_nt^-_n}{n}\bigg)\sum_{m=0}^\infty\int_{\U(m|m)}\d \cW\,(\mathrm{Ber}\,\cW)^N\\
\times\exp\bigg(\sum_{n=1}^\infty\frac{t_n^-}{n}\,\Str\,\cW^n - \sum_{n=1}^\infty\frac{t_n^+}{n}\,\Str\,\cW^{-n}\bigg)\,.
\end{multline}
We insert $k$ copies of this expansion, one for each $\til Z_{N_a}$, into \eqref{eq:ehhhh}. We collect the
supertraces into a $1\times k$ row vector $\mathbf{s}^+_n$ and a $k\times 1$ column vector $\mathbf{s}^-_n$:
\begin{align}
(\mathbf{s}^+_n)_a = \Str\,\cW_a^n \, , \qquad (\mathbf{s}^-_n)_a=\Str\,\cW_a^{-n}  \, .
\end{align}
We then perform the Gaussian integrals
\begin{align}
\int\frac{\d^k{\bf t}^+_n\,\d^k{\bf t}^-_n}{n^{k}\det(\mathbf{f}_n)}\,\exp\left(-\frac{1}{n} \, \mathbf{t}_n^- (\mathbf{f}_n^{-1}-1) \mathbf{t}_n^+ + \frac{1}{n} \mathbf{t}_n^- (\mathbf{s}_n^+)^\mathrm{T} - \frac{1}{n} (\mathbf{s}_n^-)^{\mathrm{T}} \mathbf{t}_n^+ \right)  = \frac{1}{\det(\mathbf{1}-{\mathbf f}_n)} \exp\qty( \frac{1}{n} \mathbf{s}^+_n \mathbf{f}_n^! \mathbf{s}^-_n  ) \, ,
\end{align}
where
\begin{align}
    \mathbf{f}^!_n = -[\mathbf{f}_n (1-\mathbf{f}_n)^{-1}]^{\mathrm{T}}  \, .
\end{align}

This is the quiver generalization of the relation \eqref{eq:fk}.

\subsection{Integrating out the scaffolding node}
\label{app:integratingout}

Here we prove the relations provided in section \ref{ssec:murthy} between the two-node quiver MGGE \eqref{eq:2nodeMGEE} and the standard one-node MGGE \eqref{eq:1nodeMGGE}.

First, let us show how to resum the $m_0=0$ two-node quiver terms
\be
G_{N,0}^{(0,m_1)}[\mathbf{f}] = \int\limits_{\U(m_1|m_1)}\d \cW_1\,\exp\qty( \sum_{n=1}^\infty \frac{1}{n} (\mathbf{f}_n^!)_{11}\,\Str(\cW_1^n)\,\Str(\cW_1^{-n}) ) .
\ee
They are independent of $N$, therefore we denote them by $G^{(0,m_1)}_{\infty,0} := G^{(0,m_1)}_{N,0}$. They also depend only on the $11$ component of the matrix $\mathbf{f}^!$ and are therefore standard one-node MGGE integrals for the transformed single-letter data $g=(\mathbf{f}^!)_{11}$. Since $(g^!)^!=g$, applying the one-node MGGE at rank zero with input $g^!$ gives
\begin{align}
   Z_{\infty}[g^!] \sum_{m_1=0}^\infty G^{(0,m_1)}_{\infty,0}[\mathbf{f}] = 1 \, .
\end{align}
From equation \eqref{eq:Zinftyagain} it follows that
\begin{align}
    Z_\infty[g]Z_\infty[g^!] = 1 \, .
\end{align}
Therefore we get
\begin{align}
    \sum_{m_1=0}^\infty G^{(0,m_1)}_{\infty,0}[\mathbf{f}] = Z_\infty[(\mathbf{f}^!)_{11}] \, .
\end{align}
The final equation can also be seen as a version of brane/negative brane expansion conjectured in \cite{Budzik:2025vzr}.

Next, let us consider the infinite-$N$ limit of \eqref{eq:2nodeMGEE}. At least formally, the Berezinian term $(\Ber\cW_0)^N$ suppresses the $m_0>0$ terms in the $N\rightarrow\infty$ limit. We can however show the identity \eqref{eq:Zinftyresum} in an alternative way using \eqref{eq:Zinftyagain} and \eqref{Zinfinf}. One can check by direct computation that
\begin{align}
    \frac{1}{\det(\mathbf{1}-\mathbf{f})} \frac{1}{1-(\mathbf{f}^!)_{11}} = \frac{1}{1-f_{00}} \, .
\end{align}
We therefore have
\begin{align}
    Z_{\infty,\infty}[\mathbf{f}] Z_{\infty}[(\mathbf{f}^!)_{11}] = \prod_{n=1}^\infty \frac{1}{\det(\mathbf{1}-\mathbf{f}_n) \qty(1-(\mathbf{f}^!_n)_{11})} = \prod_{n=1}^\infty \frac{1}{1-f_{00,n}} = Z_{\infty}[f_{00}] \, .
\end{align}

Finally, we show how to resum over $m_1$ at each fixed $m_0>0$ using formulas from \ref{app:quivergge}. Since node $1$ has rank zero, its auxiliary factor \eqref{hubbard} is
\begin{align}
    \widetilde{Z}_0(t^+_1,t^-_1) &= 1 \, .
\end{align}
The variables $t^\pm_1$ then only occur in the exponent in \eqref{eq:ehhhh}. With the Gaussian normalization of \ref{app:quivergge}, we have
\begin{align}
    \int \frac{\dd t^+_{n,1} \dd t^-_{n,1}}{n^2\det(\mathbf{f}_n)} \, \exp\qty( -\frac{1}{n}\mathbf{t}^-_n \mathbf{f}_n^{-1} \mathbf{t}^+_n ) = \frac{1}{n f_{00,n}} \exp\qty( - \frac{t^-_{n,0}t^+_{n,0}}{n f_{00,n}} ) \, .
\end{align}
Combining this result with the zeroth node auxiliary factor $\widetilde{Z}_{N}(t^+_0,t^-_0)$ gives precisely the auxiliary integral of the one-node MGGE for the single-letter data $f_{00}$. Proceeding with the standard derivation then gives
\begin{align}
Z_{\infty,\infty}[\mathbf{f}] \sum_{m_1=0}^\infty G^{(m_0,m_1)}_{N,0}[\mathbf{f}] = Z_{\infty}[f_{00}] \, \widetilde{G}_N^{(m_0)} [f_{00}] \, .
\end{align}

Alternatively, summing over $m_1$ at fixed $m_0$ can be seen as a generalization of MGGE to adjoint, fundamental and antifundamental matter, or again as an application of the brane/negative brane expansion \cite{Budzik:2025vzr} to that same matter content. The ``effective" single-letter data we obtain by integrating out the second node in this way is
\begin{align}
(\mathbf{f}^!)_{00} + \frac{(\mathbf{f}^!)_{01} (\mathbf{f}^!)_{10}}{1-(\mathbf{f}^!)_{11}}  = -\frac{f_{00}}{1-f_{00}} = f_{00}^! \, ,
\end{align}
which is precisely the data entering the one-node Murthy expansion \eqref{eq:1nodeMGGE}.

\section{States in the presence of backreaction} 
\label{app:backreaction}

As promised in section \ref{ssec:bcov4}, let us give a sample computation of how the BCOV states found there deform in the presence of $N$ units of D5-brane flux. This is not needed to follow the rest of the paper. We will leave a general computation of all such deformed states to future studies.

As our example, let us study a state that happens to be dual to the single-trace operator $\Tr\Phi_{IJ}$ inserted at $x=0$. To compute this, we begin with the following state pre-backreaction:
\be
\begin{split}
    \xi_{5,2} &= \p\big(\bar\delta^2(\mu)\,W_{[I}\d W_{J]}\,\D^3Z\big)\\
    &= \bar\delta^2(\mu)\,\big(\d W_I\,\d W_J\,\D^3Z - 2\,W_{[I}\d W_{J]}\,\d^4Z\big)\,.
\end{split}
\ee
This can be guessed by matching Lorentz and R-charges: both $\Tr\Phi_{IJ}$ and this state have fugacity $x_Ix_J/z_1z_2$. It is also easy to verify that this state is projective. To compute its deformation, we will find it useful to dualize it to a polyvector field,
\be\label{22state}
\beta_{2,2} = \bar\delta^2(\mu)\,\veps_{IJKL}\frac{\p}{\p W_K}\frac{\p}{\p W_L}\,.
\ee
It may be checked that $\beta_{2,2}\,\lrcorner\,\Omega_X = \xi_{5,2}$ up to numerical factors. Such polyvectors are only defined up to shifts along the Euler vector field. Additionally, the $\p$-closure constraint translates to the constraint that $\beta_{2,2}$ be divergence-free with respect to the volume form $\Omega_X$. This is manifest for our state \eqref{22state}.

The backreaction \eqref{xi43} dualizes to the polyvector field
\be
\beta_N = \frac{N\,\D^3\br W}{\|W\|^8}\,\frac{\veps^{ABCD}A_A}{A\cdot Z}\frac{\p}{\p Z^B}\frac{\p}{\p Z^C}\frac{\p}{\p Z^D}
\ee
in the patch $A_A Z^A\neq0$. We can cover $\PT = \P^3-\{\lambda_\al=0\}$ by the two patches $\{\lambda_0\neq0\}$ and $\{\lambda_1\neq0\}$. Let us work with the former. In this patch, $\beta_N$ induces a deformation of the dbar operator on $X$ into the operator
\be
\dbar_N = \dbar + \beta_N = \dbar + \frac{N\,\D^3\br W}{\|W\|^8}\,\frac{1}{\lambda_0}\frac{\p}{\p\lambda_1}\frac{\p}{\p\mu^{\dot0}}\frac{\p}{\p\mu^{\dot1}}\,.
\ee
The linearized field equation of a perturbation $\beta_{2,2}$ in the presence of this flux becomes
\be
\dbar_N\beta_{2,2} \equiv \dbar\beta_{2,2} + [\beta_N,\beta_{2,2}] \overset{!}{=} 0\,,
\ee
where $[-,-]$ denotes the Schouten-Nijenhuis bracket for polyvector fields.\footnote{See equation (3.12) of \cite{Costello:2018zrm} for a convenient formula for this bracket.}

Our state \eqref{22state} is $\dbar$-closed but fails $[\beta_N,\beta_{2,2}]=0$. But crucially, since $\beta_N$ is singular along $W_I=0$, now we are working on the space $\til X = X-\{W_I=0\}$ proper, and away from the locus $\mu^\dal=0$ around which we are evaluating local cohomology. Next we use the fact that, up to factors of $2\pi i$, we can write
\be
\bar\delta^2(\mu) = \dbar\left(\frac{\D\hat\mu}{\|\mu\|^4}\right)\,,
\ee
where $\D\hat\mu = \hat\mu_\dal\d\hat\mu^\dal$ and $\hat\mu_\dal$ is the hermitian conjugate of $\mu^\dal$. Then, since $\dbar\beta_N = 0$ away from $W_I=0$, we obtain
\be
\dbar_N\beta_{2,2} = \left[\beta_N,\bar\delta^2(\mu)\,\veps_{IJKL}\frac{\p}{\p W_K}\frac{\p}{\p W_L}\right] = \dbar\left[\beta_N,\frac{\D\bar\mu}{\|\mu\|^4}\,\veps_{IJKL}\frac{\p}{\p W_K}\frac{\p}{\p W_L}\right]\,.
\ee
It is easily checked that the operation $[\beta_N,-]$ squares to zero because of the high antiholomorphic form degree of $\beta_N$ in the $\br W^I$-directions. Hence, we may freely replace $\dbar$ on the right hand side of this equation by $\dbar_N$ and combine it with the left hand side. This leads to the following backreaction-corrected state dual to $\Tr\Phi_{IJ}(0)$,
\begin{multline}
\beta_{2,2}^{(N)} = \bar\delta^2(\mu)\,\veps_{IJKL}\frac{\p}{\p W_K}\frac{\p}{\p W_L} + 2N\veps_{IJKL}\,\frac{\D^3\br W}{\|W\|^8}\frac{\D\hat\mu}{\|\mu\|^6}\frac{\hat\mu^\dal}{\lambda_0}\frac{\p}{\p\mu^\dal}\frac{\p}{\p\lambda_1}\frac{\p}{\p W_K}\frac{\p}{\p W_L}\\
+8N\veps_{IJKL}\br W^K\,\frac{\D^3\br W}{\|W\|^{10}}\frac{\D\hat\mu}{\|\mu\|^4}\frac{1}{\lambda_0}\frac{\p}{\p\lambda_1}\frac{\p}{\p\mu^{\dot0}}\frac{\p}{\p\mu^{\dot1}}\frac{\p}{\p W_L}\,.
\end{multline}
When one acts on this by $\dbar_N$, the $\dbar$ is understood to only produce a delta function at the $\mu^\dal=0$ singularity in the second and third terms, and not at the singularity at $W_I=0$. By construction, we find $\dbar_N\beta_{2,2}^{(N)}=0$.

Similarly, the state dual to $\Tr\Phi_{IJ}(x)$ can be obtained by replacing $\mu^\dal$ with $\mu^\dal-x^{\al\dal}\lambda_\al$. This procedure gives a general principle to construct BCOV states: start with a projective and $\p$-exact combination of the flat space states \eqref{xitildekl}, and then perturbatively correct it as a formal series in $N$. A similar procedure has been developed in greater detail for twisted holography in section 8 of \cite{Costello:2018zrm} and is key to matching holographic correlators. For our purposes, we only use the fact that the backreaction effects do not appear to change the signed count of states or their fugacity assignments, but only deform their detailed structure. So the BCOV graded partition functions we computed in section \ref{sec:bcov} are not expected to be affected by such deformations. We do not prove this rigorously, but this expectation will be bolstered by the fact that we will match the result \eqref{Zbcov4} for $Z_\text{BCOV}$ to a computation in the dual gauge theory in equation \eqref{Zbcov4match}.

\bibliographystyle{JHEP}
\bibliography{gge}

@article{Sharma:2025ntb,
    author = "Sharma, Atul and Skinner, David",
    title = "{Chiral holography}",
    eprint = "2512.04152",
    archivePrefix = "arXiv",
    primaryClass = "hep-th",
    month = "12",
    year = "2025"
}

@article{Sundborg:1999ue,
    author = "Sundborg, Bo",
    title = "{The Hagedorn transition, deconfinement and N=4 SYM theory}",
    eprint = "hep-th/9908001",
    archivePrefix = "arXiv",
    reportNumber = "USITP-99-06",
    doi = "10.1016/S0550-3213(00)00044-4",
    journal = "Nucl. Phys. B",
    volume = "573",
    pages = "349--363",
    year = "2000"
}

@article{Witten:1998zw,
    author = "Witten, Edward",
    editor = "Bergstrom, L. and Lindstrom, U.",
    title = "{Anti-de Sitter space, thermal phase transition, and confinement in gauge theories}",
    eprint = "hep-th/9803131",
    archivePrefix = "arXiv",
    reportNumber = "IASSNS-HEP-98-21",
    doi = "10.4310/ATMP.1998.v2.n3.a3",
    journal = "Adv. Theor. Math. Phys.",
    volume = "2",
    pages = "505--532",
    year = "1998"
}

@article{Aharony:2003sx,
    author = "Aharony, Ofer and Marsano, Joseph and Minwalla, Shiraz and Papadodimas, Kyriakos and Van Raamsdonk, Mark",
    editor = "Doebner, H. D. and Dobrev, V. K.",
    title = "{The Hagedorn - deconfinement phase transition in weakly coupled large N gauge theories}",
    eprint = "hep-th/0310285",
    archivePrefix = "arXiv",
    reportNumber = "WIS-29-03-DPP",
    doi = "10.4310/ATMP.2004.v8.n4.a1",
    journal = "Adv. Theor. Math. Phys.",
    volume = "8",
    pages = "603--696",
    year = "2004"
}

@article{Penrose:1967wn,
    author = "Penrose, R.",
    title = "{Twistor algebra}",
    doi = "10.1063/1.1705200",
    journal = "J. Math. Phys.",
    volume = "8",
    pages = "345",
    year = "1967"
}

@article{Budzik:2021fyh,
    author = "Budzik, Kasia and Gaiotto, Davide",
    title = "{Giant gravitons in twisted holography}",
    eprint = "2106.14859",
    archivePrefix = "arXiv",
    primaryClass = "hep-th",
    doi = "10.1007/JHEP10(2023)131",
    journal = "JHEP",
    volume = "10",
    pages = "131",
    year = "2023"
}

@article{Roiban:2004yf,
    author = "Roiban, Radu and Spradlin, Marcus and Volovich, Anastasia",
    title = "{On the tree level S matrix of Yang-Mills theory}",
    eprint = "hep-th/0403190",
    archivePrefix = "arXiv",
    reportNumber = "NSF-KITP-04-35",
    doi = "10.1103/PhysRevD.70.026009",
    journal = "Phys. Rev. D",
    volume = "70",
    pages = "026009",
    year = "2004"
}

@article{Costello:2022wso,
    author = "Costello, Kevin and Paquette, Natalie M.",
    title = "{Celestial holography meets twisted holography: 4d amplitudes from chiral correlators}",
    eprint = "2201.02595",
    archivePrefix = "arXiv",
    primaryClass = "hep-th",
    doi = "10.1007/JHEP10(2022)193",
    journal = "JHEP",
    volume = "10",
    pages = "193",
    year = "2022"
}

@article{Adamo:2017qyl,
    author = "Adamo, Tim",
    title = "{Lectures on twistor theory}",
    eprint = "1712.02196",
    archivePrefix = "arXiv",
    primaryClass = "hep-th",
    doi = "10.22323/1.323.0003",
    journal = "PoS",
    volume = "Modave2017",
    pages = "003",
    year = "2018"
}

@article{Woodhouse:1985id,
    author = "Woodhouse, N. M. J.",
    title = "{Real methods in twistor theory}",
    doi = "10.1088/0264-9381/2/3/006",
    journal = "Class. Quant. Grav.",
    volume = "2",
    pages = "257--291",
    year = "1985"
}

@article{Ward:1977ta,
    author = "Ward, R. S.",
    title = "{On self-dual gauge fields}",
    doi = "10.1016/0375-9601(77)90842-8",
    journal = "Phys. Lett. A",
    volume = "61",
    pages = "81--82",
    year = "1977"
}

@article{Costello:2018zrm,
    author = "Costello, Kevin and Gaiotto, Davide",
    title = "{Twisted Holography}",
    eprint = "1812.09257",
    archivePrefix = "arXiv",
    primaryClass = "hep-th",
    month = "12",
    year = "2018"
}

@article{Costello:2023hmi,
    author = "Costello, Kevin and Paquette, Natalie M. and Sharma, Atul",
    title = "{Burns space and holography}",
    eprint = "2306.00940",
    archivePrefix = "arXiv",
    primaryClass = "hep-th",
    doi = "10.1007/JHEP10(2023)174",
    journal = "JHEP",
    volume = "10",
    pages = "174",
    year = "2023"
}

@article{Mason:2005zm,
    author = "Mason, L. J.",
    title = "{Twistor actions for non-self-dual fields: A Derivation of twistor-string theory}",
    eprint = "hep-th/0507269",
    archivePrefix = "arXiv",
    doi = "10.1088/1126-6708/2005/10/009",
    journal = "JHEP",
    volume = "10",
    pages = "009",
    year = "2005"
}

@article{Witten:2003nn,
    author = "Witten, Edward",
    title = "{Perturbative gauge theory as a string theory in twistor space}",
    eprint = "hep-th/0312171",
    archivePrefix = "arXiv",
    doi = "10.1007/s00220-004-1187-3",
    journal = "Commun. Math. Phys.",
    volume = "252",
    pages = "189--258",
    year = "2004"
}

@article{Berkovits:2004jj,
    author = "Berkovits, Nathan and Witten, Edward",
    title = "{Conformal supergravity in twistor-string theory}",
    eprint = "hep-th/0406051",
    archivePrefix = "arXiv",
    reportNumber = "IFT-P-019-2004",
    doi = "10.1088/1126-6708/2004/08/009",
    journal = "JHEP",
    volume = "08",
    pages = "009",
    year = "2004"
}

@article{Costello:2021bah,
    author = "Costello, Kevin J.",
    title = "{Quantizing local holomorphic field theories on twistor space}",
    eprint = "2111.08879",
    archivePrefix = "arXiv",
    primaryClass = "hep-th",
    month = "11",
    year = "2021"
}

@article{Caron-Huot:2023wdh,
    author = {Caron-Huot, Simon and Coronado, Frank and M\"uhlmann, Beatrix},
    title = "{Determinants in self-dual $ \mathcal{N} $ = 4 SYM and twistor space}",
    eprint = "2304.12341",
    archivePrefix = "arXiv",
    primaryClass = "hep-th",
    doi = "10.1007/JHEP08(2023)008",
    journal = "JHEP",
    volume = "08",
    pages = "008",
    year = "2023"
}

@article{Costello:2019jsy,
    author = "Costello, Kevin and Li, Si",
    title = "{Anomaly cancellation in the topological string}",
    eprint = "1905.09269",
    archivePrefix = "arXiv",
    primaryClass = "hep-th",
    doi = "10.4310/ATMP.2020.v24.n7.a2",
    journal = "Adv. Theor. Math. Phys.",
    volume = "24",
    number = "7",
    pages = "1723--1771",
    year = "2020"
}

@article{Budzik:2023xbr,
    author = "Budzik, Kasia and Gaiotto, Davide and Kulp, Justin and Williams, Brian R. and Wu, Jingxiang and Yu, Matthew",
    title = "{Semi-chiral operators in 4d $ \mathcal{N} $ = 1 gauge theories}",
    eprint = "2306.01039",
    archivePrefix = "arXiv",
    primaryClass = "hep-th",
    doi = "10.1007/JHEP05(2024)245",
    journal = "JHEP",
    volume = "05",
    pages = "245",
    year = "2024"
}

@article{Costello:2020jbh,
    author = "Costello, Kevin and Paquette, Natalie M.",
    title = "{Twisted Supergravity and Koszul Duality: A case study in AdS$_3$}",
    eprint = "2001.02177",
    archivePrefix = "arXiv",
    primaryClass = "hep-th",
    doi = "10.1007/s00220-021-04065-3",
    journal = "Commun. Math. Phys.",
    volume = "384",
    number = "1",
    pages = "279--339",
    year = "2021"
}

@article{Gaberdiel:2021qbb,
    author = "Gaberdiel, Matthias R. and Gopakumar, Rajesh",
    title = "{String Dual to Free N=4 Supersymmetric Yang-Mills Theory}",
    eprint = "2104.08263",
    archivePrefix = "arXiv",
    primaryClass = "hep-th",
    doi = "10.1103/PhysRevLett.127.131601",
    journal = "Phys. Rev. Lett.",
    volume = "127",
    number = "13",
    pages = "131601",
    year = "2021"
}

@article{Gaberdiel:2021jrv,
    author = "Gaberdiel, Matthias R. and Gopakumar, Rajesh",
    title = "{The worldsheet dual of free super Yang-Mills in 4D}",
    eprint = "2105.10496",
    archivePrefix = "arXiv",
    primaryClass = "hep-th",
    doi = "10.1007/JHEP11(2021)129",
    journal = "JHEP",
    volume = "11",
    pages = "129",
    year = "2021"
}

@article{Bittleston:2024efo,
    author = "Bittleston, Roland and Costello, Kevin and Zeng, Keyou",
    title = "{Self-Dual Gauge Theory from the Top Down}",
    eprint = "2412.02680",
    archivePrefix = "arXiv",
    primaryClass = "hep-th",
    month = "12",
    year = "2024"
}

@article{Berkovits:2004hg,
    author = "Berkovits, Nathan",
    title = "{An Alternative string theory in twistor space for N=4 superYang-Mills}",
    eprint = "hep-th/0402045",
    archivePrefix = "arXiv",
    doi = "10.1103/PhysRevLett.93.011601",
    journal = "Phys. Rev. Lett.",
    volume = "93",
    pages = "011601",
    year = "2004"
}

@article{Boels:2006ir,
    author = "Boels, Rutger and Mason, L. J. and Skinner, David",
    title = "{Supersymmetric Gauge Theories in Twistor Space}",
    eprint = "hep-th/0604040",
    archivePrefix = "arXiv",
    doi = "10.1088/1126-6708/2007/02/014",
    journal = "JHEP",
    volume = "02",
    pages = "014",
    year = "2007"
}

@article{Costello:2012cy,
    author = "Costello, Kevin J. and Li, Si",
    title = "{Quantum BCOV theory on Calabi-Yau manifolds and the higher genus B-model}",
    eprint = "1201.4501",
    archivePrefix = "arXiv",
    primaryClass = "math.QA",
    month = "1",
    year = "2012"
}

@article{Bershadsky:1993cx,
    author = "Bershadsky, M. and Cecotti, S. and Ooguri, H. and Vafa, C.",
    title = "{Kodaira-Spencer theory of gravity and exact results for quantum string amplitudes}",
    eprint = "hep-th/9309140",
    archivePrefix = "arXiv",
    reportNumber = "HUTP-93-A025, RIMS-946, SISSA-142-93-EP",
    doi = "10.1007/BF02099774",
    journal = "Commun. Math. Phys.",
    volume = "165",
    pages = "311--428",
    year = "1994"
}

@article{Fernandez:2024tue,
    author = "Fern{\'a}ndez, V{\'\i}ctor E. and Paquette, Natalie M. and Williams, Brian R.",
    title = "{Twisted holography on AdS$_3 \times S^3 \times$ K3 {\&} the planar chiral algebra}",
    eprint = "2404.14318",
    archivePrefix = "arXiv",
    primaryClass = "hep-th",
    doi = "10.21468/SciPostPhys.17.4.109",
    journal = "SciPost Phys.",
    volume = "17",
    number = "4",
    pages = "109",
    year = "2024"
}

@article{Mason:2010yk,
    author = "Mason, L. J. and Skinner, David",
    title = "{The Complete Planar S-matrix of N=4 SYM as a Wilson Loop in Twistor Space}",
    eprint = "1009.2225",
    archivePrefix = "arXiv",
    primaryClass = "hep-th",
    doi = "10.1007/JHEP12(2010)018",
    journal = "JHEP",
    volume = "12",
    pages = "018",
    year = "2010"
}

@article{Adamo:2011dq,
    author = "Adamo, Tim and Bullimore, Mathew and Mason, Lionel and Skinner, David",
    title = "{A Proof of the Supersymmetric Correlation Function / Wilson Loop Correspondence}",
    eprint = "1103.4119",
    archivePrefix = "arXiv",
    primaryClass = "hep-th",
    doi = "10.1007/JHEP08(2011)076",
    journal = "JHEP",
    volume = "08",
    pages = "076",
    year = "2011"
}

@article{Jiang:2019xdz,
    author = "Jiang, Yunfeng and Komatsu, Shota and Vescovi, Edoardo",
    title = "{Structure constants in $ \mathcal{N} $ = 4 SYM at finite coupling as worldsheet g-function}",
    eprint = "1906.07733",
    archivePrefix = "arXiv",
    primaryClass = "hep-th",
    reportNumber = "CERN-TH-2019-093, Imperial-TP-EV-2019-01",
    doi = "10.1007/JHEP07(2020)037",
    journal = "JHEP",
    volume = "07",
    number = "07",
    pages = "037",
    year = "2020"
}

@article{Lee:2023iil,
    author = "Lee, Ji Hoon",
    title = "{Trace relations and open string vacua}",
    eprint = "2312.00242",
    archivePrefix = "arXiv",
    primaryClass = "hep-th",
    doi = "10.1007/JHEP02(2024)224",
    journal = "JHEP",
    volume = "02",
    pages = "224",
    year = "2024"
}

@article{Beem:2013sza,
    author = "Beem, Christopher and Lemos, Madalena and Liendo, Pedro and Peelaers, Wolfger and Rastelli, Leonardo and van Rees, Balt C.",
    title = "{Infinite Chiral Symmetry in Four Dimensions}",
    eprint = "1312.5344",
    archivePrefix = "arXiv",
    primaryClass = "hep-th",
    reportNumber = "YITP-SB-13-45, CERN-PH-TH-2013-311, HU-EP-13-78",
    doi = "10.1007/s00220-014-2272-x",
    journal = "Commun. Math. Phys.",
    volume = "336",
    number = "3",
    pages = "1359--1433",
    year = "2015"
}

@article{Balasubramanian:2001nh,
    author = "Balasubramanian, Vijay and Berkooz, Micha and Naqvi, Asad and Strassler, Matthew J.",
    title = "{Giant gravitons in conformal field theory}",
    eprint = "hep-th/0107119",
    archivePrefix = "arXiv",
    reportNumber = "UPR-T-943, WIS-15-01-DPP",
    doi = "10.1088/1126-6708/2002/04/034",
    journal = "JHEP",
    volume = "04",
    pages = "034",
    year = "2002"
}

@article{McGreevy:2000cw,
    author = "McGreevy, John and Susskind, Leonard and Toumbas, Nicolaos",
    title = "{Invasion of the giant gravitons from Anti-de Sitter space}",
    eprint = "hep-th/0003075",
    archivePrefix = "arXiv",
    reportNumber = "SU-ITP-00-09",
    doi = "10.1088/1126-6708/2000/06/008",
    journal = "JHEP",
    volume = "06",
    pages = "008",
    year = "2000"
}

@article{Siegel:1992xp,
    author = "Siegel, W.",
    title = "{N=2 (4) string theory is self-dual N=4 Yang-Mills theory}",
    eprint = "hep-th/9205075",
    archivePrefix = "arXiv",
    doi = "10.1103/PhysRevD.46.R3235",
    journal = "Phys. Rev. D",
    volume = "46",
    number = "8",
    pages = "R3235",
    year = "1992"
}

@article{Siegel:1992wd,
    author = "Siegel, W.",
    title = "{Selfdual N=8 supergravity as closed N=2 (N=4) strings}",
    eprint = "hep-th/9207043",
    archivePrefix = "arXiv",
    reportNumber = "ITP-SB-92-31",
    doi = "10.1103/PhysRevD.47.2504",
    journal = "Phys. Rev. D",
    volume = "47",
    pages = "2504--2511",
    year = "1993"
}

@article{Chalmers:1996rq,
    author = "Chalmers, G. and Siegel, W.",
    title = "{The Selfdual sector of QCD amplitudes}",
    eprint = "hep-th/9606061",
    archivePrefix = "arXiv",
    reportNumber = "ITP-SB-96-29",
    doi = "10.1103/PhysRevD.54.7628",
    journal = "Phys. Rev. D",
    volume = "54",
    pages = "7628--7633",
    year = "1996"
}

@article{Cachazo:2004kj,
    author = "Cachazo, Freddy and Svrcek, Peter and Witten, Edward",
    title = "{MHV vertices and tree amplitudes in gauge theory}",
    eprint = "hep-th/0403047",
    archivePrefix = "arXiv",
    doi = "10.1088/1126-6708/2004/09/006",
    journal = "JHEP",
    volume = "09",
    pages = "006",
    year = "2004"
}

@article{Ishtiaque:2018str,
    author = "Ishtiaque, Nafiz and Faroogh Moosavian, Seyed and Zhou, Yehao",
    title = "{Topological holography: The example of the D2-D4 brane system}",
    eprint = "1809.00372",
    archivePrefix = "arXiv",
    primaryClass = "hep-th",
    doi = "10.21468/SciPostPhys.9.2.017",
    journal = "SciPost Phys.",
    volume = "9",
    number = "2",
    pages = "017",
    year = "2020"
}

@article{Bonetti:2016nma,
    author = "Bonetti, Federico and Rastelli, Leonardo",
    title = "{Supersymmetric localization in AdS$_{5}$ and the protected chiral algebra}",
    eprint = "1612.06514",
    archivePrefix = "arXiv",
    primaryClass = "hep-th",
    doi = "10.1007/JHEP08(2018)098",
    journal = "JHEP",
    volume = "08",
    pages = "098",
    year = "2018"
}

@article{Eberhardt:2019ywk,
    author = "Eberhardt, Lorenz and Gaberdiel, Matthias R. and Gopakumar, Rajesh",
    title = "{Deriving the AdS$_{3}$/CFT$_{2}$ correspondence}",
    eprint = "1911.00378",
    archivePrefix = "arXiv",
    primaryClass = "hep-th",
    doi = "10.1007/JHEP02(2020)136",
    journal = "JHEP",
    volume = "02",
    pages = "136",
    year = "2020"
}

@article{Haggi-Mani:2000dxu,
    author = "Haggi-Mani, Parviz and Sundborg, Bo",
    title = "{Free large N supersymmetric Yang-Mills theory as a string theory}",
    eprint = "hep-th/0002189",
    archivePrefix = "arXiv",
    reportNumber = "USITP-00-01",
    doi = "10.1088/1126-6708/2000/04/031",
    journal = "JHEP",
    volume = "04",
    pages = "031",
    year = "2000"
}

@article{Sundborg:2000wp,
    author = "Sundborg, Bo",
    editor = "Sorokin, Dmitri P.",
    title = "{Stringy gravity, interacting tensionless strings and massless higher spins}",
    eprint = "hep-th/0103247",
    archivePrefix = "arXiv",
    doi = "10.1016/S0920-5632(01)01545-6",
    journal = "Nucl. Phys. B Proc. Suppl.",
    volume = "102",
    pages = "113--119",
    year = "2001"
}

@article{Sezgin:2001zs,
    author = "Sezgin, E. and Sundell, P.",
    title = "{Doubletons and 5-D higher spin gauge theory}",
    eprint = "hep-th/0105001",
    archivePrefix = "arXiv",
    reportNumber = "CTP-TAMU-13-01, UG-01-27",
    doi = "10.1088/1126-6708/2001/09/036",
    journal = "JHEP",
    volume = "09",
    pages = "036",
    year = "2001"
}

@article{Mikhailov:2002bp,
    author = "Mikhailov, Andrei",
    title = "{Notes on higher spin symmetries}",
    eprint = "hep-th/0201019",
    archivePrefix = "arXiv",
    reportNumber = "NSF-ITP-01-181, ITEP-TH-66-01",
    month = "1",
    year = "2002"
}

@article{Sezgin:2002rt,
    author = "Sezgin, E. and Sundell, P.",
    title = "{Massless higher spins and holography}",
    eprint = "hep-th/0205131",
    archivePrefix = "arXiv",
    reportNumber = "CTP-TAMU-08-02, UU-01-10",
    doi = "10.1016/S0550-3213(02)00739-3",
    journal = "Nucl. Phys. B",
    volume = "644",
    pages = "303--370",
    year = "2002",
    note = "[Erratum: Nucl.Phys.B 660, 403--403 (2003)]"
}

@article{Gaberdiel:2018rqv,
    author = "Gaberdiel, Matthias R. and Gopakumar, Rajesh",
    title = "{Tensionless string spectra on AdS$_{3}$}",
    eprint = "1803.04423",
    archivePrefix = "arXiv",
    primaryClass = "hep-th",
    doi = "10.1007/JHEP05(2018)085",
    journal = "JHEP",
    volume = "05",
    pages = "085",
    year = "2018"
}

@article{Eberhardt:2018ouy,
    author = "Eberhardt, Lorenz and Gaberdiel, Matthias R. and Gopakumar, Rajesh",
    title = "{The Worldsheet Dual of the Symmetric Product CFT}",
    eprint = "1812.01007",
    archivePrefix = "arXiv",
    primaryClass = "hep-th",
    doi = "10.1007/JHEP04(2019)103",
    journal = "JHEP",
    volume = "04",
    pages = "103",
    year = "2019"
}

@article{Berkovits:2007rj,
    author = "Berkovits, Nathan and Vafa, Cumrun",
    editor = "Edelstein, Jose D. and Grandi, Nicolas and Nunez, Carmen A. and Schvellinger, Martin",
    title = "{Towards a Worldsheet Derivation of the Maldacena Conjecture}",
    eprint = "0711.1799",
    archivePrefix = "arXiv",
    primaryClass = "hep-th",
    reportNumber = "IFT-P-018-2007",
    doi = "10.1088/1126-6708/2008/03/031",
    journal = "JHEP",
    volume = "03",
    pages = "031",
    year = "2008"
}

@article{Berkovits:2007zk,
    author = "Berkovits, Nathan",
    title = "{A New Limit of the AdS$_5\times S^5$ Sigma Model}",
    eprint = "hep-th/0703282",
    archivePrefix = "arXiv",
    reportNumber = "IFT-P-06-2007",
    doi = "10.1088/1126-6708/2007/08/011",
    journal = "JHEP",
    volume = "08",
    pages = "011",
    year = "2007"
}

@article{Berkovits:2025xok,
    author = "Berkovits, Nathan",
    title = "{Topological A-Model for $AdS_5\times S^5$ Superstring and the Maldacena Conjecture}",
    eprint = "2506.10907",
    archivePrefix = "arXiv",
    primaryClass = "hep-th",
    month = "6",
    year = "2025"
}

@article{Alday:2007hr,
    author = "Alday, Luis F. and Maldacena, Juan Martin",
    title = "{Gluon scattering amplitudes at strong coupling}",
    eprint = "0705.0303",
    archivePrefix = "arXiv",
    primaryClass = "hep-th",
    reportNumber = "SPIN-07-16, ITP-UU-07-24",
    doi = "10.1088/1126-6708/2007/06/064",
    journal = "JHEP",
    volume = "06",
    pages = "064",
    year = "2007"
}

@article{Grisaru:2000zn,
    author = "Grisaru, Marcus T. and Myers, Robert C. and Tafjord, Oyvind",
    title = "{SUSY and goliath}",
    eprint = "hep-th/0008015",
    archivePrefix = "arXiv",
    reportNumber = "MCGILL-00-21, BRX-TH-472",
    doi = "10.1088/1126-6708/2000/08/040",
    journal = "JHEP",
    volume = "08",
    pages = "040",
    year = "2000"
}

@article{McStay:2024dtk,
    author = "McStay, N. M. and Reid-Edwards, R. A.",
    title = "{String theory in twistor space and minimal tension holography}",
    eprint = "2411.08836",
    archivePrefix = "arXiv",
    primaryClass = "hep-th",
    doi = "10.1007/JHEP10(2025)073",
    journal = "JHEP",
    volume = "10",
    pages = "073",
    year = "2025"
}

@article{Budzik:2025vzr,
    author = "Budzik, Kasia",
    title = "{Supergroup Invariants and the Brane/Negative Brane Expansion}",
    eprint = "2509.20451",
    archivePrefix = "arXiv",
    primaryClass = "hep-th",
    month = "9",
    year = "2025"
}

@article{Murthy:2022ien,
    author = "Murthy, Sameer",
    title = "{Unitary matrix models, free fermions, and the giant graviton expansion}",
    eprint = "2202.06897",
    archivePrefix = "arXiv",
    primaryClass = "hep-th",
    doi = "10.4310/PAMQ.2023.v19.n1.a12",
    journal = "Pure Appl. Math. Quart.",
    volume = "19",
    number = "1",
    pages = "299--340",
    year = "2023"
}

@article{Imamura:2021ytr,
    author = "Imamura, Yosuke",
    title = "{Finite-N superconformal index via the AdS/CFT correspondence}",
    eprint = "2108.12090",
    archivePrefix = "arXiv",
    primaryClass = "hep-th",
    reportNumber = "TIT/HEP-686",
    doi = "10.1093/ptep/ptab141",
    journal = "PTEP",
    volume = "2021",
    number = "12",
    pages = "123B05",
    year = "2021"
}

@article{Gaiotto:2021xce,
    author = "Gaiotto, Davide and Lee, Ji Hoon",
    title = "{The giant graviton expansion}",
    eprint = "2109.02545",
    archivePrefix = "arXiv",
    primaryClass = "hep-th",
    doi = "10.1007/JHEP08(2024)025",
    journal = "JHEP",
    volume = "08",
    pages = "025",
    year = "2024"
}

@article{borodin2000fredholm,
  title={A Fredholm determinant formula for Toeplitz determinants},
  author={Borodin, Alexei and Okounkov, Andrei},
  journal={Integral Equations and Operator Theory},
  volume={37},
  number={4},
  pages={386--396},
  year={2000},
  publisher={Springer}
}

@article{Liu:2022olj,
    author = "Liu, James T. and Rajappa, Neville Joshua",
    title = "{Finite N indices and the giant graviton expansion}",
    eprint = "2212.05408",
    archivePrefix = "arXiv",
    primaryClass = "hep-th",
    reportNumber = "LCTP-22-16",
    doi = "10.1007/JHEP04(2023)078",
    journal = "JHEP",
    volume = "04",
    pages = "078",
    year = "2023"
}

@article{Eniceicu:2023uvd,
    author = "Eniceicu, Dan Stefan",
    title = "{Comments on the Giant-Graviton Expansion of the Superconformal Index}",
    eprint = "2302.04887",
    archivePrefix = "arXiv",
    primaryClass = "hep-th",
    month = "2",
    year = "2023"
}

@article{Kinney:2005ej,
    author = "Kinney, Justin and Maldacena, Juan Martin and Minwalla, Shiraz and Raju, Suvrat",
    title = "{An Index for 4 dimensional super conformal theories}",
    eprint = "hep-th/0510251",
    archivePrefix = "arXiv",
    doi = "10.1007/s00220-007-0258-7",
    journal = "Commun. Math. Phys.",
    volume = "275",
    pages = "209--254",
    year = "2007"
}

@article{Choi:2022caq,
    author = "Choi, Sunjin and Kim, Seok and Lee, Eunwoo and Park, Jaemo",
    title = "{The shape of non-graviton operators for SU(2)}",
    eprint = "2209.12696",
    archivePrefix = "arXiv",
    primaryClass = "hep-th",
    reportNumber = "KIAS-P22052",
    doi = "10.1007/JHEP09(2024)029",
    journal = "JHEP",
    volume = "09",
    pages = "029",
    year = "2024"
}

@article{Choi:2023znd,
    author = "Choi, Sunjin and Kim, Seok and Lee, Eunwoo and Lee, Siyul and Park, Jaemo",
    title = "{Towards quantum black hole microstates}",
    eprint = "2304.10155",
    archivePrefix = "arXiv",
    primaryClass = "hep-th",
    doi = "10.1007/JHEP11(2023)175",
    journal = "JHEP",
    volume = "11",
    pages = "175",
    year = "2023",
    note = "[Erratum: JHEP 03, 091 (2025)]"
}

@article{Choi:2023vdm,
    author = "Choi, Jaehyeok and Choi, Sunjin and Kim, Seok and Lee, Jehyun and Lee, Siyul",
    title = "{Finite N black hole cohomologies}",
    eprint = "2312.16443",
    archivePrefix = "arXiv",
    primaryClass = "hep-th",
    reportNumber = "SNUTP23-002, KIAS-P23070, LCTP-23-20, SNUTP23-002; KIAS-P23070; LCTP-23-20;",
    doi = "10.1007/JHEP12(2024)029",
    journal = "JHEP",
    volume = "12",
    pages = "029",
    year = "2024"
}

@article{Chang:2022mjp,
    author = "Chang, Chi-Ming and Lin, Ying-Hsuan",
    title = "{Words to describe a black hole}",
    eprint = "2209.06728",
    archivePrefix = "arXiv",
    primaryClass = "hep-th",
    doi = "10.1007/JHEP02(2023)109",
    journal = "JHEP",
    volume = "02",
    pages = "109",
    year = "2023"
}

@article{Chang:2024zqi,
    author = "Chang, Chi-Ming and Lin, Ying-Hsuan",
    title = "{Holographic covering and the fortuity of black holes}",
    eprint = "2402.10129",
    archivePrefix = "arXiv",
    primaryClass = "hep-th",
    month = "2",
    year = "2024"
}

@article{Lee:2022vig,
    author = "Lee, Ji Hoon",
    title = "{Exact stringy microstates from gauge theories}",
    eprint = "2204.09286",
    archivePrefix = "arXiv",
    primaryClass = "hep-th",
    doi = "10.1007/JHEP11(2022)137",
    journal = "JHEP",
    volume = "11",
    pages = "137",
    year = "2022"
}

@article{Arai:2020qaj,
    author = "Arai, Reona and Fujiwara, Shota and Imamura, Yosuke and Mori, Tatsuya",
    title = "{Schur index of the ${\cal N}=4$ $U(N)$ supersymmetric Yang-Mills theory via the AdS/CFT correspondence}",
    eprint = "2001.11667",
    archivePrefix = "arXiv",
    primaryClass = "hep-th",
    reportNumber = "TIT/HEP-677",
    doi = "10.1103/PhysRevD.101.086017",
    journal = "Phys. Rev. D",
    volume = "101",
    number = "8",
    pages = "086017",
    year = "2020"
}

@article{Arai:2019xmp,
    author = "Arai, Reona and Imamura, Yosuke",
    title = "{Finite $N$ Corrections to the Superconformal Index of S-fold Theories}",
    eprint = "1904.09776",
    archivePrefix = "arXiv",
    primaryClass = "hep-th",
    reportNumber = "TIT-HEP-671",
    doi = "10.1093/ptep/ptz088",
    journal = "PTEP",
    volume = "2019",
    number = "8",
    pages = "083B04",
    year = "2019"
}

@article{Arai:2019wgv,
    author = "Arai, Reona and Fujiwara, Shota and Imamura, Yosuke and Mori, Tatsuya",
    title = "{Finite $N$ corrections to the superconformal index of orbifold quiver gauge theories}",
    eprint = "1907.05660",
    archivePrefix = "arXiv",
    primaryClass = "hep-th",
    reportNumber = "TIT/HEP-674",
    doi = "10.1007/JHEP10(2019)243",
    journal = "JHEP",
    volume = "10",
    pages = "243",
    year = "2019"
}

@article{Arai:2019aou,
    author = "Arai, Reona and Fujiwara, Shota and Imamura, Yosuke and Mori, Tatsuya",
    title = "{Finite $N$ corrections to the superconformal index of toric quiver gauge theories}",
    eprint = "1911.10794",
    archivePrefix = "arXiv",
    primaryClass = "hep-th",
    reportNumber = "TIT/HEP-675",
    doi = "10.1093/ptep/ptaa023",
    journal = "PTEP",
    volume = "2020",
    number = "4",
    pages = "043B09",
    year = "2020"
}

@article{Eleftheriou:2026yct,
    author = "Eleftheriou, Giorgos and Ji, Ziming and Murthy, Sameer",
    title = "{Fermionic trace relations and supersymmetric indices at finite $N$}",
    eprint = "2605.08085",
    archivePrefix = "arXiv",
    primaryClass = "hep-th",
    month = "5",
    year = "2026"
}

@article{Budzik:2023vtr,
    author = "Budzik, Kasia and Murali, Harish and Vieira, Pedro",
    title = "{Following Black Hole States}",
    eprint = "2306.04693",
    archivePrefix = "arXiv",
    primaryClass = "hep-th",
    month = "6",
    year = "2023"
}

@article{deMelloKoch:2024pcs,
    author = "de Mello Koch, Robert and Kim, Minkyoo and Kim, Seok and Lee, Jehyun and Lee, Siyul",
    title = "{Brane-fused black hole operators}",
    eprint = "2412.08695",
    archivePrefix = "arXiv",
    primaryClass = "hep-th",
    reportNumber = "SNUTP24-004",
    doi = "10.1007/JHEP07(2025)216",
    journal = "JHEP",
    volume = "07",
    pages = "216",
    year = "2025"
}

@article{Kim:2025vup,
    author = "Kim, Seok and Lee, Jehyun and Lee, Siyul and Oh, Hyunwoo",
    title = "{BPS phases and fortuity in higher spin holography}",
    eprint = "2511.03105",
    archivePrefix = "arXiv",
    primaryClass = "hep-th",
    month = "11",
    year = "2025"
}

@article{Chang:2024lxt,
    author = "Chang, Chi-Ming and Chen, Yiming and Sia, Bik Soon and Yang, Zhenbin",
    title = "{Fortuity in SYK models}",
    eprint = "2412.06902",
    archivePrefix = "arXiv",
    primaryClass = "hep-th",
    doi = "10.1007/JHEP08(2025)003",
    journal = "JHEP",
    volume = "08",
    pages = "003",
    year = "2025"
}

@article{Chang:2025wgo,
    author = "Chang, Chi-Ming and Zhang, Haoyu",
    title = "{Fortuity and R-charge concentration in the D1-D5 CFT}",
    eprint = "2511.23294",
    archivePrefix = "arXiv",
    primaryClass = "hep-th",
    month = "11",
    year = "2025"
}

@article{Chang:2025mqp,
    author = "Chang, Chi-Ming and Lin, Ying-Hsuan",
    title = "{Violation of S-duality in classical $Q$-cohomology}",
    eprint = "2510.24008",
    archivePrefix = "arXiv",
    primaryClass = "hep-th",
    month = "10",
    year = "2025"
}

@article{Chen:2025sum,
    author = "Chen, Yiming",
    title = "{Fortuity with a single matrix}",
    eprint = "2511.00790",
    archivePrefix = "arXiv",
    primaryClass = "hep-th",
    month = "11",
    year = "2025"
}

@article{Behan:2025hbx,
    author = "Behan, Connor and de Gioia, Leonardo Pipolo",
    title = "{Two roads to fortuity in ABJM theory}",
    eprint = "2512.23603",
    archivePrefix = "arXiv",
    primaryClass = "hep-th",
    month = "12",
    year = "2025"
}

@article{Belin:2025hsg,
    author = "Belin, Alexandre and Singh, Palash and Vadala, Rita and Zaffaroni, Alberto",
    title = "{Fortuity in ABJM}",
    eprint = "2512.04146",
    archivePrefix = "arXiv",
    primaryClass = "hep-th",
    month = "12",
    year = "2025"
}

@article{Imamura:2022aua,
    author = "Imamura, Yosuke",
    title = "{Analytic continuation for giant gravitons}",
    eprint = "2205.14615",
    archivePrefix = "arXiv",
    primaryClass = "hep-th",
    reportNumber = "TIT/HEP-689",
    doi = "10.1093/ptep/ptac127",
    journal = "PTEP",
    volume = "2022",
    number = "10",
    pages = "103B02",
    year = "2022"
}

@article{Chen:2024cvf,
    author = "Chen, Yiming and Mahajan, Raghu and Tang, Haifeng",
    title = "{Giant graviton expansion from eigenvalue instantons}",
    eprint = "2407.08155",
    archivePrefix = "arXiv",
    primaryClass = "hep-th",
    doi = "10.1007/JHEP06(2026)147",
    journal = "JHEP",
    volume = "06",
    pages = "147",
    year = "2026"
}

@article{Eleftheriou:2023jxr,
    author = "Eleftheriou, Giorgos and Murthy, Sameer and Rossell{\'o}, Mart{\'\i}",
    title = "{The giant graviton expansion in $AdS_5 \times S^5$}",
    eprint = "2312.14921",
    archivePrefix = "arXiv",
    primaryClass = "hep-th",
    doi = "10.21468/SciPostPhys.17.4.098",
    journal = "SciPost Phys.",
    volume = "17",
    number = "4",
    pages = "098",
    year = "2024"
}

@article{Deddo:2024liu,
    author = "Deddo, Evan and Liu, James T. and Pando Zayas, Leopoldo A. and Saskowski, Robert J.",
    title = "{Giant Graviton Expansion from Bubbling Geometry: Discreteness from Quantized Geometry}",
    eprint = "2402.19452",
    archivePrefix = "arXiv",
    primaryClass = "hep-th",
    reportNumber = "LCTP-24-04",
    doi = "10.1103/PhysRevLett.132.261501",
    journal = "Phys. Rev. Lett.",
    volume = "132",
    number = "26",
    pages = "261501",
    year = "2024"
}

@article{Ezroura:2024wmp,
    author = "Ezroura, Nizar and Liu, James T. and Rajappa, Neville Joshua",
    title = "{Analytic continuation and the giant graviton expansion}",
    eprint = "2408.02759",
    archivePrefix = "arXiv",
    primaryClass = "hep-th",
    reportNumber = "LCTP-24-12",
    doi = "10.1007/JHEP01(2025)028",
    journal = "JHEP",
    volume = "01",
    pages = "028",
    year = "2025"
}

@article{Imamura:2024lkw,
    author = "Imamura, Yosuke",
    title = "{Giant Graviton Expansions for the Line Operator Index}",
    eprint = "2403.11543",
    archivePrefix = "arXiv",
    primaryClass = "hep-th",
    reportNumber = "TIT/HEP-699",
    doi = "10.1093/ptep/ptae084",
    journal = "PTEP",
    volume = "2024",
    number = "6",
    pages = "063B03",
    year = "2024"
}

@article{Berkovits:2026ifu,
    author = "Berkovits, Nathan",
    title = "{Instanton Solutions of the Topological A-model Dual to Free Super-Yang-Mills}",
    eprint = "2602.19985",
    archivePrefix = "arXiv",
    primaryClass = "hep-th",
    month = "2",
    year = "2026"
}

@article{Lee:2024hef,
    author = "Lee, Ji Hoon and Stanford, Douglas",
    title = "{Bulk thimbles dual to trace relations}",
    eprint = "2412.20769",
    archivePrefix = "arXiv",
    primaryClass = "hep-th",
    month = "12",
    year = "2024"
}

@article{Caputa:2025ikn,
    author = "Caputa, Pawel and de Mello Koch, Robert",
    title = "{Gauge invariants at arbitrary N and trace relations}",
    eprint = "2509.05834",
    archivePrefix = "arXiv",
    primaryClass = "hep-th",
    reportNumber = "YITP-25-137",
    doi = "10.1007/JHEP12(2025)165",
    journal = "JHEP",
    volume = "12",
    pages = "165",
    year = "2025"
}

@article{Choi:2025bhi,
    author = "Choi, Jaehyeok and Lee, Eunwoo",
    title = "{Konishi lifts a black hole}",
    eprint = "2511.09519",
    archivePrefix = "arXiv",
    primaryClass = "hep-th",
    month = "11",
    year = "2025"
}

@article{Eberhardt:2017pty,
    author = "Eberhardt, Lorenz and Gaberdiel, Matthias R. and Li, Wei",
    title = "{A holographic dual for string theory on AdS$_{3}${\texttimes}S$^{3}${\texttimes}S$^{3}${\texttimes}S$^{1}$}",
    eprint = "1707.02705",
    archivePrefix = "arXiv",
    primaryClass = "hep-th",
    doi = "10.1007/JHEP08(2017)111",
    journal = "JHEP",
    volume = "08",
    pages = "111",
    year = "2017"
}

@article{Eberhardt:2019qcl,
    author = "Eberhardt, Lorenz and Gaberdiel, Matthias R.",
    title = "{String theory on AdS$_3$ and the symmetric orbifold of Liouville theory}",
    eprint = "1903.00421",
    archivePrefix = "arXiv",
    primaryClass = "hep-th",
    doi = "10.1016/j.nuclphysb.2019.114774",
    journal = "Nucl. Phys. B",
    volume = "948",
    pages = "114774",
    year = "2019"
}

@article{Dei:2019osr,
    author = "Dei, Andrea and Eberhardt, Lorenz and Gaberdiel, Matthias R.",
    title = "{Three-point functions in AdS$_{3}$/CFT$_{2}$ holography}",
    eprint = "1907.13144",
    archivePrefix = "arXiv",
    primaryClass = "hep-th",
    doi = "10.1007/JHEP12(2019)012",
    journal = "JHEP",
    volume = "12",
    pages = "012",
    year = "2019"
}

@article{Eberhardt:2019niq,
    author = "Eberhardt, Lorenz and Gaberdiel, Matthias R.",
    title = "{Strings on $\text{AdS}_3 \times \text{S}^3 \times \text{S}^3 \times \text{S}^1$}",
    eprint = "1904.01585",
    archivePrefix = "arXiv",
    primaryClass = "hep-th",
    doi = "10.1007/JHEP06(2019)035",
    journal = "JHEP",
    volume = "06",
    pages = "035",
    year = "2019"
}

@article{Dei:2019iym,
    author = "Dei, Andrea and Eberhardt, Lorenz",
    title = "{Correlators of the symmetric product orbifold}",
    eprint = "1911.08485",
    archivePrefix = "arXiv",
    primaryClass = "hep-th",
    doi = "10.1007/JHEP01(2020)108",
    journal = "JHEP",
    volume = "01",
    pages = "108",
    year = "2020"
}

@article{Dei:2020zui,
    author = "Dei, Andrea and Gaberdiel, Matthias R. and Gopakumar, Rajesh and Knighton, Bob",
    title = "{Free field world-sheet correlators for ${\rm AdS}_3$}",
    eprint = "2009.11306",
    archivePrefix = "arXiv",
    primaryClass = "hep-th",
    doi = "10.1007/JHEP02(2021)081",
    journal = "JHEP",
    volume = "02",
    pages = "081",
    year = "2021"
}

@article{Dei:2021xgh,
    author = "Dei, Andrea and Eberhardt, Lorenz",
    title = "{String correlators on AdS$_{3}$: three-point functions}",
    eprint = "2105.12130",
    archivePrefix = "arXiv",
    primaryClass = "hep-th",
    doi = "10.1007/JHEP08(2021)025",
    journal = "JHEP",
    volume = "08",
    pages = "025",
    year = "2021"
}

@article{Dei:2021yom,
    author = "Dei, Andrea and Eberhardt, Lorenz",
    title = "{String correlators on AdS$_{3}$: four-point functions}",
    eprint = "2107.01481",
    archivePrefix = "arXiv",
    primaryClass = "hep-th",
    doi = "10.1007/JHEP09(2021)209",
    journal = "JHEP",
    volume = "09",
    pages = "209",
    year = "2021"
}

@article{Dei:2022pkr,
    author = "Dei, Andrea and Eberhardt, Lorenz",
    title = "{String correlators on $\text{AdS}_3$: Analytic structure and dual CFT}",
    eprint = "2203.13264",
    archivePrefix = "arXiv",
    primaryClass = "hep-th",
    doi = "10.21468/SciPostPhys.13.3.053",
    journal = "SciPost Phys.",
    volume = "13",
    number = "3",
    pages = "053",
    year = "2022"
}

@article{Knighton:2020kuh,
    author = "Knighton, Bob",
    title = "{Higher genus correlators for tensionless AdS$_{3}$ strings}",
    eprint = "2012.01445",
    archivePrefix = "arXiv",
    primaryClass = "hep-th",
    doi = "10.1007/JHEP04(2021)211",
    journal = "JHEP",
    volume = "04",
    pages = "211",
    year = "2021"
}

@article{Eberhardt:2020bgq,
    author = "Eberhardt, Lorenz",
    title = "{Partition functions of the tensionless string}",
    eprint = "2008.07533",
    archivePrefix = "arXiv",
    primaryClass = "hep-th",
    doi = "10.1007/JHEP03(2021)176",
    journal = "JHEP",
    volume = "03",
    pages = "176",
    year = "2021"
}

@article{Eberhardt:2021jvj,
    author = "Eberhardt, Lorenz",
    title = "{Summing over Geometries in String Theory}",
    eprint = "2102.12355",
    archivePrefix = "arXiv",
    primaryClass = "hep-th",
    doi = "10.1007/JHEP05(2021)233",
    journal = "JHEP",
    volume = "05",
    pages = "233",
    year = "2021"
}

@article{Gaberdiel:2023dxt,
    author = "Gaberdiel, Matthias R. and Guo, Bin and Mathur, Samir D.",
    title = "{Tensionless strings on AdS$_{3}$ orbifolds}",
    eprint = "2312.01348",
    archivePrefix = "arXiv",
    primaryClass = "hep-th",
    doi = "10.1007/JHEP04(2024)057",
    journal = "JHEP",
    volume = "04",
    pages = "057",
    year = "2024"
}

@article{Gaberdiel:2025rtu,
    author = "Gaberdiel, Matthias R. and Guo, Bin",
    title = "{Correlators for tensionless strings on ${\rm AdS}_3$ orbifolds}",
    eprint = "2511.07725",
    archivePrefix = "arXiv",
    primaryClass = "hep-th",
    month = "11",
    year = "2025"
}

@article{Aharony:2024fid,
    author = "Aharony, Ofer and Urbach, Erez Y.",
    title = "{Type II string theory on AdS3{\texttimes}S3{\texttimes}T4 and symmetric orbifolds}",
    eprint = "2406.14605",
    archivePrefix = "arXiv",
    primaryClass = "hep-th",
    doi = "10.1103/PhysRevD.110.046028",
    journal = "Phys. Rev. D",
    volume = "110",
    number = "4",
    pages = "046028",
    year = "2024"
}

@article{Eberhardt:2025sbi,
    author = "Eberhardt, Lorenz and Gaberdiel, Matthias R.",
    title = "{A localising AdS$_3$ sigma model}",
    eprint = "2505.09226",
    archivePrefix = "arXiv",
    primaryClass = "hep-th",
    doi = "10.21468/SciPostPhys.19.2.060",
    journal = "SciPost Phys.",
    volume = "19",
    number = "2",
    pages = "060",
    year = "2025"
}

@article{Chang:2025rqy,
    author = "Chang, Chi-Ming and Lin, Ying-Hsuan and Zhang, Haoyu",
    title = "{Fortuity in the D1-D5 system}",
    eprint = "2501.05448",
    archivePrefix = "arXiv",
    primaryClass = "hep-th",
    month = "1",
    year = "2025"
}

@article{Hughes:2025tdy,
    author = "Hughes, Marcel R. R. and Shigemori, Masaki",
    title = "{Fortuity and supergravity}",
    eprint = "2505.14888",
    archivePrefix = "arXiv",
    primaryClass = "hep-th",
    reportNumber = "YITP-25-74",
    doi = "10.1007/JHEP03(2026)130",
    journal = "JHEP",
    volume = "03",
    pages = "130",
    year = "2026"
}

@article{deMelloKoch:2025cec,
    author = "de Mello Koch, Robert and Ghosh, Animik and Van Zyl, Hendrik J. R.",
    title = "{Bosonic fortuity in vector models}",
    eprint = "2504.14181",
    archivePrefix = "arXiv",
    primaryClass = "hep-th",
    doi = "10.1007/JHEP06(2025)246",
    journal = "JHEP",
    volume = "06",
    pages = "246",
    year = "2025"
}

@article{Jarov:2025qhz,
    author = "Jarov, Seraphim",
    title = "{Twisted holography from the B-model on a 7-fold}",
    eprint = "2512.07412",
    archivePrefix = "arXiv",
    primaryClass = "hep-th",
    month = "12",
    year = "2025"
}

@article{Costello:2023vyy,
    author = "Costello, Kevin J.",
    title = "{Bootstrapping two-loop QCD amplitudes}",
    eprint = "2302.00770",
    archivePrefix = "arXiv",
    primaryClass = "hep-th",
    doi = "10.1007/JHEP08(2025)011",
    journal = "JHEP",
    volume = "08",
    pages = "011",
    year = "2025"
}

@article{Costello:2022upu,
    author = "Costello, Kevin and Paquette, Natalie M.",
    title = "{Associativity of One-Loop Corrections to the Celestial Operator Product Expansion}",
    eprint = "2204.05301",
    archivePrefix = "arXiv",
    primaryClass = "hep-th",
    doi = "10.1103/PhysRevLett.129.231604",
    journal = "Phys. Rev. Lett.",
    volume = "129",
    number = "23",
    pages = "231604",
    year = "2022"
}

@article{Dixon:2024mzh,
    author = "Dixon, Lance J. and Morales, Anthony",
    title = "{On gauge amplitudes first appearing at two loops}",
    eprint = "2407.13967",
    archivePrefix = "arXiv",
    primaryClass = "hep-th",
    reportNumber = "SLAC-PUB-17784",
    doi = "10.1007/JHEP08(2024)129",
    journal = "JHEP",
    volume = "08",
    pages = "129",
    year = "2024"
}

@article{Dixon:2024tsb,
    author = "Dixon, Lance J. and Morales, Anthony",
    title = "{Rational QCD loop amplitudes and quantum theories on twistor space}",
    eprint = "2411.10967",
    archivePrefix = "arXiv",
    primaryClass = "hep-th",
    doi = "10.1007/JHEP03(2025)188",
    journal = "JHEP",
    volume = "03",
    pages = "188",
    year = "2025"
}

@article{Morales:2025alm,
    author = "Morales, Anthony",
    title = "{Two-loop QCD amplitudes from the chiral algebra bootstrap}",
    eprint = "2510.20156",
    archivePrefix = "arXiv",
    primaryClass = "hep-th",
    doi = "10.1103/yk39-n6zq",
    journal = "Phys. Rev. D",
    volume = "113",
    number = "10",
    pages = "105017",
    year = "2026"
}

@article{Charanya:2026pnh,
    author = "Charanya, Jaazib and Morales, Anthony and Paquette, Natalie M.",
    title = "{Form factors of $\mathscr{N}=4$ self-dual Yang-Mills from the chiral algebra bootstrap}",
    eprint = "2604.21015",
    archivePrefix = "arXiv",
    primaryClass = "hep-th",
    month = "4",
    year = "2026"
}

@article{Fernandez:2024qnu,
    author = "Fern{\'a}ndez, V{\'\i}ctor E. and Paquette, Natalie M.",
    title = "{Associativity is enough: an all-orders 2d chiral algebra for 4d form factors}",
    eprint = "2412.17168",
    archivePrefix = "arXiv",
    primaryClass = "hep-th",
    doi = "10.1088/1361-6382/ae0089",
    journal = "Class. Quant. Grav.",
    volume = "42",
    number = "18",
    pages = "185005",
    year = "2025"
}

@article{Fernandez:2023abp,
    author = "Fern{\'a}ndez, V{\'\i}ctor E.",
    title = "{One-loop corrections to the celestial chiral algebra from Koszul Duality}",
    eprint = "2302.14292",
    archivePrefix = "arXiv",
    primaryClass = "hep-th",
    doi = "10.1007/JHEP04(2023)124",
    journal = "JHEP",
    volume = "04",
    pages = "124",
    year = "2023"
}

@article{Hayashi:2024aaf,
    author = "Hayashi, Hirotaka and Nosaka, Tomoki and Okazaki, Tadashi",
    title = "{M2-M5 giant graviton expansions}",
    eprint = "2409.13239",
    archivePrefix = "arXiv",
    primaryClass = "hep-th",
    doi = "10.1007/JHEP12(2024)109",
    journal = "JHEP",
    volume = "12",
    pages = "109",
    year = "2024"
}

@article{Beccaria:2023sph,
    author = "Beccaria, Matteo and Giombi, Simone and Tseytlin, Arkady A.",
    title = "{(2,0) theory on $S^5\times S^1$ and quantum M2 branes}",
    eprint = "2309.10786",
    archivePrefix = "arXiv",
    primaryClass = "hep-th",
    reportNumber = "PUPT-2648, Imperial-TP-AT-2023-05",
    doi = "10.1016/j.nuclphysb.2023.116400",
    journal = "Nucl. Phys. B",
    volume = "998",
    pages = "116400",
    year = "2024"
}

@article{Beccaria:2023cuo,
    author = "Beccaria, Matteo and Tseytlin, Arkady A.",
    title = "{Large N expansion of superconformal index of $k=1$ ABJM theory and semiclassical M5 brane partition function}",
    eprint = "2312.01917",
    archivePrefix = "arXiv",
    primaryClass = "hep-th",
    reportNumber = "Imperial-TP-AT-2023-06",
    doi = "10.1016/j.nuclphysb.2024.116507",
    journal = "Nucl. Phys. B",
    volume = "1001",
    pages = "116507",
    year = "2024"
}

@article{Chen:2026fpe,
    author = "Chen, Heng-Yu and Dorey, Nick and Moriyama, Sanefumi and Mouland, Rishi and Sanli, Canberk",
    title = "{An M2/M5 Duality from the Giant Graviton Expansion}",
    eprint = "2601.17114",
    archivePrefix = "arXiv",
    primaryClass = "hep-th",
    month = "1",
    year = "2026"
}

@article{Dorey:2025qht,
    author = "Dorey, Nick and Roehl, Paul Luis",
    title = "{Giant Gravitons, Fermionic Forms and Vertex Algebras}",
    eprint = "2511.03709",
    archivePrefix = "arXiv",
    primaryClass = "hep-th",
    month = "11",
    year = "2025"
}

@article{Chen:2024erz,
    author = "Chen, Heng-Yu and Dorey, Nick and Moriyama, Sanefumi and Mouland, Rishi and Sanli, Canberk",
    title = "{Giant gravitons and volume minimisation}",
    eprint = "2412.05357",
    archivePrefix = "arXiv",
    primaryClass = "hep-th",
    doi = "10.1007/JHEP08(2025)121",
    journal = "JHEP",
    volume = "08",
    pages = "121",
    year = "2025"
}

@article{Choi:2022ovw,
    author = "Choi, Sunjin and Kim, Seok and Lee, Eunwoo and Lee, Jehyun",
    title = "{From giant gravitons to black holes}",
    eprint = "2207.05172",
    archivePrefix = "arXiv",
    primaryClass = "hep-th",
    reportNumber = "KIAS-P22048, SNUTP22-001",
    doi = "10.1007/JHEP11(2023)086",
    journal = "JHEP",
    volume = "11",
    pages = "086",
    year = "2023"
}

@article{Purkayastha:2025whi,
    author = "Purkayastha, Souradeep and Qu, Zishen and Zahabi, Ali",
    title = "{Quiver superconformal index and giant gravitons: asymptotics and expansions}",
    eprint = "2509.12123",
    archivePrefix = "arXiv",
    primaryClass = "hep-th",
    month = "9",
    year = "2025"
}

@article{Roiban:2004vt,
    author = "Roiban, Radu and Spradlin, Marcus and Volovich, Anastasia",
    title = "{A Googly amplitude from the B model in twistor space}",
    eprint = "hep-th/0402016",
    archivePrefix = "arXiv",
    reportNumber = "NSF-KITP-04-15",
    doi = "10.1088/1126-6708/2004/04/012",
    journal = "JHEP",
    volume = "04",
    pages = "012",
    year = "2004"
}

@article{Maldacena:1998bw,
    author = "Maldacena, Juan Martin and Strominger, Andrew",
    title = "{AdS(3) black holes and a stringy exclusion principle}",
    eprint = "hep-th/9804085",
    archivePrefix = "arXiv",
    reportNumber = "HUTP-98-A016",
    doi = "10.1088/1126-6708/1998/12/005",
    journal = "JHEP",
    volume = "12",
    pages = "005",
    year = "1998"
}

@article{Evans:1998qu,
    author = "Evans, Jonathan M. and Gaberdiel, Matthias R. and Perry, Malcolm J.",
    title = "{The no ghost theorem for AdS(3) and the stringy exclusion principle}",
    eprint = "hep-th/9806024",
    archivePrefix = "arXiv",
    reportNumber = "DAMTP-98-37",
    doi = "10.1016/S0550-3213(98)00561-6",
    journal = "Nucl. Phys. B",
    volume = "535",
    pages = "152--170",
    year = "1998"
}

@article{Susskind:1998dq,
    author = "Susskind, Leonard and Witten, Edward",
    title = "{The Holographic bound in anti-de Sitter space}",
    eprint = "hep-th/9805114",
    archivePrefix = "arXiv",
    reportNumber = "SU-ITP-98-39, IASSNS-HEP-98-44",
    month = "5",
    year = "1998"
}

@article{Lunin:2000yv,
    author = "Lunin, Oleg and Mathur, Samir D.",
    title = "{Correlation functions for M**N / S(N) orbifolds}",
    eprint = "hep-th/0006196",
    archivePrefix = "arXiv",
    reportNumber = "OHSTPY-HEP-T-00-010",
    doi = "10.1007/s002200100431",
    journal = "Commun. Math. Phys.",
    volume = "219",
    pages = "399--442",
    year = "2001"
}

@article{Lunin:2001pw,
    author = "Lunin, Oleg and Mathur, Samir D.",
    title = "{Three point functions for M(N) / S(N) orbifolds with N=4 supersymmetry}",
    eprint = "hep-th/0103169",
    archivePrefix = "arXiv",
    reportNumber = "OHSTPY-HEP-T-01-005",
    doi = "10.1007/s002200200638",
    journal = "Commun. Math. Phys.",
    volume = "227",
    pages = "385--419",
    year = "2002"
}

@article{Kontsevich:1994dn,
    author = "Kontsevich, Maxim",
    title = "{Homological Algebra of Mirror Symmetry}",
    eprint = "alg-geom/9411018",
    archivePrefix = "arXiv",
    month = "11",
    year = "1994"
}

@book{Ward:1990vs,
    author = "Ward, R. S. and Wells, R. O.",
    title = "{Twistor geometry and field theory}",
    doi = "10.1017/CBO9780511524493",
    isbn = "978-0-521-42268-0, 978-0-521-42268-0, 978-0-511-86977-8",
    publisher = "Cambridge University Press",
    series = "Cambridge Monographs on Mathematical Physics",
    month = "8",
    year = "1991"
}

@inproceedings{Aspinwall:2004jr,
    author = "Aspinwall, Paul S.",
    title = "{D-branes on Calabi-Yau manifolds}",
    booktitle = "{Theoretical Advanced Study Institute in Elementary Particle Physics (TASI 2003): Recent Trends in String Theory}",
    eprint = "hep-th/0403166",
    archivePrefix = "arXiv",
    reportNumber = "DUKE-CGTP-04-04",
    doi = "10.1142/9789812775108_0001",
    pages = "1--152",
    month = "3",
    year = "2004"
}

@article{Spradlin:2004pp,
    author = "Spradlin, Marcus and Volovich, Anastasia",
    title = "{A Pendant for Polya: The One-loop partition function of N=4 SYM on R x S**3}",
    eprint = "hep-th/0408178",
    archivePrefix = "arXiv",
    reportNumber = "NSF-KITP-04-99",
    doi = "10.1016/j.nuclphysb.2005.01.007",
    journal = "Nucl. Phys. B",
    volume = "711",
    pages = "199--230",
    year = "2005"
}

@article{Giombi:2004xv,
    author = "Giombi, S. and Kulaxizi, M. and Ricci, R. and Robles-Llana, Daniel and Trancanelli, D. and Zoubos, K.",
    title = "{Orbifolding the twistor string}",
    eprint = "hep-th/0411171",
    archivePrefix = "arXiv",
    reportNumber = "YITP-SB-04-61",
    doi = "10.1016/j.nuclphysb.2005.04.037",
    journal = "Nucl. Phys. B",
    volume = "719",
    pages = "234--252",
    year = "2005"
}

@article{Eastwood:1981jy,
    author = "Eastwood, Michael G. and Penrose, R. and Wells, R. O.",
    title = "{Cohomology and Massless Fields}",
    doi = "10.1007/BF01942327",
    journal = "Commun. Math. Phys.",
    volume = "78",
    pages = "305--351",
    year = "1981"
}

@article{Bhat:2021dez,
    author = "Bhat, Faizan and Gopakumar, Rajesh and Maity, Pronobesh and Radhakrishnan, Bharathkumar",
    title = "{Twistor coverings and Feynman diagrams}",
    eprint = "2112.05115",
    archivePrefix = "arXiv",
    primaryClass = "hep-th",
    doi = "10.1007/JHEP05(2022)150",
    journal = "JHEP",
    volume = "05",
    pages = "150",
    year = "2022"
}

@article{Gaiotto:2024dwr,
    author = "Gaiotto, Davide and L{\'o}pez-Raven, Adri{\'a}n and Silverans, Hanne and Zeng, Keyou",
    title = "{Categorical {\textquoteright}t Hooft expansion and chiral algebras}",
    eprint = "2411.00760",
    archivePrefix = "arXiv",
    primaryClass = "hep-th",
    doi = "10.1007/JHEP12(2025)141",
    journal = "JHEP",
    volume = "12",
    pages = "141",
    year = "2025"
}

@article{Popov:2004nk,
    author = "Popov, Alexander D. and Wolf, Martin",
    title = "{Topological B-model on weighted projective spaces and self-dual models in four dimensions}",
    eprint = "hep-th/0406224",
    archivePrefix = "arXiv",
    reportNumber = "ITP-UH-20-04",
    doi = "10.1088/1126-6708/2004/09/007",
    journal = "JHEP",
    volume = "09",
    pages = "007",
    year = "2004"
}

@article{Katz:2002gh,
    author = "Katz, Sheldon H. and Sharpe, Eric",
    title = "{D-branes, open string vertex operators, and Ext groups}",
    eprint = "hep-th/0208104",
    archivePrefix = "arXiv",
    reportNumber = "ILL-TH-02-06, ILL-(TH)-02-06",
    doi = "10.4310/ATMP.2002.v6.n6.a1",
    journal = "Adv. Theor. Math. Phys.",
    volume = "6",
    pages = "979--1030",
    year = "2003"
}

@article{Witten:1992fb,
    author = "Witten, Edward",
    title = "{Chern-Simons gauge theory as a string theory}",
    eprint = "hep-th/9207094",
    archivePrefix = "arXiv",
    reportNumber = "IASSNS-HEP-92-45",
    journal = "Prog. Math.",
    volume = "133",
    pages = "637--678",
    year = "1995"
}

@article{Krasnov:2016emc,
    author = "Krasnov, Kirill",
    title = "{Self-Dual Gravity}",
    eprint = "1610.01457",
    archivePrefix = "arXiv",
    primaryClass = "hep-th",
    doi = "10.1088/1361-6382/aa65e5",
    journal = "Class. Quant. Grav.",
    volume = "34",
    number = "9",
    pages = "095001",
    year = "2017"
}

@article{Bittleston:2022nfr,
    author = "Bittleston, Roland and Skinner, David and Sharma, Atul",
    title = "{Quantizing the Non-linear Graviton}",
    eprint = "2208.12701",
    archivePrefix = "arXiv",
    primaryClass = "hep-th",
    doi = "10.1007/s00220-023-04828-0",
    journal = "Commun. Math. Phys.",
    volume = "403",
    number = "3",
    pages = "1543--1609",
    year = "2023"
}

@article{Budzik:2025zvu,
    author = "Budzik, Kasia and Kulp, Justin",
    title = "{Loop Corrected Supercharges from Holomorphic Anomalies}",
    eprint = "2512.07771",
    archivePrefix = "arXiv",
    primaryClass = "hep-th",
    month = "12",
    year = "2025"
}

@article{Elvang:2013cua,
    author = "Elvang, Henriette and Huang, Yu-tin",
    title = "{Scattering Amplitudes}",
    eprint = "1308.1697",
    archivePrefix = "arXiv",
    primaryClass = "hep-th",
    month = "8",
    year = "2013"
}

@article{Yoo:2025qlw,
    author = "Yoo, Philsang",
    title = "{Twists of Supersymmetric Yang{\textendash}mills Theory in Topological String Theory}",
    eprint = "2502.06647",
    archivePrefix = "arXiv",
    primaryClass = "hep-th",
    doi = "10.4134/JKMS.j250056",
    journal = "J. Korean Math. Soc.",
    volume = "62",
    number = "6",
    pages = "1539--1572",
    year = "2025"
}

@article{beilinson1978coherent,
  title="{Coherent sheaves on $\mathbb{P}^n$ and problems of linear algebra}",
  author={Beilinson, Alexander A},
  journal={Functional Analysis and Its Applications},
  volume={12},
  number={3},
  pages={214--216},
  year={1978},
  publisher={Springer}
}

@article{bondal2002generators,
  title={Generators and representability of functors in commutative and noncommutative geometry},
  author={Alexei Bondal and Michel Van den Bergh},
  year={2002},
  eprint={math/0204218},
  archivePrefix={arXiv},
  primaryClass={math.AG},
  url={https://arxiv.org/abs/math/0204218}
}

@article{Douglas:2000gi,
    author = "Douglas, Michael R.",
    title = "{D-branes, categories and N=1 supersymmetry}",
    eprint = "hep-th/0011017",
    archivePrefix = "arXiv",
    reportNumber = "RUNHETC-2000-42",
    doi = "10.1063/1.1374448",
    journal = "J. Math. Phys.",
    volume = "42",
    pages = "2818--2843",
    year = "2001"
}

@article{Aspinwall:2002hg,
    author = "Aspinwall, Paul S. and Karp, Robert L.",
    title = "{Solitons in Seiberg-Witten theory and D branes in the derived category}",
    eprint = "hep-th/0211121",
    archivePrefix = "arXiv",
    reportNumber = "DUKE-CGTP-02-10",
    doi = "10.1088/1126-6708/2003/04/049",
    journal = "JHEP",
    volume = "04",
    pages = "049",
    year = "2003"
}

@book{Loday:2012ev,
    author    = "Loday, Jean-Louis and Vallette, Bruno",
    title     = "{Algebraic Operads}",
    series    = "Grundlehren der mathematischen Wissenschaften",
    volume    = "346",
    publisher = "Springer",
    address   = "Berlin",
    year      = "2012",
    doi       = "10.1007/978-3-642-30362-3"
}

@article{Balisa:2026icv,
    author = "Balisa, Matheus and Casali, Eduardo",
    title = "{Supersymmetric twists in twistor space and holography}",
    eprint = "2607.02145",
    archivePrefix = "arXiv",
    primaryClass = "hep-th",
    month = "7",
    year = "2026"
}

@article{Arai:2020uwd,
    author = "Arai, Reona and Fujiwara, Shota and Imamura, Yosuke and Mori, Tatsuya and Yokoyama, Daisuke",
    title = "{Finite-$N$ corrections to the M-brane indices}",
    eprint = "2007.05213",
    archivePrefix = "arXiv",
    primaryClass = "hep-th",
    reportNumber = "TIT/HEP-681",
    doi = "10.1007/JHEP11(2020)093",
    journal = "JHEP",
    volume = "11",
    pages = "093",
    year = "2020"
}

@article{Kim:2024ucf,
    author = "Kim, Seunggyu and Lee, Eunwoo",
    title = "{Holographic tests for giant graviton expansion}",
    eprint = "2402.12924",
    archivePrefix = "arXiv",
    primaryClass = "hep-th",
    doi = "10.1007/JHEP04(2025)119",
    journal = "JHEP",
    volume = "04",
    pages = "119",
    year = "2025"
}

@article{Beccaria:2024szi,
    author = "Beccaria, Matteo and Cabo-Bizet, Alejandro",
    title = "{Giant graviton expansion of Schur index and quasimodular forms}",
    eprint = "2403.06509",
    archivePrefix = "arXiv",
    primaryClass = "hep-th",
    doi = "10.1007/JHEP05(2024)282",
    journal = "JHEP",
    volume = "05",
    pages = "282",
    year = "2024"
}

@article{Eleftheriou:2025lac,
    author = "Eleftheriou, Giorgos and Murthy, Sameer and Rossell{\'o}, Mart{\'\i}",
    title = "{Localization and wall-crossing of giant graviton expansions in AdS$_{5}$}",
    eprint = "2501.13910",
    archivePrefix = "arXiv",
    primaryClass = "hep-th",
    doi = "10.1007/JHEP07(2025)126",
    journal = "JHEP",
    volume = "07",
    pages = "126",
    year = "2025"
}

@article{Deddo:2025lfm,
    author = "Deddo, Evan and Jayaprakash, Sabarenath and Liu, James T. and Pando Zayas, Leopoldo A.",
    title = "{Quantized giant gravitons as the periodic table of supersymmetric states: D3, M2 and M5}",
    eprint = "2509.18252",
    archivePrefix = "arXiv",
    primaryClass = "hep-th",
    reportNumber = "LITP-25-11",
    doi = "10.1007/JHEP05(2026)065",
    journal = "JHEP",
    volume = "05",
    pages = "065",
    year = "2026"
}

@article{deMelloKoch:2025ngs,
    author = "de Mello Koch, Robert and Jevicki, Antal",
    title = "{Structure of loop space at finite N}",
    eprint = "2503.20097",
    archivePrefix = "arXiv",
    primaryClass = "hep-th",
    doi = "10.1007/JHEP06(2025)011",
    journal = "JHEP",
    volume = "06",
    pages = "011",
    year = "2025"
}

@article{deMelloKoch:2025qeq,
    author = "de Mello Koch, Robert and Kim, Minkyoo and Van Zyl, Hendrik J. R.",
    title = "{From symmetry to structure: gauge-invariant operators in multi-matrix quantum mechanics}",
    eprint = "2507.01219",
    archivePrefix = "arXiv",
    primaryClass = "hep-th",
    doi = "10.1007/JHEP01(2026)031",
    journal = "JHEP",
    volume = "01",
    pages = "031",
    year = "2026"
}

@article{deMelloKoch:2025rkw,
    author = "de Mello Koch, Robert and Jevicki, Antal",
    title = "{Hilbert space of finite N multi-matrix models}",
    eprint = "2508.11986",
    archivePrefix = "arXiv",
    primaryClass = "hep-th",
    doi = "10.1007/JHEP11(2025)145",
    journal = "JHEP",
    volume = "11",
    pages = "145",
    year = "2025"
}

@article{deMelloKoch:2025eqt,
    author = "de Mello Koch, Robert and Jevicki, Antal and Kemp, Garreth and Rudra, Anik",
    title = "{Collective theory at finite-N: reduction of the emergent Hilbert space}",
    eprint = "2510.22525",
    archivePrefix = "arXiv",
    primaryClass = "hep-th",
    doi = "10.1007/JHEP04(2026)048",
    journal = "JHEP",
    volume = "04",
    pages = "048",
    year = "2026"
}

@article{deMelloKoch:2026pck,
    author = "de Mello Koch, Robert and Rudra, Anik and Mahu, Augustine Larweh",
    title = "{Overcrowding and the Finite-$N$ Hilbert Space}",
    eprint = "2607.23025",
    archivePrefix = "arXiv",
    primaryClass = "hep-th",
    month = "7",
    year = "2026"
}

@article{Aharony:2005bq,
    author = "Aharony, Ofer and Marsano, Joseph and Minwalla, Shiraz and Papadodimas, Kyriakos and Van Raamsdonk, Mark",
    title = "{A First order deconfinement transition in large N Yang-Mills theory on a small S**3}",
    eprint = "hep-th/0502149",
    archivePrefix = "arXiv",
    reportNumber = "WIS-03-05-JAN-DPP",
    doi = "10.1103/PhysRevD.71.125018",
    journal = "Phys. Rev. D",
    volume = "71",
    pages = "125018",
    year = "2005"
}

@article{Bershadsky:1993ta,
    author = "Bershadsky, M. and Cecotti, S. and Ooguri, H. and Vafa, C.",
    editor = "Greene, B. and Yau, Shing-Tung",
    title = "{Holomorphic anomalies in topological field theories}",
    eprint = "hep-th/9302103",
    archivePrefix = "arXiv",
    reportNumber = "HUTP-93-A008, RIMS-915",
    doi = "10.1016/0550-3213(93)90548-4",
    journal = "Nucl. Phys. B",
    volume = "405",
    pages = "279--304",
    year = "1993"
}

@article{Aspinwall:2001pu,
    author = "Aspinwall, Paul S. and Lawrence, Albion E.",
    title = "{Derived categories and zero-brane stability}",
    eprint = "hep-th/0104147",
    archivePrefix = "arXiv",
    reportNumber = "SLAC-PUB-8812, DUKE-CGTP-01-07, SU-ITP-01-18",
    doi = "10.1088/1126-6708/2001/08/004",
    journal = "JHEP",
    volume = "08",
    pages = "004",
    year = "2001"
}

@article{berele2025invariants,
  title={Invariants of Superalgebras as Complex Integrals},
  author={Berele, Allan},
  journal={arXiv preprint arXiv:2511.19361},
  year={2025}
}

@article{Marino:2022rpz,
    author = "Marino, Marcos and Schiappa, Ricardo and Schwick, Maximilian",
    title = "{New Instantons for Matrix Models}",
    eprint = "2210.13479",
    archivePrefix = "arXiv",
    primaryClass = "hep-th",
    month = "10",
    year = "2022"
}

@article{Schiappa:2023ned,
    author = "Schiappa, Ricardo and Schwick, Maximilian and Tamarin, Noam",
    title = "{All the D-branes of resurgence}",
    eprint = "2301.05214",
    archivePrefix = "arXiv",
    primaryClass = "hep-th",
    doi = "10.21468/SciPostPhys.20.5.135",
    journal = "SciPost Phys.",
    volume = "20",
    number = "5",
    pages = "135",
    year = "2026"
}

@article{Kristensson:2020nly,
    author = "Kristensson, Alexander T. and Wilhelm, Matthias",
    title = "{From Hagedorn to Lee-Yang: partition functions of $ \mathcal{N} $ = 4 SYM theory at finite N}",
    eprint = "2005.06480",
    archivePrefix = "arXiv",
    primaryClass = "hep-th",
    doi = "10.1007/JHEP10(2020)006",
    journal = "JHEP",
    volume = "10",
    pages = "006",
    year = "2020"
}

@article{Cherman:2020zea,
    author = "Cherman, Aleksey and Dhumuntarao, Aditya",
    title = "{Confinement and graded partition functions for $\mathcal{N}=4$ SYM}",
    eprint = "2012.12341",
    archivePrefix = "arXiv",
    primaryClass = "hep-th",
    doi = "10.1103/PhysRevD.103.066013",
    journal = "Phys. Rev. D",
    volume = "103",
    number = "6",
    pages = "066013",
    year = "2021"
}

@article{Imamura:2024zvw,
    author = "Imamura, Yosuke and Sei, Akihiro and Yokoyama, Daisuke",
    title = "{Giant graviton expansion for general Wilson line operator indices}",
    eprint = "2406.19777",
    archivePrefix = "arXiv",
    primaryClass = "hep-th",
    reportNumber = "TIT/HEP-702",
    doi = "10.1007/JHEP09(2024)202",
    journal = "JHEP",
    volume = "09",
    pages = "202",
    year = "2024"
}

@article{Fujiwara:2023bdc,
    author = "Fujiwara, Shota and Imamura, Yosuke and Mori, Tatsuya and Murayama, Shuichi and Yokoyama, Daisuke",
    title = "{Simple-Sum Giant Graviton Expansions for Orbifolds and Orientifolds}",
    eprint = "2310.03332",
    archivePrefix = "arXiv",
    primaryClass = "hep-th",
    reportNumber = "TIT/HEP-696",
    doi = "10.1093/ptep/ptae006",
    journal = "PTEP",
    volume = "2024",
    number = "2",
    pages = "023B02",
    year = "2024"
}

@article{Ahn:2026ohj,
    author = "Ahn, Byoungjoon and de Mello Koch, Robert and Garg, Vatsal and Jevicki, Antal and Yoon, Junggi",
    title = "{Finite-$N$ Operator Algebras and the Hilbert Space of Bilocal Holography}",
    eprint = "2607.04143",
    archivePrefix = "arXiv",
    primaryClass = "hep-th",
    month = "7",
    year = "2026"
}

@article{Lee:2025veh,
    author = "Lee, Ji Hoon and Li, Wei",
    title = "{AdS$_3$ Quantum Gravity and Finite $N$ Chiral Primaries}",
    eprint = "2511.00636",
    archivePrefix = "arXiv",
    primaryClass = "hep-th",
    month = "11",
    year = "2025"
}

@article{deMelloKoch:2026plq,
    author = "de Mello Koch, Robert and Kim, Minkyoo and Mahu, Augustine Larweh and Rudra, Anik",
    title = "{A pedagogical introduction to invariant theory and finite-$N$ holography}",
    eprint = "2608.22952",
    archivePrefix = "arXiv",
    primaryClass = "hep-th",
    month = "8",
    year = "2026"
}

@article{Yamada:2006rx,
    author = "Yamada, Daiske and Yaffe, Laurence G.",
    title = "{Phase diagram of N=4 super-Yang-Mills theory with R-symmetry chemical potentials}",
    eprint = "hep-th/0602074",
    archivePrefix = "arXiv",
    doi = "10.1088/1126-6708/2006/09/027",
    journal = "JHEP",
    volume = "09",
    pages = "027",
    year = "2006"
}

@article{Rouquier2008,
  author        = {Rouquier, Rapha{\"e}l},
  title         = {Dimensions of triangulated categories},
  journal       = {Journal of K-Theory},
  volume        = {1},
  number        = {2},
  pages         = {193--256},
  year          = {2008},
  eprint        = {math/0310134},
  archivePrefix = {arXiv},
  primaryClass  = {math.CT}
}

@article{Kim:2026rnx,
    author = "Kim, Seok and Kim, Seongmin and Lee, Siyul and Park, Jiyoo",
    title = "{Quantum black hole cohomologies}",
    eprint = "2606.27955",
    archivePrefix = "arXiv",
    primaryClass = "hep-th",
    month = "6",
    year = "2026"
}

@article{Beilinson1996KoszulDP,
  title={Koszul Duality Patterns in Representation Theory},
  author={Alexander Beilinson and Victor Ginzburg and Wolfgang Soergel},
  journal={Journal of the American Mathematical Society},
  year={1996},
  volume={9},
  pages={473-527},
  url={https://api.semanticscholar.org/CorpusID:653354}
}

@article{Dhar:2003fi,
    author = "Dhar, Avinash and Mandal, Gautam and Wadia, Spenta R.",
    title = "{String bits in small radius AdS and weakly coupled N=4 superYang-Mills theory. 1.}",
    eprint = "hep-th/0304062",
    archivePrefix = "arXiv",
    reportNumber = "TIFR-TH-03-09",
    month = "4",
    year = "2003"
}

@article{Bianchi:2003wx,
    author = "Bianchi, Massimo and Morales, Jose F. and Samtleben, Henning",
    title = "{On stringy AdS(5) x S**5 and higher spin holography}",
    eprint = "hep-th/0305052",
    archivePrefix = "arXiv",
    reportNumber = "ROM2F-2003-08, ITPUU-03-22, SPIN-03-12",
    doi = "10.1088/1126-6708/2003/07/062",
    journal = "JHEP",
    volume = "07",
    pages = "062",
    year = "2003"
}

@article{Beisert:2004di,
    author = "Beisert, Niklas and Bianchi, Massimo and Morales, Jose F. and Samtleben, Henning",
    title = "{Higher spin symmetry and N=4 SYM}",
    eprint = "hep-th/0405057",
    archivePrefix = "arXiv",
    reportNumber = "AEI-2004-025, ROM2F-04-08, DESY-04-058",
    doi = "10.1088/1126-6708/2004/07/058",
    journal = "JHEP",
    volume = "07",
    pages = "058",
    year = "2004"
}

@article{Gopakumar:2003ns,
    author = "Gopakumar, Rajesh",
    title = "{From free fields to AdS}",
    eprint = "hep-th/0308184",
    archivePrefix = "arXiv",
    doi = "10.1103/PhysRevD.70.025009",
    journal = "Phys. Rev. D",
    volume = "70",
    pages = "025009",
    year = "2004"
}

@article{Gopakumar:2004qb,
    author = "Gopakumar, Rajesh",
    title = "{From free fields to AdS. 2.}",
    eprint = "hep-th/0402063",
    archivePrefix = "arXiv",
    doi = "10.1103/PhysRevD.70.025010",
    journal = "Phys. Rev. D",
    volume = "70",
    pages = "025010",
    year = "2004"
}

@article{Gopakumar:2005fx,
    author = "Gopakumar, Rajesh",
    title = "{From free fields to AdS: III}",
    eprint = "hep-th/0504229",
    archivePrefix = "arXiv",
    doi = "10.1103/PhysRevD.72.066008",
    journal = "Phys. Rev. D",
    volume = "72",
    pages = "066008",
    year = "2005"
}

@article{Furuuchi:2005qm,
    author = "Furuuchi, Kazuyuki",
    title = "{From free fields to AdS: Thermal case}",
    eprint = "hep-th/0505148",
    archivePrefix = "arXiv",
    doi = "10.1103/PhysRevD.72.066009",
    journal = "Phys. Rev. D",
    volume = "72",
    pages = "066009",
    year = "2005"
}

@article{Alday:2005kq,
    author = "Alday, Luis F. and David, Justin R. and Gava, Edi and Narain, K. S.",
    title = "{Towards a string bit formulation of N=4 super Yang-Mills}",
    eprint = "hep-th/0510264",
    archivePrefix = "arXiv",
    reportNumber = "ITP-UU-05-48, SPIN-05-33",
    doi = "10.1088/1126-6708/2006/04/014",
    journal = "JHEP",
    volume = "04",
    pages = "014",
    year = "2006"
}

@article{Aharony:2006th,
    author = "Aharony, Ofer and Komargodski, Zohar and Razamat, Shlomo S.",
    title = "{On the worldsheet theories of strings dual to free large N gauge theories}",
    eprint = "hep-th/0602226",
    archivePrefix = "arXiv",
    reportNumber = "WIS-03-06-FEB-DPP",
    doi = "10.1088/1126-6708/2006/05/016",
    journal = "JHEP",
    volume = "05",
    pages = "016",
    year = "2006"
}

@article{Harmark:2006ta,
    author = "Harmark, Troels and Orselli, Marta",
    title = "{Matching the Hagedorn temperature in AdS/CFT}",
    eprint = "hep-th/0608115",
    archivePrefix = "arXiv",
    doi = "10.1103/PhysRevD.74.126009",
    journal = "Phys. Rev. D",
    volume = "74",
    pages = "126009",
    year = "2006"
}

@article{Aharony:2007fs,
    author = "Aharony, Ofer and David, Justin R. and Gopakumar, Rajesh and Komargodski, Zohar and Razamat, Shlomo S.",
    title = "{Comments on worldsheet theories dual to free large N gauge theories}",
    eprint = "hep-th/0703141",
    archivePrefix = "arXiv",
    reportNumber = "SLAC-PUB-12405, WIS-03-07-MAR-DPP",
    doi = "10.1103/PhysRevD.75.106006",
    journal = "Phys. Rev. D",
    volume = "75",
    pages = "106006",
    year = "2007"
}

@article{Berkovits:2008qc,
    author = "Berkovits, Nathan",
    title = "{Perturbative Super-Yang-Mills from the Topological AdS(5) x S**5 Sigma Model}",
    eprint = "0806.1960",
    archivePrefix = "arXiv",
    primaryClass = "hep-th",
    reportNumber = "IFT-P-011-2008",
    doi = "10.1088/1126-6708/2008/09/088",
    journal = "JHEP",
    volume = "09",
    pages = "088",
    year = "2008"
}

@article{Bae:2016hfy,
    author = "Bae, Jin-Beom and Joung, Euihun and Lal, Shailesh",
    title = "{On the Holography of Free Yang-Mills}",
    eprint = "1607.07651",
    archivePrefix = "arXiv",
    primaryClass = "hep-th",
    doi = "10.1007/JHEP10(2016)074",
    journal = "JHEP",
    volume = "10",
    pages = "074",
    year = "2016"
}

@article{Bae:2017spv,
    author = "Bae, Jin-Beom and Joung, Euihun and Lal, Shailesh",
    title = "{One-loop free energy of tensionless type IIB string in AdS$_5\times$S$^5$}",
    eprint = "1701.01507",
    archivePrefix = "arXiv",
    primaryClass = "hep-th",
    doi = "10.1007/JHEP06(2017)155",
    journal = "JHEP",
    volume = "06",
    pages = "155",
    year = "2017"
}

@article{Harmark:2017yrv,
    author = "Harmark, Troels and Wilhelm, Matthias",
    title = "{Hagedorn Temperature of AdS$_5$/CFT$_4$ via Integrability}",
    eprint = "1706.03074",
    archivePrefix = "arXiv",
    primaryClass = "hep-th",
    doi = "10.1103/PhysRevLett.120.071605",
    journal = "Phys. Rev. Lett.",
    volume = "120",
    number = "7",
    pages = "071605",
    year = "2018"
}

@article{Harmark:2018red,
    author = "Harmark, Troels and Wilhelm, Matthias",
    title = "{The Hagedorn temperature of AdS$_5$/CFT$_4$ at finite coupling via the Quantum Spectral Curve}",
    eprint = "1803.04416",
    archivePrefix = "arXiv",
    primaryClass = "hep-th",
    doi = "10.1016/j.physletb.2018.09.033",
    journal = "Phys. Lett. B",
    volume = "786",
    pages = "53--58",
    year = "2018"
}

@article{Berkovits:2019ulm,
    author = "Berkovits, Nathan",
    title = "{Sketching a Proof of the Maldacena Conjecture at Small Radius}",
    eprint = "1903.08264",
    archivePrefix = "arXiv",
    primaryClass = "hep-th",
    doi = "10.1007/JHEP06(2019)111",
    journal = "JHEP",
    volume = "06",
    pages = "111",
    year = "2019"
}

@article{Gopakumar:2024jfq,
    author = "Gopakumar, Rajesh and Kaushik, Rishabh and Komatsu, Shota and Mazenc, Edward A. and Sarkar, Debmalya",
    title = "{Strings from Feynman diagrams}",
    eprint = "2412.13397",
    archivePrefix = "arXiv",
    primaryClass = "hep-th",
    doi = "10.1007/JHEP05(2026)054",
    journal = "JHEP",
    volume = "05",
    pages = "054",
    year = "2026"
}

@article{Hawking:1982dh,
    author = "Hawking, S. W. and Page, Don N.",
    title = "{Thermodynamics of Black Holes in anti-De Sitter Space}",
    reportNumber = "PRINT-83-0019 (CAMBRIDGE)",
    doi = "10.1007/BF01208266",
    journal = "Commun. Math. Phys.",
    volume = "87",
    pages = "577",
    year = "1983"
}

@article{Keller:2011xi,
    author = "Keller, Christoph A.",
    title = "{Phase transitions in symmetric orbifold CFTs and universality}",
    eprint = "1101.4937",
    archivePrefix = "arXiv",
    primaryClass = "hep-th",
    doi = "10.1007/JHEP03(2011)114",
    journal = "JHEP",
    volume = "03",
    pages = "114",
    year = "2011"
}

@inproceedings{Shifman:1999mv,
    author = "Shifman, Mikhail A. and Vainshtein, Arkady I.",
    title = "{Instantons versus supersymmetry: Fifteen years later}",
    eprint = "hep-th/9902018",
    archivePrefix = "arXiv",
    reportNumber = "TPI-MINN-99-08-T, UMN-TH-1743-99",
    pages = "485--647",
    month = "2",
    year = "1999"
}

@article{Basar:2014jua,
    author = "Basar, Gokce and Cherman, Aleksey and McGady, David A.",
    title = "{Bose-Fermi Degeneracies in Large $N$ Adjoint QCD}",
    eprint = "1409.1617",
    archivePrefix = "arXiv",
    primaryClass = "hep-th",
    reportNumber = "PUPT-2466, UMN-TH-3352-14, FTPI-MINN-14-26",
    doi = "10.1007/JHEP07(2015)016",
    journal = "JHEP",
    volume = "07",
    pages = "016",
    year = "2015"
}

@article{Harmark:2007px,
    author = "Harmark, Troels and Kristjansson, Kristjan R. and Orselli, Marta",
    title = "{Decoupling limits of N=4 super Yang-Mills on $\mathbb{R}\times S^3$}",
    eprint = "0707.1621",
    archivePrefix = "arXiv",
    primaryClass = "hep-th",
    reportNumber = "NORDITA-2007-21",
    doi = "10.1088/1126-6708/2007/09/115",
    journal = "JHEP",
    volume = "09",
    pages = "115",
    year = "2007"
}

@article{Unsal:2007fb,
    author = "Unsal, Mithat",
    title = "{Phases of N = infinity QCD-like gauge theories on S**3 x S**1 and nonperturbative orbifold-orientifold equivalences}",
    eprint = "hep-th/0703025",
    archivePrefix = "arXiv",
    reportNumber = "SLAC-PUB-12376",
    doi = "10.1103/PhysRevD.76.025015",
    journal = "Phys. Rev. D",
    volume = "76",
    pages = "025015",
    year = "2007"
}

@article{Basar:2013sza,
    author = {Basar, Gokce and Cherman, Aleksey and Dorigoni, Daniele and {\"U}nsal, Mithat},
    title = "{Volume Independence in the Large $N$ Limit and an Emergent Fermionic Symmetry}",
    eprint = "1306.2960",
    archivePrefix = "arXiv",
    primaryClass = "hep-th",
    reportNumber = "FTPI-MINN-13-20, UMN-TH-3210-13, DAMTP-2013-31",
    doi = "10.1103/PhysRevLett.111.121601",
    journal = "Phys. Rev. Lett.",
    volume = "111",
    number = "12",
    pages = "121601",
    year = "2013"
}

@article{Basar:2015asd,
    author = {Ba{\c{s}}ar, G{\"o}k{\c{c}}e and Cherman, Aleksey and Dienes, Keith R. and McGady, David A.},
    title = "{Modularity and 4D-2D spectral equivalences for large-N gauge theories with adjoint matter}",
    eprint = "1512.07918",
    archivePrefix = "arXiv",
    primaryClass = "hep-th",
    reportNumber = "INT-PUB-15-073",
    doi = "10.1007/JHEP06(2016)148",
    journal = "JHEP",
    volume = "06",
    pages = "148",
    year = "2016"
}

@article{Cherman:2015rpc,
    author = "Cherman, Aleksey and McGady, David A. and Yamazaki, Masahito",
    title = "{Spectral sum rules for confining large-N theories}",
    eprint = "1512.09119",
    archivePrefix = "arXiv",
    primaryClass = "hep-th",
    reportNumber = "IPMU-15-0224, INT-PUB-15-077",
    doi = "10.1007/JHEP06(2016)095",
    journal = "JHEP",
    volume = "06",
    pages = "095",
    year = "2016"
}

@article{Cherman:2018mya,
    author = {Cherman, Aleksey and Shifman, Mikhail and {\"U}nsal, Mithat},
    title = "{Bose-Fermi cancellations without supersymmetry}",
    eprint = "1812.04642",
    archivePrefix = "arXiv",
    primaryClass = "hep-th",
    reportNumber = "FTPI-MINN-18-20, UMN-TH-3803/18",
    doi = "10.1103/PhysRevD.99.105001",
    journal = "Phys. Rev. D",
    volume = "99",
    number = "10",
    pages = "105001",
    year = "2019"
}

@article{Cherman:2019ecx,
    author = {Cherman, Aleksey and Kamata, Syo and Sch{\"a}fer, Thomas and {\"U}nsal, Mithat},
    title = "{Flow of Hagedorn singularities and phase transitions in large $N$ gauge theories}",
    eprint = "1910.12312",
    archivePrefix = "arXiv",
    primaryClass = "hep-th",
    doi = "10.1103/PhysRevD.101.014012",
    journal = "Phys. Rev. D",
    volume = "101",
    number = "1",
    pages = "014012",
    year = "2020"
}

@article{Harmark:2006di,
    author = "Harmark, Troels and Orselli, Marta",
    title = "{Quantum mechanical sectors in thermal N=4 super Yang-Mills on R x S**3}",
    eprint = "hep-th/0605234",
    archivePrefix = "arXiv",
    doi = "10.1016/j.nuclphysb.2006.08.022",
    journal = "Nucl. Phys. B",
    volume = "757",
    pages = "117--145",
    year = "2006"
}

@article{Harmark:2021qma,
    author = "Harmark, Troels and Wilhelm, Matthias",
    title = "{Solving the Hagedorn temperature of AdS$_{5}$/CFT$_{4}$ via the Quantum Spectral Curve: chemical potentials and deformations}",
    eprint = "2109.09761",
    archivePrefix = "arXiv",
    primaryClass = "hep-th",
    doi = "10.1007/JHEP07(2022)136",
    journal = "JHEP",
    volume = "07",
    pages = "136",
    year = "2022"
}

@article{Bianchi:2006ti,
    author = "Bianchi, M. and Dolan, F. A. and Heslop, P. J. and Osborn, H.",
    title = "{N=4 superconformal characters and partition functions}",
    eprint = "hep-th/0609179",
    archivePrefix = "arXiv",
    reportNumber = "DAMTP-06-73, ROM2F-2006-21, NSF-KITP-06-80",
    doi = "10.1016/j.nuclphysb.2006.12.005",
    journal = "Nucl. Phys. B",
    volume = "767",
    pages = "163--226",
    year = "2007"
}

@article{Gromov:2010vb,
    author = "Gromov, Nikolay and Kazakov, Vladimir and Tsuboi, Zengo",
    title = "{PSU(2,2|4) Character of Quasiclassical AdS/CFT}",
    eprint = "1002.3981",
    archivePrefix = "arXiv",
    primaryClass = "hep-th",
    reportNumber = "OIQP-09-14, LPT-ENS-10-06",
    doi = "10.1007/JHEP07(2010)097",
    journal = "JHEP",
    volume = "07",
    pages = "097",
    year = "2010"
}

\end{document}